# jOsOcO

## Journal of Social Computing

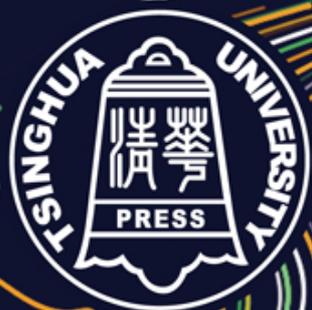

# JOURNAL OF SOCIAL COMPUTING

# CONTENTS

# CONTENTS

# Editorial

We are thrilled to present this special issue of the *Journal of Social Computing*, entitled "Technology Ethics in Action: Critical and Interdisciplinary Perspectives".

This special issue interrogates the meaning and impacts of "tech ethics": the embedding of ethics into digital technology research, development, use, and governance. In response to concerns about the social harms associated with digital technologies, many individuals and institutions have articulated the need for a greater emphasis on ethics in digital technology. Yet as more groups embrace the concept of ethics, critical discourses have emerged questioning whose ethics are being centered, whether "ethics" is the appropriate frame for improving technology, and what it means to develop "ethical" technology in practice.

This interdisciplinary issue takes up these questions, interrogating the relationships among ethics, technology, and society in action. This special issue engages with the normative and contested notions of ethics itself, how ethics has been integrated with technology across domains, and potential paths forward to support more just and egalitarian technology. Rather than starting from philosophical theories, the authors in this issue orient their articles around the real-world discourses and impacts of tech ethics—i.e., tech ethics in action. In many cases, this focus derives from the authors' own engagements with tech ethics as scholars, practitioners, and activists.

This special issue emerged from the Ethical Tech Working Group at the Berkman Klein Center for Internet & Society at Harvard. The working group was co-founded in 2017 by Mary Gray and Kathy Pham, two trailblazers in embedding ethics, responsibility, and justice into technology. Every contributor in this special issue was a regular participant in the Working Group's weekly meetings between 2017 and 2020.

After its founding, the Ethical Tech Working Group quickly blossomed into a flourishing, interdisciplinary, and welcoming community. It became a hub of trust, respect, and friendship. Weekly meetings were structured to provide everyone with a platform and to welcome newcomers. We began by having everyone in the room introduce themself and answer the question: "What is on your mind in terms of ethics and technology this week?" Our discussions explored technology ethics through lenses that include art, anthropology, communications, computer science, divinity, history, labor, law, race and gender studies, philosophy, political science, and STS (science, technology, and society). We aimed to recognize and honor everyone's perspective and knowledge, gaining an understanding across fields and viewpoints. It was a common occurrence to hear: "You just said [*this word*]. What does that word mean to you? Because it means something different to me."

The articles in this issue reflect this spirit of community and dialogue: each article represents its author's distinct perspective yet is simultaneously deeply informed by conversation with the other members of the Ethical Tech Working Group.

The articles in this special issue are split across two sections. The first section begins with the introduction to the special issue: Ben Green summarizes recent developments and challenges in tech ethics, suggesting the need to study tech ethics through a sociotechnical lens. The rest of the first section focuses on the value and limits of ethics for improving digital technology. With its connections to philosophy and connotations of moral behavior, ethics appears well-suited for improving the development and applications of digital technology. Moral philosophy indeed sheds light on the normative principles and obligations that arise in complex sociotechnical contexts. Yet the superficial and legitimizing role that ethics often plays in digital technology and other domains suggests that ethics may suffer from significant shortcomings as an organizing principle for reform. Diagnosing the contours of these limits is thus an essential task for achieving more just technologies moving forward.

Jasmine E. McNealy sets the stage for the particular frames and terms used to discuss ethics, describing several frames that provide vague and misleading promises and calling for counternarratives. Lily Hu draws parallels between ethics and the history of human rights, suggesting the need to be suspicious of moral language that evades political and ideological battles. Ben Green advocates for an explicit embrace of politics instead of ethics, articulating how the field of data science can productively evolve toward a politics of social justice. Elettra Bietti reflects on critical discourses about tech ethics, laying out a philosophically informed middle ground between "ethics-washing" and "ethics-bashing". Salomé Viljoen analyzes calls to embrace law rather than ethics, describing how the law (much like ethics) is a terrain of contestation and how the law has structured the ethical crises in tech.

*The second section (to be published in Issue 4) considers broader frames and strategies, beyond the explicit label of tech ethics, for improving digital technology. The limits of tech ethics indicate a need to expand the scope of ethical analysis of technology, not to abandon ethical analysis altogether. Doing so broadens the focus beyond technology design to consider the entire lifecycle, infrastructure, and governance of sociotechnical systems, thus opening up new terrains for contestation and action. On this view, many of the central problems related to technology ethics are less problems of technology itself than problems connected to broader social and political injustices. Exploring how these larger contexts shape technology—and how to reform them—is therefore essential to a more expansive approach to remediating the impacts of digital technologies. This section moves from interventions focused on design processes to broader reorientations of pedagogy, culture, and institutions.*

*Luke Stark introduces Apologos as a method for eliciting ethics, norms, and human values in sociotechnical design processes on a compressed time scale. Aden Van Noppen describes how tech companies can improve technology design by adopting practices of spiritual care. Joanne Cheung draws an analogy between the financialization of land and the financialization of social media, emphasizing the need to focus on business models (in addition to design decisions) and describing possibilities to steward social networks in the public interest. Jonnie Penn challenges the myth of automation, describing how the pursuit of digital automation systematically extracts human labor and proposing the corrective of "algorithmic silence". Jenny Ungbha Korn calls for a critical tech ethics that embraces critical race theory, intersectional feminist theory, and critical race feminist theory. Maya Malik and Momin M. Malik introduce and explore the process of how people from technical disciplines come to embrace more critical orientations, describing the importance of these awakenings for technology ethics and social justice. Sabelo Mhlambi critiques data colonialism and surveillance capitalism, arguing for technology development grounded in the Sub-Saharan African philosophy of Ubuntu to AI.*

We are deeply grateful to the many, many people who contributed to this special issue. All of the authors shared their wisdom and care in crafting their articles. In addition to these authors, many Ethical Tech Working Group peers have heavily influenced our perspectives, including Doaa Abu Elyounes, Kendra Albert, Bao Kham Chau, Mary Gray, Jenn Halen, Dean Jansen, Ram Kumar, Keith Porcaro, Boaz Sender, and Suchana Seth. We could not have done our work without the support of the Berkman Klein Center Staff: Carey Anderson, Sebastian Diaz, Daniel Jones, Reuben Langevin, and Ellen Popko, who made all of our events and convenings possible; and Becca Tabasky, who tirelessly built community and taught us how to effectively bring everyone to the table.

We are also grateful to the *Journal of Social Computing* for the opportunity to publish this special issue and all of the labor that went into it. Peaks Krafft provided the initial connection that made this special issue possible and managed many of the logistics, including the peer review process. The Editors-in-Chief—Xiaoming Fu, James Evans, and Jar-Der Luo—embraced the vision of the special issue and supported all of our efforts. The staff at Information Science Division, Tsinghua University Press managed the entire publication process. Finally, the many peer reviewers provided thoughtful feedback and substantially improved each article.

Collectively, the articles in this special issue provide a range of perspectives and proposals regarding the value of tech ethics and paths forward for improving digital technologies. These varied perspectives embody our commitment to honoring all forms of knowledge and expertise, particularly those forms that technologists all too often ignore. While many of the articles reflect common viewpoints, there is also disagreement across articles—in some cases explicit, in others implicit—regarding questions such as the proper role for "ethics" and strategies for interacting with the technology industry. Our goal is not to produce precisely defined answers. Instead, we aim to explore the contours of debate and action related to technology ethics, in the service of a more just society. We hope that this special issue will help you to do the same.

In solidarity,

Ben Green (Special Issue Editor)

Kathy Pham (Berkman Klein Center Ethical Tech Working Group Co-Founder)

# The Contestation of Tech Ethics: A Sociotechnical Approach to Technology Ethics in Practice

Ben Green*

**Abstract:** This article introduces the special issue "Technology Ethics in Action: Critical and Interdisciplinary Perspectives". In response to recent controversies about the harms of digital technology, discourses and practices of "tech ethics" have proliferated across the tech industry, academia, civil society, and government. Yet despite the seeming promise of ethics, tech ethics in practice suffers from several significant limitations: tech ethics is vague and toothless, has a myopic focus on individual engineers and technology design, and is subsumed into corporate logics and incentives. These limitations suggest that tech ethics enables corporate "ethics-washing": embracing the language of ethics to defuse criticism and resist government regulation, without committing to ethical behavior. Given these dynamics, I describe tech ethics as a terrain of contestation where the central debate is not whether ethics is desirable, but what "ethics" entails and who gets to define it. Current approaches to tech ethics are poised to enable technologists and technology companies to label themselves as "ethical" without substantively altering their practices. Thus, those striving for structural improvements in digital technologies must be mindful of the gap between ethics as a mode of normative inquiry and ethics as a practical endeavor. In order to better evaluate the opportunities and limits of tech ethics, I propose a sociotechnical approach that analyzes tech ethics in light of who defines it and what impacts it generates in practice.

**Key words:** technology ethics; AI ethics; ethics-washing; Science, Technology, and Society (STS); sociotechnical systems

## 1 Introduction: A Crisis of Conscience

If digital technology production in the beginning of the 2010s was characterized by the brash spirit of Facebook's motto "move fast and break things" and the superficial assurances of Google's motto "do not be evil", digital technology toward the end of the decade was characterized by a "crisis of conscience"[1]. While many have long been aware of digital technology's harms, an influx of stories about salient harms led to widespread critique of digital technology. The response was the "techlash": a growing public animosity toward major technology companies. In 2018, Oxford Dictionaries and the Financial Times both deemed techlash to be one of the words of the year[2, 3].

Consider just a few of the controversies that prompted this crisis of conscience within tech and the associated techlash:

**Disinformation:** Throughout the 2016 US presidential election between Donald Trump and Hillary Clinton, social media was plagued with fraudulent stories that went viral[4, 5]. In turn, numerous commentators—including Hillary Clinton—blamed Facebook for Donald Trump's presidential election victory[6–9]. Later reporting revealed that Facebook's leadership has actively resisted taking strong measures to curb disinformation, instead prioritizing the company's business strategies[10, 11].

**Cambridge Analytica:** In 2018, *The New York Times*

• Ben Green is with the Society of Fellows and the Gerald R. Ford School of Public Policy, University of Michigan, Ann Arbor, MI 48109, USA. E-mail: bzgreen@umich.edu.
∗ To whom correspondence should be addressed.

and *The Guardian* reported that the voter-profiling firm Cambridge Analytica had harvested information from millions of Facebook users, without their knowledge or permission, in order to target political ads for Donald Trump's 2016 presidential campaign[12, 13]. Cambridge Analytica had acquired these data by exploiting the sieve-like nature of Facebook's privacy policy.

**Military and ICE Contracts**: In 2018, journalists revealed that Google was working with the US Department of Defense (DoD) to develop software that analyzes drone footage[14]. This effort, known as Project Maven, was part of a ＄7.4 billion investment in AI by the DoD in 2017[14] and represented an opportunity for Google to gain billions of dollars in future defense contracts[15]. Another story revealed that Palantir was developing software for Immigration and Customs Enforcement (ICE) to facilitate deportations[16].

**Algorithmic Bias:** In 2016, ProPublica revealed that an algorithm used in criminal courts was biased against Black defendants, mislabeling them as future criminals at twice the rates of white defendants[17]. Through popular books about the harms and biases of algorithms in settings such as child welfare, online search, and hiring[18−20], the public began to recognize algorithms as both fallible and discriminatory.

These and other tech-related controversies were a shock to many, as they arrived in an era of widespread (elite) optimism about the beneficence of technology. Yet these controversies also brought public attention to what scholars in fields such as Science, Technology, and Society (STS), philosophy of science, critical data and algorithm studies, and law have long argued: technology is shaped by social forces, technology structures society often in deleterious ways, and technology cannot solve every social problem. Broadly speaking, these fields bring a "sociotechnical" approach to studying technologies, analyzing how technologies shape, are shaped by, and interact with society[21−24]. As tech scandals mounted, a variety of sociotechnical insights, long ignored by most technologists and journalists, were newly recognized (or in some form recreated).

Many in the tech sector and academia saw the harms of digital technology as the result of an inattention to ethics. On this view, unethical technologies result from a lack of training in ethical reasoning for engineers and a dearth of ethical principles in engineering practice[1, 25−28]. In response, academics, technologists, companies, governments, and more have embraced a broad set of goals often characterized with the label "tech ethics": the introduction of ethics into digital technology education, research, development, use, and governance. In the span of just a few years, tech ethics has become a dominant discourse discussed in technology companies, academia, civil society organizations, and governments.

This article reviews the growth of tech ethics and the debates that this growth has prompted. I first describe the primary forms of tech ethics in practice. I focus on the people and organizations that explicitly embrace the label of "tech ethics" (and closely related labels, such as AI ethics and algorithmic fairness). I then summarize the central critiques made against these efforts, which call into question the effects and desirability of tech ethics. Against the backdrop of these critiques, I argue that tech ethics is a terrain of contestation: the central debate is not whether ethics is desirable but what ethics entails and who has the authority to define it. These debates suggest the need for a sociotechnical approach to tech ethics that focuses on the social construction and real-world effects of tech ethics, disambiguating between the value of ethics as a discipline and the limits of tech ethics as a practical endeavor. I introduce this approach through four frames: objectivity and neutrality, determinism, solutionism, and sociotechnical systems.

## 2　The Rise of Tech Ethics

Although some scholars, activists, and others have long considered the ethics of technology, attention to digital technology ethics has rapidly grown across the tech industry, academia, civil society, and government in recent years. As we will see, tech ethics typically involves applied forms of ethics such as codes of ethics and research ethics, rather than philosophical inquiry (i.e., moral philosophy). For instance, one common treatment of tech ethics is statements of ethical principles. One analysis of 36 prominent AI principles documents shows the sharp rise in these statements, from 2 in 2014 to 16 in 2018[29]. These documents tend to cover the themes of fairness and non-discrimination, privacy, accountability, and transparency and explainability[29]. Many documents also reference human rights, with some taking international human rights as the framework for ethics[29].

## 2.1   Tech industry

The most pervasive treatment of tech ethics within tech companies has come in the form of ethics principles and ethics oversight bodies. Companies like Microsoft, Google, and IBM have developed and publicly shared AI ethics principles, which include statements such as "AI systems should treat all people fairly" and "AI should be socially beneficial"[30−32]. These principles are often supported through dedicated ethics teams and advisory boards within companies, with such bodies in place at companies including Microsoft, Google, Facebook, DeepMind, and Axon[33−37]. Companies such as Google and Accenture have also begun offering tech ethics consulting services[38, 39].

As part of these efforts, the tech industry has formed several coalitions aimed at promoting safe and ethical artificial intelligence. In 2015, Elon Musk and Sam Altman created OpenAI, a research organization that aims to mitigate the "existential threat" presented by AI, with more than ＄1 billion in donations from major tech executives and companies[40]. A year later, Amazon, Facebook, DeepMind, IBM, and Microsoft founded the Partnership on AI (PAI), a nonprofit coalition to shape best practices in AI development, advance public understanding of AI, and support socially beneficial applications of AI[41, 42].①

## 2.2   Academia

Computer and information science programs at universities have rapidly increased their emphasis on ethics training. While some universities have taught computing ethics courses for many years[44−46], the emphasis on ethics within computing education has increased dramatically in recent years[47]. One crowdsourced list of tech ethics classes contains more than 300 courses[48]. This plethora of courses represents a dramatic shift in computer science training and culture, with ethics becoming a popular topic of discussion and study after being largely ignored by the mainstream of the field just a few years prior.

Research in computer science and related fields has also become more focused on the ethics and social impacts of computing. This trend is observable in the recent increase in conferences and workshops related to computing ethics. The ACM Conference on Fairness, Accountability, and Transparency (FAccT) and the AAAI/ACM Conference on AI, Ethics, and Society (AIES) both held their first annual meetings in February 2018 and have since grown rapidly. There have been several dozen workshops related to fairness and ethics at major computer science conferences[49]. Many universities have supported these efforts by creating institutes focused on the social implications of technology. 2017 alone saw the launch of the AI Now Institute at NYU[50], the Princeton Dialogues on AI and Ethics[51], and the MIT/Harvard Ethics and Governance of Artificial Intelligence Initiative[52]. More recently formed centers include the MIT College of Computing[53]; the Stanford Institute for Human-Centered Artificial Intelligence[54]; and the University of Michigan Center of Ethics, Society, and Computing[55].

## 2.3   Civil society

Numerous civil society organizations have coalesced around tech ethics, with strategies that include grantmaking and developing principles. Organizations such as the MacArthur and Ford Foundations have begun exploring and making grants in tech ethics[56]. For instance, the Omidyar Network, Mozilla Foundation, Schmidt Futures, and Craig Newmark Philanthropies partnered on the Responsible Computer Science Challenge, which awarded ＄3.5 million between 2018 and 2020 to support efforts to embed ethics into undergraduate computer science education[57]. Many foundations also contribute to the research, conferences, and institutes that have emerged in recent years.

Other organizations have been created or have expanded their scope to consider the implications and governance of digital technologies. For example, the American Civil Liberties Union (ACLU) has begun hiring technologists and is increasingly engaged in debates and legislation related to new technology. Organizations such as Data & Society, Upturn, the Center for Humane Technology, and Tactical Tech study the social implications of technology and advocate for improved technology governance and design practices.

Many in civil society call for engineers to follow an ethical oath modeled after the Hippocratic Oath (an ethical oath taken by physicians)[20, 58−60]. In 2018, for instance, the organization Data for Democracy partnered

---

① Although PAI also includes civil society partners, these organizations do not appear to have significant influence. In 2020, the human rights organization Access Now resigned from PAI, explaining that "there is an increasingly smaller role for civil society to play within PAI" and that "we did not find that PAI influenced or changed the attitude of member companies"[43].

with Bloomberg and the data platform provider BrightHive to develop a code of ethics for data scientists, developing 20 principles that include "I will respect human dignity" and "It is my responsibility to increase social benefit while minimizing harm"[61]. Former US Chief Data Scientist DJ Patil described the event as the "Constitutional Convention" for data science[58]. A related effort, produced by the Institute for the Future and the Omidyar Network, is the Ethical OS Toolkit, a set of prompts and checklists to help technology developers "anticipate the future impact of today's technology" and "not regret the things you will build"[62].

## 2.4    Government

Many governments developed commissions and principles dedicated to tech ethics. In the United States, for example, the National Science Foundation formed a Council for Big Data, Ethics, and Society[63]; the National Science and Technology Council published a report about AI that emphasized ethics[64]; and the Department of Defense adopted ethical principles for AI[65]. Elsewhere, governing bodies in Dubai[66], Europe[67], Japan[68], and Mexico[69], as well as international organizations such as the OECD[70], have all stated principles for ethical AI.

## 3    The Limits of Tech Ethics

Alongside its rapid growth, tech ethics has been critiqued along several lines. First, the principles espoused by tech ethics statements are too abstract and toothless to reliably spur ethical behavior in practice. Second, by emphasizing the design decisions of individual engineers, tech ethics overlooks the structural forces that shape technology's harmful social impacts. Third, as ethics is incorporated into tech companies, ethical ideals are subsumed into corporate logics and incentives. Collectively, these issues suggest that tech ethics represents a strategy of technology companies "ethics-washing" their behavior with a façade of ethics while largely continuing with business-as-usual.

### 3.1    Tech ethics principles are abstract and toothless

Tech ethics codes deal in broad principles[71]. In 2016, for example, Accenture published a report explicitly outlining "a universal code of data ethics"[72]. A 2019 analysis of global AI ethics guidelines found 84 such documents, espousing a common set of broad principles: transparency, justice and fairness, non-maleficence, responsibility, and privacy[73]. Professional computing societies also present ethical commitments in a highly abstract form, encouraging computing professionals "to be ever aware of the social, economic, cultural, and political impacts of their actions" and to "contribute to society and human well-being"[74]. Ethics codes in computing and information science are notably lacking in explicit commitments to normative principles[74].

The emphasis on universal principles papers over the fault lines of debate and disagreement that spurred the emergence of tech ethics in the first place. Tech ethics principles embody a remarkable level of agreement: two 2019 reports on global AI ethics guidelines noted a "global convergence"[73] and a "consensus"[29] in the principles espoused. Although these documents tend to reflect a common set of global principles, the actual interpretation and implementation of these principles raise substantive conflicts[73]. Furthermore, these principles have been primarily developed in the US and UK, with none from Africa or South America[73]. The superficial consensus around abstract ideals may thus hinder substantive deliberation regarding whether the chosen values are appropriate, how those values should be balanced in different contexts, and what those values actually entail in practice.

The abstraction of tech ethics is particularly troubling due to a lack of mechanisms to enact or enforce the espoused principles. When framed at such a high level of abstraction, values such as fairness and respect are unable to guide specific actions[75]. In companies, ethics oversight boards and ethics principles lack the authority to veto projects or require certain behaviors[76, 77]. Similarly, professional computing organizations such as the IEEE and ACM lack the power to meaningfully sanction individuals who violate their codes of ethics[75]. Moreover, unlike fields such as medicine, which has a strong and established emphasis on professional ethics, computing lacks a common aim or fiduciary duty to unify disparate actors around shared ethical practices[75]. All told, "Principles alone cannot guarantee ethical AI"[75].

### 3.2    Tech ethics has a myopic focus on individual engineers and technology design

Tech ethics typically emphasizes the roles and responsibilities of engineers, paying relatively little attention to the broader environments in which these individuals work. Although professional codes in computing and related fields assert general commitments to the public, profession, and one's employer, "the morality of a profession's or an employer's motives are not scrutinized"[74]. Similarly, ethics within computer science curricula tends to focus on ethical decision making for individual engineers[78].

From this individualistic frame comes an emphasis on appealing to the good intentions of engineers, with the assumption that better design practices and procedures will lead to better technology. Ethics becomes a matter of individual engineers and managers "doing the right thing" "for the right reasons"[79]. Efforts to provide ethical guidance for tech CEOs rest on a similar logic: "if a handful of people have this much power—if they can, simply by making more ethical decisions, cause billions of users to be less addicted and isolated and confused and miserable—then, is not that worth a shot?"[1]. The broader public beyond technical experts is not seen as having a role in defining ethical concerns or shaping the responses to these concerns[71].

Tech ethics therefore centers debates about how to build better technology rather than whether or in what form to build technology (let alone who gets to make such decisions). Tech ethics follows the assumption that artificial intelligence and machine learning are "inevitable", such that "'better building' is the only ethical path forward"[71]. In turn, tech ethics efforts pursue technical and procedural solutions for the harmful social consequences of technology[79]. Following this logic, tech companies have developed numerous ethics and fairness toolkits[80–84].

Although efforts to improve the design decisions of individual engineers can be beneficial, the focus on individual design choices relies on a narrow theory of change for how to reform technology. Regardless of their intentions and the design frameworks at their disposal, individual engineers typically have little power to shift corporate strategy. Executives can prevent engineers from understanding the full scope of their work, limiting knowledge and internal dissent about controversial projects[85, 86]. Even when engineers do know about and protest projects, the result is often them resigning or being replaced rather than the company changing course[60, 85]. The most notable improvements in technology use and regulation have come from collective action among activists, tech workers, journalists, and scholars, rather than individual design efforts[87, 88].

More broadly, the emphasis on design ignores the structural sources of technological harms. The injustices associated with digital technologies result from business models that rely on collecting massive amounts of data about the public[89, 90]; companies that wield monopolistic power[91, 92]; technologies that are built through the extraction of natural resources and the abuse of workers[93–96]; and the exclusion of women, minorities, and non-technical experts from technology design and governance[97, 98].

These structural conditions place significant barriers on the extent to which design-oriented tech ethics can guide efforts to achieve reform. As anthropologist Susan Silbey notes, "while we might want to acknowledge human agency and decision-making at the heart of ethical action, we blind ourselves to the structure of those choices—incentives, content, and pattern—if we focus too closely on the individual and ignore the larger pattern of opportunities and motives that channel the actions we call ethics"[78]. To the extent that it defines ethical technology in terms of individual design decisions, tech ethics will divert scrutiny away from the economic and political factors that drive digital injustice, limiting our ability to address these forces.

### 3.3   Tech ethics is subsumed into corporate logics and incentives

Digital technology companies have embraced ethics as a matter of corporate concern, aiming to present the appearance of ethical behavior for scrutinizing audiences. As Alphabet and Microsoft noted in recent SEC filings, products that are deemed unethical could lead to reputational and financial harms[99]. Companies are eager to avoid any backlash, yet do not want to jeopardize their business plans. An ethnography of ethics work in Silicon Valley found that "performing, or even showing off, the seriousness with which a company takes ethics becomes a more important sign of ethical practices than real changes to a product"[79]. For instance, after an effort at Twitter to reduce online harassment stalled, an external researcher involved in the effort noted, "The impression I came away with from this experience is that Twitter was more sensitive to deflecting criticism than in solving the problem of harassment"[100].

Corporate tech ethics is therefore framed in terms of its direct alignment with business strategy. A software engineer at LinkedIn described algorithmic fairness as being profitable for companies, arguing, "If you are very biased, you might only cater to one population, and eventually that limits the growth of your user base, so from a business perspective you actually want to have everyone come on board, so it is actually a good business decision in the long run"[101]. Similarly, one of the people behind the Ethical OS toolkit described being motivated to produce "a tool that helps you think through societal consequences and makes sure what you are designing is good for the world and good for your longer-term bottom line"[102].

Finding this alignment between ethics and business is an important task for those charged with promoting ethics in tech companies. Recognizing that "market success trumps ethics", individuals focused on ethics in Silicon Valley feel pressure to align ethical principles with corporate revenue sources[79]. As one senior researcher in a tech company notes, "the ethics system that you create has to be something that people feel adds value and is not a massive roadblock that adds no value, because if it is a roadblock that has no value, people literally will not do it, because they do not have to"[79]. When ethical ideals are at odds with a company's bottom line, they are met with resistance[1].

This emphasis on business strategy creates significant conflicts with ethics. Corporate business models often rely on extractive and exploitative practices, leading to many of the controversies at the heart of the techlash. Indeed, efforts to improve privacy and curb disinformation have led Facebook and Twitter stock values to decline rapidly[103, 104]. Thus, even as tech companies espouse a devotion to ethics, they continue to develop products and services that raise ethical red flags but promise significant profits. For example, even after releasing AI ethics principles that include safety, privacy, and inclusiveness[31] and committing not to "deploy facial recognition technology in scenarios that we believe will put democratic freedoms at risk"[105], Microsoft invested in AnyVision, an Israeli facial recognition company that supports military surveillance of Palestinians in the West Bank[106]. Similarly, several years after Google withdrew from Project Maven due to ethical concerns among employees, and then created AI ethics guidelines, the company began aggressively pursuing new contracts with the Department of Defense[107].

In sum, tech ethics is being subsumed into existing tech company logics and business practices rather than changing those logics and practices (even if some individuals within companies do want to create meaningful change). This absorption allows companies to take up the mantle of ethics without making substantive changes to their processes or business strategies. The goal in companies is to find practices "which the organization is not yet doing but is capable of doing"[79], indicating an effort to find relatively costless reforms that provide the veneer of ethical behavior. Ethics statements "co-opt the language of some critics", taking critiques grounded in a devotion to equity and social justice and turning them into principles akin to "conventional business ethics"[71]. As they adopt these principles, tech companies "are learning to speak and perform ethics rather than make the structural changes necessary to achieve the social values underpinning the ethical fault lines that exist"[79].

These limits to corporate tech ethics are exemplified by Google's firings of Timnit Gebru and Meg Mitchell. Despite Gebru's and Mitchell's supposed charge as co-leads of Google's Ethical AI team, Google objected to a paper they had written (alongside several internal and external co-authors) about the limitations and harms of large language models, which are central to Google's business[108]. Google attempted to force the authors to retract the paper, claiming that they failed to acknowledge recent technical advances that mitigate many of the paper's concerns[108]. Soon after, journalists revealed that this incident reflected a larger pattern: Google had expanded its review of papers that discuss "sensitive topics", telling researchers, for instance, to "take great care to strike a positive tone" regarding Google's technologies and products[109]. Thus, even as Google publicly advertised its care for ethics, internally the company was carefully reviewing research to curtail ethical criticisms that it deemed threatening to its core business interests.

### 3.4 Tech ethics has become an avenue for ethics-washing

As evidence of tech ethics' limitations has grown, many have critiqued tech ethics as a strategic effort among technology companies to maintain autonomy and profits.

This strategy has been labeled "ethics-washing" (i.e., "ethical white-washing"): adopting the language of ethics to diminish public scrutiny and avoid regulations that would require substantive concessions[110–112]. As an ethnography of ethics in Silicon Valley found, "It is a routine experience at 'ethics' events and workshops in Silicon Valley to hear ethics framed as a form of self-regulation necessary to stave off increased governmental regulation"[79]. This suggests that the previously described issues with tech ethics might be features rather than bugs: by focusing public attention on the actions of individual engineers and on technical dilemmas (such as algorithmic bias), companies perform a sleight-of-hand that shifts structural questions about power and profit out of view. Companies can paint a self-portrait of ethical behavior without meaningfully altering their practices.

Thomas Metzinger, a philosopher who served on the European Commission's High-Level Expert Group on Artificial Intelligence (AI HLEG), provides a particularly striking account of ethics-washing in action[110]. The AI HLEG contained only four ethicists out of 52 total people and was dominated by representatives from industry. Metzinger was tasked with developing "Red Lines" that AI applications should not cross. However, the proposed red lines were ultimately removed by industry representatives eager for a "positive vision" for AI. All told, Metzinger describes the AI HLEG's guidelines as "lukewarm, short-sighted, and deliberately vague" and concludes that the tech industry is "using ethics debates as elegant public decorations for a large-scale investment strategy"[110].

Tech companies have further advanced this "ethics-washing" agenda through funding academic research and conferences. Many of the scholars writing about tech policy and ethics are funded by Google, Microsoft, and other companies, yet often do not disclose this funding[113, 114]. Tech companies also provide funding for prominent academic conferences, including the ACM Conference on Fairness, Accountability, and Transparency (FAccT); the AAAI/ACM Conference on Artificial Intelligence, Ethics, and Society (AIES); and the Privacy Law Scholars Conference (PLSC). Even if these funding practices do not directly influence the research output of individual scholars, they allow tech companies to shape the broader academic and public discourse regarding tech ethics, raising certain voices and conversations at the expense of others.[②]

In December 2019, then-MIT graduate student Rodrigo Ochigame provided a particularly pointed account of ethics-washing[119]. Describing his experiences working in the Media Lab's AI ethics group and collaborating with the Partnership on AI, Ochigame articulated how "the discourse of 'ethical AI' was aligned strategically with a Silicon Valley effort seeking to avoid legally enforceable restrictions of controversial technologies". Ochigame described witnessing firsthand how the Partnership on AI made recommendations that "aligned consistently with the corporate agenda" by reducing political questions about the criminal justice system to matters of technical consideration. A central part of this effort was tech companies strategically funding researchers and conferences in order to generate a widespread discourse about "ethical" technology. Finding that "the corporate lobby's effort to shape academic research was extremely successful", Ochigame concluded that "big tech money and direction proved incompatible with an honest exploration of ethics".

Ochigame's article prompted heated debate about the value and impacts of tech ethics. Some believed that Ochigame oversimplified the story, failing to acknowledge the many people behind tech ethics[120–122]. On this view, tech ethics is a broad movement that includes efforts by scholars and activists to expose and resist technological harms. Yet many of the people centrally involved in those efforts see their work as distinct from tech ethics. Safiya Noble described Ochigame's article as "All the way correct and worth the time to read"[123]. Lilly Irani and Ruha Benjamin expressed similar sentiments, noting that "AI ethics is not a movement"[124] and that "many of us do not frame our work as 'ethical AI'"[125]. On this view, tech ethics represents the narrow domain of efforts, typically promulgated by tech companies, that explicitly embrace the label of "tech ethics".

The debate over Ochigame's article exposed the fault lines at the heart of tech ethics. The central question is what tech ethics actually entails in practice. While some frame tech ethics as encompassing broad societal debates about the social impacts of technology, others define tech ethics as narrower industry-led efforts to

---

② The integrity of academic tech ethics has been further called into question due to funding from other sources beyond tech companies[115–117]. A related critique of academic tech ethics institutes is the lack of diversity within their leadership[118].

explicitly promote "ethics" in technology. On the former view, tech ethics is an important and beneficial movement for improving digital technology. On the latter view, tech ethics is a distraction that hinders efforts to achieve more equitable technology.

## 4 The Contestation of Tech Ethics

The debates described in the previous section reveal that the central question regarding tech ethics is not whether it is desirable to be ethical, but what "ethics" entails and who gets to define it. Although the label of ethics carries connotations of moral philosophy, in practice the "ethics" in tech ethics tends to take on four overlapping yet often conflicting definitions: moral justice, corporate values, legal risk, and compliance[126]. With all of these meanings conflated in the term ethics, superficially similar calls for tech ethics can imply distinct and even contradictory goals. There is a significant gap between the potential benefits of applying ethics (as in rigorous normative reasoning) to technology and the real-world effects of applying ethics (as in narrow and corporate-driven principles) to technology.

As a result, tech ethics represents a terrain of contestation. The contestation of tech ethics centers on certain actors attempting to claim legitimate authority over what it means for technology to be "ethical", at the expense of other actors. These practices of "boundary-work"[127] enable engineers and companies to maintain intellectual authority and professional autonomy, often in ways that exclude women, minorities, the Global South, and other publics[128–130]. We can see this behavior in technology companies projecting procedural toolkits as solutions to ethical dilemmas, computer scientists reducing normative questions into mathematical metrics, academic tech ethics institutes being funded by billionaires and led primarily by white men, and tech ethics principles being disseminated predominantly by the US and Western Europe. Furthermore, many of the most prominent voices regarding tech ethics are white men who claim expertise while ignoring the work of established fields and scholars, many of whom are women and people of color[131, 132].

Two examples of how ethics has been implemented in other domains—science and business—shed light on the stakes of present debates about tech ethics.

### 4.1 Ethics in science

Many areas of science have embraced ethics in recent decades following public concerns about the social implications of emerging research and applications. Despite the seeming promise of science ethics, however, existing approaches fail to raise debates about the structure of scientific research or to promote democratic governance of science.

Rather than interrogating fundamental questions about the purposes of research or who gets to shape that research, ethics has become increasingly institutionalized, instrumentalized, and professionalized, with an emphasis on filling out forms and checking off boxes[133]. Science ethics bodies suffer from limited "ethical imaginations" and are often primarily concerned with "keeping the wheels of research turning while satisfying publics that ethical standards are being met"[133]. "Ethical analysis that does not advance such instrumental purposes tends to be downgraded as not worthy of public support"[133].

In turn, "systems of ethics play key roles in eliding fundamental social and political issues" related to scientific research[134]. For instance, efforts to introduce ethics into genetic research throughout the 1990s and 2000s treated ethics "as something that could be added onto science—and not something that was unavoidably implicit in it"[134]. The effort to treat ethics as an add-on obscured how "ethical choices inhered in efforts to study human genetic variation, regardless of any explicit effort to practice ethics"[134]. As a result, these research projects "bypassed responsibility for their roles in co-constituting natural and moral orderings of human difference, despite efforts to address ethics at the earliest stages of research design"[134].

The turn to ethics can also entail an explicit effort among scientists to defuse external scrutiny and to develop a regime of self-governance. In the 1970s, frightened by calls for greater public participation in genetic engineering, biologists organized a conference at the Asilomar Conference Center in California[135]. The scientific community at Asilomar pursued two, intertwined goals. First, to present a unified and responsible public image, the Asilomar organizers restricted the agenda to eschew discussions of the most controversial applications of genetic engineering (biological warfare and human genetic engineering).

Second, to convince the American public and politicians that allow biologists could self-govern genetic engineering research, the Asilomar attendees "redefined the genetic engineering problem as a technical one" that only biologists could credibly discuss[135]. Although Asilomar is often hailed as a remarkable occasion of scientific self-sacrifice for the greater good, accounts from the conference itself present a different account. "Self-interest, not altruism, was most evident at Asilomar", as not making any sacrifices and appearing self-serving would have invited stringent, external regulation[135].

Tech ethics mirrors many of these attributes in scientific ethics. As with ethics in other fields of science, tech ethics involves a significant emphasis on institutionalized design practices, often entailing checklists and worksheets. Mirroring ethics in genetic research, the emphasis on ethical design treats ethics as something that can be added on to digital technologies by individual engineers, overlooking the epistemologies and economic structures that shape these technologies and their harms. Just like the molecular biologists at Asilomar, tech companies and computer scientists are defining moral questions as technical challenges in order to retain authority and autonomy.③ The removal of red lines in the European Commission's High-Level Expert Group on AI resembles the exclusion of controversial topics from the agenda at Asilomar.

## 4.2   Corporate ethics and co-optation

Codes of ethics have long been employed by groups of experts (e.g., doctors and lawyers) to codify a profession's expected behavior and to shore up the profession's public reputation[137, 138]. Similarly, companies across a wide range of sectors have embraced ethics codes, typically in response to public perceptions of unethical behavior[139].

Yet it has long been clear that the public benefits of corporate ethics codes are minimal. While ethics codes can help make a group appear ethical, they do little to promote a culture of ethical behavior[139]. The primary goal of business ethics has instead been the "inherently unethical" motivation of corporate self-preservation: to reduce public and regulatory scrutiny by promoting a visible appearance of ethical behavior[139, 140]. Ethics codes promote corporate reputation and profit by making universal moral claims that "are extremely important as claims but extremely vague as rules" and emphasizing individual actors and behaviors, leading to a narrow, "one-case-at-a-time approach to control and discipline"[137]. Ethics codes in the field of information systems have long exhibited a notable lack of explicit moral obligations for computing professionals[74, 141].

Business ethics is indicative of the broader phenomenon of co-optation: an institution incorporating elements of external critiques from groups such as social movements—often gaining the group's support and improving the institution's image—without meaningfully acting on that group's demands or providing that group with decision-making authority[142–144]. The increasing centrality of companies as the target of social movements has led to a particular form of co-optation called "corporatization", in which "corporate interests come to engage with ideas and practices initiated by a social movement and, ultimately, to significantly shape discourses and practices initiated by the movement"[145]. Through this process, large corporate actors in the United States have embraced "diluted and deradicalized" elements of social movements "that could be scaled up and adapted for mass markets"[145]. Two factors make movements particularly susceptible to corporatization: heterogeneity (movement factions that are willing to work with companies gain influence through access to funding) and materiality (structural changes get overlooked in favor of easily commodifiable technological "fixes"). By participating in movement-initiated discourses, companies are able to present themselves as part of the solution rather than part of the problem, and in doing so can avoid more restrictive government regulations.

Tech ethics closely resembles corporate ethics. Abstract and individualized tech ethics codes reproduce the virtue signaling and self-preservation behind traditional business ethics. In a notable example of co-optation and corporatization, technology companies have promoted tech ethics as a diluted and commoditized version of tech-critical discourses that originated among activists, journalists, and critical scholars. Because societal efforts to improve technology are often aimed at companies and include both heterogeneity and materiality, it is particularly vulnerable to

③ In an ironic parallel, the Future of Life Institute organized an Asilomar Conference on Beneficial AI in 2017, leading to the development of 23 "Asilomar AI Principles"[136].

corporatization. Through corporatization, tech companies use ethics to present themselves as part of the solution rather than part of the problem and use funding to empower the voices of certain scholars and academic communities. In doing so, tech companies shore up their reputation and hinder external regulation. The success of tech ethics corporatization can be seen in the expanding scope of work that is published and discussed under the banner of "tech ethics". Even scholars who do not embrace the tech ethics label are increasingly subsumed into this category, either lumped into it by others or compelled into it as opportunities to publish research, impact policymakers, and receive grants are increasingly shifting to the terrain of "tech ethics".

### 4.3    The stakes of tech ethics

These examples of ethics in science and business suggest two conclusions about tech ethics. First, tech ethics discourse enables technologists and technology companies to label themselves as "ethical" without substantively altering their practices. Tech ethics follows the model of science ethics and business ethics, which present case studies for how ethics-washing can stymie democratic debate and oversight. Continuing the process already underway, tech companies and technologists are poised to define themselves as "ethical" even while continuing to generate significant social harm. Although some individuals and groups are pursuing expansive forms of tech ethics, tech companies have sufficient influence to promote their narrow vision of "tech ethics" as the dominant understanding and implementation.

Second, those striving for substantive and structural improvements in digital technologies must be mindful of the gap between ethics as normative inquiry and ethics as a practical endeavor. Moral philosophy is essential to studying and improving technology, suggesting that ethics is inherently desirable. However, the examples of ethics in technology, science, and business indicate that ethics in practical contexts can be quite distinct from ethics as a mode of moral reasoning. It is necessary to recognize these simultaneous and conflicting roles of ethics. Defenders of ethics-as-moral-philosophy must be mindful not to inadvertently legitimize ethics-as-superficial-practice when asserting the importance of ethics. Meanwhile, critics who would cede ethics to tech companies and engineers as a denuded concept should be mindful that ethics-as-moral-philosophy has much to offer their own critiques of ethics-as-superficial-practice.

Attending to these porous and slippery boundaries is essential for supporting efforts to resist oppressive digital technologies. As indicated by the responses to Ochigame's critique of ethics-washing, many of the more radical critics of digital technology see themselves as outside of—if not in opposition to—the dominant strains of tech ethics. Activists, communities, and scholars have developed alternative discourses and practices: refusal[85, 146, 147], resistance[148], defense[149, 150], abolition[150, 151], and decentering technology[152]. Although some may see these alternative movements as falling under the broad umbrella of tech ethics, they embody distinct aspirations from the narrow mainstream of tech ethics. Labeling these burgeoning practices as part of tech ethics risks giving tech ethics the imprimatur of radical, justice-oriented work even as its core tenets and practices eschew such commitments.

## 5    A Sociotechnical Approach to Tech Ethics

Rather than presenting a unifying and beneficent set of principles and practices, tech ethics has emerged as a central site of struggle regarding the future of digital architectures, governance, and economies. Given these dynamics of contestation surrounding tech ethics, ethics will not, on its own, provide a salve for technology's social harms. In order to better evaluate the opportunities and limits of tech ethics, it is necessary to shift our focus from the value of ethics in theory to the impacts of ethics in practice.

This task calls for analyzing tech ethics through a sociotechnical lens. A sociotechnical approach to technology emphasizes that artifacts cannot be analyzed in isolation. Instead, it is necessary to focus on technology's social impacts and on how artifacts shape and are shaped by society. Similarly, a sociotechnical approach to tech ethics emphasizes that tech ethics cannot be analyzed in isolation. Instead, it is necessary to focus on the social impacts of tech ethics and on how tech ethics shapes and is shaped by society. If "technologies can be assessed only in their relations to the sites of their production and use"[22], then so too, we might say, tech ethics can be assessed only in relation to the sites of its conception and use. With this aim in mind, it is fruitful to consider tech ethics through the lens of four sociotechnical frames: objectivity and neutrality, determinism, solutionism, and sociotechnical systems.

## 5.1   Objectivity and neutrality

A sociotechnical lens on technology sheds light on how scientists and engineers are not objective and on how technologies are not neutral. It makes clear that improving digital technologies requires grappling with the normative commitments of engineers and incorporating more voices into the design of technology[153, 154]. Similarly, it is necessary to recognize that the actors promoting tech ethics are not objective and that tech ethics is not neutral. Currently, the range of perspectives reflected in ethics principles is quite narrow and ethics is treated as an objective, universal body of principles[29, 71, 73]. In many cases, white and male former technology company employees are cast to the front lines of public influence regarding tech ethics[131, 132]. As a result, the seeming consensus around particular ethical principles may say less about the objective universality of these ideals than about the narrow range of voices that influence tech ethics. Thus, rather than treating tech ethics as a body of objective and universal moral principles, it is necessary to grapple with the standpoints and power of different actors, the normative principles embodied in different ethical frameworks, and potential mechanisms for adjudicating between conflicting ethical commitments.

## 5.2   Determinism

A central component of a sociotechnical approach to technology is rejecting technological determinism: the belief that technology evolves autonomously and determines social outcomes[155, 156]. Scholarship demonstrates that even as technology plays a role in shaping society, technology and its social impacts are also simultaneously shaped by society[21, 23, 157, 158]. Similarly, it is necessary to recognize the various factors that influence the impacts of tech ethics in practice. Currently, ethics in digital technology is often treated through a view of "ethical determinism", with an underlying assumption that adopting "ethics" will lead to ethical technologies. Yet evidence from science, business, and digital technology demonstrates that embracing "ethics" is typically not sufficient to prompt substantive changes. As with technology, ethics does not on its own determine sociotechnical outcomes. We therefore need to consider the indeterminacy of tech ethics: i.e., how the impacts of tech ethics are shaped by social, political, and economic forces.

## 5.3   Solutionism

Closely intertwined with a belief in technological determinism is the practice of technological solutionism: the expectation that technology can solve all social problems[159]. A great deal of sociotechnical scholarship has demonstrated how digital technology "solutions" to social problems not only typically fail to provide the intended solutions, but also can exacerbate the problems they are intended to solve[160−163]. Similarly, it is necessary to recognize the limits of what tech ethics can accomplish. Currently, even as tech ethics debates have highlighted how technology is not always the answer to social problems, a common response has been to embrace an "ethical solutionism": promoting ethics principles and practices as the solution to these sociotechnical problems. A notable example (at the heart of many tech ethics agendas) is the response to algorithmic discrimination through algorithmic fairness, which often centers narrow mathematical definitions of fairness but leaves in place the structural and systemic conditions that generate a great deal of algorithmic harms[164, 165]. Efforts to introduce ethics in digital technology function similarly, providing an addendum of ethical language and practices on top of existing structures and epistemologies which themselves are largely uninterrogated. Thus, just as technical specifications of algorithmic fairness are insufficient to guarantee fair algorithms, tech ethics principles are insufficient to guarantee ethical technologies. Ethics principles, toolkits, and training must be integrated into broader approaches for improving digital technology that include activism, policy reforms, and new engineering practices.

## 5.4   Sociotechnical systems

A key benefit of analyzing technologies through a sociotechnical lens is expanding the frame of analysis beyond the technical artifact itself. Rather than operating in isolation, artifacts are embedded within sociotechnical systems, such that the artifact and society "co-produce" social outcomes[21]. Similarly, it is necessary to view tech ethics as embedded within social, economic, and legal environments, which shape the uses and impacts of tech ethics. Currently, efforts to promote ethical technology typically focus on the internal characteristics of tech ethics—which principles to promote, for instance—with little attention to the impacts of these efforts when integrated into a tech company or computer science curriculum. In turn, tech ethics has had limited effects on technology production and has played a legitimizing role for technology companies. Attempts to promote more equitable technology must instead consider the full context in which tech ethics is embedded. The impacts of tech ethics are shaped by the beliefs and actions of engineers, the economic incentives of companies, cultural and political pressures, and regulatory environments. Evaluating tech ethics in light of these factors can generate better predictions about how particular efforts will fare in practice. Furthermore, focusing on these contextual factors can illuminate reforms that are more likely to avoid the pitfalls associated with tech ethics.

## 6   Conclusion

A sociotechnical lens on tech ethics will not provide clear answers for how to improve digital technologies. The technological, social, legal, economic, and political challenges are far too entangled and entrenched for simple solutions or prescriptions. Nonetheless, a sociotechnical approach can help us reason about the benefits and limits of tech ethics in practice. Doing so will inform efforts to develop rigorous strategies for reforming digital technologies.

That is the task of this special issue: "Technology Ethics in Action: Critical and Interdisciplinary Perspectives". The articles in this issue provide a range of perspectives regarding the value of tech ethics and the desirable paths forward. By interrogating the relationships between ethics, technology, and society, we hope to prompt reflection, debate, and action in the service of a more just society.

## Acknowledgment

B. Green thanks Elettra Bietti, Anna Lauren Hoffmann, Jenny Korn, Kathy Pham, and Luke Stark for their comments on this article. B. Green also thanks the Harvard STS community, particularly Sam Weiss Evans, for feedback on an earlier iteration of this article.

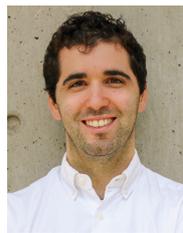

**Ben Green** is a postdoctoral scholar in the Society of Fellows and an assistant professor in the Gerald R. Ford School of Public Policy, University of Michigan. He received the PhD degree in applied math (with a secondary field in STS) from Harvard University and the BS degree in mathematics & physics from Yale College in 2020 and 2014, respectively.

# Framing and Language of Ethics: Technology, Persuasion, and Cultural Context

Jasmine E. McNealy*

**Abstract:** What are the consequences of the language we use for technology, and, how we describe the frameworks regarding technology and its creation, use, and deployment? The language used to describe technology has the possibility to deceive and be abusive. How language is used demonstrates what can occur when one party is able to assert linguistic power over another. The way in which organizations frame their relationships with technology is one such power asymmetry. This article examines the complications of the imagery used for ethics in technology. Then, the author offers a brief overview of how language influences our perceptions. The frames used to describe phenomena, including ethical frameworks and technology, allow for the creation of heuristics, or shortcuts that are "good enough" for understanding what is being described and for decision-making. Therefore, descriptions matter for relaying meaning and constructing narratives related to ethical uses of technical systems. After this, the author investigates what we mean by ethics and the codes that corporate, governmental, and other organizations use to depict how they understand their relationship to the technology they create and deploy. The author explores three examples of frames of ethics and descriptions of technology, which though appearing progressive, once understood holistically, fail to adequately describe technology and its possible impact. The author ends this article with a discussion of the complexity of describing and communicating ethical uses of technology.

**Key words:** language; framing; ethics; technology; culture

*"…metaphors give rise to technical models, which inform design processes, which in turn shape knowledges and politics…"*

– Shannon Mattern[1]

## 1 Introduction

In her piece, "The City is not a Computer", is noted by scholar of media and spaces Shannon Mattern[1]. She described the error in analogizing a city to a computational device. Theorists and technologists used this analogy in discussions of emerging smart cities and the implementation of civic technology. Mattern argued that the representation of a city as a computer was inaccurate; cities were neither programmable nor rational. The description of a city as a computer was also inadequate — life in cities did not always follow specific aims and plans, and urban environments were not simply "apparati for record-keeping and information management"[1]. More happens in cities than simple information processing and storage. And the image does more than fail to correctly describe the environment, it sets the stage for the implementation of policy and the reification of structural issues that can be harmful. It also ignores the complexity of describing urban society. Viewing data as the fuel upon which a city runs disregards other kinds of information flowing throughout a city that cannot be reduced to 1 s and 0 s; it overlooks the voices of people and history.

The repeated use of certain language to describe

- Jasmine E. McNealy is with the College of Journalism and Communications, University of Florida, Gainesville, FL 32601, USA. E-mail: jmcnealy@ufl.edu.
* To whom correspondence should be addressed.

something informs how we construct, or imagine, that thing. Therefore, if we view urban spaces as simply parts of an information processor, we may ignore or miss the vital social, political, and cultural activities that need to be considered in designing things like parks, social services, and policy.

The same can be said for the kind of language and descriptions used in talking about ethics in technology. Considerations of what it means for technology, and the firms that create and implement it, to be ethical are critically important in a world in which the fruit of failing to consider the implications of technology is manifesting in amplified ways. This includes how we discuss technology in general, as how we describe something influences how we understand its purpose and what it can do.

Take, for instance, the controversy over DeepNudes, an app that uses machine learning to make photos appear as though the women in them were naked. The app is an example of the deepfakes phenomenon, in which images, videos, and sound can be altered by using cheap artificial intelligence to produce products that look and sound real[2]. The creators of DeepNudes, who pulled down the service after receiving widespread negative attention, claimed they created it for entertainment purposes[3]. This framing of such a potent technology as for mere amusement, like framing a city as a computer, ignores the ramifications of creating a system that targets women for abuse. What's more, the "entertainment" designation makes it appear as though legitimate uses of the app exist.

What, then, are the consequences of the language we use for technology, and, how we describe the frameworks regarding technology and its creation, use, and deployment? This article examines the complications of the imagery used for ethics in technology. In Section 2, the author offers a brief overview of how language influences our perceptions. The frames used to describe phenomena, including ethical frameworks and technology, allow for the creation of heuristics, or shortcuts, that are "good enough" for understanding what is being described and for decision-making[4, 5]. Therefore, descriptions matter for relaying meaning and constructing narratives related to ethical uses of technical systems. This section, though acknowledging that framing theory has been used in several different fields, focuses on framing from the perspective of mass communication scholars. In Section 3, the author investigates what we mean by ethics and the codes that corporate, governmental, and other organizations use to depict how they understand their relationship to the technology they create and deploy. Section 4 explores three examples of frames of ethics and descriptions of technology, which though appearing progressive, once understood holistically, fail to adequately describe technology and its possible impact. The author ends this article with a discussion of the complexity of describing and communicating ethical uses of technology in Section 5.

## 2   Constructing Descriptions

In June 2019, Google announced the construction of its third subsea cable stretching from Portugal to South Africa[6, 7]. The cable would connect Africa to Europe and increase Google's cloud infrastructure. The significance of a new, private, and international telecommunications cable project from the mega-corporation was appreciable alone. But what garnered noteworthy attention was Google's choice in name for the project, "Equiano". This name, chosen by the large global tech company, may have appeared benign to some, but others noted the historical context.

Equiano was the surname of Olaudah Equiano, also known as Gustavas Vassa, a formerly enslaved man who became a writer and abolitionist in the late 18th century. According to his own memoir, Equiano was kidnapped as a child from his home in what is now Nigeria and sold into slavery[8]. As an adult Equiano was able to purchase his freedom and then moved to London where he was instrumental in founding the Sons of Africa, an abolitionist group composed of formerly enslaved Africans. He spent the rest of his life advocating for poor Black people in London. Therefore, when Google chose to use Equiano as the name for its underwater cable, it evoked connections to colonialism, imperialism, and human subjugation.

Equiano, like all words, has meaning. How we define words is linked to social and cultural context[9]. And meanings change, and words may have multiple meanings that are created and adapted for specific situations. Meanings are situated, or assembled out of many different features, while at the same time, being created from what linguistics and literacy scholar James Paul Gee calls cultural models[9]. Cultural models, also called explanatory theories, help to clarify patterns that emerge in sense-making, or interpreting everyday life[10]. These theories are based in sociocultural practices that include beliefs about the meanings. What, then, is the sociocultural context of Equiano and how is meaning assembled for this word in relation to the construction of an underwater cable connecting Europe to Africa?

Within context, it is possible to view Google's name choice as a mirror image of the extractive historical relationship that Europe, representing Western or Global North countries, have had with Africa and other Global South countries. Countries of the Global South offer raw material and resources — food, minerals, oil, and people — of use for firms and governments in the North, and these resources have historically been exploited. In a more modern sense, the data from Internet users who will be connected to this cable are the raw material awaiting exploitation — Google is not the only organization creating data infrastructure in an attempt to connect with Africa; Facebook, too, has plans for an underwater cable to be named "Simba"[11] and Microsoft and Amazon are opening data centers on the continent[12]. The data flows embody all five dimensions of anthropologist of media and communication Arjun Appadurai's global cultural flows — ethnoscapes, mediascapes, technoscapes, financescapes, and ideoscapes — which explain the complexity of the global economy as being the result of "fundamental disjunctures between economy, culture, and politics"[13].

Google's choice in name, of course, could be viewed as an ode to Equiano; the company has another private subsea cable off the coast of Brazil named "Curie" and a third that runs from Virginia Beach in the United States to France called "Dunant". All three cables would be regarded as being named in recognition of an individual of significant achievement. But this view would be based in cultural context that ignores the history of extraction and exploitation. A cable named after Marie or Pierre Curie, both French scientists, off the coast of South America, a continent to which they may only have traveled is not the same as using the name of a man for cable laid off the coast from which he was stolen.

How people interpret the use of the name, or any word or idea, is based on the frames used. Culture — along with the communicator, the text, and the receiver — is a location of frames in the process of communication[14, 15]. Frames are created when a communicator chooses to make certain aspects of an idea more salient than others[14, 15]. According to mass communication scholar Vreese[13], framing involves a source of information presenting and defining an issue. Frames may also identify the origin of an issue, evaluate the causes and possible impacts, and offer remedies[15, 16]. Frames are, by definition, selective descriptions — highlighting certain information and downplaying or omitting others[17, 18]. Frames offer a construction of reality to the audience and, therefore, exert power[15]. Put another way, frames communicated to an audience can affect how that audience understands the subject[19, 20].

Framing theory has a long history and has been examined in related several different academic fields, the author focuses on framing theory as it has been explored in mass communication or media studies research. Mass communication is relevant as it is the communications of a message to a large audience[21]. Scholarship has identified two kinds of frames: equivalency frames and emphasis frames[22−24]. While equivalency frames tend to be associated most with media effect because they "involve manipulating the presentation of logically equivalent information", emphasis frames involve the manipulation of content[22]. According to Cacciatore et al.[22], emphasis framing is sociologically oriented — it focuses on the messages the audience receives, and emphasizes one set of considerations over another.

It is appropriate, then, to reexamine Google's own announcement of the cable Equiano, then, with emphasis framing analysis in mind. The blog post mentions that the cable is named for the abolitionist and notes that he had been enslaved as a child. The announcement further references that all three of its subsea cables have been named for "historical luminaries"[6]. Yet, the blog post omits the history European exploitation and extraction. It offers no further explanation of slavery, Africa, or Equiano's life. What is made salient — highlighted as most important — is that the company was creating private infrastructure connecting Europe to Africa, and that the cable is named after an important historical person. The power in this construction of reality is that the use of the name is promoted as a tribute, while the historical context—connection, infrastructure, colonialism, and subjugation — is omitted or ignored. Google may have done a great thing in honoring Equiano, but the company also fails to note the larger picture, and the reality constructed for the audience is one that omits important background issues. In doing this, the mega-corporation creates an assemblage of frames that center the honorific while ignoring the extraction.

Likewise, how we communicate about ethics and technology may omit or ignore important sociocultural and historical contexts, offering a construction of reality that is both inaccurate and inadequate for assessing our relationship to technology. In Section 3, the author considers what it is we mean by ethics, and how organizations and individuals are framing the discussion of ethics and technology.

## 3    Communicating Ethics

Discussions about the ethics of the creation and use of certain technology have proliferated[25]. Yet, we are often confronted with a lack of a definitive or unified definition. Studies of business ethics have found several different descriptions of ethics in the textbooks used to teach business majors at universities[26], and ethics are at times conflated with social responsibility[27]. It is important, then, to define what it is we mean when we say ethics.

Ethics have been defined as a system for determining what is right and proscribing what is wrong[28]. This system is based in rationality and must be applied in a consistent manner to be valuable. Further, the decision of whether an act is right or wrong is situational, based on the context of the action or policy. But even this definition does not provide us with a universal description of the kinds of behaviors or actions that will be considered ethical. This may be because there are many different ethical systems, some of which conflict.

Deontological ethics, for instance, frame certain standards of conduct as being right based on normative ideas of duty and morality[29]. It is a view that no matter how good the effects of a choice are, some choices are wrong based on specific values. Therefore, if a choice is considered wrong, an individual should not make it, even if the outcome of that choice is positive. As an illustration, imagine a society in which the normative belief system says that the destruction of forestland, for any reason, is morally wrong. Say then, a particular forest is between one town and another larger city where a large trauma medicine center is located. Ambulances and others must navigate around the forest, spending more time than if a road were to be cut directly through the trees. Under a system of deontological ethics, although a consequence of destroying part of nature in this case would mean achieving the good of being able to reach emergency and other health services faster, the act of clearing the land would still be considered wrongful.

Contrast deontological ideas of ethics with a consequential ethical system like utilitarianism. Consequential ethics, as it sounds, looks at the effects of an action or policy[30]. In a consequential ethics system like act-based utilitarianism, the ethics of an action are judged based on the consequences as they relate to outcomes like happiness or welfare. In rule-based utilitarian systems, rules are only created if they result in overwhelming benefits[31]. In the forest scenario above, under a consequentialist ethical system, we would first examine the effects of clearing a path in the forest for a road, a significant one of which would be the decrease in time for emergency health services, among other things, between the two towns. These outcomes would be considered a benefit to society, most-likely outweighing the costs of losing some of the forested area.

So far only two kinds of ethical systems have been identified, both of which come from a Western-centric philosophy. Many other systems of ethics and morality exist across the globe[32]. The various indigenous systems, for example, have dramatically different ideas about what should be considered right or wrong and how society should deal with non-conformance. How, then, can we know what we mean when we talk about ethics, particularly ethics as applied to technology and the organizations that create technology? According to Ellwood[33], how a society functions — the purview of the social sciences — furnishes the "raw material" for the creation and study of ethics. Social sciences investigate humans and their relationships. Therefore, the author defines ethics here as describing the relationship that individuals and organizations have with the "thing" at issue, in this case technology, in making determinations about right and wrong.

The explicit language of an organization's code of ethics or how it talks about its relationship to its technology, in theory, tells us about how an organization perceives certain products and behaviors. Codes of ethics and ethical statements are public-facing expressions of organizational standards[34]. Codes express to others how an organization views its product or service within the context of societal ideas of what is right or wrong. And these codes can also be a way for organizations to appear ethical to the public[35]. Codes, though expressing statements of how an organization perceives behavior, are not self-enforcing, nor do all codes even consider or express the consequences of noncompliance. According to Wood and Rimmer[35], a code by itself is only a façade for an organization to behave ethically and could be considered deceptive.

But codes are the steps that tech companies and organizations employing technology are being used to express their understanding of the relationship between their use or creation of a product or service and the greater society. Three kinds of ethics codes exist: regulatory, aspirational, and educational[36, 37]. Regulatory codes use language that express imperatives. Behaviors and activities are expressly prescribed according to the specified rules of the organization. An organization enforces these codes through monitoring and bad actors can be sanctioned for failure to comply. An example of a regulatory code of ethics is that like the professional responsibility rules that lawyers admitted to a bar in the United States must follow. Failure to follow the rules of professional responsibility could mean suspension from practicing law, and sometimes, loss of law license.

Aspirational codes are those expressing an organization's ideals. These are statements of levels to which an organization would like to reach, but behaviors are not mandated. Therefore, an aspirational code for a tech-related organization could contain a clause stating that the company will "strive to recognize the humanity of all people". This certainly reads like an important way for an organization to behave. Language centering humanity would persuade some that the organization is doing something with respect to how it will treat people in relation to its use or creation of technology. But the statement does not offer an articulation of what "recognize the humanity of all people" means, nor does it include a description or inference as to the kind of actions that do not meet this standard. It certainly does not provide any indication of what will happen when it fails to meet this code, if that ever could happen under such a vague standard.

Lastly, educational codes are those that may offer proscriptions, but also provide commentary, with the goal of offering the reader an understanding of its interpretation of the language used. The rules of professional responsibility mentioned above, those to which attorneys in the US must adhere, are often annotated to provide commentary and example scenarios by which readers can judge the ethics of their actions.

Many organizational, professional, and societal ethical codes for tech organizations are aspirational. Organizations may provide vague statements and are allowed to self-police. According to Stark and Hoffmann[38], these codes "elide granular attention" to actual actions. And these ethical codes are also represented in the pronouncements organizations make about themselves and their products. Google, for instance, for a long time used the assertion "Do not be evil". Certainly, staying away from building and using technology for bad acts was a laudable goal, but the mega-corporation offered no definition of what it considered evil, nor did it provide a way for the public to hold it accountable for failing to meet this standard. Significantly, Google removed "Do not be evil" from its code of conduct in 2018[39, 40]. Section 4 considers some of the other aspirational language used not only by organizations, but by others in considering ethics in technology.

## 4 Framing Ethics and Technology

How an organization or individual describes its relationship, its ethics, to technology has implications for how that technology — its creation, use, and deployment — is understood by society. Several different ways of expressing this relationship exist. This section explores three common frames of emphasis for describing this relationship to technology and the implications of how these frames construct reality surrounding technology. To do this, the author investigates how organizations communicate messages about technology to the public and by examining how they define and identify important ideas and issues and by evaluating which descriptions are made salient or omitted. These frames are neutrality, property, and user-centeredness.

### 4.1 Neutrality

An enduring frame of technology is that a tool, system, or structure is neutral, or not programmed with biases or values. A popular topic in philosophy of science and technology studies, this framing of the relationship with technology, value neutrality in technology, can be interpreted in three ways: that the technology is neutral because it has many different purposes, that the technology is neutral before it is used, and that the technology is only an application of "a scientific, mathematized, and value-free view of nature"[41]. The first and second value-neutrality interpretations align as they focus on the use and purposes of technology and lead to the principal question of whether a "thing" can be created without having any inherent values or biases. Though this interpretation of value-neutrality is still popular, the evidence, both empirical and otherwise, demonstrates that the answer is "no". Recent scholarship by Safiya Umoja Noble[42] and Virginia Eubanks[43] has illustrated the danger in believing in, and relying on, the neutrality of algorithms used for search and to implement civic policy. Technology, broadly defined, has been shown to be endowed with the "politics", or the viewpoints, of its creator in both its use and the impact[44].

The third interpretation of value-neutrality in technology, too, is popular in that it denies the existence of politics in technology by pointing to laws of nature, science, and math. This imagining of value-neutrality constructs technology as reflecting only what occurs naturally and outside of human or social control. This reflects technological determinism, a theory that technology evolves by itself and has the power to shape society[42, 43]. Of course, the dispute between social constructivists — who believe that humans shape technology — and determinists has been ongoing[45]. But to ignore that our understanding of science and technology, itself, is grounded in societal context is to ignore that what we think of as scientific and technological laws are based on interpretation by humans.

Facial recognition technology (FRT) provides an emergent technology for the exploration of frames of value-neutrality. FRT, which scans the human face for supposedly unique characteristics to create a map akin to a fingerprint, has been framed as a neutral technology by some technologists[46, 47]. FRT systems have been touted as systems for good in law enforcement, anti-terrorism, and finding missing persons[48]. Amazon has created its own FRT, Rekognition, which it frames as a system that allows the user to "detect, analyze, and compare faces for a wide variety of user verification, people counting, and public safety use cases"[49]. In promoting its system, the company focuses on the capability of the product, calling it fast and accurate. Beyond simply facial recognition, Amazon promotes Rekognition as useful for six other services, including facial analysis, celebrity recognition, and unsafe content detection. Rekognition is also described as offering benefits to the user like low costs, and real-time analysis. Lastly, Amazon employs images as part of its description of the technology. These photos are used to demonstrate how the system works.

As with Google's use of Equiano for its subsea cable, Amazon's framing of its FRT fails to provide context for its technology. In fact, the description provided for the Rekognition makes it appear as though the technology is neutral when evidence has proven that FRT is anything but. Research published by computer scientists Joy Buolamwini and Timnit Gebru found that because FRT is trained on biased datasets, in this case datasets in which the majority of faces in the training data were of lighter skin, the systems could not provide accurate results, especially for women with darker skin[50]. Inaccurate results from FRT disparately impact individuals from already marginalized communities, particularly Black people[51].

In describing its relationship, then, with this technology, Amazon omits the significant consequences of the uses of FRT. Of course, negative impacts are not often selling-points for any product. But in failing to provide any explanation of how the system is trained, and the ramifications of that training, the organization presents FRT as though the technology were devoid of any inherent values.

## 4.2  Property

Another common emphasis frame used in discussing our relationship with technology is that of property. Property-centered language is used to describe our data and the rights we may assert over another individual or organization attempting to access, use, or control personal information. Examples of property-based language can be found in the discussion of privacy, and legal scholars have found that "the right of privacy originates in property-based ideas, whereas one of the functions of property law is to protect private interests"[52].

But property is a creation of society. This allows the ownership of a "thing", be it land or intangible information[53]. Therefore property, and the rights associated with the ownership of property, can be restrained. What is often used in relation to property is the metaphor of a "bundle or rights", signifying the constraints on the property owner[52, 53]. The bundle of rights has been interpreted as conceptualizing property as the relationships between people and not between people and a thing[54, 55]. Within this bundle is the ability to possess, use, and sell the property. But these rights can be overcome for reasons of public policy. For example, although we technically own our own bodies, it is illegal in the US to sell your organs.

But property ownership has never, historically, been a right allowed to all people, nor has everyone been allowed to have the "sticks" in the bundle of rights respected. In the US, for instance, property rights have been found to be "rooted in racial domination"[56]. Native Americans had their rights to land stripped by US Supreme Court decision[57]; African Americans were deemed property and denied rights related to self-fulfillment[58]. Other groups, too, have been denied rights in property or possession in both land and self.

And the language of property conflicts with that of humanity. To look upon a human as property allows individuals and organizations to behave toward that person in ways that would not be sanctioned in relation to others. The same can be said for property language in connection to personal data. Discussions of data, broadly defined, use property language and the rhetoric of ownership, control, and access in relation to personal data, thereby creating a definition of data divorced from the individual and ignoring the very real harms of personal data aggregation. When it is just data being collected, the consequences of that collection can be ignored; when those data are more closely attached to humans, the possible harms become more tangible.

Detroit's Project Greenlight provides a case for exploration. The City's own website describes the project as partnership between the Detroit Police Department and businesses with the aim of fighting crime[59]. Businesses and other organizations involved in the partnership must install surveillance cameras and high-speed internet. The videos, the data, from the cameras are streamed to the DPD for analysis. Instead of using officers to watch the stream, the DPD employs "civilian crime analysts" tasked with identifying and finding crime suspects[60]. The DPD page fails to provide any mention that the department uses FRT to analyze the video[61].

Viewing the people who may appear in the video streams as points of data used to solve crime has ramifications. Although city officials have denied it, it is possible that FRT may be used to identify people who are not necessarily suspected of any crime[62, 63]. Video surveillance provides more than just points of data; it can offer a construction of the life of a private individual who is not breaking any law. Points of data, then, are more than simple places for analysis; these observations are what makes individuals unique and identifiable[64]. Therefore, the DPD framing of its Project Greenlight omits the possible consequences of the technology to personal privacy.

### 4.3 User-centeredness

Lastly, a common emphasis frame used by organizations explaining their relationship to technology is that of being "user-centered". The user-centered approach to design calls for involving the user in the process of design[65]. The process focuses on ensuring usability by understanding how users may interact with a product or service[66]. Users participate with designers throughout the development of the item and the system evolves in iterations. Throughout the development, designers are supposed to make "explicit and conscious design" choices[66].

From the outset, there are various reasons why the framing of product design as "user-centered" is fraught. To make their process truly user-centered, the designers and researcher would have to be able to consider all the possible kinds of users, their abilities, and the social and cultural systems in which they will interact with a product. User-centered language, as currently implemented, may ignore that there are many kinds of users, each with their own needs, wants, and desires. It may further ignore that different users have different abilities and experiences that are important for consideration in design. For the most part, creators design for what is considered the default — white and male[67, 68] — thereby, ignoring the diverse sets of individuals that may use, come in contact, or be impacted by the products they create[69].

User-centered framing also ignores that a product or service may impact non-users. For the most part, then, user-centered design is a customer-centered approach[70]. User experience researchers create personas that imagine how someone adopting a product may use it and what their needs may be. These are usually constructed out of the imaginary of the prototypical product user[71]. But there may be several reasons that someone may not adopt technology, including lack of access and complete rejection.

The City of Boston's neighborhood resources site provides an example of a technology for analysis. In April 2019, the City announced that it was working to port over its old "My Neighborhood Resource" tool to its new Boston.gov website[72]. The My Neighborhood page provides access to information on properties within the City including city services and resources, landmarks, information on voting, and political representation. In describing the revisions to the site, the City's announcement states that the team tasked with the makeover "wanted to make sure they were creating a user focused application", and that it wanted "to make sure we always keep our users' needs in focus"[72]. The announcement goes on to describe the various tests conducted in revising the app and provides interview data from some of the users.

This framing of a user-centered civic technology, on the surface, appears to be a great approach for a municipality attempting to provide the services that its constituents need. Certainly, a city providing ways for its residents to access information and services efficiently is laudable and conducting research to ensure that it was best serving the people who may use the technology should also be commended. This framing of the tool as user-friendly, however, ignores the residents and others who may not have access to the information available on the app.

Civic technology aimed at connecting city residents to information and services has proliferated[73]. But not all possible users can take advantage of these systems. At the very least using these resources requires a phone, in the case of 311 numbers — which allow residents to report issues to government departments — or an internet connection for those who want to access online services. The users that a city focuses on, then, are those who can afford these connections. Further, some residents may consciously choose to reject a civic service for fear of surveillance. Therefore, although a user-centered approach to civic technology, like Boston's My Neighborhood, appears to target all users, it ignores the impediments to adoption of the technology.

## 5 Reframing Relationship with Technology

In her 1992 book, *Talking Power*, linguist Robin Tolmach Lakoff asserted, "Language is powerful; language is power. Language is a change-creating force and therefore to be feared and used, if at all, with great care, not unlike fire"[74]. Like Mattern, who argued that how we imagine a city had the power to influence design and policy, Lakoff, too, focused on how language was used to seek power and was, therefore, always political. For Lakoff, language was always used to persuade, but of particular interest was the possibility of deception and abuse that can occur when one party is able to assert linguistic power over another.

The ways in which organizations frame their relationships with technology is one such power asymmetry. Emphasis frames like neutrality, property, and user-centeredness offer surface-level interpretations of technology that appear benign. Sometimes, as in the case of user-centered, the language used constructs an imaginary of a progressive way of thinking about how we should create, use, and implement technology. Organizations prioritize to the public narratives that appear advanced or enlightened, like naming a telecommunication cable after a formerly enslaved abolitionist. But these frames, and many others, fall short of offering express and complete descriptions and explanations of the technology and the organization's relationship with it. They are reductive, oversimplified ways of viewing impactful systems and relationships. At most, these frames provide aspirational goals for organizations to reach. More realistically, these frames offer vague and inaccurate views of the possible implications of the technology for which an organization is responsible. Like the story told about the choice of Equiano as a name for the subsea cable, these explanations hinder progress by obstructing true examinations of power dynamics through framing.

Frames can be composed of four possible elements: a definition, an identified origin, an evaluation, and remediation. The three frames explored — neutrality, property, and user-centered — were missing elements of the frames that might produce alternative understandings, thereby demonstrating the creation of salience relevant to one aspect of a discussion. In the neutrality frame, both the evaluation of causes and possible impacts and the remedies were lacking. For neutrality, omitted in the promotion of facial recognition technology was an explanation of the underlying bias in how the technology is trained, as well as the possible impacts to already marginalized populations. For property, omitted was an evaluation of the consequences of not viewing data collected as representative of people. Finally, for user-centered, omitted was an evaluation of the impediments to use and how this might affect who an organization includes when designing a technology.

The influence of these emphasis frames on public understanding of the implications of various technology is significant. As both Lakoff and Mattern assert, language has the power to shape our perspectives[1, 74]. These frames, along with others used by organizations and individuals to explain relationships with technology, shape our interpretations of how or whether technology should be designed and used. DeepNudes, the service that used AI to allow users to alter photos to make it appear as though the women in them were naked, offers an example for consideration. In their attempt at explaining their relationship with the app, the creators framed it as being created for entertainment[75]. As a frame, the label entertainment colored how people initially understood the service. Entertainment, usually, supposes amusement and light-hearted fun. But this frame ignores the consequences of virtually disrobing unsuspecting women without their consent.

The example technology and the language used in connection identified throughout this article offer some inferences. Perhaps the primary conclusion is that talking is hard[76]. Language is constructed of words and descriptions situated in culture, that helps us form explanations for ideas and phenomena. Consequently, it is important for organizations to understand the various cultural models that may arise with the descriptions of technology. Although Google's Equiano announcement did note that the abolitionist had been a slave, it did not consider the context of extraction and imperialism that could change the tenor of the statement for some.

Context matters. The audiences for these statements will have divergent views based on their understandings and experience with both the organization and the subject technology. How a Black Detroiter understands the police department's Project Greenlight may be very different than how a white business owner interprets the initiative. They arrive at their understandings of the project based in their experiences and histories. It is important, then, for government, corporate, and other organizations to reckon with the context of the technology they create and deploy.

No simple solution exists ensuring that organizations use adequate and accurate descriptions for their relationships with technology. This is not to say that there is one, definitive depiction or phrasing that would cover the entirety of the history and implications of technology. But current portrayals are woefully insufficient. And these representations are political and persuasive[77]. It would be beneficial, then, for organizations to reconsider how they approach creation and use of these systems. This may require changing codes of ethics from aspirational to more educational as well as reexamining the frames used in public statements. More importantly, it requires a rethinking of the power that certain organizations are allowed to amass with respect to technology. This power is partially derived from how firms are able to persuade the public[78]. A way to shift power, perhaps, may be in modifying how organizations promote their product or service to the public. It will also take continued vigilance in ensuring that counternarratives revealing the risks of technology are exposed to the public through advocacy. In the end, language matters.

At the same time, expecting organizations, including government organizations, to agree with shifts in burdens and power seems too simplistic of a resolution to a complex issue. Further, under this idea the onus would remain on individuals and communities to protect themselves from manipulative message tactics used by organizations. It is important, then, for regulators to bring reforms to how these messages are communicated to audiences.

Organizational messages are commercial messages made to persuade individuals into believing ideas that benefit the firm. In the United States regulations related to the kinds of claims made in organizational ethics codes already exist, especially as these claims are public-facing and material to whether individuals choose to adopt software or services. When companies fail to measure up to their claims, or an individual acting rationally could be deceived, regulators can step in to punish organizations for these deceptive or unfair practices. The author leaves the specifics to future research, but regulators have the power to prohibit and punish commercial messages that are otherwise deceptive. This could be necessary step in having organizations rethink the language they use in connection with ethics and impacts.

## Acknowledgment

J. E. McNealy would like to acknowledge the collective wisdom of the Ethical Tech Collective and those who participate in the Ethical Tech Working Group.

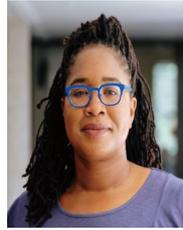

**Jasmine E. McNealy** is an associate professor in College of Journalism and Communications, University of Florida. Her research focuses on emerging technology, with an emphasis on privacy, surveillance, and data governance. She has published in internationally and nationally recognized journals.

# Tech Ethics: Speaking Ethics to Power, or Power Speaking Ethics?

Lily Hu*

**Abstract:** In recent years, tens of product teams, research institutes, academic conferences, and college courses—the list goes on—have cropped up under the banner of tech ethics to grapple with the social and political impact of technology. For some, an orientation around ethics indicates a moment of humility in an industry characterized by hubris. Now even major tech corporations are seeking expertise outside of the technical sphere. In speaking tech ethics, we speak ethics to power. For others, the outlook is less rosy. Critical observers take tech ethics to just be the latest tool in the same-old corporate toolshed—new rhetoric in service of old interests. Tech ethics is a wolf in sheep's clothing. It is power speaking ethics. But debate about tech ethics concerns more than descriptive analyses of current efforts as such. The capacities of ethical tech as a political movement are also up for scrutiny. What is the political payoff of anyone speaking ethics at all? In this article, the author approaches the question by drawing on a critical history of another moral-turned-political movement. A critical inquiry into the ascendency of human rights, the author suggests, elucidates the multiple functions of moral reasoning and rhetoric in political movements and lends insight into how they may ultimately bear on political efficacy. The 20th century history of human rights gives reason to be suspicious of moral language that is evasive of engaging political and ideological battles. However, it also points to the possibility that long-standing moral ideals may be renewed and refashioned into new claims. Tech ethics may yet play such a role: placing explicitly moral demands on those typically taken to be exempt from moral standards. This demand reaches beyond what the specialized moniker of "tech ethics" suggests.

**Key words:** ethics of technology; political movements; human rights

## 1 Introduction

Every year in late-April and early-May, thousands of tech enthusiasts gather in convention center auditoriums, usually in the San Francisco and Seattle Areas, to watch the industry's biggest names unveil their companies' latest innovations. The sequential late-spring slate of developer conferences—Google I/O, Facebook F8, and the Apple Worldwide Developers Conference—are like a techie's West Coast Met Gala: celebrities don signature outfits and dazzle star-struck fans; press and critics report on who best captured the zeitgeist; at the center

• Lily Hu is with the Applied Mathematics and Philosophy Departments, Harvard University, Cambridge, MA 02138, USA. E-mail: lilyhu@g.harvard.edu.
* To whom correspondence should be addressed.

of the events, the products themselves shine (in the case of most digital devices, literally emitting light) as though beaming at the crowds.

But unlike the Met Gala, which showcases reactions to a theme announced ahead of time, developer conferences also set an agenda to come. Like all things in the tech world, conferences are about the future. We are shown snippets of our soon-to-be world—if the tech companies get their way, that is—a world of fancy wrist devices that "watch" much more than time, of cylindrical home assistants that serve as home stage lighting directors, of phones that unlock at a glance. The futures imagined by Silicon Valley are idealized visions of the human-technology partnership: technologies are our tools. They help us do what we want. The more we develop, the better off we will be in attaining what we want. The questions that follow these assumptions—What do we want? What should we want? To what extent are technologies really mere tools?—are best left to be pondered by others outside the conference center.

It was striking, then, that Microsoft CEO Satya Nadella's opening keynote at his company's 2018 Build conference looked not to the future but to the past—first, to the Industrial Revolution by way of the economist Robert Gordon's anti-techno-optimist book *The Rise and Fall of American Growth*, and then, even more surprisingly, to the mid-century existentialist philosopher of technology Hans Jonas. Nadella recounted Jonas's claim that acting responsibly is to "act so that the effects of your action are compatible with the permanence of genuine life". Transitioning to a discussion of Jonas's 1973 essay "Technology and Responsibility", Nadella continued: "That is something that we need to reflect on, because he was talking about the power of technology being such that it far outstrips our ability to completely control it, especially its impact even on future generations." Nadella then segued from Jonas's words on responsibility to outlining the three core pillars he claims will guide Microsoft's plans for the future: privacy, cybersecurity, and ethical artificial intelligence. Technology might well still be a tool, but it is also something that needs to be controlled and even constrained.

It is notable that this dual challenge of the contemporary tech moment found its way into Microsoft's biggest publicity event of the year by way of philosophy. What are we to make of Nadella's choice—which we now see repeated in the rhetoric of many institutions grappling with the rapidly growing social and political impact of technology—to adopt the language of ethics in response to tech's crisis of legitimacy? It is likely, of course, that the invocation of Jonas and Nadella's entreaty to his fellow engineers to consider "not only what computers can do but what computers should do" was mere publicity stunts aimed at humanizing both Microsoft and the tech industry more widely. But granting, for the sake of argument, his commitment to the matter, we might still ask what good it would serve? What can philosophy and ethics do in the harsh technological realities of our present world?

## 2   A Turn Toward, and Away From, Tech Ethics

While discourse in and about tech continues to be largely ruled by a spirit of optimism, a tempered tone, tales of caution, and attention to societal risks and harms have become features of the industry's narrative, too. Concern about such wide-ranging matters as the role of social media in our civic landscape to a digital economy built on surveillance to automation-driven joblessness has technology companies under greater public scrutiny. Increasing attention to the ways technology makes and remakes society has been followed by demands for oversight, regulation, and more generally for a reassertion of values into the discussion of what we build.

Latter values-focused approaches to technology's implications for society, often centering around the language of moral ideals and principles, appear under the broad umbrella term of "tech ethics", and their rise both inside and outside of the companies themselves has been accompanied by two kinds of responses. On one view, to the extent that philosophy can be useful at all for building a just society, some amount of moral theorizing needs to make its way down from the heavens to affect the practices and politics of our earthly institutions. If we can come to an agreement on the content of certain shared moral ideals (a tough proposition to be sure, but one that is not impossible), public declarations infused with ethical language can give shape to those moral ideals in the real world and give directedness to actions aimed at achieving them. "Ethics" can force a shift in companies' normative orientations, from their own bottom lines to the roles they play in society: the duties and obligations they owe to a broader public. In doing so, thinking morally can help companies avoid potential future missteps and their accompanying social consequences.

Others see tech's adoption of ethical language as serving less honorable purposes. Skeptics not only doubt the extent to which ethics can transform tech's practices but have questioned whether ethics, as deployed, is meant to even serve those purposes at all. Commentators such as Ben Wagner have decried the recent onslaught of company principles, frameworks, and guidelines, and as mere "ethics-washing", aimed at masking deeper structural critiques and preventing regulatory actions[1]. In an industry ridden with scandals and rapidly losing public trust, critics wonder why ethics has been chosen as the rallying cry. For companies that have as of yet been largely unconstrained by state and legal forces, why have ethical frameworks, promises, and principles appeared as safe policies to embrace? The tech ethics cynic sees the easy co-optation of ethics language as, to use a popular phrase in tech, "features, not bugs" of the ideas themselves: vague claims to center human values, consider the social good, and avoid bias and unfairness. In his book *Radical Technologies*, Adam Greenfield characterizes messaging like Nadella's as a "fig leaf of 'ethical development'", allowing corporations to carry on with business as usual, so long as they assure the public of their attention to various ethical considerations[2].

But critics like Wagner and Greenfield are skeptical more generally of the tech ethics program, even when formulated by seemingly independent tech advocacy groups. Their reason for suspicion is clear once you follow the money. Besides those initiatives that are official company efforts, many organizations that forward an ethics-centric agenda—the Association of Computing Machinery's Fairness, Accountability, and Transparency (FAccT) conference, the Good Technology Collective think tank, and the Center for Humane Technology, to name a few—are financially backed by Big Tech. Partnership on AI is a non-profit collaborative effort between several of the most prominent tech firms (Amazon, Apple, Facebook, Google, IBM, and Microsoft). OpenAI, similarly, is sponsored by Amazon and Microsoft, and supported by Elon Musk and Peter Thiel. Even the ivory tower, often caricatured as fetishizing separation from the concerns of reality in favor of high-minded independent inquiry, has sought a slice of the tech ethics pie and the money guaranteed to come along with it. The Stanford Institute for Human-Centered Artificial Intelligence (HAI), an endeavor set on incorporating human values into technology design and policy, is advised by a roster of Silicon Valley and Wall Street executives; meanwhile, the Technical University of Munich Institute for Ethics in Artificial Intelligence has received $7.5 million in funding from Facebook.

As a practical matter, this relationship between "independent" research and corporate cash is par for our neoliberal course. Ours is an era of unprecedented slashes to the public financing of non-profit organizations, as well as some of our most important democratic institutions: elections, libraries, universities, and public service broadcasting. Whereas ethics ventures might have received public support in the past, shrinking budgets in funding agencies such as the offices of the National Endowment for the Humanities have left institutions increasingly reliant on the graces, whims, and self-interest of private philanthropy, both corporate and individual. Thus, given that programs and promises to be ethical need funding, institutions are left with little place to go but to the standard stock of elite private donors. Recent unveilings of colleges and university centers dedicated to the "social good" demonstrate shocking cases of short-term memory loss: Stephen Schwarzman, known in part for allying with Saudi Crown Prince Mohammed Bin Salman, will have his name forever emblazoned on an "ethical" College of Computing at Massachusetts Institute of Technology (MIT) in exchange for a financial setback of $350 million, while Henry Kissinger spoke at the MIT College's inaugural festivities and attended the HAI launch. It is clear why these collaborations are also in the interest of their patrons: for the rich, famous, and morally dubious, paying for tech ethics buys a seat at the table and an opportunity to eclipse the more unsavory parts of one's history. Tech ethics are indulgences; universities gladly sell.

Much of the ethics-washing discourse has well identified the at-best-amoral coffers of tech ethics initiatives, which allow corporations to maintain oversight and even steer the public conversation about their growing power. On this view, Silicon Valley, with its long financial strings, plays the tech ethics marionette; "ethics" is a show, and they know it.

But while an ethics-washing story that centers corporate control over the terms of political conversation captures one important aspect of tech ethics, it underplays another critical feature of the dynamic between ethics ventures and tech companies in today's movement. The author wants to suggest that there is a much deeper dependence than the mere financial one between tech ethics initiatives and the corporations they attempt to keep in check—one that is relatively less explored and lies in the political rationality of tech ethics. This dependency is mutual. Just as Big Tech needs "ethics" on its side to maintain public goodwill, "ethics" ventures need Big Tech for their own legitimacy. It is an uncomfortable fact that however much external advisory boards and universities claim to be "third parties", ethical tech institutions are in fact parasitic on the continual moral failures and disappointments of a hegemonic tech industry. These groups and efforts survive only because Big Tech has chosen to engage the ethics discourse while it has blocked most other political movement-building. Up to now, the tech ethics discourse has only been able to make headway to the extent that corporate power has remained largely intact.

This mutual dependency, however, also suggests that applying the common ethics-washing critique is less straightforward than we might expect. All sorts of tech critiques now appear in the language of ethics for a variety of reasons—some might take on "ethics" as a convenient label that now happens to hold sway with companies; others might masquerade as "ethics" simply to survive in the space; still others might intentionally choose to reinsert "ethics" in our political discourse. Still, the endorsement of ethics by corporate board members and organizing tech workers alike is unexpected and also unsettling. How should we understand such a multifaceted movement that lies at the convergence of so many different political motives and ideologies? Does the mutual dependency between tech ethics efforts and tech corporations expose the minimal political capacities of the movement? Or is it evidence of shifting tides in the public's expectations of corporate behavior? How can we interpret and update the ethics-washing critique in light of the highly varied nature of the tech ethics landscape and of the political moment in which it sits? The author considers these questions through a lens that focuses on the place of moral rhetoric in political movements. What are the political affordances of the tech ethics movement's self-conscious orientation around the language of ethics?

The author wants to note from the outset that her investment in these questions is not that of a disinterested onlooker. As a researcher, the author has worked in the broader tech ethics area. The author has participated in conferences, organized workshops, and even taught classes on the field. This article is equal parts personal and academic interest. On one hand, self-reflection and anxiety about the author's own experiences and relationship to this burgeoning tech ethics space. On the other hand, diagnosis and analysis of what tech ethics does and can do as a trend, a practical strategy, and a field of study. The exercise here is an attempt at scrutinizing a movement and community of which the author is a part—recognizing all the limitations of theorizing without remove.

## 3   Moral Ideals and Political Movements

Moral ideals occupy a delicate position within political programs. The capacity for a moral political campaign to achieve democratic victories is highly contingent on its surrounding political conditions. Interpreting the political capacities of the tech ethics movement requires an analysis of both the material and ideological conditions under which such ideas and activism are able to flourish today. But tech ethics is not the first inspiring political movement to self-consciously center moral ideals. In this article, the author looks to another moral-turned-political human rights, which rose to become the lingua franca of global justice in the latter half of the 20th century, as a frame through which to analyze the contemporary tech ethics moment. The author shows that in the cases of both tech ethics and human rights, there arises a mutual dependency between the movements and their moral ideals on one hand and reigning institutions and their logic on the other. Just as reigning institutions appeal to higher moral ideals to bolster claims of legitimacy, both the present-day tech ethics and the 20th century human rights movements rely on the power of sponsor institutions to ensure their continued political relevance. The tension at the nexus of moral ideal, political practice, and institutional instrumentalization is a central feature of the history of human rights and one that the author argues is crucial to interpreting "tech ethics" as a contemporary phenomenon.

Second, the author reads the tech ethics movements, both the corporate and tech worker movement one, as part of broader projects that look to (re)claim the role of moral reasoning and language in our political sphere. If the author is right, then the stakes of the movement are much greater than the specialized title "tech ethics" suggests. Here again, the history of the rise of human rights has something to offer. If the political demands we make are at-bottom moral claims about living in a just society, what factors influence the fate of these moral arguments? Knowing an answer to this question can help us assess our current ethical movement—has it been irredeemably captured by tech industry elites? Or does it have political potential? In looking to the post-World War II development of human rights, we gain a new perspective on ethics-washing charges and can in turn, better evaluate the opportunities and risks of today's tech ethics efforts.

### 3.1   Human rights: Moral or political?

Human rights, those rights we are entitled to simply by virtue of being human, project an unconditional moral objectivity, justifying their priority over the more contingent facts of our worldly existence: what leaders we might have, what government happens to rule us, and what political system we currently live under. Though human rights are genealogically descendant of a natural rights tradition that reaches back centuries, their rapid international ascendancy in the 1970s spawned a new orientation to global justice that emphasized individual rights separate from those entitled by citizenship. Humans rights claimed higher moral ground than those enshrined by positive law; hence, Sen's description that they are often seen as "parents of law"[3]. By emphasizing rights outside of, and indeed above, governance structures, the modern-day human rights movement did not have to confront perennial challenges of political organization.

Some scholars who study the political circumstances surrounding human rights see a less rosy picture of their international prominence. Moyn's account of the 20th century history of human rights locates their ascendency at a time of exhaustion with ambitious egalitarian visions[4, 5]. Moyn sees such timing as evidence of the compatibility of modern-day human rights activism with whatever dominant ideological order happens to reign. This is perhaps a first hazard of relying on moral language—even moral language that we more or less "all agree with"—as political speech. Far from offering a stable moral lens through which to appraise the well-beings of humans and their rights, the concept of "human rights" has always functioned as a political tool, to be folded into, rather than to destabilize, the reigning geopolitical calculus of those who choose to wield it, be it watchdog NGOs, international political bodies, or nation-state governments.

Consider, as example, the state of US foreign policy before and after Jimmy Carter's famously human rights-centric inaugural address in 1977. In the decades leading up to the "golden era" of human rights in the 1970s, the US amassed a remarkable record of toppling regimes and replacing them with right-wing military dictatorships—most notably in Latin America with Guatemala in 1954, Brazil in 1964, Chile in 1973, and Argentina in 1976. But to believe that foreign policy principles and strategies were fundamentally altered after 1977 is to fail to appreciate the fundamentally politically-embedded and instrumental nature of moral discourse. Moralistic human rights language could also be easily incorporated into pre-existing interstate allegiances and conflicts. "Good" human rights-focused foreign policies became entangled with the more morally-ambiguous ideal of "democracy promotion". Interventions originally justified in the name of the former were frequently later defended by reference to the latter. Human rights rhetoric reached new heights of dark irony in the 1980s when the Reagan Administration embarked on its bloody foreign policy strategy in Central America that left dead hundreds of thousands of civilians, much of which was pursued under the direction of Assistant Secretary of State for Human Rights and Humanitarian Affairs, Elliott Abrams. Abrams repeatedly upheld the human rights record of the right-wing military junta in El Salvador that was responsible for an estimated seventy-thousands civilian deaths during the course of the country's civil war.① He even continued to push for more US aid to the Salvadoran government, explicitly saying, "The purpose of our aid is to permit people who are fighting on our side to use more violence"[10]. The real human rights mission was to protect American-style democracy, and on Abrams' view, the junta were "freedom fighters", so the moral choice was clear.②

Such blatantly self-serving rhetoric remained so much a feature of the Reagan Administration's human rights-centric foreign policy that in 1985, advocacy groups explicitly accused Abrams of developing and articulating a "human rights ideology which complements and justifies Administration policies"[11]. Funnily enough, the same charge has often been levied against the entire realm of human rights practice and politics itself. From their United Nations declaration in 1948 to Carter's human rights inaugural speech to Amnesty International's Nobel Peace Prize in the 1977, human rights have always relied on the approval of the reigning Western political bodies for the legitimacy of its moral force. It is for this reason that the contemporary human rights agenda has retained a largely liberal approach to justice, eschewing the broad egalitarian economic concerns that have been at the center of other notable 20th

---

① On Elliott Abrams' human rights offenses and defenses of the US foreign policy in Central America[6−9].

② The descriptor "freedom fighters" was oft used in the Reagan Administration. President Reagan used the term to refer to anti-Communist insurgents everywhere in his first State of the Union of his second term in 1985. Elliot Abrams adopted the term to refer both to the Contras in Nicaragua and to the insurgents in El Salvador.

century political movements, such as socialism. Reliance on institutional endorsement has thus limited the extent to which human rights can stand independently of the larger animating political ideals of the dominant powers that be, let alone challenge them. As the US's war in Iraq so devastatingly showed, "humanitarian" campaigns have proven compatible with a diverse set of political frameworks and agendas. Without their own positive independent vision for global justice, human rights, even when pursued earnestly as a guide to moral political action, have been continually subordinated to more assertive ideologies—in the case of the US, ideologies of neoconservatism, of imperial expansion, and of global capitalism.

## 3.2  Tech ethics: Moral or political?

If there are lessons to draw from this recent history of human rights for the purpose of understanding tech ethics, this transition from moral theory to institutional political instrumentalization is a good place to start. Just as causes of all sorts have marched under the banner of human rights, so we see the same in conversations about tech ethics: Google's capacity to bring high-quality information to people across the globe becomes a social responsibility to augment its user base. At the WIRED25 Summit, Sundar Pichai portrayed the business decision to expand into global markets as an urgent moral choice, saying, "Today, people either get fake cancer treatments, or they actually get useful information"[12]. Following this line of reasoning—in which Google withholds life-saving information when it fails to service populations—Pichai arrived at the conclusion that Google is in fact ethically "obliged" to consider how it can expand its services to the 1.4 billion people in state-censored China. For Apple's Tim Cook, taking ethics seriously means calling the business model of ad tech what it is: platforms built on exploitation and surveillance. What is the solution to this "data industrial complex"[13]? Ensuring strong protections against personal data extraction via hardware solutions—luxury good devices that feature premium encryption for users. Fortunately for Google and Apple, doing ethics-aligned business is not so hard after all. The business instrumentalization of tech ethics follows the same pattern as that of the state's deployment of human rights rhetoric: enlisted to complement and justify more fundamental strategies that protect political and economic interests.

Ethics-washing critics have called attention this corporate ethics charade, but as the author has suggested, the incorporation of ethical language into business pitches represents only one modality of the tech industry and tech ethics interdependence. While firms might refer to ethics to stave-off greater public scrutiny, the legitimacy of tech ethics as a viable political program also in part depends on the recognition that the effort is awarded by corporations. Ethics-washing critics have much less noted this second type of reliance. Beyond the material consequences that tech ethics groups would face if they issued a genuine challenge to tech power, many mainstream organizations adhere to a theory of change that requires corporate approval—a dependency on institutional heavyweights that, as the author has shown, echoes the logic and geopolitical power relations of the human rights political landscape. In the case of tech ethics, proposals to be ethical can only remain relevant if tech firms choose to endorse them. Ironically then, tech ethics groups become reliant on a certain sweet-spot of crisis: enough to sustain their sense of purpose and urgency, but not too much to spur calls for a rejection of industry elites and a radical revision of our institutions. That is, a deeper ethics-washing charge may in fact cut both ways—corporations use ethics as a diversion that distracts from meeting more substantial political demands; independent tech advocacy groups use ethics to bolster their own relevance as institutional changemakers.

This joint convergence on a weak political program is no surprise to critical scholars of human rights. We see the same with modern advocates of "human rights" who envision a global community of watchdogs for abuse but rarely ask whether the baseline from which urgent crises deviate is itself morally and politically acceptable. Much of human rights appears now so obvious to the Western public that the moral consensus seems to justify a movement that retreats from the political sphere. Of course, the de facto reliance of human rights on dominant political powers and their governing ideologies continues to demonstrate political allegiance, albeit a silent one. The failure of human rights and humanitarian organizations to see their work as politically inflected simply serves to naturalize these dominant political conditions and ideology. Here we notice a superficial but rather telling trend in how tech ethics institutions are named. Many of their names emphasize an alignment with "humanity"— the Stanford Institute for Human-Centered Artificial Intelligence, Center for Human-Compatible AI, All Tech is Human, and the Center for Humane Technology, for example. The obviously-good alignment with "humans" provides groups cover for failing to commit to more specific political projects. Tech ethics proposals have thus existed mainly as the negative of crisis moments: every breach of our privacy and revelation of biased technological design is fodder for ethics watchdogs, which can then prompt (gentle) intervention to correct the aberration.

One such example of how the ethics-washing charges may indict all do-good tech organizations who push the mainstream tech ethics agenda is well illustrated by the activism pursued by the Center for Humane Technology (CHT), an organization which boldly declares on its website that, "Technology is hijacking our minds and society". Its ethical concerns have primarily cashed out in the form of advocating for more conscientious consumption of technology and greater emphasis on the design of applications that allow users to better monitor their digital activity. Tristan Harris, co-founder of CHT and former Google Design Ethicist, sees the roll-out of recent phone use limiting features built into Apple's iOS 12 and Google's Well-being tool as encouraging responses to CHT's "Time Well Spent" campaign against "attention-hacking". Although Harris acknowledges that such apps represent only baby steps in a larger battle, he sees "Time Well Spent" as flipping a switch, triggering a "race to the top for who can care more about the fabric of society"[14]. On Harris's view, then, profit-driven market interactions still operate as the fixed point of institutional behavior with which ethical aspirations must align. A movement that takes this tack can hardly see a role for tech beyond serving as either our harvester or our caretaker.

Interdisciplinary tech ethics-adjacent research ventures like the Fairness, Accountability, and Transparency (FAccT) conference illustrate a more specialized form of mutual dependency in which the tech ethics academic discourse feeds on the shortcomings of Big Tech, while Big Tech bolsters the legitimacy of tech ethics by engaging the ethics discourse. While FAccT as an academic venue does shine light on important normative, technical, and critical inquiry in the fields of computer science, law, and tech-concerned social sciences, it is also likely the case that without the support of large tech companies, the field would not be seen as urgent, impactful, and generally as "hot" of a research area as it is today. Applied research of this sort greatly benefits when large tech companies adopt their proposed "more fair" technical practices or ethical guidelines. FAccT researchers are, generally-speaking, not shouting into the void; quite the opposite, many are in fact meeting at post-conference corporate-sponsored cocktail parties to discuss collaborations across institutions and interests. In environments like these, it is easy for considerations about making a real-world positive impact to become considerations about how companies can be convinced to adopt such reforms. Sadly, this thought process effectively subordinates questions about what justice requires to questions about what companies will likely find agreeable. The scope of the tech ethics discourse can thus be easily hemmed by the naturalization of corporate logic.

It bears noting that the limitations of CHT, FAccT, and similar organizations are not specific to the groups themselves; they have arisen due to a general shift in our political economy, in which the realm of the economic increasingly shapes and even displaces the realm of the political. Just as the ascension of human rights cannot be understood absent the parallel dawn of the neoliberal age, tech ethics efforts must also be situated within this greater political context. Mainstream ethics efforts fill a vacuum of institutional political activism in an area that exists due to a variety of factors: successful political capture, insufficiencies of collective action, a significant structural advantage of Big Tech in the economy, and a genuine uncertainty among both policymakers and the general public about the harms and benefits of technology. This stalemate, along with the chilling effect of financial sponsorship, limits the extent to which ethics groups are willing and able to agitate for more ambitious structural change. What remains is the narrow ability to challenge those impacts and behaviors that organizations view as clearly morally objectionable— hence the language of ensuring "humane" tech solutions—in order to ameliorate those particular ills.

## 4   Inevitability and Contingency in the Politics of Tech Ethics

In pointing out the mutual dependency that underlies much of the mainstream tech ethics movement today, the author does not intend to immediately undercut the critical value and independent integrity of all such ventures of research. The interdependency does, however, bring to the fore important questions about the politics and morality of conducting "ethical" research in an area that is shot through with neoliberal logic. As a researcher who has participated in FAccT, the author finds these conflicting desires exceedingly difficult to negotiate. On one hand, the author has an interest in producing work that speaks with courage and honesty to her normative political commitments; on the other hand, the author has an interest in being accepted by a larger community of scholars, many of whom reside at Big Tech, and the author carries a (faint) hope that tech companies will consider her scholarship in a way that destabilizes unjust yet profitable business practices. On one hand, the author has an interest in scholarship that dispels with the siren song of political neutrality on the most urgent questions of ethical tech; on the other hand, the author has an interest in the community's continual appeal to Big Tech, which allows it to persist as a model of productive discourse between academia and industry. In ideal conditions of practical discourse, perhaps these two visions would be reconcilable. But such a rosy interpretation refuses to confront the necessity of political struggle in a sphere well overdue for it.

The problem, then, is that the success of FAccT's constructive cross-sector exchange cuts both ways. It proves that tech companies' products and processes can be shaped by thoughtful ethics-adjacent research, but it also shows how a symbiotic relationship between tech firms and tech ethics can obscure the fundamental fact of political contestation undergirding the ethical issues at stake. This latter consequence is what the author finds to be most worrying about tech ethics collaborations today. If the story of ethical technology has, up to now, been one of effective assimilation under corporate influence, then we may have to face up to a great potential irony of tech ethics: that pursuing the ethical movement we most need would actually compel us to immediately cast many of our current campaigns into obsolescence. This, in fact, is the ethics-washing charge at its strongest: a claim about the use and norms of tech ethics in a corporatized language-game inimical to our dire need for a genuine redistribution of power.

The strong ethics-washing claim that the political virility of tech ethics language has been doomed from the start shares notable similarities with another influential idea in the scholarship on human rights. The view that an appeal to "ethical technology" undermines the larger political project parallels historian Lynn Hunt's "logic of rights" account of how the inexorable cascade of natural rights philosophizing led to the current wide acceptance of human rights[15]. On Hunt's view, once human rights were born in the 18th century America and France, it was only a matter of time before they would develop into a full-fledged form as they did in the latter half of the 20th century. Whereas Hunt claims that rights language could only lead to an earnest commitment to their undergirding moral principles by the powers that be, the tech ethics cynic sees that ethics language in tech could only lead to a full absorption of such principles into corporate logic. The two perspectives share a belief about inevitability, though their conclusions are diametrically opposed.

Hunt's account, however, sees only continuities in the intellectual history of natural rights stretching to human rights practice today, overlooking broader political context as a force shaping the course of the movement. An "ethics-washing" tale about the inevitability of corporate capture of ethical language in our current moment commits a similar oversight and fails to account for the significance of historical contingency to all intellectual and political movements. In *The Last Utopia*, Samuel Moyn reminds us that a more complete history of human rights is not a tale of ripening—a slow but sure coming into being—but a tale about the breakdown of political alternatives: a national sovereignty mission toward social democracy accompanied by a decolonization project toward a more egalitarian international order. Neither were these projects doomed from the start. The New International Economic Order (NIEO), proposed in 1974, sought to upend the reigning global economic order by calling for redistributive justice and an international body in which every nation-state, regardless of its size or economic power, would be given one vote in matters of global import. Leaders of new nations in the Global South were especially focused on gaining the ability to override the liberal notions of free trade and economic ownership that had been taken as central in matters of international governance. They asserted a "right to development", a collective claim by former colonized people against their colonizers in the North to both take their national fates into their own hands and to a fundamental equality on the international stage. Alas, in the late 1970s, when a political future like that proposed by NIEO was seen no longer as viable, a limp moral individualism dressed up as human rights was left to take up the mantle of global justice. It is important, however, to recognize that the NIEO did not fail of its own accord—politics are always operating beneath the surface. Rather, elite neoliberals who feared the effects that runaway democracy would have on the reach of property and capital undertook a concerted effort to make known the great danger that the NIEO posed to Western civilization. General Agreement on Tariffs and Trade (GATT) chief economist Jan Tumlir scoffed at the audacity of the Global South's attempt to restructure international politics to achieve redistribution, speaking about NIEO with a sneer, "Not only do nations claim to be determining their own future within a global order; now that order itself is to be transcended"[16].

The human rights of the past fifty years must be read in light of a shattered NIEO. The present movement, in contrast, has not sought such bold plans as restructuring international governance. It has largely defended a minimalist conception of global justice, aimed at mitigating the harms of famine, severe poverty, and those reprehensible political leaders who starve, torture, and kill. Transformations of the social, economic, and political order within the nation, along with aspirations of solidarity and egalitarianism at the international level, have been left behind.

## 5   An Outlook on Tech Ethics

The cynic who sees human rights as descendent of an Anglo-American tradition of liberal individualism interprets this to be an unavoidable outcome of moral ideals with inherently impoverished political capacities. But this conclusion is wrong. Alternative histories and origins of human rights can be found, even within the narrow confines of the Enlightenment.③ In the nearby French tradition, human rights were closely tied with egalitarian (though, it should be noted, still largely exclusionary) ideas of democratic self-rule and participatory government. There is no reason that a 20th century human rights practice built on these tenets could never have flourished. Nevertheless, few critics of human rights now hold out hope for this possibility: the thin moral individualist capture of human rights has proven too successful. It is better to pursue other ways forward now.

With this framing in mind, the question for our own movement is simple: has the corporate capture of tech ethics proven too successful as well? Commitments to ethics and social responsibility now sit comfortably within a corporation's standard stock of business-speak, while even the nominally-independent-but-flush-with-corporate-cash tech ethics sphere can only plea for decency. What role now remains for ethical language to play in a movement that wishes for a genuine challenge to corporate power?

Returning to the importance of political and historical contingency to the development of human rights practice is instructive. Even if the global justice affordances of the human rights project have more or less been settled, the same question about the capacity for justice within the tech ethics movement has not been. If Moyn is right that the fate of a movement is as much dictated by the fate of alternatives, then declaring the larger fight for tech ethics as dead on arrival is premature.

First, there is good reason to believe that the happy illusion of consensus enforced by steady economic growth and Third Way politics is coming to an end. The 2000s have already brought startling revelations that the United States (and capitalist liberal democracies more broadly) is neither economically nor politically stable. Reform in the form of technocratic tinkering is no longer the horizon of our mainstream political imaginary. If the ascent of a sufficientarian human rights program could only sit comfortably once egalitarian internationalisms had lost out to a rising neoliberal agenda, then the return of politics means a resurgence of ideological debate—and a potential overthrow of previously reigning conceptions of justice. Perhaps tech corporations will no longer be able to smooth over their crisis of legitimacy with good ethics messaging. The ever-louder ethics-washing chorus itself demonstrates the mounting challenges that corporations face in trying to assert their own visions of ethics. The public is increasingly keeping their eyes on the capture and subversion of our ideals.

Second, the decision to place ethical language at the center of a promise of better behavior is not a risk-free strategy. Companies that choose to do so make the explicit and important concession that their conduct

③ For two recently published books that look elsewhere for origins of human rights[17, 18].

should be held accountable to normative principles and demands from the public. In this renewed era of political mobilization, it is possible that—or one can only hope that—attempts to pervert ethical language for business purposes represent such a clear transgression against the urgency of reevaluating our society's moral commitments that the tech ethics strategy can backfire: companies might find themselves unable to tame demands for ethical tech and instead need to commit to them in earnest.

Whether this will in fact happen will of course be determined by a variety of factors, but there is reason for cautious hope. Tech workers protest against their companies' unethical practices have already been a surprising instance of collective mobilization in direct response to the hypocrisy of tech ethics: for example, Google employees successfully pressured their employer to cancel its multiple bids for government military contracts that would contribute to more effective killing operations[19, 20], as well as to retract a controversial external advisory board on ethics that included a member with anti-LGBTQ, anti-immigrant, and climate denialist views, mere days after it was announced[21]. In their activism, tech workers are increasingly recognizing the role that ethics language has served for companies up to now, but rather than cede the conceptual ground, they have continued to insist on an ethics that, in the words of legal scholar Rashida Richardson, serves as a "moral compass" rather than "just another rubber stamp"—an ethics that refuses to be controlled by tech but instead seeks to holds power within it[22]. As an ideological transformation beyond just a policy one, neoliberalism expunges our social world of ethical commitments to anything other than private economic interests. Rejecting neoliberalism—and preserving democratic politics—requires this exact kind of struggle to reclaim ethics from those who attempt to redefine its meaning and possibility.

One can recognize the historical contingency of ideas and the performativity of words while also still acknowledging that some bannered slogans will be more effective than others in achieving a political vision. Choosing language is a task of political strategizing. But in the end, no words, even the most carefully selected and perfectly suited, predispose a movement to victory. A belief in the inherent lack of certain concepts and the superior natures of others can mask the fact that political efforts are never descriptive; they are always aspirational. Taking ethical principles and language to always be deployed as speech acts should help us to re-interpret our current tech ethics moment as a failure of deeds, not only a failure of words. Moral principles, be that of human rights or of ethical tech, communicate a political end that we insist on. Their assimilation under other logics is dangerous precisely because they risk redefining not only the words themselves but the terms of the larger political project. Their successful capture disciplines our ambitions for a better world.

Asserting a tech ethics that insists on the moral commitments between us and our institutions, each of us to each other, is political work that can never be carried out by corporations and the elite, orchestrating conduct from above, but only by all of us from below, collectively building and agitating for a future that is fully our own.

## Acknowledgment

L. Hu would like to thank Will Holub-Moorman for his invaluable feedback on this article and Ben Green for his hard work and patience in corralling together this selection of articles and making this special issue possible. L. Hu would also like to acknowledge support from the National Science Foundation and the Jain Family Institute.

**Lily Hu** is a PhD candidate in applied mathematics and philosophy at Harvard University. She received the AB degree in mathematics from Harvard College in 2015. She will start as an assistant professor of philosophy at Yale University in 2022. Her work is presently supported by a fellowship from the Edmond J. Safra Center for Ethics and the Jain Family Institute.

# Data Science as Political Action: Grounding Data Science in a Politics of Justice

Ben Green*

**Abstract:** In response to public scrutiny of data-driven algorithms, the field of data science has adopted ethics training and principles. Although ethics can help data scientists reflect on certain normative aspects of their work, such efforts are ill-equipped to generate a data science that avoids social harms and promotes social justice. In this article, I argue that data science must embrace a political orientation. Data scientists must recognize themselves as political actors engaged in normative constructions of society and evaluate their work according to its downstream impacts on people's lives. I first articulate why data scientists must recognize themselves as political actors. In this section, I respond to three arguments that data scientists commonly invoke when challenged to take political positions regarding their work. In confronting these arguments, I describe why attempting to remain apolitical is itself a political stance—a fundamentally conservative one—and why data science's attempts to promote "social good" dangerously rely on unarticulated and incrementalist political assumptions. I then propose a framework for how data science can evolve toward a deliberative and rigorous politics of social justice. I conceptualize the process of developing a politically engaged data science as a sequence of four stages. Pursuing these new approaches will empower data scientists with new methods for thoughtfully and rigorously contributing to social justice.

**Key words:** data science; ethics; politics; social justice; social change; social good; pedagogy

## 1 Introduction

The field of data science has entered a period of reflection and reevaluation.[①] Alongside its rapid growth in both size and stature in recent years, data science has become beset by controversies and scrutiny. Machine learning algorithms that guide decisions in areas such as hiring, healthcare, criminal sentencing, and welfare are often biased, inscrutable, and proprietary[1−6]. Algorithms that drive social media feeds manipulate people's emotions[7], spread misinformation[8], and amplify political extremism[9]. Facilitating these and other algorithms are massive datasets, often gained illicitly or without meaningful consent, that reveal sensitive and intimate information about people[10−13].

Many individuals and organizations responded to these controversies by advocating for a focus on ethics in computing training and practice[14]. Universities have created new courses that train students to consider the ethical implications of computer science[15−18]; one crowdsourced list includes more than 300 such classes[19]. Former US Chief Data Scientist D. J. Patil has argued that data scientists need a code of ethics[20]. The Association for Computing Machinery (ACM), the world's largest educational and scientific computing society, updated its Code of Ethics and Professional Conduct in 2018 for the first time since 1992[21]. The broad motivation behind these efforts is the assumption that, if only data scientists were more attuned to the

● Ben Green is with the Society of Fellows and the Gerald R. Ford School of Public Policy, University of Michigan, Ann Arbor, MI 48109, USA. E-mail: bzgreen@umich.edu.
∗ To whom correspondence should be addressed.

① Throughout this article, "data science" encompasses the use of computational methods (including artificial intelligence and machine learning) to derive patterns from data in order to make predictions about the future. In this sense, a data scientist is anyone who works with data and algorithms in these settings. My particular focus is on the application of data science methods to social and political contexts.

ethical implications of their work, many harms associated with data science could be avoided[14].

Although emphasizing ethics is an important step in data science's development toward greater socially responsible, it is an insufficient response to the broad issues of social justice that are implicated by data science.② As described in the introductory article for this special issue, technology ethics as applied in practice suffers from four significant limitations[14]. First, technology ethics principles are abstract and lack mechanisms to ensure that engineers follow ethical principles. Second, technology ethics has a myopic focus on individual engineers and on technology design, overlooking the structural sources of technological harms. Third, technology ethics is subsumed into corporate logics and practices rather than substantively altering behavior. All told, the rise of technology ethics often reflects a practice dubbed "ethics-washing": tech companies deploying the language of ethics to resist more structural reforms that would curb their power and profits.

Thus, while ethics provides useful frameworks to help data scientists reflect on their practice and the impacts of their work, these approaches are insufficient for generating a data science that avoids social harms and that promotes social justice. The normative responsibilities of data scientists cannot be managed through to a narrow professional ethics that lacks normative weight and supposes that, with some reflection and a commitment to best practices, data scientists will make the "right" decisions that lead to "good" technology. Instead of relying on vague moral principles that obscure the structural drivers of injustice, data scientists must engage in politics: the process of negotiating between competing perspectives, values, and goals.

In other words, we must recognize data science as a form of political action. Data scientists must recognize themselves as political actors engaged in normative constructions of society. In turn, data scientists must evaluate their efforts according to the downstream impacts on people's lives.

By politics and political, I do not refer directly to partisan or electoral debates about specific parties and candidates. Instead, I invoke these terms in a broader sense that transcends activity directly pertaining to the government, its laws, and its representatives. Two aspects of politics are paramount. First, politics is everywhere in the social world. As defined by politics professor Adrian Leftwich, "politics is at the heart of all collective social activity, formal and informal, public and private, in all human groups, institutions, and societies"[23]. Second, politics has a broad reach. Political scientist Harold Lasswell describes politics as "who gets what, when, how"[24]. The "what" here could mean many things: money, goods, status, influence, respect, rights, and so on. Understood in these terms, politics comprises any activities that affect or make claims about the who, what, when, and how in social groups, both small and large.

Data scientists are political actors in that they play an increasingly powerful role in determining the distribution of rights, status, and goods across many social contexts. As data scientists develop tools that inform important social and political decisions—who receives a job offer, what news people see, where police patrols—they shape social outcomes around the world. Data scientists are some of today's most powerful (and obscured) political actors, structuring how institutions conceive of problems and make decisions.

This article will justify and develop the notion of data science as political action. My argument raises two questions: (1) Why must data scientists recognize themselves as political actors? and (2) How can data scientists ground their practice in a politics of social justice? The two primary sections of this article will take up these questions in turn.

My aim is to support data science toward playing a more productive role in promoting equity and social justice. I do not intend to stop data science in its tracks, critique individual practitioners, or discourage data scientists from working on social problems. The path ahead does not require data scientists to abandon their technical expertise, but it does require data scientists to expand their notions of what problems to work on and how to engage with society. This process may involve an uncomfortable period of change. But I am confident that exciting new areas for research and practice will emerge, producing a field that can contribute to a more egalitarian and just society.

---

② In *Black Feminist Thought*, Patricia Hill Collins defines a "social justice project" as "an organized, long-term effort to eliminate oppression and empower individuals and groups within a just society". Oppression, she writes, is "an unjust situation where, systematically and over a long period of time, one group denies another group access to the resources of society"[22].

## 2 Why Must Data Scientists Recognize Themselves as Political Actors?

The first part of this article will attempt to answer this question in the form of a dialogue with a well-intentioned skeptic. I will respond to three arguments that are commonly invoked by data scientists when they are challenged to take political stances regarding their work. These arguments have been expressed in a variety of public and private settings and will be familiar to anyone who has engaged in discussions about the social responsibilities of data scientists.

These are by no means the only arguments proffered in this larger debate, nor do they represent any sort of unified position among data scientists. In practice, computer scientists are "diverse and ambivalent characters"[25] who engage in "nuanced, contextualized, and reflexive practices"[26]. Some computer science subfields (such as CSCW[27]) have long histories of engaging with sociotechnical practices and normative implications, while others (such as the ACM Conference on Fairness, Accountability, and Transparency (FAccT)) are actively developing such approaches. Nonetheless, in my experience, the three positions considered here are the most common and compelling arguments made against a politically oriented data science. Any promotion of a more politically engaged data science must contend with them.

### 2.1 Argument 1: "I am just an engineer"

This first argument represents a common attitude among engineers. In this view, although engineers develop new tools, their work does not determine how a tool will be used. Artifacts are seen as neutral objects that lack any inherent normative character and that can simply be used in good or bad ways. By this logic, engineers bear no responsibility for the impacts of their creations.

It is common for data scientists to argue that the impacts of technology are unknowable. As one computer scientist who faced criticism for developing facial recognition software argued in defense of his work, "Anything can be used for good. Anything can be used for bad"[28]. Similarly, during a 2019 NeurIPS workshop, in which two panelists highlighted the harmful impacts of AI on communities of color, several computer scientists in the audience countered that it is impossible to know what the impacts of research will be or to prevent others from misusing products[29].

By articulating their limited role as neutral researchers, data scientists provide themselves with an excuse to abdicate responsibility for the social and political impacts of their work. When a paper that used neural networks to classify crimes as gang-related was challenged for its potentially harmful effects on minority communities, a senior author on the paper deflected responsibility by arguing, "It's basic research"[30].

Although it is common for engineers to see themselves as separate from politics, many scholars have thoroughly articulated how technology embeds politics and shapes social outcomes. As political theorist Langdon Winner describes, "technological innovations are similar to legislative acts or political foundings that establish a framework for public order that will endure over many generations. For that reason, the same careful attention one would give to the rules, roles, and relationships of politics must also be given to such things as the building of highways, the creation of television networks, and the tailoring of seemingly insignificant features on new machines. The issues that divide or unite people in society are settled not only in the institutions and practices of politics proper, but also, and less obviously, in tangible arrangements of steel and concrete, wires and semiconductors, and nuts and bolts"[31].

Even though technology does not conform to conventional notions of politics, it often shapes society in much the same way as laws, elections, and judicial opinions. In this sense, "the scientific workplace functions as a key site for the production of social and political order"[32]. Thus, as with many other types of scientists, data scientists possess "a source of fresh power that escapes the routine and easy definition of a stated political power"[33].

There are many examples of engineers developing and deploying technologies that, by structuring behavior and shifting power, shape aspects of society. As one example, Winner famously (and controversially[34, 35]) describes how Robert Moses designed the bridges over the parkways on Long Island, New York with low overpasses[31]. Moses purportedly did this to prevent buses (which predominantly carried lower-class and non-white urban residents) from navigating these parkways and accessing the parks to which they led.

Another historical example similarly demonstrates how the design of traffic technologies can have social and political ramifications. As historian Peter Norton describes, when automobiles were introduced onto city streets in the 1920s, they created chaos and conflict in the existing social order[36]. Many cities turned to traffic engineers as "disinterested experts" whose scientific methods could provide a neutral and optimal solution. But the engineers' solution contained unexamined assumptions and values, namely, that "traffic efficiency worked for the benefit of all". As traffic engineers changed the timings of traffic signals to enable cars to flow freely, their so-called solution "helped to redefine streets as motor thoroughfares where pedestrians did not belong". These actions by traffic engineers helped shape the next several decades of automobile-focused urban development in US cities.

Although these particular outcomes could be chalked up to unthoughtful design, any decisions that the traffic engineers made would have had some such impact: determining how to time streetlights requires judgments about what outcomes and whose interests to prioritize. Whatever they and the public may have believed, traffic engineers were never "just" engineers optimizing society "for the benefit of all". Instead, they were engaged in the process—via formulas and signal timings—of defining which street uses should be supported and which should be constrained. The traffic engineers may not have decreed by law that streets were for cars, but their technological intervention assured this outcome by other means.

Data scientists today risk repeating this pattern of designing tools with inherently political characters yet largely overlooking their own agency and responsibility. By imagining an artificially limited role for themselves, engineers create an environment of scientific development that requires few moral or political responsibilities. But this conception of engineering has always been a mirage. Developing any technology contributes to the particular "social contract implied by building that technological system in a particular form"[31].

Of course, we must also resist placing too much responsibility on data scientists. The point is not that, if only they recognized their social impacts, engineers could themselves solve social issues. Technology is at best just one tool among many for addressing complex social problems[37]. Nor should we uncritically accept the social influence that data scientists have. Having unelected and unaccountable technical experts make core decisions about governance away from the public eye imperils essential notions of how a democratic society ought to function. As Science, Technology, and Society (STS) scholar Sheila Jasanoff argues, "The very meaning of democracy increasingly hinges on negotiating the limits of the expert's power in relation to that of the publics served by technology"[38].

Nonetheless, the design and implementation of technology does rely, at some level, on trained practitioners. This raises several questions that animate the rest of this article. What responsibilities should data scientists bear? How must data scientists reconceptualize their scientific and societal roles in light of these responsibilities?

## 2.2 Argument 2: "Our job is not to take political stances"

Data scientists adhering to this second argument likely accept the response to Argument 1 but feel stuck, unsure how to appropriately act as more than "just" an engineer. "Sure, I am developing tools that impact people's lives", they may acknowledge, before asking, "But is not the best thing to just be as neutral as possible?"

Although it is understandable how data scientists come to this position, their desire for neutrality suffers from two important failings. First, neutrality is an unachievable goal, as it is impossible to engage in science or politics without being influenced by one's background, values, and interests. Second, striving to be neutral is not itself a politically neutral position. Instead, it is a fundamentally conservative one.③

An ethos of objectivity has long been prevalent among scientists. Since the nineteenth century, objectivity has evolved into a set of widespread ethical and normative scientific practices. Conducting good science—and being a good scientist—meant suppressing one's own perspective so that it would not contaminate the interpretations of observations[39].

Yet this conception of science was always rife with contradictions and oversights. Knowledge is shaped and bounded by the social contexts that generated it. This insight forms the backbone of standpoint theory, which articulates that "nothing in science can be protected from cultural influence—not its methods, its research technologies, its conceptions of nature's fundamental ordering principles, its other concepts, metaphors,

③ I use conservative here in the sense of maintaining the status quo rather than in relation to any specific political party or movement.

models, narrative structures, or even formal languages"[40]. Although scientific standards of objectivity account for certain kinds of individual subjectivity, they are too narrowly construed: "methods for maximizing objectivism have no way of detecting values, interests, discursive resources, and ways of organizing the production of knowledge that first constitute scientific problems, and then select central concepts, hypotheses to be tested, and research designs"[40].

These processes make the supposedly objective scientific "gaze from nowhere" nothing more than "an illusion"[41]. Every aspect of science is imbued with the characteristics and interests of those who produce it. This does not invalidate every scientific finding as arbitrary, but points to science's contingency and reliance on its practitioners: all research and engineering are developed within particular institutions and cultures and with particular problems and purposes in mind.

Just as it is impossible to conduct science in any truly neutral way, there is no such thing as a neutral (or apolitical) approach to politics. As philosopher Roberto Unger writes, political neutrality is an "illusory and ultimately idolatrous goal" because "no set of practices and institutions can be neutral among conceptions of the good"[42].

Instead of being neutral and apolitical, attempts to be neutral and apolitical embody an implicitly conservative politics. Because neutrality does not mean value-free—it means acquiescence to dominant social and political values, freezing the status quo in place. Neutrality may appear to be apolitical, but that is only because the status quo is taken as a neutral default. Anything that challenges the status quo—which efforts to promote social justice must do by definition—will therefore be seen as political. But efforts for reform are no more political than efforts to resist reform or even the choice simply to not act, both of which preserve existing systems.

Although surely not the intent of every scientist and engineer who strives for neutrality, broad cultural conceptions of science as neutral entrench the perspectives of dominant social groups, who are the only ones entitled to legitimate claims of neutrality. For example, many scholars have noted that neutrality is defined by a masculine perspective, making it impossible for women to be seen as objective or for neutral positions to consider female standpoints[40, 43–45]. The voices of Black women are particularly subjugated as partisan and anecdotal[22]. Because of these perceptions, when people from marginalized groups critique scientific findings, they are cast off as irrational, political, and representing a particular perspective[41]. In contrast, the practices of science and the perspectives of the dominant groups that uphold it are rarely considered to suffer from the same maladies.

Data science exists on this political landscape. Whether articulated by their developers or not, machine learning systems already embed political stances. Overlooking this reality merely allows these political judgments to pass without scrutiny, in turn granting data science systems with more credence and legitimacy than they deserve.

Predictive policing algorithms offer a particularly pointed example of how striving to remain neutral entrenches and legitimize existing political conditions. The issue is not simply that the training data behind predictive policing algorithms are biased due to a history of overenforcement in minority neighborhoods. In addition, our very definitions of crime and how to address it are the product of racist and classist historical processes. Dating back to the eras of slavery and reconstruction, cultural associations of Black men with criminality have justified extensive police forces with broad powers[46]. The War on Drugs, often identified as a significant cause of mass incarceration, emerged out of an explicit agenda by the Nixon administration to target people of color[47].④ Meanwhile, crimes like wage theft are systemically underenforced by police and do not even register as relevant to conversations about predictive policing.⑤

Moreover, predictive policing rests on a model of policing that is itself unjust. Predictive policing software could exist only in a society that deploys vast punitive resources to prevent social disorder, following "broken

---

④  As Nixon's special counsel John Ehrlichman explained years later, "We knew we could not make it illegal to be either against the war or black. But by getting the public to associate the hippies with marijuana and blacks with heroin, and then criminalizing both heavily, we could disrupt those communities. We could arrest their leaders, raid their homes, break up their meetings, and vilify them night after night on the evening news. Did we know we were lying about the drugs? Of course we did."[48]

⑤  Wage theft occurs when employers deny their employees the wages or benefits to which they are legally entitled (e.g., not paying employees for overtime work). Wage theft steals more value than all other kinds of theft (such as burglaries) combined, typically carried out by business owners against low-income workers[49].

windows" tactics. Policing has always been far from neutral: "the basic nature of the law and the police, since its earliest origins, is to be a tool for managing inequality and maintaining the status quo"[50]. The issues with policing are not flaws of training or methods or "bad apple" officers, but are endemic to policing itself[46, 50].

Against this backdrop, choosing to develop predictive policing algorithms is not neutral. Accepting common definitions of crime and how to address it may seem to allow data scientists to remove themselves from politics, but instead upholds historical politics of social hierarchy.

Although predictive policing represents a notably salient example of how data science cannot be neutral, the same could be said of all applied data science. Biased data are certainly one piece of the story, but so are existing social and political conditions, definitions and classifications of social problems, and the set of institutions that respond to those problems. None of these factors are neutral and removed from politics. And while data scientists are of course not responsible for creating these aspects of society, they are responsible for choosing how to interact with them. Neutrality in the face of injustice only reinforces that injustice. When engaging with aspects of the world steeped in history and politics, in other words, it is impossible for data scientists to not take political stances.

I do not mean to suggest that every data scientist should share a singular political vision—that would be wildly unrealistic. It is precisely because the field (and world) hosts a diversity of normative perspectives that we must surface political debates and recognize the role they play in shaping data science practice. Nor is my argument meant to suggest that articulating one's political commitments is a simple task. Normative ideals can be complex and conflicting, and one's own principles can evolve over time. Data scientists need not have precise answers about every political question. However, they must act in light of articulated principles and grapple with the uncertainty that surrounds these ideals.

## 2.3　Argument 3: "We should not let the perfect be the enemy of the good"

Following the responses to Arguments 1 and 2, data scientists asserting this third argument likely acknowledge that their creations will unavoidably have social impacts and that neutrality is not possible. Yet still holding out against a thorough political engagement, they fall back on a seemingly pragmatic position: because data science tools can improve society in incremental but important ways, we should support their development rather than argue about what a perfect solution might be.

Despite being the most sophisticated of the three arguments, this position suffers from several underdeveloped principles. First, data science lacks robust theories regarding what "perfect" and "good" actually entail. As a result, the field typically adopts a superficial approach to reform that involves making vague (almost tautological) claims about what social conditions are desirable. Second, this argument fails to articulate how to evaluate or navigate the relationship between the perfect and the good. Efforts to promote social good thus tend to take for granted that technology-centric incremental reform is an appropriate strategy for social progress. Yet, considered from a perspective of substantive equality and anti-oppression, many data science efforts to do good are not, in fact, consistently doing good.

### 2.3.1　Data science lacks a thorough definition of "social good"

Across the broad world of data science, from academic institutes to conferences to companies to volunteer organizations, "social good" (or just "good") has become a popular term. Numerous universities across the United States and Europe have hosted the Data Science for Social Good Summer Fellowship.⑥ Several major computer science conferences have hosted AI for Social Good workshops,⑦ and in 2014 the theme of the entire ACM SIGKDD Conference on Knowledge Discovery and Data Mining (KDD) was "Data Mining for Social Good".⑧ Since 2014, the company Bloomberg has hosted an annual Data for Good Exchange.⑨ The non-profit Delta Analytics strives to promote "Data-driven solutions for social good".⑩

While this energy to do good among the data science community is both commendable and exciting, the field has not developed (nor even much debated) any working definitions of the term "social good" to guide its efforts. Instead, the field seems to operate on a "know it when

---

⑥　http://www.dssgfellowship.org
⑦　https://aiforsocialgood.github.io/
⑧　https://www.kdd.org/kdd2014/
⑨　https://www.bloomberg.com/company/d4gx/
⑩　http://www.deltanalytics.org

you see it" approach, relying on rough proxies such as crime = bad, poverty = bad, and so on. The term's lack of precision prompted one of Delta Analytics' founders to write that "'data for good' has become an arbitrary term to the detriment of the goals of the movement"[51]. The notable exception is Mechanism Design for Social Good (MD4SG), which articulates a clear research agenda "to improve access to opportunity, especially for communities of individuals for whom opportunities have historically been limited"[52].

In fact, the term "social good" lacks a thorough definition even beyond the realm of data science. It is not defined in dictionaries like Merriam-Webster, the Oxford English Dictionary, and Dictionary.com, nor does it have a page on Wikipedia.① To find a definition one must look to the financial education website Investopedia, which defines social good as "something that benefits the largest number of people in the largest possible way, such as clean air, clean water, healthcare, and literacy"[54]. There is, of course, extensive literature (spanning philosophy, STS, and other fields) that considers what is socially desirable, yet data science efforts to promote "social good" rarely reference this literature.

This lack of definition leads to "data science for social good" projects that span a wide range of conflicting political orientations. For example, some work under the "social good" umbrella is explicitly developed to enhance police accountability and promote non-punitive alternatives to incarceration[55, 56]. In contrast, other work under the "social good" label aims to enhance police operations. One such paper aimed to classify gang crimes in Los Angeles[30, 57]. This project involved taking for granted the legitimacy of the Los Angeles Police Department's gang data—a notoriously biased type of data[58] from a police department that has a long history of abusing minorities in the name of gang suppression[50]. That such politically disparate and conflicting work could be similarly characterized as "social good" should prompt a reconsideration of the core terms and principles. When the term encompasses everything, it means nothing.

The point is not that there exists a single optimal definition of "social good", nor that every data scientist should agree on one set of principles. Instead, there is a multiplicity of perspectives that must be openly acknowledged to surface debates about what "good" actually entails. Currently, however, the field lacks the language and perspective to sufficiently evaluate and debate differing visions of what is "good". By framing their notions of "good" in such vague and undefined terms, data scientists get to have their cake and eat it too: they can receive praise and publications based on broad claims about solving social challenges, while avoiding substantive engagement with social and political impacts.

Most dangerously, data science's vague framing of social good allows those already in power to present their normative judgments about what is "good" as neutral facts that are difficult to challenge. As discussed in Section 2.2, neutrality is an impossible goal and attempts to be neutral tend to reinforce the status quo. Thus, if the field does not openly debate definitions of "perfect" and "good", the assumptions and values of dominant groups will tend to win out. Projects that purport to enhance social good but fail to reflexively engage with the political context are likely to reproduce the exact forms of social oppression that many working towards "social good" seek to dismantle.②

### 2.3.2 Pursuing an incremental "good" can reinforce oppression

Even if data scientists acknowledge that "social good" is often poorly defined, they may still adhere to the argument that "we should not let the perfect be the enemy of the good". "After all", they might say, "is not some solution, however imperfect, better than nothing?" As one paper asserts, "we should not delay solutions over concerns of optimal" outcomes[60].

At this point the second failure of Argument 3 becomes clear: it tells us nothing about the relationship between the perfect and the good. Data science has thus far not developed any rigorous methodology for considering the relationship between algorithmic interventions and social impacts. Although data scientists generally acknowledge that data science cannot provide perfect solutions to social problems, the field typically takes for granted that incremental reforms using data science contribute to the "social good". On this logic, we should applaud any attempts to alleviate issues such as crime, poverty, and discrimination. Meanwhile, because "the perfect" represents an

---

① Searching Wikipedia for "social good" automatically redirects to the page for "common good", a term similarly undefined in data science parlance[53].

② Reflexivity refers to the practice of treating one's own scientific inquiry as a subject of analysis[59].

unrealizable utopia we should not waste time and energy debating the ideal solution.

Although efforts to promote "social good" using data science can be productive,[33] pursuing such applications without a rigorous theory of social change can lead to harmful consequences. A reform that seems desirable from a narrow perspective focused on immediate improvements can be undesirable from a broader perspective focused on long-term, structural reforms. Understood in these terms, the dichotomy between the idealized "perfect" and the incremental "good" is a false one: articulating visions of an ideal society is an essential step for developing and evaluating incremental reforms. In order to rigorously conceive of and compare potential incremental reforms, we must first debate and refine our conceptions of the society we want to create; following those ideals, we can then evaluate whether potential incremental reforms push society in the desired direction. Because there is a multiplicity of imagined "perfects", which in turn suggest an even larger multiplicity of incremental "goods", reforms must be evaluated based on what type of society they promote in both the short and long term. In other words, rather than treating any incremental reform as desirable, data scientists must recognize that different incremental reforms can push society down drastically different paths.

When attempting to achieve reform, an essential task is to evaluate the relationship between incremental changes and long-term agendas for a more just society. As social philosopher André Gorz proposes, we must distinguish between "reformist reforms" and "non-reformist reforms"[61]. Gorz explains, "A reformist reform is one which subordinates its objectives to the criteria of rationality and practicability of a given system and policy." In contrast, a non-reformist reform "is conceived not in terms of what is possible within the framework of a given system and administration, but in view of what should be made possible in terms of human needs and demands".

Reformist and non-reformist reforms are both categories of incremental reform, but they are conceived through distinct processes. Reformist reformers start within existing systems, looking for ways to improve them. In contrast, non-reformist reformers start beyond existing systems, looking for ways to achieve emancipatory social conditions. Because of the distinct ways that these two types of reforms are conceived, the pursuit of one versus the other can lead to widely divergent social and political outcomes.

The solutions proposed by data scientists are almost entirely reformist reforms. The standard logic of data science—grounded in accuracy and efficiency—tends toward accepting and working within the parameters of existing systems. Data science interventions are therefore typically proposed to improve the performance of a system rather than to substantively alter it. And while these types of reforms have value under certain conditions, such an ethos of reformist reforms is unequipped to identify and pursue the larger changes that are necessary across many institutions. This approach may even serve to entrench and legitimize the status quo. From the standpoint of existing systems, it is impossible to imagine alternative ways of structuring society—when reform is conceived in this way, "only the most narrow parameters of change are possible and allowable"[62].

In this sense, data science's dominant strategy of pursuing a reformist, incremental good resembles a greedy algorithm: at every point in time, the strategy is to make immediate improvements in the local vicinity of the status quo. Although a greedy strategy can be useful for simple problems, it is unreliable in complex search spaces: we may quickly find a local maximum but will never reach a further-afield terrain of far better solutions. Moves that are immediately beneficial can be counterproductive for finding the global optimum. Similarly, although reformist reforms can lead to certain improvements, a strategy limited to reformist reforms cannot guide robust responses to complex political problems. Reforms that appear desirable within the narrow scope of a reformist strategy can be counterproductive for achieving structural reforms. Even though the optimal political solution is rarely achievable (and is often subject to significant debate), it is necessary to fully characterize the space of possible reforms and to evaluate how reliably different approaches can generate more egalitarian outcomes.

The US criminal justice system, a domain where data scientists are increasingly striving to do good, exemplifies the limits of a reformist mindset. Because criminal justice reform can be "superficial and deceptive"[63], it is necessary to couch reform efforts within a broader vision of long-term, non-reformist change. This is the approach taken by the movement for police and prison abolition. Notably, prison abolitionists object to reforms that "render criminal law administration more humane, but fail to substitute alternative institutions or approaches to realize social order maintenance goals"[64]. Instead, abolitionists pursue only reforms that reduce or replace carceral responses to social disorder.

In contrast with this abolitionist ethos, most data science efforts to contribute "good" are grounded in the existing practices of the criminal justice system. A notable example is pretrial risk assessments. Even if they lead to incremental improvements, these tools legitimize policies that drive racial injustice and mass incarceration[65]. Meanwhile, an entirely separate incremental reform—an abolitionist and non-reformist (and non-technological) one—is possible: ending cash bail and pretrial detention. Recent surveys show public support for such reforms[66, 67].

Adopting pretrial risk assessments and abolishing pretrial detention appear to respond to the same problems, suggesting that these two reforms are aligned. However, these reforms derive from conflicting visions of the "perfect". Reformers supporting risk assessments accept pretrial detention as part of criminal justice system, aiming merely to improve the means by which people are selected for pretrial detention. Meanwhile, reformers aiming to abolishing pretrial detention reject pretrial detention, aiming to abolish the practice altogether. In other words, the debate about risk assessments hinges on political questions about how the criminal justice system should be structured. It is only by articulating our imagined perfects that we can even recognize the underlying tension between these two incremental reforms, let alone properly debate which one to pursue.

The point is not that data science is incapable of improving society. However, data science interventions must be evaluated against alternative reforms as just one of many options, rather than compared merely against the status quo as the only possible reform. There should not a default presumption that machine learning provides an appropriate reform for every problem.

In sum, attempts by data scientists to avoid politics overlook technology's social impacts, privilege the status quo, and narrow the range of possible reforms. The field of data science will be unable to meaningfully advance social justice without accepting itself as political. The question that remains is how it can do so.

## 3 How Can Data Scientists Ground Their Practice in Politics?

The first part of this article argued that data scientists must recognize themselves as political actors. Yet several questions remain: What would it look like for data science to be explicitly grounded in a politics of social justice? How might the field evolve toward this end?

I conceptualize the process of incorporating politics into data science as following four stages, with reforms at both the individual and the institutional/cultural levels. Stage 1 (Interest) involves data scientists becoming interested in working directly on addressing social issues. In Stage 2 (Reflection), the data scientists involved in that work come to recognize the politics that underlie these issues and their attempts to address them.⁴⁴ This leads to Stage 3 (Applications), in which data scientists direct the methods at their disposal toward new problems. Finally, Stage 4 (Practice) involves the long-term project of developing new methods and structures that orient data science around a politics of social justice.

I discuss each stage in more detail below. While not every person or project will follow this precise trajectory, it presents a possible path for data scientists to incorporate politics into their practice. In fact, many data scientists already are following some version of these stages toward a politically informed data science.

### 3.1 Stage 1: Interest

The first step toward infusing a deliberate politics into data science is for data scientists to orient their work around addressing social issues. Such efforts are already well underway, from "data for good" programs to civic technology groups to the growing numbers of data scientists working in governments and non-profits. Although they may not have an articulated vision of "social good", many data scientists are eager to apply

⁴⁴ Some might argue that the order of Stages 1 and 2 should be reversed: data scientists should reflect first, then act to address social issues. This would be the most responsible approach and is the practice that data scientists should follow in the long term. In my experience, however, data scientists' engagements with politics tend to begin with an interest in addressing social challenges, which then leads to reflection on the politics of data science. New pedagogical approaches could merge these two stages. For instance, a "public interest tech" program could integrate reflection on the political nature of data science into its efforts to apply data science in practice.

their work to pressing societal challenges.

However, relative to the excitement around such work, there is a dearth of opportunities for data scientists to apply their skills to an articulated vision of social benefit. Many academic departments and conferences tend not to consider such work to be valid research, companies can find more profit elsewhere, and governments and non-profits have few internal data science roles. Thus, many data scientists who want to do socially impactful work often settle for more traditional research or jobs, in which technical contributions and profit provide the primary imperatives.

Data science programs should work towards a model of "public interest technology" that trains data scientists to address social issues. This involves not simply adopting this label, but also providing methods, pathways, and a broader culture of support for data scientists to improve society. For example, data science programs should develop clinics where students provide technical and policy assistance to "clients" such as activists and government agencies. Programs should also provide funding and guidance for students to find internships and jobs focused on social impact.⑤

It is essential that "social good" and "public interest tech" programs prioritize social and political reforms over deploying technology. The driving goal should be to positively impact society rather than to develop sophisticated tools. This requires an attitude of agnosticism: "approaching algorithms instrumentally, recognizing them as just one type of intervention, one that cannot provide the solution to every problem"[68]. The more that data scientists work directly with governments, communities, and service providers (rather than on abstract technology problems), the more thoroughly they will come to see technology as an imperfect means rather than as an end in itself. Without this technology-agnostic focus on social impacts, efforts to apply data science to social problems will reproduce the issues described in Section 2.3 and will prevent progression to the following stages.

### 3.2   Stage 2: Reflection

As they work on data science for social good projects, data scientists will encounter the political nature of both the issues at hand and their own efforts to address these issues. To the extent that they maintain an open-minded and critical approach grounded in impact, data scientists will begin to reflect on political questions.

We have seen this process play out most clearly with respect to algorithmic bias and fairness. Where just a few years ago it was common to hear claims that data represents "facts" and that algorithms are "objective"[69, 70], today it is widely acknowledged within data science that data contains biases and that algorithms can discriminate. In addition to the annual ACM Conference on Fairness, Accountability, and Transparency (FAccT), there have been numerous workshops dedicated to these issues at major computer science conferences[71]. Moreover, there is also an emerging literature that articulates the limitations and politics of common approaches to studying and promoting algorithmic fairness[72−74].

Over time, data scientists must expand this critical and reflexive lens to increasingly interrogate how all aspects of their work are political. For example, returning to the discussion of predictive policing from Section 2.2, it is not sufficient to develop algorithms just with a recognition that crime data are biased. It is necessary to also recognize that our definitions of crime, the set of institutions that are tasked with responding to it, and the interventions that those institutions provide are all the result of historical political processes laden with discrimination.

Reflection of this sort is propelled by approaching research with an open mind and honoring the expertise of other disciplines, policymakers, and affected communities. Such reflection will be particularly enhanced by fluency in fields such as STS and critical algorithm studies. Exposure to these fields should become central to data science training programs, particularly those with an emphasis on applications of data science for social good. For data scientists hoping to improve society, familiarity with STS and related fields is just as essential as knowledge of databases and statistics.

### 3.3   Stage 3: Applications

In the short term, the insights provided in Stage 2 are not likely to shake the fundamental structures and practices of data science. Instead, these insights will empower data scientists to seek new applications for how existing data science methods can address injustice and shift power.

---

⑤ See e.g., a list of job boards and other resources that I have compiled: https://www.benzevgreen.com/jobs/.

These effects will demonstrate how incorporating a political perspective into data science produces new directions for research and applications rather than a dead end.

Several frameworks can guide data scientists in these efforts. For example, André Gorz's schema of non-reformist reforms and the framework of prison abolition provide conceptual tools for moving beyond the false dichotomy between incremental and radical reform[61, 64]. The notion of "critical design" embodies a similar approach: in contrast to "affirmative design, which "reinforces how things are now", "critical design provides a critique of how things are now through designs that embody alternative social, cultural, technical, or economic values"[75]. A related framework is "anti-oppressive design", which provides "a guide for how best to expend resources, be it the choice of a research topic, the focus of a new social enterprise, or the selection of clients and projects, rather than relying on vague intentions or received wisdom about what constitutes good"[76].

At each stage of the research and design process, data scientists should evaluate their efforts according to these frameworks: Should the design of this algorithm be affirmative or critical? Would the implementation of this model represent a reformist or non-reformist reform? Would empowering our project partner with this system challenge or entrench oppression? Such analyses can help data scientists interrogate their notions of "good" to engage in non-reformist, critical, and anti-oppressive data science. These approaches can also help data scientists recognize situations in which non-technological reforms are more desirable than technological ones[37, 77].

This ethos of pursuing different, politically motivated data science applications can inform work in areas such as policing. One dimension of this shift involves a critical and anti-oppressive approach to selecting project partners. For example, some researchers explicitly articulate an intention to work with community groups and social service providers rather than with law enforcement, recognizing that the latter tend to contribute to structural oppression[55, 78, 79]. Another dimension of this shift involves orienting the analytic gaze away from individuals and towards institutions. One example of this work used machine learning to predict which police officers will be involved in adverse events such as racial profiling or inappropriate use of force[56]. Others have used new algorithmic methods to find evidence of racial bias in police behavior[80, 81].

Although Stage 3 represents a significant evolution of data science toward politics, it suffers from three notable shortcomings. First, it is possible to operate in Stage 3 without ever articulating an explicit politics. Although not raising a project's political motivations may enable some projects to pass without scrutiny, it does little to provide language or direction for other data scientists. The field will not evolve if political debates remain shrouded. Moreover, only relatively minor reforms could be successfully promoted in this covert manner: more significant reforms will likely be challenged and will advance only if they can be explicitly defended.

Second, existing data science methods have a limited ability to promote social justice. Because of data science's adherence to mathematical formalism, current methods are incapable of rigorously representing and reasoning about social contexts and political impacts[68]. Thus, even well-intentioned and seemingly well-designed data science tools can promote injustice[74].

Third, merely directing data science toward new applications remains fundamentally undemocratic: it allows data scientists to shape society without deliberation or accountability. In this frame, a cadre of data scientists—no matter their intentions or actions—retain an outsized power to shape institutions and decision-making processes. Even when their actions are grounded in anti-oppressive ideals, the efforts of data scientists can serve coercive functions if they are not grounded in the needs and desires of the communities supposedly being served. In order to promote long-term structural change and social justice, larger shifts in data science practice are necessary.

### 3.4 Stage 4: Practice

The final stage is to develop new modes for what it means to practice data science. Achieving changes along these lines requires developing new epistemologies, methodologies, and cultures for data science. While the path ahead remains somewhat speculative, several broad directions are clear.

#### 3.4.1 Participatory data science

Data scientists must abandon their desire for a removed objectivity in favor of participation and deliberation among diverse perspectives. STS scholar Donna

Haraway argues for a new approach centered on "situated knowledges": she articulates the need "for a doctrine and practice of objectivity that privileges contestation and deconstruction", one that recognizes that every claim emerges from the perspective of a particular person or group of people[41]. Following this logic, the "neutral" data scientist who attempts to minimize position-taking must be replaced by a data science of situated values—a "participatory counterculture of data science"[82]. This perspective highlights the importance of groups such as Black in AI,⑯ LatinX in AI,⑰ Queer in AI,⑱ and Women in Machine Learning,⑲ all of which work to increase the presence of underrepresented groups in the field of artificial intelligence. Given that data science is influenced by practitioners' perceptions of problems and of how to address them, it is essential to encourage greater diversity in data science[83].

Complementing this participatory approach is for data science to focus more directly on "designing with" rather than "designing for" affected communities and social movements. Data scientists must develop procedures for incorporating a multitude of public voices into their work. When engineers privilege their own perspectives and fail to consider the multiplicity of needs and values across society, they tend to erase and subjugate those who are already marginalized[84−90]. To avoid participating in these oppressive (even if inadvertent) acts, data scientists must center affected communities in their work. One approach toward this end is the principle of "Nothing about us without us", which has been invoked in numerous social movements (in particular, among disability rights activists in the 1990s) to signify that no policies should be developed without direct participation from the people most directly affected by those policies[91]. The Design Justice Network articulates a powerful enactment of these values, with its commitments to "center the voices of those who are directly impacted" and to "look for what is already working at the community level"[92].

This type of approach represents a notable departure from traditional data science practice and values—efficiency and convenience—toward democracy and empowerment. A great deal of work in recent years has exemplified this approach[79, 93−100]. Mechanisms for participatory design and decision making—such as charrettes, participatory budgeting, and co-production—present further models of designing with communities. Any participatory practices should entail not just the design of an algorithm, but also broader questions such as whether an algorithm should be developed in the first place and how it should be used. Additionally, an essential component of developing a more democratic data science is to bring data scientists, technology companies, and governments within the ambit of democratic oversight and accountability[101].

### 3.4.2 New methods and cultures

Adapting data science to a political orientation and to participatory practices will require new methods. Broadly speaking, data science must move toward a "critical technical practice" that rejects "the false precision of mathematical formalism" to engage with the political world in its full complexity and ambiguity[102]. It is necessary to expand the bounds of algorithmic reasoning, shifting from the dominant method of "algorithmic formalism" to the alternative method of "algorithmic realism" that better accounts for the realities of social life and the impacts of algorithmic interventions[68].

As a central component of this evolution, the field should change its internal structures to incentivize greater attention to the implementation and impacts of data science. To embrace justice and tackle the most pressing social issues related to algorithms, data science must take a more expansive approach to research contributions that looks for more than technical contributions. Actually improving people's lives with data science requires far more than just developing a technical tool—it also requires thoughtfully adapting data science methods to the needs of a particular organization or community[37]. If data scientists are to contribute to improving society, they need a more rigorous methodology for ensuring that data science tools produce beneficial impacts when implemented in real-world contexts. New workshops, conferences, and journals will be essential mechanisms for fostering novel methods that blend technical and nontechnical approaches.

Along these lines, data scientists must also adopt a reflexive political standpoint that grounds their efforts in

---

⑯ https://blackinai.github.io/
⑰ http://www.latinxinai.org/
⑱ https://sites.google.com/view/queer-in-ai/
⑲ https://wimlworkshop.org

rigorous evaluations of downstream social and political consequences. What ultimately matters is not how an algorithm performs in the abstract, but what impacts an algorithm has when introduced into complex sociopolitical environments. Data scientists cannot be expected to perfectly predict the impacts of their work—the entanglements between technology and society are far too complex. However, through collaborations with communities and with scholars from other fields, well-grounded analyses are possible. Just as data scientists would demand rigor in claims that one algorithm is superior to another, they should also demand rigor in claims that a technology will have any particular impacts. Toward this end, one necessary direction for future research is to develop interdisciplinary frameworks that will help data scientists consider the downstream impacts of their interventions. This requires being mindful of the various forms of "indeterminacy" that may lead an algorithm to generate different impacts than its developers expect[68].

As one example of a reform that emphasizes impacts as a central concern, in 2018 the ACM Future of Computing Academy proposed that peer reviewers should consider the potential negative implications of submitted work and that conducting "anti-social research" should factor negatively into promotion and tenure cases[103]. Just two years later, the Neural Information Processing Systems Conference (NeurIPS)—one of the world's top AI conferences—announced that every paper at the 2020 conference must include a "broader impact" section that discusses the positive and negative social consequences of the research[104].

### 3.4.3 Engaging with the broader political context
Of course, shifts in data science practice do not occur in a vacuum. Shifts in data science practice require broader structural reforms that contribute to a more just society. As historian Elizabeth Fee notes, "we can expect a sexist society to develop a sexist science; equally, we can expect a feminist society to develop a feminist science"[105]. Similarly, we can expect a militarized society of economic inequality to produce a militarized and unequal data science[106, 107].

Data scientists committed to social justice must work toward more structural reforms against the harms of digital technologies. For instance, building solidarity and power among workers can shift the development of data science away from the most harmful applications. In recent years, tech workers have organized against their companies' partnerships with the United States Departments of Defense and Homeland Security. Rather than perceiving themselves as "just an engineer", these technologists recognize their position within larger sociotechnical systems, recognize the connection between their work and its social ramifications, and hold themselves (and their companies) accountable for these impacts. Building on this movement, thousands of computer science students from more than a dozen US universities pledged in 2019 that they will not work for Palantir due to its partnerships with Immigration and Customs Enforcement (ICE)[108]. Data scientists should also provide support for communities and activists organizing in opposition to oppressive algorithms.

Data scientists alone cannot be held responsible for promoting social and political progress. They are just one set of actors among many. The task of data scientists is not to eradicate social challenges on their own, but to act as thoughtful and productive partners in broad coalitions and social movements striving for a more just society.

## 4 Conclusion

The field of data science must abandon its self-conception of being neutral to recognize how, despite not being engaged in what is typically seen as political activity, data science logics, methods, and technologies shape society. Restructuring the values and practices of data science around a political vision of social justice will not be easy or immediate, but it is necessary. Given the political stakes of algorithms, it is not enough to have good intentions—data scientists must ground their efforts in clear political commitments and rigorous evaluations of the consequences.

As a form of political action, data science can no longer be separated from broader analyses of social structures, public policies, and social movements. Instead, the field must debate what impacts are desirable and how to promote those outcomes—thus prompting rigorous evaluations of the issues at hand and openness to the possibility of non-technological alternatives. Such deliberation needs to occur not just among data scientists, but also with scholars from other fields, policymakers, and communities affected by data science systems.

Recognizing data science as a form of political action

  

will empower and enlighten data scientists with new frameworks to improve society. By deliberating about political goals and strategies and by developing new methods and norms, data scientists can more rigorously contribute to social justice.

## Acknowledgment

B. Green is grateful to the Berkman Klein Center Ethical Tech Working Group for fostering his thinking on matters of technology, ethics, and politics. B. Green also thanks Catherine D'Ignazio, Anna Lauren Hoffmann, Lily Hu, Momin Malik, Dan McQuillan, Luke Stark, Salomé Viljoen, and the reviewers for providing helpful discussions and suggestions.

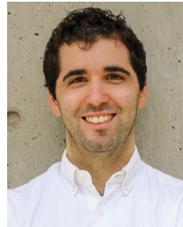

**Ben Green** is a postdoctoral scholar in the Society of Fellows and an assistant professor in the Gerald R. Ford School of Public Policy, University of Michigan . He received the PhD degree in applied math (with a secondary field in STS) from Harvard University and the BS degree in mathematics & physics from Yale College in 2020 and 2014, respectively.

# From Ethics Washing to Ethics Bashing: A Moral Philosophy View on Tech Ethics

Elettra Bietti*

**Abstract:** Weaponized in support of deregulation and self-regulation, "ethics" is increasingly identified with technology companies' self-regulatory efforts and with shallow appearances of ethical behavior. So-called "ethics washing" by tech companies is on the rise, prompting criticism and scrutiny from scholars and the tech community. The author defines "ethics bashing" as the parallel tendency to trivialize ethics and moral philosophy. Underlying these two attitudes are a few misunderstandings: (1) philosophy is understood in opposition and as alternative to law, political representation, and social organizing; (2) philosophy and "ethics" are perceived as formalistic, vulnerable to instrumentalization, and ontologically flawed; and (3) moral reasoning is portrayed as mere "ivory tower" intellectualization of complex problems that need to be dealt with through other methodologies. This article argues that the rhetoric of ethics and morality should not be reductively instrumentalized, either by the industry in the form of "ethics washing", or by scholars and policy-makers in the form of "ethics bashing". Grappling with the role of philosophy and ethics requires moving beyond simplification and seeing ethics as a mode of inquiry that facilitates the evaluation of competing tech policy strategies. We must resist reducing moral philosophy's role and instead must celebrate its special worth as a mode of knowledge-seeking and inquiry. Far from mandating self-regulation, moral philosophy facilitates the scrutiny of various modes of regulation, situating them in legal, political, and economic contexts. Moral philosophy indeed can explainin the relationship between technology and other worthy goals and can situate technology within the human, the social, and the political.

**Key words:** ethics; technology; artificial intelligence; big tech; ethics washing; law; regulation; moral philosophy; political philosophy

## 1 Introduction

On May 26th, 2019, Google announced that it would put in place an external advisory council for the responsible development of AI, the Advanced Technology External Advisory Council (ATEAC).[1] Following a petition signed by 2556 Google workers demanding the removal of one of the body's board members, anti-LGBT advocate Kay Coles James, the advisory body was

• Elettra Bietti is with the Harvard Law School, Harvard University, Cambridge, MA 02138, USA. E-mail: ebietti@sjd.law.harvard.edu.
∗ To whom correspondence should be addressed.

withdrawn approximately one week after its announcement.[2, 3] On December 3rd, 2020, Timnit Gebru, a Google AI researcher, was abruptly fired for sending an internal letter to Google employees which discussed her superiors' questionable resistance to the publication of a research paper she co-authored.[4−6] Her Tweet produced a wave of reactions in academia and beyond, with many Google employees subsequently quitting.[7] These episodes and the backlash they produced provide a salient illustration of the tensions around the corporate use of "ethics" language in technology circles. Corporate and policy instrumentalization and misuse of such language in technology policy have taken two forms.

On one hand, the term has been used by companies as an acceptable façade that justifies deregulation, self-regulation or market driven governance, and is increasingly identified with technology companies' self-interested adoption of appearances of ethical behavior. Such growing instrumentalization of ethical language by tech companies has been called "ethics washing".[8] Beyond AI ethics councils or AI Ethics researchers, the ethics washing critique extends to corporate practices that have tended to co-opt the value of ethical work: the hiring of in-house moral philosophers who have little power to shape internal company policies; the careful selection of employees that will not question the status quo; the focus on humane design—e.g., nudging users to reduce time spent on apps—that does not address the risks inherent in tech products themselves;[9] the funding of "fair" machine learning systems combined to the defunding of work on algorithmic systems that questions the broader impacts of those systems on society.[10, 11]

On the other hand, the technology community's criticism and scrutiny of instances of ethics washing, when imprecise, have sometimes bordered into the opposite fallacy, which the author calls "ethics bashing". This is a tendency, common amongst non-philosophers, to simplify the issues around tech "ethics" and "moral philosophy" either by drawing a sharp distinction between ethics and law and defining ethics as that which operates in the absence of law[12] or by conflating all forms of moral inquiry with routine politics, for instance by merging or drawing artificial separations between the frameworks of "ethics", "justice", and "political action".[13, 14] Distinguishing between "law" and "ethics" is a common legal positivist move, configurable within a long philosophical tradition that sees the practice of making, interpreting, and applying law as processes whose existence and relevance are distinct and separable from their moral and societal implications.[15] The relation between "ethics", "justice", and "political action" instead is complex. Understanding ethics and moral inquiry as either a mode of political action or a discrete, individual-centric, and particularized exercise that is easily instrumentalized and is unsuited to tackling political and institutional questions is misleading yet frequent. As described by Jacob Metcalf, Emanuel Moss, and Danah Boyd, the distinction between narrow "ethics" and capacious "justice" became a central focus of discussions during the 2019 ACM Conference on Fairness, Accountability and Transparency.[13]

Equating serious engagement in moral argument with the social and political dynamics within ethics boards or understanding ethics as a methodological stance that is antithetic to—instead of complementary to and inherent in—serious engagement in law-making and democratic decision-making, is a frequent and dangerous fallacy. The misunderstandings underlying the broad trend of ethics bashing are at least three-fold: (1) philosophy is either confused with "self-interested politics" or understood in opposition to law, justice, political representation, and social organizing; (2) philosophy and "ethics" are seen as a formalistic methodology, vulnerable to instrumentalization and abuse, and thus ontologically flawed; and (3) engagement in moral philosophy is downplayed and portrayed as mere "ivory tower" intellectualization of complex problems that need to be dealt with through alternative and more practical methodologies.

Grappling with the role of ethics in tech policy requires moving beyond both ethics washing and ethics bashing and seeing ethics as a mode of inquiry that informs work in law, policy, and technological design alike in emancipatory directions. Policy-makers, lawyers, technologists, corporates, and academics do moral theorizing all the time. Asking whether a corporate ethics council can improve internal policy-making, whether a given machine learning system can lead to fairer criminal justice enforcement, or whether a given corporate decision to fire a researcher or ban facial recognition is acceptable in context involves asking moral questions that, if properly framed, can lead to a better understanding of these phenomena and also to better policies. Awareness of the ubiquity of morality would enable all actors in the technological and AI space to contextualize their work with greater subtlety, at several levels of abstraction, and to more rigorously assess the legitimacy of corporate self-regulation and other ethics initiatives.

One aim of this article is to distinguish between what ethics is often thought to be (a neutral and context-independent methodology, a self-interested corporate rhetoric) and what ethics could be (a principled methodology for evaluating political disagreements around technology). To understand that distinction, another distinction must be captured between the intrinsic and the instrumental value of ethics. The intrinsic perspective sees ethics as a mode of inquiry which is independently valuable as an aspirational process, particularly for those engaging in it. The instrumental perspective instead sees the value of ethics as lying in its results. The value of ethics understood in this way depends on its end-results, ethics' causal role in bringing about desired results, such as reputation, innovation, and profit. Intrinsic and instrumental perspectives on ethics and moral inquiry are not mutually exclusive. One can understand ethics as an intrinsically valuable process with valuable results. However, distinguishing facial appearances of ethics from approaches that emphasize ethics' potential entails emphasizing intrinsic value over instrumental value. The author will argue that the more the process of engaging in ethics is motivated by outcomes independent of the process itself—the less ethics is taken as an intrinsically valuable process—the weaker its moral value becomes for society. Ethics washing and ethics bashing are instrumental understandings of ethics, in that both positions or tendencies envision or experience ethics as a means to an end and nothing more.

What is at stake in recent controversies around the weaponization of "ethics" rhetoric are also competing moral conceptions of technology companies' role. Corporate-friendly conceptions benefit from inserting ethical work within larger communications and public relations strategies.[13, 16−18] Critical conceptions reject these corporate efforts and prefer participatory democracy and activism.[11, 19] Yet both corporations and their critics obscure the potential role that moral inquiry can and must play in developing a thicker conception of technology politics. There is no neutral perspective "outside morality" from which the normative implications of technology can be teased out. It should thus be possible to maintain a critical outlook on the instrumentalization of ethics in technology settings, while also recognizing the special value and centrality of moral inquiry to expanding horizons.

This article has two goals. First, it aims to articulate the weaknesses of both the ethics washing and ethics bashing fallacies, explaining why both are impoverished views of the relationship between technology and ethics. Second, it aims to clarify the potential of moral philosophy in debates about the impact of new technologies on society and thereby to dissipate misunderstandings of moral philosophy as either too abstract to inform concrete policy or as a red herring that prevents proper focus on political and social action. Far from constituting a barrier to appropriate governance, moral philosophy enables us to seriously scrutinize the future of technology governance, law, and policy, and to understand what humans need from new technologies and innovation from a unique vantage point.

The article is structured as follows. In Sections 2−4, the article begins by explaining the function and meaning of ethics and moral philosophy, some common criticisms of moral philosophy, and what it is for. Section 5 of the article then provides background on the rise of ethics in tech and the advent of so-called "ethics washing". In Section 6 it explains the limits of existing critiques of ethics washing, identifying "ethics bashing" as a fallacious depiction of ethics as opposed to law, politics, or justice. In Sections 7 and 8, adopting a view internal to moral philosophy, the author engages in a moral argument and shows that commitment to moral principles and engagement in moral reasoning also leads to the conclusion that corporate ethics efforts are by and large wrong and that ethics is antithetical to what happens inside corporate settings. Finally, Section 9 of this article suggests a way forward that moves beyond both ethics washing and ethics bashing, that adopts a less instrumentalist position on ethics, and that requires developing governance frameworks that enable the emergence of renewed moral, political, and legal thinking and action outside corporate settings.

## 2 Ethics and Moral Philosophy

The English word "ethics" is derived from the ancient Greek words ēthikós and êthos which refer to character and moral nature.[20] Morality comes from the Latin moralis which means manner, character, and proper behavior. Both "ethics" and "morality" thus refer to the study of good and bad character, appropriate behavior, and virtue. The two terms are often employed interchangeably but have slightly distinct uses and connotations. Morality is often associated with etiquette and rules of appropriate social behavior, whereas ethics has instead a more personal connotation. Ethics pertains to the cultivation of individual virtue abstracted from society and is sometimes used to refer to personal and professional standards of behavior embodied in "codes of ethics". In Confucian philosophy, morality is about respecting the family and pursuing social harmony and stability through virtues including altruism, loyalty, and piety.[21]

In the discussion to follow, the term "ethics" will refer to the rhetoric of morality employed in technology circles, and "moral philosophy" will instead refer to the philosophical discipline that investigates questions around human agency, freedom, responsibility, blame, and the relationships between individuals, amongst other questions. The author adopts a primarily Anglo-American liberal approach to the practice and understanding of moral philosophy[22] but the author's perspective is by no means intended to close the door to alternative approaches to moral philosophy and ethics. According to some accounts, moral philosophy's scope is limited to relationships between humans and ethics extends instead beyond humans to animals and nature. Some would also distinguish moral from political philosophy while others such as Ronald Dworkin see them as interconnected.[23] Like Ronald Dworkin, the author construes the "moral" widely as consisting of the domain of "value", i.e., an evaluative mode of inquiry which is distinguishable from scientific or descriptive modes of inquiry, which focus on facts.[23, 24] The domain of "value" is the specific domain of inquiry of moral philosophers.

To better illustrate what moral philosophy is, consider the example of surveillance. Let us ask: what is wrong or unethical about big data and certain forms of surveillance? Disparate arguments can be offered to show that big data and surveillance are wrong in some respects or worth carrying out in other respects. Different persons will likely have different views on which of these arguments are strongest. As philosophers might put it: the morality of surveillance is an evaluative matter, i.e., a matter on which reasonable people disagree because they hold competing moral interpretations of what is at stake. Numerous lines of reasoning support the wrongness of surveillance and business models that rely on data extraction. Surveillance is objectionable on self-development and virtue ethics grounds because it incentivizes self-censorship, reducing human beings' ability to develop themselves or to engage in other valuable causes for fear that these actions will be held against them. Another argument focuses on harm: some surveillance and big data activities cause harm to individuals (e.g., they lead to unjustified and stereotype-enhancing discriminatory treatment, they create asymmetries of knowledge and power, they perpetuate pre-existing and unjustified inequalities). A third line of reasoning focuses on equal dignity and respect for persons: some forms of data processing and surveillance fail to treat individuals as equally worthy of respect because they are covert and because some people are surveilled more than others.

Each line of argument entails a different way of evaluating policy. For instance, if someone considers that surveillance inhibits the pursuit of worthy behavior or individuality, they might be satisfied with aspects of big data and surveillance practices that enhance the pursuit of certain worthy life goals, including certain targeted and personalized work opportunities, as long as they are empowering and equally distributed. On the other hand, if one believes that the core problem is that the data collected can cause unintended harm to individuals, they might advocate for solutions that minimize discriminatory impacts and ensure that harms are reduced. Finally, someone who believes that surveillance and the opacity of big data activities are denials of respect for the persons surveilled might be keen to ban surveillance completely or to reduce any tolerable surveillance to a de minimis threshold.

Which reasons we find most weighty is a matter of commitment and deliberation on how to actualize moral values such as autonomy, equality, and human flourishing. The process of weighing some reasons against others allows us to overcome the intuitive belief that "surveillance feels creepy",[25] and to instead ground or re-evaluate one's commitment to privacy or its limitation based on carefully weighed argument on how different forms of surveillance and data extraction might interact with autonomy, dignity, equality, and human flourishing. Identifying the drawbacks of surveillance business models and their morally unacceptable core also facilitates the design of nuanced concrete strategies for addressing them.

This process of revising and refining moral beliefs through philosophical inquiry is what John Rawls has called reflective equilibrium.[26] What Rawls' methodology and other analogous modes of moral evaluation have in common is that they provide a lens through which to interpret issues of societal importance, to locate them within existing debates, consider them from all relevant standpoints, and evaluate which angle or way of approaching them is capable of shedding the most valuable light on the issues themselves. When engaging in this process, the broader the spectrum of considerations that are taken into account in moral theorizing, the more interesting, capacious, and morally significant are the outcomes, and the more inspiring and valuable are its practical implications.

It is also important to emphasize that moral philosophy and ethics can mean different things as part of different fields of study and intellectual traditions. The above is intended to capture only a glimpse of a larger roadmap of possible uses of the terminology of ethics and moral philosophy in technology governance and policy. It is not intended to fix the meaning of these rich and complex modes of inquiry.

## 3   What Moral Philosophy Is For

A key question is what ethics and moral philosophy are for and what they can contribute to existing technology policy debates. In asking this question, The author focuses on the reflexive value of engaging in moral reasoning from the perspective of those engaging in it, i.e., "from within". In the technology policy context, moral and other philosophical work is valuable in at least four ways for those who pursue it.

First, philosophical reasoning and deliberation can provide a meta-level perspective from which to consider any disagreement relating to the governance of technology. Instead of taking arguments narrowly, intuitively, or personally, philosophical reasoning provides a framework for stepping back, situating any problem within its broader context and understanding it within or in relation to other relevant or analogous debates. As such, the practice or method of engaging in moral argument allows us to broaden our perspective and to look at a debate from a wider lens, overcoming confusions, filling in gaps, correcting inconsistencies, and drawing clarifying distinctions. In debates on the acceptability or necessity of facial recognition technologies, for instance, a philosophical method can help us rethink our reasons for rejecting or promoting existing technologies, clarify points of agreement between a variety of opponents to these technologies, and focus on where disagreements lie and what they entail in practice: what freedom, equality, and human flourishing require in an era of structural surveillance and systemic inequality. Otherwise put, philosophy is a good antidote to knee-jerk reactions: it can help reduce unbridgeable value conflicts and make agreement possible by moving discussions between different levels of specificity or abstraction. This is not to say that ideology and value conflicts are unimportant, but merely to recognize the importance of philosophy as a method aimed at overcoming or clarifying those conflicts.

A second, related, contribution of moral philosophy to tech debates is that it adds rigor principled thinking to value-laden, emotional, or subjective discussions. Moral philosophy should be understood as an explanatory mode of inquiry which requires us to set out the justifications and reasons for advancing one view and not a different one. By centering attention on the explanation and the justification for a position, philosophy enables a dialectic to take place, a Socratic dialogue which we can have internally with ourselves or externally with others, that sheds light on blind spots and enables fluid and iterative repositioning. Winning the argument is not as important as laying all its facets on the table. Such principled and disinterested inquiry is frequently absent in technology policy and governance discussions for at least two reasons. The first is that current policy debates are instinctive, emotional, polarizing and inimical to measured reflection. The second is that many of these debates are mediated by platforms whose corporate incentives are difficult to align with disinterested reflection on societal impacts.[27, 28]

Third, a normative philosophical lens can substantively move us beyond a narrow focus on procedural fairness, diversity, and representation in technology governance, and towards substantive goal evaluation. As explained in more detail below, the problem is not just whether an AI ethics board's members have diverse perspectives and backgrounds, but also whether the board's decisions can actually constrain Google's profit-motivated actions. Similarly, the question is not just whether a facial recognition algorithm properly recognizes black faces, but whether such algorithm is deployed in circumstances where it can harm black people. A capacious moral philosophy approach can help us move beyond checklists and proceduralism to question whether an existing or future structural governance framework and its substantive outcomes are morally acceptable and worth pursuing.

Fourth, far from obscuring ideological conflicts and structural divisions[19, 29] engaging in moral philosophy can facilitate dialogue, encourage the building of common ground, and provide a basis for collaborative and participatory approaches to policy-making capable of bridging divides in a polarized landscape. An important drawback of critical work that centers on power, value conflicts, and unbridgeable ideological divides is that it renders dialogue between people holding different views or occupying different social positions more difficult. Pursuing such strategies has its advantages but it can also lead to fragmentation in an already polarized and emotions-driven public sphere. Understanding philosophy as a dialectic discipline that enables empathy and grounds methodology in the aspirational possibilities of commonality, justification, and conflict resolution can instead help navigate fragmentation and polarization today. The many "embedded ethics" initiatives at computer science and philosophy departments in the United States and beyond are fostering greater debates and have been shown to promote the building of common ground across disciplinary boundaries.[30–33]

Still, while acknowledging the important contributions of Western philosophy to the promotion of an inclusive and discursive public sphere, awareness of how power and inequality manifest within such discursive public sphere is key. Not every person has the same voice and the same ability to be heard.[34] Equalizing a space in the face of structural inequality must thus be one of the first considerations when building spaces for dialogue and "ethical" reflection. Contemporary approaches that embed ideology and structural power asymmetries within normative philosophical inquiry[19, 29, 35] account for the advantages of a discursive methodology while expanding the horizon of philosophical inquiry to include issues of structural inequality, power, domination, and ideological entrenchment.

## 4　How to Criticize Ethics and Moral Philosophy

Work in moral philosophy and ethics has a number of limitations. Before turning to the rise of ethics discourse in technology and the fallacies associated with that trend, here are six ways of criticizing moral philosophy that are targeted at moral philosophy as a reflexive exercise and as a methodology. By addressing these important criticisms, my aim is to shed light on moral philosophy as a critical method, showing that it can channel change, re-assessment, and revision of commonly held beliefs.

First, philosophy can be criticized for being abstract and for not being accessible to large audiences. This makes philosophical work often unsuited to advocacy or activism or to making provocative contributions to time-sensitive issues. Philosophy is also rarely suited to opeds, for example, or to those who aim at quick and easy policy fixes. Yet depth and abstraction are also one of the discipline's advantages: engaging in philosophical work prompts us to pause and think, to shield our thinking from pragmatic pressures, to enlarge the temporal and geographical scope of our research scope. As we engage in this process, our intuitions change, we extend our thoughts or revise them so that they can connect with and make sense of other problems, we learn how to think slower, to think with more depth and more systematically. To achieve meaningful cultural and social renewal in the technology industry, countering a technological culture of fast-paced permissionless innovation driven by an ethos of "move fast and break things", slowness needs to be taken more seriously.[36]

Second, some work in moral philosophy, particularly in its connections with technology, is seen as not going far enough prescriptively or as doing harm in practice. Recent work in social science, for example, has attempted to rely on the philosophical heuristic of the trolley problem[37] to address the regulation of Autonomous Vehicles (AVs), with scarce practical success and generating significant controversy. The Moral Machines experiment at MIT,[38, 39] a large-scale experiment that gamifies the trolley problem to extrapolate aggregate data and then guidelines for programming AVs, has been criticized for simplifying, scaling, and misusing a case-specific and contextual philosophical mode of reasoning.[40] Similarly, Basl and Behrends argued that attempts at applying trolley problem insights directly to AV policy are flawed because they fail to take into account the complexity and contextuality of machine learning development.[41]

More generally, entrenching high level principles for ethical AI in Codes[42] also arguably remains too abstract to guide individuals and policy-makers' actions in practice on AI questions.[11, 43, 44] In the absence of a deep understanding of context, focusing on the trolley problem or outlining high level theoretical principles for ethical AI appears unlikely to lead to workable and morally compelling regulatory strategies. These examples leave us perplexed: much philosophical work seems irrelevant or unsuited to resolving pressing problems in technological contexts. What is needed however is not less philosophical work, but more thinking on what moral principles can do in practice, and what they mean contextually. Helen Nissenbaum's work on contextual privacy is an important example of how thoroughly articulating the contextual implications of abstract privacy norms can impactfully guide the work of communities of practice.[45]

Third, the application of philosophical work can have effects in practice that sometimes contradict the philosopher's motivations. Hegel and Nietzsche's philosophical ideas have been instrumentalized by the German Nazi regime to pursue inhumane ends, an instrumentalization that had little connection to what these philosophers were actually doing or thinking.[46, 47] More concretely, philosophers frequently understand reflection and engagement with the politics and context of their work as corrupting, and thereby fail to prevent misuses of their ideas for unworthy ends. The hiring of moral philosophers by technology companies is but one instance in which philosophical ideas need to be scrutinized in context; such work cannot be taken at face value just because they are the ideas of a trained philosopher. Philosophers are hired, and then their skills are subordinated to the commercial goals of their employers. In this way, work that might have seemed apolitical in an academic setting acquires a new politics. This work can become harmful if it hides under the appearance of neutral thinking allowing the legitimation of controversial states of affairs, such as the secrecy of algorithms and their control by private companies. As important as it is, this criticism however should not be seen as fatal to the kind of work philosophers do. The emergence of in-house philosophers means philosophical work must be scrutinized with greater care, must be publicly accountable, and philosophers must exercise an enhanced level of caution regarding the context and consequences of what they do. Importantly, the funding of philosophical work in the technology and governance field must be disclosed and discussed more openly.

Fourth, work in ethics can be understood as normalizing, as an attempt to discipline social life by devising and applying universally applicable norms of conduct that entrench existing power dynamics by placing them outside the realm of contestation.[48] Marxist critics of moral philosophy have also argued that capitalist incentives can influence philosophical work in directions that favor the interests of businesses and elites.[49] Ethnographers speak of "ordinary ethics" as the descriptive way ethics and morality structure routine social interaction.[13] Zigon however emphasizes the importance of distinguishing routine and unconscious moral claims from conscious ethical claims that arise during "breakdown" moments and are aimed at changing a culture and at "returning to the unreflective mode of everyday moral dispositions".[50] While Zigon's anthropological perspective on morality and ethics captures the pivotal role played by moments of breakdown and moral dilemma, he still sees morality and ethics as fundamentally about the need to return to unreflected normality, to revise beliefs so they can be fixed, routinized, and remain unchallenged once again. For philosophers, instead, morality and ethics are centrally about reflectiveness, conscious revising of beliefs and constant changes to the status quo. Contrary to anthropologists and ethnographers, moral philosophers and ethicists are only marginally concerned with the normalization of moral beliefs. For a philosopher, the task is indeed to engage in direct moral questioning about these beliefs and to bring them to the foreground of our consciousness, instead of emphasizing their regularities and embeddedness in social norms and cultural contexts.

Fifth, philosophical theorizing is frequently criticized for creating an appearance of principled reasoning, neutrality, and objectivity when much of what is at play are a philosopher's subjective views.[19, 51] There is some validity to this criticism, but it is less powerful than it first appears. Good normative philosophical work does not attempt to convey an appearance of absolute objectivity. Quite the contrary, such work is very clear regarding the uncertain bases on which it stands. A large share of Anglo-American moral philosophy follows Rawls' reflective equilibrium or a similar method, to progressively match intuitions and beliefs to considered judgments. This iterative process is one of many approaches that Anglo-American philosophers use to formulate normative conclusions. Although any philosophical conclusion necessarily originates in a thinker's subjective intuitions and beliefs, it is also the product of structured and iterative revisions. It gives conclusions a normative weight or subtlety that raw intuitions do not have. Far from presenting ultimate and final words on a subject, good philosophical work is rigorous yet porous and open to scrutiny: its aim is to broaden perspectives, allowing us to see the limits of the existing and to constantly revise our beliefs.

Finally, sociologists have argued, often rightly, that philosophy is not sufficiently from a gender and racial perspective in particular, dominated instead by Western male figures.[52]

These criticisms are grounded in the idea that moral philosophy can be a worthy enterprise but that its objective appearance or moral weight too often leads philosophers in the wrong direction. Philosophers and theorists interested in the potential of ethical reflection in technology should not only be aware of these vulnerabilities but must also combat them by embedding inclusion and resistance to the exploitation and instrumentalization of moral inquiry into their very methodologies and practices.

As shown, moral philosophy is a reflexive pursuit that is valuable as a process for those who engage in it in view of making sense of the world around them with caution and empathy. Moral philosophy in this sense is not a synonym of the ethical initiatives that occur within corporate settings which are mostly self-centered and instrumental;[18] it is an exercise that, if construed radically as an inclusive emancipatory methodology, is in inherent tension with industry players' profit logics. In Section 5, the author explains the development and rise of technology ethics and its entrenchment within private companies, a trend often aimed at reputational enhancement which has been called "ethics washing".[8]

## 5   The Rise of Tech Ethics and Ethics Washing

In an important essay in 1980, Winner showed that artifacts have politics in two important ways: technologies embed and express the biases and power relations of the society and people who design them, and the deployment and use of these artifactual affordances in turn change and shape the politics and power relations in society.[53] The rise and promise of machine learning and artificial intelligence technologies have brought about a renewed urgency to the debate on the political nature of technology and its ethical implications. A number of prominent books and articles on the subject have shown that the deployment of artificial intelligence can have significant consequences for privacy, human dignity, equality and non-discrimination, gender, social, racial, and economic justice.[54−61] The growing awareness of AI's societal implications and political nature, and a significant "techlash",[62] have led companies involved in developing AI systems to pay attention to the ethical implications of data science and artificial intelligence.

In the last few years technology ethics has grown in popularity and been adopted and endorsed in a multitude of overlapping forms.[43] High-level statements of principled artificial intelligence have been created or endorsed by private companies, civil society, governments, as well as transnational and multi-stakeholder entities.[42] Ethics training has been developed and embedded in the computer science curriculum of a growing number of universities.[30−32, 63] The growing research field of AI and the growing body of research around its ethical and societal implications has led to the creation of a number of new conferences and dedicated research institutes.[42]

Private companies have been involved in these efforts at each level: developing and publicly sharing statements of AI principles,[42] hiring in-house ethicists,[64] forming ethics councils and bodies,[3] and putting in place ethics and diversity trainings and structures for their employees.[18] As regards principles, Google, for instance, has published principles emphasizing the need for AI applications to be socially beneficial, to avoid creating or reinforcing bias, to be safe and accountable.[65] Microsoft and IBM have also engaged in codifying principles and procedures for safe and trustworthy AI.[66, 67] Microsoft's website states the need to move beyond principles and toward implementation of ethical AI through ad hoc internal bodies:

We put our responsible AI principles into practice through the Office of Responsible AI (ORA) and the AI, Ethics, and Effects in Engineering and Research (Aether) Committee. The Aether Committee advises our leadership on the challenges and opportunities presented by AI innovations. ORA sets our rules and governance processes, working closely with teams across the company to enable the effort.[67]

When they do not engage directly in crafting statements of principles and setting up internal ethics boards, private companies sponsor AI conferences, research institutes and efforts that shape the research agenda and discourse around the societal impact of AI.[68] The Partnership on AI, a non-profit established to study and formulate best practices on AI technologies, was founded by Amazon, Facebook, Google, DeepMind, Microsoft, and IBM, and is entirely funded by industry stakeholders. Palantir, Google, and Facebook frequently fund major law, computer science, and privacy conferences.[18, 43] In turn, AI ethics is becoming a business, with consultancy firms and law firms developing AI ethics expertise to assist tech companies in their compliance efforts.[69, 70]

As these instances show, companies such as Google, Facebook, Microsoft, and Palantir are concerned about their ethical reputation in the face of new technological developments in data science and beyond. Their efforts to promote and arguably build more trustworthy and ethical AI indicate a calculative stance, a method for preempting financial and reputational risk, more than a recognition of the political nature of AI and its implications.[13, 14, 16] Even though it might be argued that the intentions behind these initiatives are good, the practices themselves are too limited and opportunistic to be in line with a conception of morality and ethics as reflexive capacious exercises that can foster disinterested selfless change. Overall, speaking of AI "ethics" instead of AI "politics" can be seen as a way to depoliticize and normalize the impacts of company efforts in this space,[14] allowing companies to "ethics wash" their reputations and to narrow the space for real debate and change in AI.[8, 71]

# 6   Critiques of Ethics Washing: Merits and Limits

Efforts such as embedding ethicists or ethical guidelines within industry practices and creating codes of ethical principles aimed at more responsible and trustworthy technological design have been criticized by scholars for normalizing and depoliticizing data science and AI (Green, this issue). They have been criticized for bringing about a performative "transformation of ethics and design into discourses about ethics and design",[11] a routinized checklist approach to ethics that is powered by capitalist logics and a technosolutionist mindset.[13] Companies are "learning to speak and perform ethics rather than make the structural changes necessary to achieve the social values underpinning the ethical fault lines that exist".[13] For Greene, Hoffmann, and Stark, these practices are both too focused on technical tweaks, blinded by technical concerns about how to embed fairness and accountability within machine learning systems and neglectful of structural injustice, and are universalist projects "justified by reference to a hazy biological essentialism".[11] For human rights experts such as Paul Nemitz[12] and Phillip Alston who jokingly said at a 2018 AI Now conference that he wanted to "strangle ethics",[13] technology ethics is seen as a substitute or an alternative to more adequate human rights laws.[16]

As argued further below, these critiques ought to be taken seriously. They shed light on the politics of AI and on crucial blind spots that are performatively and voluntarily obscured by corporate ethics practices. Yet they are at their weakest when, instead of understanding that legal and technological governance are necessarily embedded in ethical and moral thinking, they draw sharp dichotomies between "ethics" and "law", between "ethics" and "justice", as if these were incompatible alternatives and they often misconstrue the relation between "ethics" and "politics" failing to take them as all ingredients playing complementary roles in a desirable understanding of technology governance. The author calls ethics bashing the reduction and dismissal of ethics as a simplistic alternative to law or justice, and the lazy conflation of moral thinking and inquiry with a politics of neutral thinking and with appearances of "ethics" that are hardly in line with what morality requires. The author identifies three fallacies that characterize ethics bashing positions.

First, Nemitz has drawn sharp distinctions between ethics and law as separable and discrete practices: the key question, writes Nemitz, is "which of the challenges of AI can be safely and with good conscience left to ethics, and which challenges of AI need to be addressed by rules which are enforceable and based on democratic process, thus laws".[12] Such distinctions operate on the positivist assumption that law—its making, interpretation, and application—are institutional facts whose existence and relevance are entirely distinct and separable from its societal and moral implications. Positivists, frequently relying on a Humean separation of "is" and "ought", or fact and value, argue that law belongs to the realm of positive facts while morality is completely distinct and belongs to the realm of moral value and of the "ought".[72] An understanding of law as conceptually separate from morality obscures how law is constructed—written, interpreted, and applied—in ways that embed certain moral and political commitments. As Dworkin understood and theorized, law has no factual existence other than the existence we give it through the principled moral and political commitments we express as we interpret and apply it.[24] Consequently, the task of understanding, applying, and re-making law is inseparable from engagement in the internal reflexive exercise of moral commitment and ethical evaluation. Instead of saying that law is superior to ethics, we might want to respond to obtuse corporate ethics efforts by saying that a capacious understanding of morality and ethics is incompatible with ethics washing and extensive self-regulation and that morality instead requires effective laws and robust external checks and accountability mechanisms on machine learning systems, especially when they affect vulnerable populations.[73]

The second and third fallacies, the conflation of "ethics" and "self-interested politics" and the distinction between "ethics" and "social justice", are connected. Both attitudes are grounded in a relatively narrow understanding of moral inquiry as a discrete, individual-centric, and particularized exercise whose politics and impact lie in its separateness from broader political and institutional questions. As described by Metcalf et al., the distinction between narrow "ethics" and capacious "justice" became a central focus of discussions during the 2019 ACM Conference on Fairness, Accountability and Transparency.[13] However, justice and morality are inseparably intertwined. Critics are right to argue that the focus on design and on embedding fairness in machine learning is too narrow to address more urgent questions around these technical systems' political dimensions and effects on structural inequality, capitalist exploitation, surveillance, disinformation, and environmental degradation.[10, 13, 14] However, responding to narrow and techno-solutionist corporate approaches on "ethics" is not exhaustively done by arguing somewhat simplistically that justice is superior to ethics, whatever that means, or that ethics has a flawed politics. It must be done by showing that any meaningful understanding of ethics (or politics) must include concerns about structural inequality, capitalist extraction, and environmental justice, or else it is an empty exercise that has little to do with the ethics, justice, and politics of new technologies and their societal impacts.

The answer to instrumentalized ethics is not to draw simplistic dichotomies, but to provide a richer account of how ethics, politics, and law are connected and can work together to enable a better understanding of AI's shortcomings and to foster political and other change. By addressing ethics from the outside, as a discrete practice that does not include them, critics of corporate ethics often fail to recognize that ethics is something they also engage in and that existing corporate practices are in fact morally flawed. The task is therefore to change the way we collectively engage in moral inquiry, equipping ourselves with a better understanding of injustice, inequality, and other digital harms. Corporate logics of profit, expanding production, capitalist exploitation, and so on are often incompatible with a capacious view of morality.

In the remainder of this article, the author articulates what the role of moral philosophy should be in technology policy debates and how a view that takes the reflexive internal exercise of moral inquiry as valuable can shed light on the "ethics washing" debate. The author then concludes with what ethics in technology must look like going forward.

## 7  The Moral Limits of Corporate Ethics and Self-Regulation

Equipped with a richer understanding of what ethics and moral philosophy are and can do, the question now is what role moral philosophy can play in informing technology policy and particularly the question of what makes ethics-based efforts as practiced in corporate tech settings particularly problematic from a moral philosophical perspective. Moral philosophy can provide a lens to evaluate the moral wrongness of some of these efforts.

As described above, companies such as Google, Apple, Microsoft, OpenAI, Palantir, and Facebook are increasingly making efforts to consider an ethical standpoint. The intentions behind their proactive efforts are often presented as good, but the practices remain driven by market incentives and techno-centric perspectives and motivated primarily by the need to avoid financial and other company risk.[11, 13] Notwithstanding good intentions, therefore, embedding philosophers or ethicists within technology companies appears to be a façade that is frequently used to legitimate certain pre-existing practices and to shield companies from measures more protective of consumers. This is true of corporate settings but also of public institutions. Taylor and Dencik for example have described the political dynamics within the European Commission's High Level Expert Group on AI, showing that instead of having outcomes guided by processes of reflection and philosophical principles, ethical reflections are often designed to produce pre-determined instrumental outcomes.[18] They state that after months of discussion around "red lines" on the use of AI, corporate participants in the High Level Group stated: "the word 'red lines' cannot be in this document … at any point … and the word 'non-negotiable' has to be out of this document."[18] As Taylor and Dencik point out, "if the possibility of delineating meaningful boundaries for technology … is off the table, then so is an important part of the task of ethics."[18]

As we assess these ethics initiatives, we are therefore pulled in two directions. On one hand, we are tempted to welcome some of these developments as positive. On the other hand, we are moved to criticize these efforts for the opportunism they represent. Where we stand on this spectrum will often be informed by our situated perspective, our training, by who pays us, etc. What moral philosophy as a method enables us to do is to take a step back, to consider these attitudes along a spectrum of nuanced positions on companies' ethical behavior, and to evaluate our reasons for supporting or resisting initiatives such as a corporate ethics council or an AI Panel of Experts at EU level. It allows us to suspend our intuitive reactions and take a less polarized perspective on the question: What is wrong with the instrument-alization of ethics language? And what is wrong with ethics boards and self-regulation?

As seen, much of the debate has centered on ethics as a self-regulatory modality of governance and an alternative to law and government regulation. As Javier Ruiz is reported to have stated, "a lot of the data ethics debate is really about how … we avoid regulation. It is about saying this is too complex, regulation cannot capture it, we cannot just tell people what to do because we do not really know the detail."[18] Self-regulation and self-publicity at first both seem benign. Self-regulation in certain cases is not only tolerable but actually welcome, for instance where regulatory interference by a public agency is unlikely to be effective and where a self-regulatory approach can lead to substantive policy improvements for individuals and society. Further, in principle it does not seem morally objectionable to fund and develop initiatives that foster a positive image of one's business, nor does it seem wrong for a business to engage in self-publicity and self-advocacy. However, when looking further the reality is more complex.

To use an example, let us focus on the case of self-regulation in relation to online content moderation on Facebook. In the United States, governmental regulation of online speech is seen with suspicion.[74, 75] The solution to the regulation of online speech on Facebook has consequently materialized in the form of an internal Facebook Oversight Board (FOB), a quasi-judicial body set-up internally but composed of external experts to adjudicate on the acceptability of controversial user content on the platform.[76] The body has been praised as "one of the most ambitious constitution-making projects of the modern era",[77] and is seen as a workable and promising approach for taming Facebook's power over online content in the face of First Amendment restrictions on government regulation.[78] Nonetheless, while the Board may bring about needed transparency and an appearance that content moderation is being tackled fairly, we must look beyond Facebook's messaging to find its shortcomings, procedural and otherwise. In spite of its carefully crafted set-up and the well-intentioned messaging around its existence, it is likely that the FOB will serve the interests of Facebook more than those of users. First, it provides a way to shield Facebook from other forms of regulation and scrutiny on matters of content moderation and community guidelines, including the intervention of national or international courts but also the formulation and enforcement of legislative redlines and constraints. Second, by centering attention on content moderation and community guidelines, it allows Facebook to continue developing its News Feed algorithms as it pleases, and to continue showing individuals lucrative content, without interference from regulators or courts. Thus, far from addressing all questions of online speech harms, the FOB seems to divert attention toward some issues and away from the most pressing concerns around misinformation and political propaganda.[79]

The case of facial recognition technologies is analogous. In the United States, much state regulation of private technology firms is made difficult by the First Amendment.[80] The solution to making facial recognition more ethical was thus for some time believed to be something that must originate within the proprietary walls of tech companies and not something that can be initiated by government entities or the Federal Trade Commission (FTC). But things are changing. Following activist efforts, companies like IBM, Amazon, and Microsoft have scaled back on their offering of general purpose facial recognition software.[81, 82] More recently Facebook has declared that it will cease to use facial recognition.[83] Earlier, company ethics boards themselves, such as Axon's, recognized the importance of public oversight on these technologies.[84] In spite of litigation by tech companies to defend their self-regulatory immunities, it seems that the turning point in the relationship between state power and self-regulatory power in this space.

Self-regulatory and ethics washing initiatives such as the FOB, Google's ATEAC Board or Axon's Report on facial recognition technologies should prompt us to look beyond appearances and ask whether their very existence, in spite of appearing useful and a step forward, might in fact performatively obscure more pressing problems and risk long-term harm.

# 8   A Critique of Ethics Washing from Within Moral Philosophy

To explore the moral limits of these internal corporate efforts superficially aimed at developing more ethical artificial intelligence, we must again turn to moral philosophy. At least three moral arguments can be raised against initiatives that co-opt ethics language and self-regulation for selfish corporate purposes that include profits and reputation.

First, the type of ethics work carried out within companies or ethics boards more often than not seems to lack instrumental value: it does not have beneficial effects on individuals and society, because it is undertaken under conditions that deny these beneficial effects. Second, these practices also seem to lack much of the intrinsic, or independent, value associated with philosophical inquiry insofar as they do not seem to be undertaken in ways that value the process itself and with the aim of achieving overall justice. Third, even if these ethics-based practices were carried out in absolute good faith and in pursuit of justice, and thus maintained both their instrumental and intrinsic value, instrumentalizing ethics reasoning and language to reach company goals entails a specific kind of epistemic concern. Indeed, it seems that the performative role of ethics language remains problematic even where, as the cases of the Facebook Oversight Board or the Axon Ethics Board have illustrated, these efforts are intended to address real issues and in fact could have positive effects. This happens where, in spite of having some instrumental value, these efforts instrumentalize ethics for the sake of other selfish or less valuable ends yet are presented as panaceas that serve the public interest. In what follows I explore these three arguments.

The first critique of self-regulation and company ethics is an argument grounded in the poor instrumental value, or small positive impact, of ethical work performed within a company. Ethics bodies or in-house philosophers are purportedly set up and hired to make a difference to a company's social impact. Yet as long as philosophical inquiry is mandated and funded by a company, and carried out within closed corporate proprietary walls, its primary function is to benefit companies and fulfill their pre-existing mandates, and cannot be to benefit society at large and lead to social renewal. Internal AI ethics practices are frequently put in place for compliance purposes, to pre-empt reputational and financial risk.[13] They are subjected to internal limits, subordinated to the endorsement of high management, and dependent on company funding. This dependency on the company's control renders ethics rhetoric inadequate for addressing serious cases of company misconduct and also unfit for achieving societal change.

The narrow impact of ethics-based efforts carried out within tech companies is due in part to formal limitations on employee-philosophers' or ethics boards' mandates and in part to more diffuse pressures that companies exert on technological discourse and context. Formally, for example, Apple's philosopher in residence Joshua Cohen has been forbidden from making public appearances since he started working for the company and Microsoft's AI ethics board does not disclose the reasons for its decisions.[85] The firing of former Google employees Timnit Gebru and Margaret Mitchell for writing allegedly controversial papers and pushing for a prosocial AI agenda inside the company illustrates companies' power to formally police internal ethics efforts.[6, 7] It also however shows the potentially strong instrumental value of social media backlash following these episodes.[4] Less visibly, companies also exert diffuse influence on the broader discourse around technological innovation and ethics by funding research and policy initiatives that favor their agendas and selecting people to engage with (and whose ideas to highlight), including the people these companies choose to have as part of their ethics-based initiatives.[68, 86]

These internal pressures in turn shape the substance and conservative nature of resulting ethics-based work. Strong pushes for data protection guarantees, data minimization mandates, redlines on the use of AI in credit scoring, policing, criminal procedure, or antitrust enforcement can hardly be initiated by a company's ethics board or in-house philosopher. Their role remains confined to steering, reviewing, and advising on policies and product launches within the confines of existing business models, so as to preserve those business models. For example, in June 2020, IBM publicly announced it would stop offering general purpose facial recognition or analysis software.[81] This move, which was a significant departure from IBM's long-standing position on facial recognition and was followed by similar announcements by Amazon and Microsoft, came as a result of external political pressures in the wake of George Floyd's death in Minneapolis, not as a result of the company's internal ethical compliance processes.[82]

Yet it is precisely at moments of political and moral breakdown, where a company's activities and general goals clearly come into conflict with the interests of society, that ethics can acquire central importance[13] and can provide a fruitful lens for evaluating and deciding the way forward. In most cases, instead, the breakthrough potential of ethics as a mechanism for learning from and facing dilemmas and contradictions is missed. As long as the ultimate decision-maker on any given AI policy is the company itself, as long as internal ethics programs are focused on rhetoric more than on substance, these initiatives will keep benefiting the industry more than users and their instrumental value for society is limited.

The second critique of so-called ethics washing looks at the act of engaging in these efforts by philosophers-in-residence, or members of ethics boards, and examines the intrinsic or independent value of these people's engagement in moral thinking. Moral philosophy as a practice has value when followed in pursuit of independently valuable goals such as truth, justice, or the well-being of society. To be intrinsically valuable, engaging in moral argument must be done to a substantial extent out of commitment to moral principle, in the belief that it can lead to a better understanding of moral questions. If instead it is undertaken for the sake of earning money, pleasing employers, or obtaining honors and recognitions, it loses some of its special worth.

We might think that this critique is about the actual motivations of the philosophers and experts that engage in the exercise. When looking at cases of philosophers-in-residence, ethics boards, or academics who work closely with these companies, there are doubtless some individuals who do it to raise their profile or create connections that can lead to further work in the field, or even to obtain promotions, honors, or greater impact and salience for their work. Yet many also do it simply because they believe that their involvement might lead to a positive overall impact or in the hope of getting insights into how the company works. It is tempting to focus on these people's intentions and blame their shortsighted mindsets, but focusing on intentions seems unhelpful: the road to hell is paved with good intentions.

To better characterize the independent value of ethics-based work, we must look beyond intentions and instead at scope: actual commitment to moral principle requires questioning what an employer requires. Philosophical thinking must have the potential to reach beyond the limits imposed by companies in corporate settings. For example, saying that a facial recognition algorithm should be reviewed because it systematically identifies white people more accurately than black people seems right but is not sufficient. Rectifying bias requires more than acknowledging that the algorithm needs "fixing". It requires making sure that the algorithm is not deployed in settings where it might cause irreparable harm to black people. It also possibly involves thinking about preventing the use of such algorithms by the police, or by society at large, and replacing them with human decision-making.[10, 56] To the extent an ethics board or in-house philosopher engages in moral argument with a view to correcting the algorithm yet is prevented from considering or voluntarily ignores these other considerations, their moral inquiry seems to lack substantive independent value. Philosophical inquiry achieves its full potential only when it comes with full and unrestricted substantive commitment to moral principle and justice.

Third and finally, even if these efforts did have intrinsic and/or instrumental value, the expression "ethics washing" denotes a particular epistemic function of the activities in question which requires distinct analysis. Ethics rhetoric, as it is funded and constructed in academic and corporate circles, may have the effect of freezing popular imagination and of preventing the emergence of valuable alternatives.[68] It may promote and reinforce a narrow and confined vision of the possibilities for regulatory and societal change.

It can, for example, mislead the public into believing that previously contested policies have now become acceptable, thus creating a legitimacy buffer for objectionable corporate action. Immunizing corporate action from public scrutiny is dangerous for more than one reason: apathy strengthens corporations and weakens activists, it shifts the burden of policing new technologies from deep-pocketed security and defense departments and private companies to poorly funded activist groups and other marginalized stakeholders. It can also discredit awareness-enhancing efforts and narrow the spectrum of contestation and debate. Self-regulatory efforts, such as the example of the FOB provided above, tend to narrow the scope of a debate, marginalizing questions of structural injustice or disruptive change and instead centering attention on procedural fairness and fixable tweaks. This—predictably—ends up favoring incumbents. Although the performative dimensions of ethics washing are hardly visible by a majority of consumers, they are in fact crucial to a comprehensive analysis of corporate and governmental stakeholders' strategies in this space and of the moral value and acceptability of their efforts.

Overall, an analysis from the perspective of moral philosophy confirms the view of many critics of ethics washing efforts. It helps us see many of these in-house ethics initiatives as lacking significant instrumental and intrinsic value and also as playing a performative function that can negatively affect persons. There are no doubt exceptions of companies really working to ensure that internal ethical work is independent and valuably contributes to a more just society. However, in general policymakers should not overlook the salience and weight of these critiques of ethics as a self-interested rhetoric. Many existing internal efforts to construct a corporate ethics, particularly around AI, largely remain a façade.

## 9   Avoiding Ethics Bashing

If the reasons for criticizing and resisting ethics washing are ones found within moral philosophy, where does this leave us on the role of moral philosophy? How should we understand corporate ethics? Two main fallacies seem at play in overbroad critiques of ethics that see ethics as distinct from law, politics, justice or social organizing: a linguistic misunderstanding, that is to say the conflation of instrumentalized ethics washing efforts with moral philosophy as a reflexive exercise, and ignorance of or resistance to the possibilities and importance of moral philosophy as a discipline and method.

The linguistic misunderstanding is due to what the author has described above as companies' cooptation of the language and performative function of "ethics" to pursue self-promotional goals. Instrumentalized and emptied of its instrumental and intrinsic value, what remains of "ethics" is an empty construct trapped between meanings and signifying timid instances of self-regulation, static and finite lists of guiding principles, and other forms of narrow and conservative regulative "fixes". None of these embodied instances of the practice of ethics are actually likely to be fully morally defensible, but as the word quickly gains traction, it gets defended or criticized at face value by corporations and critics alike. These dynamics further entrench the misuse and instrumentalization of ethics language. In policy circles, the word becomes a red herring, a mode of governance or a communications strategy to dismiss. Yet the misunderstanding at bottom is this: what is called "ethics" may have nothing "ethical" in it. It may have no intrinsic value for those who perform it and may have instrumental value only for those who commission it and not for society at large.

Much of the ink used to bash "ethics" was perhaps justified but it could have been used more wisely by distinguishing corporate ethics, or ethics washing, from the practice of moral philosophy. We too frequently neglect that "ethics" can and must encompass more than what companies make of it: that properly contextualized, ethics can be a valuable methodology for rethinking the competing or complementary merits of different kinds of regulation, including self-regulation and other forms of law and policy-making.

A richer critique of corporate self-regulatory efforts therefore demands that we operate at two levels: be critical of ethics washing, while also being aware that our very critique positions ourselves distinctly within moral philosophy. In other words, when criticizing certain practices we necessarily adopt a distinct moral stance that is within moral philosophy—not outside of it. We must thus be ready to engage more thoroughly with the flaws of narrow approaches to ethics and to accept that defending more capacious ethical stances is related to a better understanding and awareness of moral philosophy's potential—not a blank rejection of it as a language, practice, discipline, and mode of inquiry. This requires a deep societal reckoning with the values and limits of moral philosophy.

To change tech ethics, it is urgent to rethink the way technology ethics comes to exist and is talked about. Since ethics washing is broadly antithetic to meaningful and capacious ethics, it is important for policy change to originate primarily outside formal and informal corporate settings. To be effective, the role of philosophers, boards, and other formalized bodies concerned to bring about ethical AI must be re-imagined, their scope of action and mandate must extend outside the corporate walls of companies such as Google or IBM, they cannot be exclusively or primarily funded by companies such as Facebook or Palantir, they must to the extent possible safeguard themselves from opportunistic corporate discourse around "ethical AI". A deep reinvention of the structures, processes and modes of governance through which technological impacts on society are evaluated is urgent. At their core, these processes must facilitate the moral evaluation, questioning, and constant re-assessment of technological developments. Far from treating technological developments as moments of ethical breakdown, technology as a whole must be seen as a system that endemically tends toward societal breakdown, and therefore requires constant reflexive reconsideration, revision, and re-imagination.

Criticized as complex, abstract, apolitical, and misleadingly neutral or objective, philosophy is frequently dismissed in areas such as technology policy which are fast moving, full of ideological conflicts, and in need of quick and effective responses. However, it is clear that quick and effective fixes are not the answer. Ideological conflicts and the pace of innovation are not barriers to doing more impactful and valuable philosophical work in this sector. Indeed, the current technological zeitgeist of strong resistance to surveillance capitalism; new data privacy laws; the complicated relationship between big tech, big oil, and climate justice; tech employee movements and whistleblowing; COVID-19 and Black Lives Matter suggests that something within technology is changing, and that it is time we adopt new tools and modes of thinking to fight technological injustice. What the tech ecosystem is in greatest need of today, in fact, seems to be a slower, richer, more comprehensive investigation of what various technology companies and stakeholders owe to humans, to animals, and to the planet. New technologies are also making us reinvestigate and question the commitments we humans owe to each other, as well as to other beings and to the global planet ecosystem. This is precisely what moral philosophy is for. We may want to stop bashing it and instead invest in re-imagining it.

## 10  Conclusion

This article has argued that ethics washing and ethics bashing are both reductive tendencies that rely on a limited understanding of what ethics actually entails. Ethical reasoning or moral inquiry can have intrinsic value as a process and instrumental value as a means to the achievement of other valuable outcomes. The author has argued that the more ethics is used in tech circles as a performative façade and the more it is instrumentalized and voided of its intrinsic reflexive value, the less value ethics can have overall as a practice and mode of inquiry. Adopting a perspective internal to moral philosophy helps us see the limits and actual similarities of what seem like polar opposites—ethics washing and ethics bashing—as two instances of instrumentalized ethics language.

The way to combat ethics washing, therefore, is not to instrumentalize, reduce, and then dispose of ethical language, but rather to distinguish performative and instrumentalized forms of ethics from valuable commitments to moral principle that promote advancements in self-knowledge, understanding, and social change. Although philosophers might never fully adapt their methodology to fast-paced and politicized technology environments, we cannot disregard the immense depth and richness that philosophy can bring to any debate, not least ones about technology governance.

We all ask moral questions as part of our daily pursuits. Technology scholars and policymakers should embrace moral philosophy and value its porous, principled, and open-ended richness, yet resist its instrumentalization or reduction to a performative ethics. Moral philosophers should take on the difficult task of rethinking how new technologies interact with humans so as to provide answers to questions in urgent need of theorization. We all ask moral questions as part of our daily pursuits. To avoid falling into reductive epistemic and ideological traps, it is everyone's duty to nourish curiosity for ethics' and moral philosophy's role in tech and beyond. However, before we can re-center attention on technology ethics, value it in our daily pursuits, and renew interest in the interconnections between moral philosophy, justice, politics, and law, it is urgent to de-center the structures for engaging in theoretical and ethical thinking from corporate settings. Making a commitment to moral principle in technology is impossible without a new governance framework that ensures that ethics in technology remains independent and capacious.

## Acknowledgment

E. Bietti thanks Jeff Behrends, Yochai Benkler, Brian Berkey, Reuben Binns, Mark Budolfson, Urs Gasser, Ben Green, Lily Hu, Lucas Stanczyk, Luke Stark, Jonathan Zittrain, and some anonymous reviewers for their valuable input on this article.

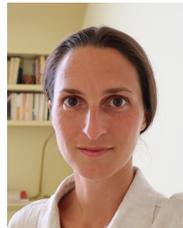

**Elettra Bietti** is pursuing the PhD degree at Harvard Law School. She is an incoming joint postdoctoral fellow at NYU Law's Information Law Institute and at the Digital Life Initiative, Cornell Tech. She is affiliated to Harvard's Berkman-Klein Center, Harvard's Weatherhead Center and Yale Law School's Information Society Project. Prior to her doctorate, she was a competition and intellectual property lawyer in London and Brussels, handling corporate transactions and patent disputes. She received the LLB degree in law from University College London, the LLM degree from Harvard Law School, and the professional diploma in intellectual property law and practice from Oxford University in 2011, 2012, and 2016, respectively, and is admitted to practice law in New York, NY, USA and England and Wales, UK.

# The Promise and Limits of Lawfulness: Inequality, Law, and the Techlash

Salomé Viljoen*

**Abstract:** In response to widespread skepticism about the recent rise of "tech ethics", many critics have called for legal reform instead. In contrast with the "ethics response", critics consider the "lawfulness response" more capable of disciplining the excesses of the technology industry. In fact, both are simultaneously vulnerable to industry capture and capable of advancing a more democratic egalitarian agenda for the information economy. Both ethics and law offer a terrain of contestation, rather than a predetermined set of commitments by which to achieve more democratic and egalitarian technological production. In advancing this argument, the essay focuses on two misunderstandings common among proponents of the lawfulness response. First, they misdiagnose the harms of the techlash as arising from law's absence. In fact, law mediates the institutions that it enacts, the productive activities it encases, and the modes and myths of production it upholds and legitimates. Second, this distinction between law's absence and presence implies that once law's presence is secured, the problems of the techlash will be addressed. This concedes the legitimacy of the very regimes currently at issue in law's own legitimacy crisis, and those that have presided over the techlash. The twin moment of reckoning in tech and law thus poses a challenge to those looking to address discontent with technology with promises of future lawfulness.

**Key words:** law; technology; ethics; tech ethics; inequality; regulation

"Laws have to determine what is legal, but you can not ban technology. Sure, that might lead to a dystopian future or something, but you can not ban it."

−David Scalzo, Kirenaga Partners[1]

"Ferment is abroad in the law."

−K. N. Llewellyn[2]

## 1 The Techlash

In the past several years, the prevailing role of Silicon Valley's California Ideology as the source of hope and inspiration for the "Western capitalist imaginary" has begun to falter[3]. No longer does the tech industry stand for the propositions of inclusive capitalism and technological progress that benefit all. In the wake of Facebook's Cambridge Analytica scandal the technology industry has been the focus of increased public distrust, civil and worker activism, and regulatory scrutiny—a collective curdling of goodwill referred to as the "techlash".①

The techlash is remarkable for its depth of field. The 2020 Edelman Trust Barometer noted a continued decline in trust both globally and in the U.S. in technology and a significant distrust of artificial intelligence[4], both linked to increased numbers of people who believe these sectors should be regulated. A 2019 study conducted by the Pew Research Center

• Salomé Viljoen is with Columbia Law School, New York City, NY 10019, USA. E-mail: sviljoen@law.columbia.edu.
∗ To whom correspondence should be addressed.

① The techlash—and the inequality it is a response to—is a global phenomenon. However, this piece will predominantly draw its examples and focus its analysis on the United States. This is in part a reflection of the author's expertise as a US legal scholar (law is a jurisdiction-specific discipline; particularly in the United States). But it is also due to the Essay's extensive engagement with—and analytical reliance on—the particularities of the U.S. common law judicial system. The role of courts in the U.S. system, as well as the specific legal intellectual tradition in and about U.S. law, informs, constrains, and limits much of the discussion below.

found that from 2015 to 2019, the number of Americans who held a positive view of technology fell by 21 percentage points[5]. In 2018, a majority of Americans (55%) said tech companies have too much power and influence[5]. Former executives have spoken out against their company's actions[6–8], and senior engineers and civil society groups have called for moratoriums or outright bans on facial recognition technology, especially for police and immigration enforcement[9]. Student groups at universities have protested or banned companies like Palantir recruiting at their schools[10]. Community groups have pushed to dismantle and delegitimize the close ties between law enforcement and surveillance technology companies[11]. The technology industry has been the site of increased worker activism from Amazon warehouses workers[12], Uber and Lyft drivers[13], line engineers at Google[14], and the tech industry writ large[15, 16]. Digital rights' activists have pressured companies about their policies and labor practices on everything ranging from content moderation, polarization, lack of diversity, surveillance, and manipulative and extractive data collection practices.

Alongside the popular backlash, technology's harmful social effects have become the subject of increased academic inquiry. Scholars seek to diagnose and address the worst excesses of industry harm, and to develop technical methods and fields of practice less conducive to committing them. These methods produce systems that are normatively relevant to the areas of life they govern: they can amplify and reproduce inequality and entrench unjust means of social ordering. Scholars of scientific method (science and techonlony studies, history of science, philosophy of science, and critical digital studies), as well as computer scientists have highlighted methodological limits in how algorithms are developed and the need for interventions better attuned to the social causes and effects in which such systems are entangled[17]. Increased attention to engineering pedagogy has placed renewed attention on need to educate future data scientists and engineers about the ethical and social dimensions of their work[18].

The techlash involves significant political stakes. Growing worker activism and agitation at companies like Google and Amazon have led to these companies firing senior engineers[19]. Oppressive and biased technologies such as facial recognition and the capacity of social media to manufacture dis- and misinformation campaigns are being used by authoritarian regimes abroad and reactionaries at home[20, 21]. Companies like AirBnB and Uber erode workers' rights and redistribute significant surplus wealth away from local renters and workers[22, 23]. The dominance of a handful of large technology companies (Facebook, Amazon, Apple, Microsoft, and Google) is spurring renewed debates over market concentration and monopoly. The pervasive data collection, processing, and analytic practices that undergird controversial technologies continue to erode our collective privacy (and contribute to the oppressive power of autonomous surveillance systems) amidst an industry-wide gold rush for data[24].

Digital activism is not new—in the United States, groups like the Electronic Frontier Foundation and the American Civil Liberties Union have long advocated for civil rights protections online. Yet these organizations have traditionally focused on civil libertarian concerns over privacy, strong free speech protections, and government overreach. As a result, their advocacy efforts focused on issues like the Edward Snowden revelations over extensive US security surveillance programs, the Digital Millennium Copyright Act and Section 230 of the Communications Decency Act. In each instance, digital advocates defended free speech (and the absence of government surveillance necessary for free speech to thrive) of users and content creators online. This strain of digital advocacy emphasized protecting individuals' online freedom but did not typically focus on other forms of injustice, such as the wealth accumulation that motivated corporate advertising-based surveillance practices or on the distributive or relational effects of the digital economy writ large. In short, while there is a long history of concern over surveillance online, this tradition of digital activism did not historically focus on the social problems of inequality that arise because of surveillance-based economic activity.

The techlash, on the other hand, evinces marked egalitarian concerns over the highly unequal distributions of wealth and power within the digital economy. It expresses a rejection of the tech industry's justificatory narrative for the inequality it generates: that technological progress on its terms will, in the long-run, benefit everyone. There is growing skepticism over technological advancement as a project of shared prosperity and a growing understanding of the technology political economy as one that works to the benefit of the few to the detriment of the many[5, 25].

Critics of digital technology firms argue that their technological progress relies upon extractive practices and oppressive purposes. This begs the question of how to achieve an alternative result, and what role (if any) "tech ethics" will play in achieving it.

## 2   Ethics Response and Lawfulness Response: Troubling the Distinction

In the ensuing public debate, some have advocated for tech to become more "human", and more "ethical"[18]. Others suspect that appeals to traditions of ethics and humanism have less to do with the moral lessons such traditions offer, and more to do with their rhetorical and public-relations capacity to forestall legal and regulatory action[26−32]. Such debates set off second order debates over whether appeals to "ethics" negate rather than require regulatory action[33, 34]. These in turn spawn tertiary debates over what such appeals substantively or materially entail, under what conditions appeals become demands, and who gets to decide what ethical practice means for the technology industry[18, 35, 36].

This initial emphasis on "responsible", "humane", "human-centered", or "ethical" technology and the resulting set of discursive moves are all part of what I call the *ethics response*. The ethics response has real power to marshal bureaucratic and material resources. The call for more ethical technology has spawned a series of ethics boards, company-funded corporate wellness and social responsibility initiatives, the rise of "ethical AI" consultancy practices, and a flurry of publications that outline ethical AI principles for industry[18, 37, 38]. This response has received much attention and been the subject of considerable debate.

Alongside an increased emphasis on ethics, a risk-averse, law-abiding modus operandi pervades the C-suites of Silicon Valley that recalls a certain attitude among banks post-2008 crisis: a patina of cowed mea culpa alongside assurances that lessons have been learned. This second response is marked by an initial commitment from executives that the era of "move fast and break things" is over, and that the strictest interpretation of legal protections will be followed. Like the ethics response, this legalistic mode coalesces from a particular set of discursive moves. Critics call for legal investigations, lawsuits, or new regulation. Companies seek to comply with these calls or proactively offer alternatives as simultaneous signal of compromise and seriousness. Like the ethics response, this response can marshal resources for meaningful new regulatory agendas. Companies change corporate governance and business practices[39, 40], embrace regulatory agendas they had previously fought[41, 42], and even join activist calls for increased oversight and regulation[43, 44]. This attitude marks another strain of response to the techlash that I call the *lawfulness response*.

The lawfulness response is often positioned in contrast to the ethics response as a more serious alternative[31−33]. While critics view the ethics response as ineffectual (or even a harmful distraction), the lawfulness response is often advanced as more capable of disciplining the excesses of the technology industry: "we do not need ethics, we need regulation." And indeed, the lawfulness response generally accompanies companies' acquiescence to a more significant regulatory agenda. Depending on how such demands were articulated and then negotiated by industry actors, the lawfulness response may result in private regulation—a change in corporate governance or firm policy, often in response to threatened or actual litigation—or legislative action, with companies joining advocates in calling for industry regulation. Where the ethics response is viewed as either too vague or too readily co-opted to provide a meaningful form of discipline, the lawfulness response appears to offer a more robust vehicle for realizing the social demands of the techlash.

Despite this perception, the ethics and lawfulness responses function quite similarly. Like the ethics response, the lawfulness response may also yield anti-egalitarian results. Cynical actors may appeal to law to seek moral cover for instituting (and then complying with) with a low standard of behavior. But well-meaning critics may also appeal to legal solutions that inadvertently legitimize the very business practices they seek to reform. Similar to the ethics response, the capacity for the lawfulness response to discipline the technology industry depends on its capacity to express and enforce egalitarian demands.

Two examples of the lawfulness response are instructive.

The first involves Uber. In its early rise to prominence, Uber gained considerable notoriety and begrudging admiration for operating at the edge of legality in pursuit of rapid and aggressive growth[45, 46]. In 2017, this strategy appeared finally to be catching up with Uber. In that year alone, Uber faced a federal criminal investigation into its Project Greyball[②] became embroiled in a legal fight with Waymo over its alleged theft of self-driving car technology, and was experiencing growing backlash from drivers over low pay and poor working conditions[47–49]. In addition, the company was embroiled in allegations of sexual harassment and a toxic work culture for women and minorities[50]. Many commentators thought this collection of scandals marked the end of the company—a fate that longstanding critics of Uber welcomed.

Focusing on the workplace culture allegations, the company's board of directors promptly hired former U.S. Attorney General Eric Holder (then at Covington & Burling) to conduct an internal investigation and issue a report, a high-profile step that was extensively covered in the media. The report resulted in the board adopting a series of corporate governance practices and ultimately firing then-CEO Travis Kalanick. This change in leadership and attendant set of institutional changes were generally understood to end the company's "wild west days" and to usher in a new era of a law-abiding Uber focused on "ensur[ing] a tone of support and a culture of compliance"[40, 51]. In line with this new culture, Uber dropped many of its more openly aggressive tactics, such as Project Greyball.

Uber's lawfulness response was an impressive display of threading the needle: it addressed the public attitude of Uber (as a deviant and morally suspect company) by signaling legal seriousness, while keeping intact a business model that was also a primary subject of critique[51]. Focusing its response on workplace culture allegations at its headquarters, Uber drew fire away from its continued use of pricing manipulations and other techniques to squeeze profit from drivers.

A second, more proactive example of the lawfulness response is Microsoft's approach to developing facial recognition technology. As questionable business practices of facial recognition companies have come to light[1], the social pressure to ban[52] or place a moratorium[53] on facial recognition technologies has

grown—even Alphabet CEO Sundar Pichai has suggested a temporary moratorium on facial recognition technologies may be needed[43].

Microsoft has called for legalistic restraint as one way to temper concerns while continuing development. The company is publicly refusing to sell their technology to California police (citing Fourth Amendment concerns), endorsing federal regulation, such as the Commercial Facial Recognition Privacy Act, and introducing its six principles for facial recognition software that include "lawful surveillance" and prohibitions against use for "unlawful discrimination"[54]. This middle path appeals to the restraint of law to narrow public critique of facial recognition to its most egregious (and, it is proposed, unlawful) applications, while preserving other areas of application intact. Microsoft's chief legal officer Brad Smith likened a wholesale ban to "try[ing] to solve a problem with a meat cleaver" when a "scalpel" is required to "enable good things to get done and bad things to stop happening"[9]. The lawfulness response provides precisely such a scalpel-like approach: a cautious-yet-optimistic program of continued development of facial recognition technology under the guiderails of existing law. As Smith notes, "This is young technology. It will get better. But the only way to make it better is actually to continue developing it. And the only way to continue developing it actually is to have more people using it"[9].

As these two examples show, the turn to lawfulness during moments of popular backlash serves an important role for companies. In the case of Uber, bringing in a high-profile legal investigator like Eric Holder—the embodiment of a trusted form of lawful authority, the Obama Justice Department—shifted perception of the company from lawless adolescence to reformed and responsible corporate adulthood, while preserving its core business model. In the case of Microsoft, faced with a far more aggressive regulatory alternative in the form of bans or moratoriums, the company emphasized the importance of continued, yet responsible, development of the technology. This tack grants Microsoft the ability to craft through law a basis for its own legitimacy: by proceeding with its business under the imprimatur of law, the company may reap the financial benefits of the technology without suffering reputational harm. In both examples, the lawfulness response is marshaled to chart a middle path, softening calls for abolition—of an entire

---

② Beginning around 2014, Uber used a program called Greyball. It operated this scheme in cities like Boston, Paris, Portland, and countries like Australia and China—all places Uber had been restricted or banned—to evade detection by using geo-fencing around government buildings and "greyballing" users identified as law enforcement or city officials[47]. While approved by Uber's general counsel at the time, other legal experts thought the program may constitute a violation of the Computer Fraud and Abuse Act or an act of intentional obstruction of justice, and a federal criminal investigation into the company's misleading tactics with local regulators soon followed.

business model or a technology—into steps for continuation, just in a more procedurally robust and accountable manner.

The lawfulness response offers companies a pathway to regain or retain legitimacy for their business in the face of accusations of injustice. It does so in part by collapsing the distinction between lawfulness and legitimacy in the company's actions. This separates out unlawful/illegitimate actions from lawful/legitimate ones—an important separation that distances those practices that are of central importance to a company's business from those that are not. By dealing seriously with the unlawful/illegitimate practices, the category distinction between these practices and the rest of the business is reinforced. This reinforced separation has significant material stakes. In the case of Uber, the lawfulness response undergirds an all-important distinction for the company: that sexual harassment at work is illegal, whereas harsh contracting terms for independent contractors are not. In the case of Microsoft, this distinction is proactive—a campaign to disambiguate the illegitimate/unlawful uses of facial recognition (backroom deals with law enforcement, warrantless searches), from the legitimate/lawful ones (a category the company argues requires further exploration). The unlawful actions thus identified and addressed, the company's remaining actions regain or retain legitimacy.

## 3 Lawfulness As Anti-Regulatory

Lending credence to the lawfulness response is that a corollary version of it—what I call the *legalist-reform response*—is accepted and even championed among some of big tech's fiercest critics. When such critics emphasize the *lawlessness* of company actions, it sets up technology companies to reply credibly to popular frustrations with the lawfulness response.

The legalist-reform response suffers from two limitations as a strategy for democratic egalitarian reform. First, it misdiagnoses the role of law in current processes of technological production as one of absence. Second, and more importantly, by invoking an absence of law or a failure to comply with existing law, such responses concede the status of such law as capable of expressing the particular demands of justice in the techlash. Such responses thus concede the legitimacy of lawfulness responses without specifying the substantive and normative commitments such an intervention should aim to secure and upon which legitimacy would seem to be contingent. Legalist-reform responses may thus articulate a claim that "compliance" or "regulation" is needed, but do not, in and of themselves, provide substantive or conceptual specificity regarding what such law should achieve or enact in order to be satisfactory.

Both limitations combine to make this form of critique conceptually vulnerable to anti-egalitarian agendas. Critics advancing legalist-reform agendas risk misdiagnosing the role of law and conceding the legitimacy of law. This then allows companies to defend exploitative business models as lawful and therefore legitimate, particularly by applying the "scalpel" of legal intervention to separate and excise the worst instances of abuse while preserving the core business practices that give rise to them. Both invoke a popular imagination of the role of law that is quite distinct from the role that law in fact plays.

### 3.1 Law as absent, law as present

Some of big tech's fiercest critics propose legalist-reform solutions. For example, Zuboff[55] reserves a key role for data protection and greater transparency in averting the disasters of surveillance capitalism. Her critique focuses on the lawless and un-governed "dark data continent of… inner life" that, absent any regulatory protection against plunder, is "summoned into the light for others' profit". She cites the General Data Protection Regulation (GDPR) as a significant positive force that may help make "the life of the law … move against surveillance capitalism". In her account, such laws provide a way to turn the interrogatory spotlight back onto tech companies. Others have similarly advocated the need for applying existing law, particularly fundamental rights protections, as "able, agile, and flexible"[56] when used against technology companies to "shape, apply, and enforce" data rights[57].

The enormity of injustice catalogued by these critics appears at odds with the solutions they propose in response to them. Indeed such proposals suggest that once companies do comply with laws like the GDPR— once the law has trained a spotlight on these companies' inner workings—they may credibly claim to engage in an "acceptable form" of surveillance capitalism: a transparent and compliant version. Legalist-reform responses concede the essential legitimacy of the legal frameworks that bind these companies, and in so doing concede the essential legitimacy of the business models that have developed within those frameworks. Under this account, the problem is not whether such technology—a platform optimized to exploit drivers, a technology designed for at-scale personal surveillance—should exist at all, but simply one of law's absence in ensuring its use is "up to code". Once companies achieve this standard of compliance, the problem is addressed.

Faith in a new regulatory regime to fill tech's legal lacuna can be misplaced, as companies actively work to shape such regimes and use them to further their ends. For instance, both critics and industry executives expected companies like Facebook and Google to come under harsh penalties and increased scrutiny for new attempts at aggressive data extraction under the GDPR. But enforcement has been largely absent, as under-resourced European authorities struggle to build complex investigations against wealthy international companies (though defenders would rightly point out that enforcement has picked up as of last year). More troublingly, companies have used the GDPR's consent rules to re-introduce technologies previously banned in the region[58]. In the U.S., state attempts to pass privacy legislation have come under heavy scrutiny from industry lobbyists; in Virginia, Amazon increased political donations tenfold over four years before successfully getting lawmakers to pass an industry-friendly privacy bill that Amazon itself drafted[59]. In Washington, Amazon lobbyists negotiated to have language inserted verbatim in the state's pioneering biometrics bill that meant the law, when it passed in 2017, would have "little, if any, direct impact on Amazon's services"[59]. Companies do not just advance new business-friendly regulatory regimes, but also shape existing doctrines into shields from accountability, distorting the doctrines of trade secrecy and commercial speech protections to protect valuable data assets[24, 60].

Legal observers have long understood that injustice is rarely a matter of law being absent. Instead, claims of injustice often arise from the ways that existing law structures patterns of exchange and establishes a particular distribution of power among actors[61, 62].

Katharina Pistor provides a compelling example in her account of the role law plays in facilitating contemporary capitalism by encoding global capital using certain well-trodden legal properties[63]. Her account makes clear that global inequality does not arise due to the capacity of assets and their owners to escape the law, but instead through their ability to *use* the law (and, by extension, the state) to distribute risk and reward in maximally beneficial ways. In his history of global neoliberalism, Quinn Slobodian further troubles the easy supposition of law's absence from the neoliberal justificatory narrative. He shows how the policy package of "privatization, deregulation, and liberalization" associated with the neoliberal mode of governance was at its core a project of legal institution building that embraced, rather than shrank from, active re-working of global projects of governance[64]. Britton-Purdy and Grewal[65] provided a similar account of law's active role in furthering and bolstering a neoliberal form of market-style governance. Cohen's[24] account of how law and technology shape one another in the emergence of informational capitalism similarly refutes the simple account of law as a powerful yet regrettably absent tool for disciplining the information economy. Instead, she shows how the formation of informational capitalism was as much a product of legal innovation as technical innovation.

What these analyses make clear is that law is a terrain of contestation for the regulatory arrangements that structure any social process—including our technology economy. Just as companies actively shape the ethics response to enhance their interests and shield them from accountability, so too does the daily business of informational capitalism actively rely on specific theories and forms of law. The problem is not law's absence from the technology industry, the digital marketplace or platform, or informational capitalism. The problem is precisely how existing law mediates the institutions that it enacts, the productive activities it encases, and the modes and myths of production it upholds and legitimates.

## 3.2   Conceding law's democratic legitimacy

The second (and perhaps more conceptually significant) limitation of the lawfulness/legalist-reform response is that it concedes the democratic legitimacy of law absent any interrogation of why such legitimacy may or may not be warranted, or under what conditions it may not hold.

Invoking law as a backstop against the harms of technology relies on the premise that law enacts our popular will regarding such harms. In other words, the lawfulness response implicitly or explicitly relies on the view that law: (1) can express our democratic will, (2) does express our democratic will, and therefore (3) offers a legitimate democratic response to the popular frustration of the techlash and social egalitarian claims that arise from it. The legalist-reform response appeals to law's role as a moral floor on what we owe one another: we may not trust technology companies, but we can trust the laws to which they are beholden.

This tees up corporate interests to invoke the lawfulness response as a way to trade on the authority and legitimacy of law itself. Where law is proposed and then invoked as moral cover, it serves to justify the patterns of wealth accumulation or technological development that law itself facilitates. Pistor notes that "strategic and well-resourced actors" quietly push for change outside the limelight of the public sphere; they couple such efforts with "claims to the authority of law to fend of critique and legitimize success"[66]. Indeed, few claims to legitimacy are more powerful at present than that something is "legal"[63].

Such normative appeals to law only warrant the legitimacy they invoke insofar as the law itself is widely accepted as a (sufficiently) legitimate expression of our social code of conduct and thus a viable channel for enforcing collective accountability. Yet a gap persists between the moral standing the lawfulness response means to invoke and the obligations its invocation in fact incurs—law's actual response to claims of injustice. This gap complicates how one evaluates the political purpose of the lawfulness response as well as the political limitations of its legalist-reform corollary. As a result, the lawfulness response (like the ethics response) may also be anti-regulatory, albeit in a more complex way.

To understand how this reliance on the legitimacy of law may be in tension with the project of democratizing technological progress, we need to turn from the techlash to a parallel phenomenon: the growing legitimacy crisis of law. The discipline of law itself is in foment over the normative gap between (1) the political ideals that form the basis of law's legitimacy and (2) how the law actually serves to bind and obligate agents to such ideals. This poses a significant challenge to the normative and political appeal of the lawfulness response. What does it mean to address the crisis of legitimacy in tech with the tools of law at a time when law is undergoing its own growing legitimacy crisis?

# 4 Law's Legitimacy Crisis

In near parallel with the emergence of the techlash, ferment is once again abroad in the law (to paraphrase Llewellyn[2]). This ferment has engulfed a broad swathe of legal regimes and institutions, but for the purposes of illustration, a focus on the Supreme Court is instructive. The Court is the paradigmatic institution of U.S. law. It enjoys cultural significance as a stand-in for the legal system more generally, and debates regarding the Court can plausibly be read to reflect broader political sentiment towards the legal system writ large. The Court is not just a cultural talisman; due to the practice of (and current standard for) judicial review, it has immense importance for the substance of U.S. law: how lawyers and regulators practice, interpret, and implement the law.

Many who once looked to the law as the primary means by which progressive justice is advanced have lost confidence that the Third Branch provides fruitful terrain on which to champion progress[67, 68]. Though still in its early days, this shift is noteworthy. The liberal-legalist mythos of the Supreme Court and liberal Justices as champions of progressive change has persisted for decades. This is despite the general trend over the last 40-odd years of the Court (and the justice system over which it presides) prioritizing the constitutional rights of corporate entities over human citizens[69, 70], eroding protections erected against discrimination[71–74], diminishing democratic governance at work and restricting employee and consumer access to recourse[75–78]. As recently as the spring of 2016, the Supreme Court was widely celebrated for providing progressive wins like *Obergefell* (2015)[79] and *Whole Women's Health* (2016)[80]. Liberal Justices, most notably Ruth Bader Ginsburg, were fêted as icons of the progressive movement, and many observed with optimism the gradual leftward drift of Justice Kennedy, the moderate swing-vote of the bench, on issues of free speech and criminal justice reform[81]. Yet four years later, the appointment of Brett Kavanaugh (despite the testimony of Christine Blasey Ford and mass protests in the wake of #MeToo) prompted popular liberal dismay at the inability of the justice system to hold itself above, let alone discipline, the political turmoil of our time. Kavanaugh's appointment marked, for many, a turning point in coming to terms with the politics—conservative

politics—of not just this Court, but the Court[82].

Of course, most reasonably sophisticated observers have always acknowledged that politics play some role in judicial reasoning and the workings of law. But the explanatory power of this role tended (in the "correct" account of both legal scholars and mainstream observers of the long 1990s) to be downplayed. On this view, while there is some partisan flavor to the judiciary, this has less to do with vulgar partisanship and far more to do with different theories of constitutional and statutory interpretation among judges that happen to fall along ideological boundaries③. On the whole, the prevailing sense was—and in notable swathes of the legal academy, still is—that there exists a meaningful "residual" in judicial reasoning once ideological affinity has been accounted for, a space that may be won through appeals to reason and precedent. For liberal-legal political reformers of the long 1990's this "residual" comprised a primary terrain of major progressive political campaigns such as the fight for LGBTQ rights, disabilities rights, and reproductive justice.

Yet in the span of a few years, political ideology—while still far from a dominant view—has become an ascendant explanans of judicial decisionmaking, as presumptions of apolitical judicial reasoning decline. On this account, the judiciary is not above and immune from politics; instead, it plays an active and willing role in conservative power consolidation. Three recent developments strengthen this alternative account. First, the mass appointment of under-qualified (by the old standards of the elite bar) partisan Trump appointees to the federal bench. Second, the failure of liberal-legalist tactics to discipline the excesses of Trump White House (e.g., the Mueller investigation and the Impeachment proceedings). Third, the willingness of the judiciary to play a deciding role in hotly-contested and highly political issues[83].

This turning point in ideological understanding coincides with the emergence of a community of legal scholars interested in methodological interventions in law. These aim to promote (1) a renewed sociological turn in jurisprudence[84, 85], (2) a greater attentiveness to the role law has played in facilitating inequality and excessive private power, and (3) a renewed ideological commitment to law's role in addressing these challenges.

③ Living Constitutionalism being a progressive or liberal theory and Originalism being prominent among the conservative judiciary.

Loosely grouped under the banner of "Law and Political Economy (LPE)", this methodological agenda unsettles the neat analytic separation between the economic considerations in private law and the political considerations in public law. LPE traces a methodological lineage to Legal Realism, a tradition that was itself closely allied with progressive aims. Like their Legal Realist forebears, LPE scholars largely share a commitment to social democratic or democratic socialist political reform, expanding the terrain on which legal reasoning and decision-making should be judged, and incorporating a more complete accounting of law's social consequences and structuring capacities.

Similar to reformers responding to the techlash, these legal reform projects aim to produce methodological interventions and agendas to develop and advance egalitarian and democratizing projects in legal scholarship and legal pedagogy.

Progressive critique of the anti-democratic nature of law is not new. The judicial branch has long been viewed as anti-majoritarian and operating at a technocratic remove from popular politics. Democrats as far back as Bentham have attacked the undue power of courts, recognizing the ideological power concealed in the judicial power to decide "what the law is"[68, 86].

In the U.S., progressives once similarly viewed the courts as the enemies of democracy. The American tradition of using "judges as secret agents of political transformation" has its roots in conservative, rather than progressive, fears of the majority[67, 68, 86]. In 1885, Englishman Sir Henry James Sumner Maine "sang the praises of the U.S. Supreme Court, as one of the many 'expedients' in the U.S. Constitution that would allow the 'difficulties' of any country 'transforming itself' into a democracy to be 'greatly mitigated' or 'altogether overcome'"[86]. American conservatives of the era, fearing the effects of mass suffrage, revived the then-obscure case *Marbury v. Madison* (1803)[87] to establish the constitutionality of judicial review over Congressional legislation (a reading of the case in contrast to how it was interpreted in its own time), and judges used this newfound power to invalidate progressive legislation. It "took the strife of the Great Depression, and fear of Franklin Roosevelt" to force the Supreme Court into granting many of the most significant pieces of legislation of that era, and which form the basis of the modern U.S. state. While the Progressives ultimately prevailed, FDR noted in 1937 that victory came at a "terrible cost"[86].

This antagonistic history makes the more recent progressive embrace of the Court all the more unusual. These critiques, both long-standing and renewed, are not for nothing. As the emerging crisis in law makes clear, the progressive embrace of legalist strategies to secure democratic agendas has produced meager results. The Warren court (the high point of progressive power on the Court) undoubtedly achieved victories for popular justice. Yet it "is worth asking whether the courts were necessary to the outcomes"—and whether it was worth expanding the political prominence of an antidemocratic power that "the right has now turned against progressives"[86].

The most prominent progressive victories in the court—de-segregation, voting rights, and legalizing abortion—have all been subjects of sustained erosion.④ By achieving these political goals as legal wins, their strength became subject to, and conditioned upon, the interpretative methods of judicial review—a method that in some sense marks the limits of these reforms. As the liberal character of the court waned and these victories have been reinterpreted ever more narrowly, the result has been to enshrine formal protections of these legal victories even as the functional social forms of injustice they were meant to prevent gain new purchase.

To take school de-segregation as one prominent example, more than sixty years after *Brown v. Board of Education* (1954)[88], functional segregation thrives even while being formally prohibited⑤. Despite this landmark judicial victory, more than half of American schoolchildren are in racially concentrated districts where over 75 percent of students are either white or nonwhite[89]. Even the districts most committed to integration have experienced notable re-segregation following successful court challenges from white parents[90].

The courts' dubious record presents a puzzle: should the project of democratizing tech and reviving an egalitarian spirit in law be to reclaim or reduce the power of the legal system over the substantive conditions of political wins and losses? If law is terrain on which the struggles of the techlash must take place, is this terrain we should seek to shield from the vicissitudes of political life or to expose further to popular accountability, access, and rule? Such questions go to the heart of longstanding debates regarding the emancipatory potential of the legal system and force us to contend with the limits of articulating the demands of justice in the language of courts, judges, and lawyers.

## 5    Democratizing Tech, Democratizing Law: Rescuing What Law May Offer

Despite the shortcomings of the lawfulness response, law will nevertheless play a key role in addressing the harms of the techlash. Yet doing so in line with egalitarian political aims will require re-invigorating the possibility of law to channel and enact democratic will rather than serving as a means for powerful interests to circumvent that will.

As discussed above, the processes of wealth extraction and social oppression at issue in the techlash exist by virtue of their encasement in law. The lawfulness response offers moral cover to continue engaging in these practices; the legalist-reform response either misdiagnoses these processes as occurring in the absence of law or appeals to existing legal tools incapable of addressing them. Instead, technology reformers can recast the problems of the technology's failure *as* problems of law's failure. Two clarifying reformulations of the twin crises of law and technology arise as a result.

First, this makes clear that both the crisis of law and the crisis in technology are part of a larger egalitarian political response to growing social inequality. Both legal and technical institutions structure (and drive) economic exchange, and thus serve to distribute power and resources. Both also enforce and enact the hierarchical relations that give shape to the social and cultural experience of contemporary life. Thus, both play a role in institutionalizing the current "justificatory narrative" of "property, entrepreneurship, and meritocracy" that informs how enduring inequalities are justified[25]. As this justificatory narrative grows more

---

④ The 2015 decision upholding constitutional protection of gay marriage undoubtedly ranks among the key progressive victories for the Court. Unlike the other examples noted here, the constitutional and statutory protections won in 2015 for members of the LGBTQ community have simply not been enshrined in law long enough to endure the sustained, decades-long legal attack that other progressive victories face. It remains an open question therefore whether these protections will face a similar fate of strong formal, negative protection, while the positive conditions required to obtain and exercise such freedoms remain out of reach for many.

⑤ It is worth noting that *Brown* is as much a legislative and democratic victory as a judicial one. Though decided in 1954, school integration in the South did not genuinely begin until a full ten years later, precisely because it ultimately required federal legislative action to enforce.

fragile and contestable, so too, do the legal and technical methods that encode and enact it. The role of both law and technology in facilitating this narrative informs how people evaluate our technology-based economy and our legal system.

That inequality has grown should come as no surprise—the hypercapitalist, neoliberal, or radical neo-propertarian ideology that gained prominence during the past several decades espouses the view that inequality is a necessary byproduct of freer markets. Under this view, inequality is required to produce a more efficient allocation of goods and to increase overall productivity (and thus overall wealth). Yet this has not turned out to be the case. Socioeconomic inequality has increased in all regions of the world since the 1980s and identitarian violence has accompanied the faith in market action and efficient allocation[25]. Inequality has had particularly pernicious effects in the US. While the top decile's share of income (not wealth, where differences are even more pronounced) has risen almost everywhere, in the US it rose from 35% to 48% of total national income. This increase for those at the top "has come at the expense of the bottom 50 percent" of the population, which as of 2018, commanded only 10% of the total national income[25] (emphasis my own).

In response to increasing inequality and its harmful social and political effects, reformers of law and technology share a broad methodological commitment to expanding the epistemic capacity of technical or legal methods to recognize and act on inequality and a broad political agenda of reforming technology or law to further social justice goals. Both express the growing democratic and egalitarian response to the challenges of rising inequality and social oppression.

Second, and perhaps of more importance for any positive legal and political agenda, we may reformulate the crisis of techlash as, at least in part, a crisis of the failure of law. Many of the tech's democracy problems may be reinterpreted as instances of law's democracy problem. Law has been instrumental in creating the social challenges of the techlash, and law, as a terrain upon which to create, enact, and enforce democratic reform, will be instrumental in addressing those challenges.

Both popularly and intellectually, the legal system's case for its own democratic legitimacy is increasingly thin. If the primary interests served by the law are those of the powerful against the powerless, how does such a legal system continue to justify itself in a democratic society, particularly in light of growing public egalitarian challenges against the failures of the status quo? If the legal system systematically cannot serve to correct for problems of inequality, unfairness, and oppression, or even provide basic recourse to make one's case against such social effects, then what, precisely, is it for?

Critiques of law as inherently anti-democratic suggest that one priority may be reducing the prominence of existing law (and the courts that uphold it) as the primary terrain on which we pursue the democratization of technology production, and focus instead on political battles to remake the law governing technology production. Yet even in its reduced role, law remains a primary means by which democratic will is expressed and enforced. The legal system is failing to provide its most basic function: to provide recourse and enforcement of our popular expression of justice through law. Its capacity to do so has been eroded over time and across core functions of law in ways that have, if not caused, then certainly exacerbated the crisis of democratic legitimacy in tech.

Another pathway is to embrace the terrain of law as essential to the project of democratizing technology production. This strategy, too, has a notable progressive tradition. Reflecting on E. P. Thompson's understanding of law's role in traditions of radical dissent, Gordon[62] notes that the Marxist historian was well aware of law's instrumental function as "a bag of weapons and tricks for the rich and powerful to use against the poor", but he "never succumbed to a crudely instrumental view of law". Instead, he understood law to be a "crucial element in the constitution of markets and relations of power and of production" that has the capacity to enact many different social roles and relations and is thus important terrain for radical dissent.

On this view, enacting meaningful legal institutions to discipline technology will require a democratic reinvigoration of law's capacity to express and enact popular democratic strength of will. Willy Forbath offers one robust positive vision of democratizing legal reform in form of constitutional political economy, developing a theory of constitutional law that does not ask what forms of redistribution the law permits, but instead what forms of redistribution the law requires: grounding political claims to the social and material conditions of freedom as necessary conditions for equal citizenship. These in turn produce a series of affirmative duties to secure these conditions against oligarchy[91]. Others disagree on whether a positive democratizing legal agenda needs to extend to constitutionalism, or focus instead on diminishing the power of constitutional constraints over popular legislation[92]. Yet both views hold that democratizing law will require departing from the predominant mode of *reinterpreting* law in anti-democratic courts in favor of remaking law in popular legislative political wins. These wins may occur at the local, state, or national level, take the form of new law (such as facial recognition bans or surveillance ordinances) or renewed law (such as revivals of FTC unfairness enforcement or substantive standards of merger review).

Waldron[93] notes that "a lot of what makes law worthwhile, … is that it commits us to a certain method of arguing about the exercise of public power". Situating the problems of techlash on legal terrain gives us recourse to this method, both to contend with the problems of the digital economy and to develop the democratic legal institutions in respond to them. Properly attending to the techlash and the lawfulness response will require re-politicizing "critical questions of self-governance" that have been lost as we cede democratic control of law in ways that facilitated mobility for some at the expense of the rest[66]. In other words, what we need is not technology that is more ethical, humane, or lawful. Instead, we must make our social institutions—including those of law and our tech-based economy—more democratic.

## Acknowledgment

This material is based on work undertaken at the Digital Life Initiative, supported in part by Microsoft. Many thanks to the ILI NYU fellows for their comments, as well as Elettra Bietti, Jake Goldenfein, and Ben Green.

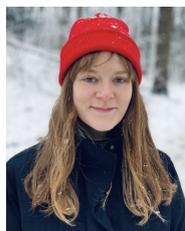

**Salomé Viljoen** is an academic fellow at Columbia Law School, and the former fellow and current affiliate at the Berkman Klein Center for Internet and Society at Harvard University. She studies how information law structures inequality in the information economy and how alternative legal arrangements might address that inequality. Her current work focuses on the political economy of social data. She is particularly interested in how (and whether) changes in the information economy create new kinds of legal claims to social and economic equality in social data production.

# Apologos: A Lightweight Design Method for Sociotechnical Inquiry

Luke Stark*

**Abstract:** While scholars involved in studying the ethics and politics flowing from digital information and communication systems have sought to impact the design and deployment of digital technologies, the fast pace and iterative tempo of technical development in these contexts, and the lack of structured engagement with sociotechnical questions, have been major barriers to ensuring values are considered explicitly in the R&D process. Here I introduce Apologos, a lightweight design methodology informed by the author's experience of the challenges and opportunities of interdisciplinary collaboration between computational and social sciences over a five-year period. Apologos, which is inspired by "design apologetics", is intended as an initial mechanism to introduce technologists to the process of considering how human values impact the digital design process.

**Key words:** values in design; values sensitive design (VSD); artificial intelligence; Values@Play; design methods; sociotechnical; ethics; values

## 1 Introduction

Human values pervade technical systems of all kinds[1], including computational technologies such as machine learning (ML) and other digital automation systems often termed artificial intelligence (AI)[2–4]. Over the past thirty years, work in fields such as science and technology studies (STS)[5, 6], social computing[7, 8], and critical studies of technology and race, gender, and sexuality[9–13] has interrogated the ways in which sociotechnical systems are conceived from and maintained in webs of normative preferences. Recent scholarship has paid particular attention to unpacking the granular technical affordances and design mechanisms in AI/ML[4, 14–19], through which particular human values are operationalized—and particular kinds of asymmetric power, injustice, and inequality maintained—in the everyday impacts of algorithmic technologies.

Alongside these academic developments, the

• Luke Stark is with the Faculty of Information and Media Studies, Western University, London, N6A 0A2, Canada. E-mail: cstark23@uwo.ca.
∗ To whom correspondence should be addressed.

increasingly obvious and deleterious societal impacts of social media platforms, artificial intelligence ventures, and other Silicon Valley firms have pushed the broad topic of "tech ethics", and the harm digital technologies do to marginalized groups, into international prominence[20–24]. Thanks to pressure from civil society groups, social justice organizations, activist scholars, and ordinary citizens, digital technology firms have been forced to begin to take responsibility for, and move to address their role in perpetuating and exacerbating social inequalities and power asymmetries.

As a result, digital technology firms have put much emphasis on high-level codes of ethical conduct around the development of technologies like AI systems, and have even begun to invest, albeit sporadically, in interdisciplinary teams of experts versed in the social impacts of computational media. However, methods and mechanisms to translate these high-level principles and diverse insights into actual decisions about products and systems are, on the whole, lacking[25]. Scholars involved in studying the ethics and politics of digital information and communication systems have long sought to have a concrete impact on the design and deployment of such artifacts in technical research and development (R&D) contexts such as academic laboratories and commercial development spaces[26−29]. However, the fast pace and iterative tempo of technical development in these contexts, and the importance of business decisions over wider societal concerns, have been major barriers to ensuring consideration of human norms and values is at play at every stage in the R&D process[30].

Urgent calls to grapple with the social, ethical, and normative implications of AI/ML and other computational systems on societies around the world make understanding and evaluating the ethics, norms, and values of all sorts of novel digital systems a necessity. Firms must move beyond lip-service to broader normative frameworks; critical and progressive responses to such technologies from lawmakers, regulators, civil society groups, and citizens in general can also benefit from such evaluations[21, 23, 31, 32].

Here, the author outlines a lightweight method for eliciting and evaluating ethics, norms, and human values in sociotechnical systems on a compressed time scale: Apologos. As a method, Apologos is inspired by the notion of "design apologetics"[33], which uses speculation to appraise technologies and their social impacts. This method seeks to present a coherent, practical, and principled approach to the problem of actively identifying norms, ethics, and values in a design process quickly. This method extends existing methodological frameworks[34−37] and draws on observations from a five-year case study of how conflicting human values intersect with the often complicated and contingent dynamics of designing digital systems[8].

The author also deploys insights from scholarship in design fiction[38], speculative design[39, 40], and the notion of "design apologetics"—thought experiments through which participants in a design process work backwards from existing or prospective artifacts to destabilize and make novel the normative notions behind digital technologies and systems. Such design apologetics ideally stimulate productive disorientation[41], and generative reflection about these technologies' possible social effects. This strategy is grounded in understanding and articulating how particular human norms, ethics, and values become incorporated in multifarious ways into the technical features of digital artifacts through design choices (both conscious and unconscious), and how such norms are made manifest in the use of technologies across diverse contexts[42].

As a method, Apologos seeks to use time itself as a visceral prompt to encourage novel thinking, and spark further in-depth reflection on and attention to the social contexts and lived realities of our everyday experience of technologies. Apologos is not intended as a panacea or replacement for more longitudinal or reflective design methods. As is a lightweight approach, it is potentially useful both in commercial settings and in broader participatory design contexts as an introduction to considering how human values are expressed in sociotechnical systems[43−45]. Apologos would be appropriate as an initial diagnostic exercise in a wide variety of digital design contexts: to begin to surface potentially confounding or complicated values tradeoffs[36]; point to spaces in the design process that might act as "values levers"[46]; and support space for novel engineering, participatory design, and refusal[47]. The theoretical and conceptual stakes of this article are thus twofold. First, what does the "nitty-gritty" of collaboration between computer scientists, social scientists, and humanists tell us about how to work across the socio-technical divide? Second, how can this experience shape an efficacious method for introducing audiences to the rich existing literature exploring how human values come to bear on the process of designing computational systems?

## 2 Lessons from Future Internet Architecture Project

Alongside already existing work on design prompts and methods such as Friedman's Valuse Sensitive Design[48] and Flanagan and Nissenbaum's Values@Play[36] methods, Apologos has been shaped by observations and insights drawn from the author's experience with the Future Internet Architecture (FIA) project, a multi-million dollar research project sponsored in two phases from 2010 to 2016 by the Computer and Network Systems (NETS) Division within the Directorate for Computer and Information Science and Engineering (CISE) of the National Science Foundation (NSF). The project involved four multi-institution research teams comprised primarily of computer scientists (henceforth referred to as the "computational teams"). These teams involved dozens of senior researchers and graduate students from more than fifteen different institutions. CISE also engaged several outside technical experts to advise and interact with the projects on an ad-hoc basis. The whole initiative was led by David Clarke of the Massachusetts Institute of Technology (MIT), a pioneer in the current Internet's early network architecture[49–52].

The FIA technical teams were asked to "explore, design, and evaluate trustworthy future Internet architectures". Braman[53] notes, "decisions about technology design and network architecture are, today, de facto social policy." Recognizing this fact, the FIA Project also included the participation of the Values in Design Council, a group "funded by NSF to involve a set of social scientists, lawyers, and economists in the FIA design process". This group of experts, for which the author were a research assistant, included more than twenty leading scholars from information science, science and technology studies, technology law, and digital media studies. The FIA-VID collaboration was a rare example of a large, ongoing, and formal attempt to bring technical expertise together with sociocultural, legal, and policy insight in the service of computational design. As such, the lessons learned from both the successes and productive failures of these collaborations—which were multiple and nuanced—have already served as the basis for insights around how interdisciplinary teams can work collaboratively to design systems with human values in mind[8, 49].

This portion of the paper is thus grounded in qualitative materials, including collaborative and individual field notes, reports, and participant observation of more than a dozen FIA Principal Investigator meetings from 2010 through 2016. The experience of the Values in Design Council demonstrated the need to get interdisciplinary collaboration right between computer science and the social sciences/humanities. It also showed the related necessity, which became ever clearer over the course of the project, to develop a tool with which to introduce and familiarize computer scientists and engineers with the notion of engaging with human values in the design process quickly and efficaciously. As Shilton[8] observes in reference to her own experience as part of the FIA-VID project, such "interventions struggled to make values reflection consistently relevant and engaging" to members of the computational teams involved in the effort. These challenging elements of the FIA project experience have shaped Apologos as a method, and suggest avenues for further refinement and elaboration. What ontological and epistemological assumptions clash in the conversations between computer scientists and social scientists/humanists? What material practices and institutional or disciplinary norms help or hinder collaborations? And how do shared aims, desires, and values motivate or inhibit working together? Answers to these questions have shaped the development of Apologos as a method.

## 2.1 Key emergent theme: Clashing technical languages & epistemologies

Perhaps unsurprisingly, each academic discipline works with and within a particular technical language. While these technical languages are often related depending on the shared history and concepts of disciplines, the same terms in each discipline can suggest not just different technical definitions, but also imply radically different epistemologies (theories of professional knowledge) that require work to be made commensurate.

Throughout the course of the FIA-VID project, one of the chief practical obstacles to collaboration between computational and STS scholars was the way in which both linguistic definitions and epistemological priors were misunderstood by project participants, at least some (if not most) of the time. Shilton[8] observed one key definitional confusion in the context of FIA-VID was the status of the term "interoperable", which for the computational teams was implicitly synonymous for technologically "neutral". Shilton observes this definitional overlap prompted a key value assumption: "the asserted belief in the neutrality of architecture was at least partially an expression of a core value: the interoperability of infrastructure"[8]. Central to Shilton's observation is that definitions, and indeed their epistemological foundations, are grounded and guided by value judgments: in this case, the historical legacy of values consensus in network engineering has made end-to-end interoperability synonymous with technical and societal impartiality (ibid.).

Debates about the definitions and the broader epistemological meanings of terms such as privacy and security also exemplify how FIA-VID participants from the computational and STS teams struggled with developing common definitions. Two FIA Principal Investigator (PI) meetings focused their agendas on the security and privacy implications of the various prototype architectures—the first, early in the project in 2011, and the second in 2015. The 2011 meeting also involved a group of outside technical security experts not attached to the four computational teams, who were tasked with advising those teams on their security and privacy plans as the projects moved forward from conceptualization to working prototypes. As privacy and security happened to be areas of considerable expertise for many Values in Design (VID) Council members, the meeting was also framed as an opportunity for the Council and technical teams to engage on an issue of common interest.

Unfortunately, it became apparent over the course of the meeting that "security" and "privacy" had different "technical" meanings for both groups. For the technical teams, "privacy" was generally articulated as a feature of their novel network architectures: something to be added as an additional layer or modification above more fundamental programming, but which was not in and of itself necessary for the functioning of the network. For the members of the VID Council, privacy was understood as a necessary outcome of network architecture, a default end state for ordinary users that ought to govern the technical teams' design decisions. One member of the VID Council noted that ordinary users of digital technologies rarely changed their default settings within their personal devices—and that privacy as a core human value ought to be integral to the teams' thinking.

In plain language, members of the technical teams wanted to know how to build systems and architectures that promoted user privacy as an actionable and imperative procedure, whereas members of the VID Council articulated privacy as an overarching end that could be enabled through a variety of different material, technological, and discursive means by the computational teams. The computational teams, a group made up primarily of computer scientists, operated primarily through what Abelson, Sussman, and Sussman term "procedural epistemology"—or in their words, "the study of the structure of knowledge from an imperative point of view, as opposed to the more declarative point of view taken by classical mathematical subjects." This epistemological logic, according to the authors, "provides a framework for dealing precisely with notions of 'how to'"[54]. This difference led to a focus on searching for concrete mechanisms through which to translate values into technical features. Differences between VID members and the computational teams thus did not seem to be grounded in a difference of underlying lived or embodied values per se. Rather, it was a disagreement around "lingo" tied to a deeper difference in epistemological constructions and prior assumptions held by the two groups.

In contrast, the members of the VID Council, as social scientists and humanists, operated through several overlapping epistemological frames. One, to borrow again from Abelson, Sussman, and Sussman, was a "declarative point of view"—which the authors associate with classical mathematical subjects, but which we here suggest is an epistemological frame that also fits with certain strands of humanistic and social scientific thought. Abelson, Sussman, and Sussman suggest that, "Mathematics provides a framework for dealing with precisely with notions of 'what is'." While the humanities and social sciences have never been accused of ontological clarity, their interest is often declarative or descriptive. Crucially, declarative epistemology is not the only flavor of knowledge construction in the human sciences—members of the VID Council also articulated discursive and critical epistemological frames through their comments.

In this particular instance of collaborative conversation, one that replayed itself around many other concepts across the life of the FIA project, both sides of the discussion were confused as to why a value each side agreed was valuable and important—privacy—seemed to nonetheless cause consternation (critically, in some contexts it became clear that privacy was not perceived as valuable by some in the conversation—a problem discussed later in the paper).

## 2.2 Key emergent theme: Conflicting disciplinary norms and incentives

Some of the most basic structural barriers to collaboration between the computational experts and VID Council members during the span of the FIA project involved the disparate and unaligned disciplinary incentives (and disincentives) around pursuing joint interdisciplinary projects. Studying interdisciplinary collaboration, particularly in the natural sciences, has become something of a cottage industry in recent years[55, 56]. However, as Callard and Fitzgerald[57] observe, this methodological focus has not extended to other academic arenas: with very few exceptions[58], the authors note, "there has not yet been any significant emergence of research on practices of interdisciplinarity within the social sciences and humanities."[57] Worse, interdisciplinary collaborations between natural and social scientists, though increasingly vital to understanding and engaging with complex sociotechnical problems, are both rare in practice and understudied in terms of their collaborative dynamics[59].

Computer science as both an academic discipline and a set of professional practice is structured quite differently than the disciplines from which members of the VID Council were drawn (chiefly law, media and information studies, and the social sciences). Moreover, the specific structure of the FIA project also introduced structural incentives that discouraged efforts at interdisciplinary collaboration between the computational teams and VID Council. These challenges around interdisciplinary collaboration were exemplified by the relative physical and temporal separation of the VID Council from the four technical teams from the outset of the project. Over the course of the FIA project, VID Council members and members of the technical teams met at a series of semi-annual two-day PI meetings hosted by various member institutions. Because of the size of both the technical teams and of the VID Council, few if any project participants were present for every meeting. Because of the length of the project, there was considerable turnover among junior researchers (doctoral students and postdoctoral researchers) involved in the technical teams. And while members of each individual technical team were bound together both by the shared content of the project and by potentially a sense of competition *vis a vis* the other technical teams, there was more physical, social, and intellectual separation between the individual teams, and between the teams and the VID Council, across the life of the project. This segmentation kept project participants in both disciplinary and project-based silos, which made communication, let alone collaboration, an ongoing challenge.

These gaps did not go unnoticed by FIA Project organizers, who eventually sought to reduce the distance between FIA Council members and the technical teams. At the May 2012 PI meeting, the NSF announced that it would provide extra funding to "embed" members of the FIA Council within particular technical teams, and several VID Council members were ultimately affiliated to a greater or lesser extent with particular technical projects[8]. However, the default case for interactions remained large, structured presentations and Q&A sessions at PI meetings. While disciplinary segmentation is well-known challenge for interdisciplinary work, the experience of the FIA project highlighted an under-noted effect of these silos: the ways in which a lack of sociality and a high degree of emotional distance hindered the project's intellectual and scholarly goals. Previous scholarship has found that sustained social bonds and shared physical space are helpful in supporting interdisciplinary inquiry by connecting individuals interpersonally as well as intellectually[60].

As Callard et al.[60] note, hierarchies within interdisciplinary collaborations can often short-circuit sustained interactions. Given the intermittent nature of most interactions between Council participants and technical team members, it is unsurprising that relatively few social bonds formed over the course of the project between the two groups. Divergent publishing conventions and expected project outputs and timelines also posed challenges for sustained cross-disciplinary collaboration across the life of the FIA project. For the computer scientists involved in the project, the core technical problems around the design of network architecture required a different collaborative cadence and pace of publication than studies by the social scientists in the VID Council. Interdisciplinary collaborations were curtailed as much by divergent professional schedules and incentives as they were by a lack of social connection between VID Council members and members of technical teams.

## 3   Apologos as Design Prompt

One important lesson of the FIA project is that many technical experts have little vocabulary or formal training in engaging with sociotechnical questions, but are eager and excited to do so if supported by appropriate research and design methods. One strategy that the Values in Design Council deployed with some success over the course of the FIA project was around design scenarios: suggesting the computational teams sketch out how their proposed systems might work in real-world conditions and what such conditions indicated about the values of the systems at hand.①
Apologos is first and foremost inspired by this

---

① Particular credit goes to VID Council members James Grimmelmann and Chris Hoofnagle for their initial championing of this approach.

experience, which could nonetheless have perhaps been improved by a standardized, lightweight set of prompts to help initiate such discussions across both the Council and computational groups.

As noted above, Apologos draws from the rich body of extant scholarship and design practice on values and human design, including how human values can be elicited or translated into technical means through particular design prompts or toolkits. Structured design prompts have become a popular method for facilitating design research and participatory co-design over the past decade in critical HCI and related fields. These prompts often feature a toolkit of playing cards or make use of other gamic elements as mechanisms to structure and vary deign outcomes[61]. Popular examples of such card-based design prompts include the *Envisioning Cards* from the University of Washington's Value-Sensitive Design (VSD) group[34, 62] among many others[63]. More broadly, the notion of "design sprints" and similar time-constrained methods to facilitate design work has proliferated in both academia and the digital technology sector[64]. Recent frameworks and toolkits for applying VSD methods to contemporary tech development environments are a salutary means of effectively integrating these design traditions to an applied context[65].

Apologos is inspired by these existing design traditions and methods. It also draws on, and is named after, the notion of "design apologetics", proposed by interface designers Nathan Shedroff and Christoper Noessel in *Make It So: Interaction Design Lessons from Science Fiction* (2012). Shedroff and Noessel borrow the term "apologetics" from theology, where it refers to reasoned argument justifying a religious doctrine. In the context of design in science fiction films and television programs, Shedroff and Noessel note they searched for ways fictional technologies "could" be explained to work, which "led to some interesting insights about the way technology should work"[33]. Other recent academic work has centered using speculative fiction as a prompt to produce novel design insights around human values and more broadly as a tool for critical design practice in computational settings[38, 66, 67].

Building on this prior work, Apologos is intended as an intervention to enable interdisciplinary groups to concisely consider a technology's sociotechnical impacts using design apologetics (as outlined in Table 1). An Apologos session is essentially a highly compressed version of the three-stage process laid out in Refs. [35, 36]: discovery of relevant values in a particular situation, implementation of those values as technical features, and verification that the assumptions made in these first two steps are broadly pertinent. Ideally, Apologos should be deployed either at the very outset of a particular design project or as a pedagogical prompt to introduce new audiences to sociotechnical analysis. Each session should be relatively brief: no less than 1 h, and potentially two hours or slightly more. The exercise is most practically tractable if undertaken by groups of 3−4 individuals; it is beneficial for larger design teams to break up into these smaller cadres and then reassemble as a larger unit for the final phase of the initial exercise as per below.

The first step in an Apologos session is for the session facilitator to ground participants in two definitions. First, ethos: a moral habit, character, disposition, or custom.

**Table 1　Apologos summary.**

| Phase | Component (total 60 min) |
| --- | --- |
| Discovery | Definitions (8 min) |
| | Brainstorming (2 min) |
| | List development (5 min) |
| Implementation | Design apologetics (2 min) |
| | Apologetics application (8 min) |
| | (Break for 5 min) |
| | Value judgment (5 min) |
| | Re-design (5 min) |
| Evaluation & follow-up | Reflection (5 min) |
| | Sharing across groups (8 min) |
| | Exercise feedback (7 min) |
| | Method transition (indeterminate) |

Second, techne: a variable and context-dependent art or craft **(approximately 8 min, in a 1 h session)**. Attentive readers will note here how the notion of ethos stands in for the broader definition of the "social" common to science and technology studies (STS) literature on sociotechnical systems. Any ethos is inherently a communal, and thus social, undertaking, but the term's normative connotation, and its emphasis on habituation, make it particularly apt for this exercise. After laying out these definitions, the facilitator should note the significant practical overlap between these two definitions around usual custom and lived concreteness—by extension the ways particular sociotechnical systems are inherently ones with unique sets of norms and values.

The participating small groups should then brainstorm three to four examples of everyday situations, scenarios, or activities where the definitions of ethics and technics already provided might be at play: a sociotechnical situation involving both aspects **(2 min in a 1 h session/time elapsed: 10 min)**. Once participants have identified a short list of scenarios or situations, such as flagging down a self-driving ride share vehicle or having one's mobile devices searched at a national border, each group should collectively develop, for just one of those scenarios, two parallel lists: one of 3−4 ethical principles or values the group associates with the situation, and one of 3−4 technical/material elements or features of the situation **(5 min in a 1 h session/time elapsed: 15 min)**. The facilitator should encourage these lists to be in parallel vertical columns on one sheet of paper. As a further prompt, the facilitator can also emphasize the utility of considering what Flanagan and Nissenbaum term "values seams"[36], or "places where multiple values are held in tension" within particular technologies or technical systems, as a means to remember that no technology is in any way "neutral".

The initial steps of an Apologos session—highlighting both the difference and potential overlaps in the definitions of *techne* and *ethos* and grounding this general discussion in particular domain contexts—are intended to immediately highlight the processual versus declarative divide in considering a value such as privacy in particular sociotechnical milieus. The group discussion intended to support these determination serves as what Shilton terms a "values lever"[46], or "practices that pry open discussions about values in design and help the team build consensus around social values as design criteria". Values levers are valuable to delicately de-lace Flanagan and Nissenbaum's "value seams", enabling collaborative examination and discussion of the multiple value perspectives within a particular sociotechnical apparatus. Through these steps, definitional and even epistemological differences can be, if not resolved, at least surfaced and recognized as such.

When these first stages of the exercise are complete, the facilitator should introduce the notion of design apologetics **(2 min in a 1 h session/time elapsed: 17 min)** to participants (as a reminder, apologetics are reasoned arguments or writings in justification of something, typically a theory or religious doctrine). Groups should not be informed about the details of the design apologetics stage before they formulate their two initial lists.

In their existing groups, participants should then perform apologetics across their lists **(8 min in a 1 h session/time elapsed: 25 min)**, speculating about or imagining reasonable ways a designer might pair the ethical principles or values discerned by the group in the first phase of the exercise with the technical features the group has picked, inasmuch as technical features can express a value or make it concrete. It is critical that participants do not change the content of either list to make this exercise easier or tidier. If members of the group cannot draw a reasonable connection between the values they first identified and the initial material or technical features, this failure should be specifically noted as another example of Flanagan and Nissenbaum's notion of "values seams". It is to encourage the exposure of such values seams that groups should not be informed about the design apologetics stage before they formulate their initial lists. This stage should be immediately followed by a stretch break for decompression and informal conversation about the exercise **(5 min in a 1 h session/time elapsed: 30 min)**.

After the break, the facilitator should focus participants on the most discordant or least convincing pairing of principles and technical features out of their list. Group members should then decide together whether they judge, in the context of the other values and features identified, whether it is the ethical value or the technical feature that is more important to the broader goals of the project, situation, or milieu **(5 min in a 1 h** **session/time elapsed: 35 min)**. For groups with little design experience, the facilitator might note that in making such a judgment, participants are expressing their own values as designers. For groups with more experience with sociotechnical analysis, the facilitator can observe beforehand that how each group understands and bounds their project or situation is itself a values judgment that inevitably shapes the decisions in this step.

After each group has made a judgment collectively, the facilitator should encourage group members to either (1) brainstorm a new technical feature that better makes concrete the value or principal the group has decided to prioritize, or (2) identify another principal or value (even a "negative" or unwanted one) suggested by the technical feature being prioritized to replace the initial ill-matched value **(5 min in a 1 h session/time elapsed: 40 min)**. Group members should then record answers to the following questions collectively **(5 min in a 1 h session/time elapsed: 45 min)**: How might this planned change interact with the other values and technical features already identified in the scenario or milieu? How might this change affect the broader parameters of the scenario or situation originally laid out?

Finally, the facilitator should quickly organize a jigsaw in which each small group splits and new groups of 2−3 people are constituted; if there is only one small group, then the following two steps can be combined. Each group member should briefly present the work of their original group to their new group mates and ask them to imagine themselves as people in the scenario the original group explored **(8 min in a 1 h session/time elapsed: 53 min).** Participants should ask their new groupmates whether there are values or technical elements of the original group's assessment with which they disagree, or that they would add or take away. Moreover, each member should ask the others how the change made by the original group around either a value or technical feature would affect them as imagined subjects in the scenario. For the last portion of the hour span **(7 min in a 1 h session/time elapsed: 60 min)**, participants should reconvene into a large group and the facilitator should ask for feedback on the design exercise, including about what was satisfying, enlightening, or useful; what was unsatisfying, frustrating, or incomplete; and if appropriate, what changes would improve the exercise for participants in a particular context.

One benefit to a lightweight design exercise like Apologos is that it can be deployed expeditiously and readily understood by teams from diverse disciplines and backgrounds. However, Apologos should by no means be the final step in a design and development process: as a final step, the group should make a plan to transition to a more fully developed method for designing with values in mind as appropriate. Frameworks such as Value Sensitive Design (VSD)[34, 48, 65], worth-centered design[68, 69], reflective design[70], adversarial design[71], and Values@ Play[35, 36], and working with values hypotheses[42] are all potential options for such a framework.

As an exercise, Apologos was designed by considering the experience of FIA participants, who faced a related set of conflicting norms grounded in the disciplinary and institutional nature of large-scale academic research. Insights from these successes and failures point to ways design exercises like Apologos can help bridge these institutional and professional gaps and to moments where broader and longer-term work is needed above and beyond any individual design technique. With both social and disciplinary challenges in mind, Apologos is designed to be a group activity that ensures a large group of disparate experts can participate in discussions around values and technologies and to be sufficiently brief that it can be deployed even at one- or two-day meetings. The social element of Apologos provides one possible scaffold for broader collaboration, or at least engagement, between members of diffuse interdisciplinary teams. Given sustained interactions and shared physical proximity are clearly ideal in this regard[60], Apologos aims to begin the process of social mixing, while building shared incentives around speculative design.

## 4 Conclusion: Disagreeing Over Values

Here, the author has proposed a lightweight design method, Apologos, intended to elicit and evaluate ethics, norms, and human values in sociotechnical systems on a compressed time scale. As already noted, Apologos is not a panacea or replacement for more comprehensive design methods or structural mechanisms to facilitate both interdisciplinary collaboration and broad, truly participatory responses to sociotechnical problems[72]. The author notes two particular limitations: the composition of the participants in an Apologos session, and the potentially limited impact of such a speculative design exercise.

As with all design, the possible outputs of an Apologos session will be bounded by the experiences, positionality, and perspectives of those in the room—as well as the shortcomings inherent in the ideology of "design" thinking itself[73]. This is of course an argument for engaging diverse teams of participants in the work of design[19]. However, it is also a warning that, as should be obvious, no single Apologos session is sufficient to adequately explore the sociotechnical terrain of any given computational artifact. The broader literature on the merits and challenges of participatory design in computing and digital media points to the shortcomings of even comprehensive methods for widespread inclusion in the design process, and Apologos should thus be understood as an introduction to this much larger and thornier area of practice. Apologos is an entry point into sociotechnical/interdisciplinary collaboration but sustaining and fostering that collaboration require much additional work.

A second limitation, related to the first, concerns how impactful the outcomes of one, or even many Apologos sessions can realistically be in changing the usual activities of a startup, corporation, or public institution[21, 74]. An Apologos session, or even several, will provide sufficient insight to change policy. However, as an entry to thinking critically about the sociotechnical landscape, Apologos is potentially useful in providing a language and framework for technologists and others who have not had a structured means to consider such questions before. Apologos should be seen thus as an introductory component of a much broader set of developments around the training, education, regulation, and design of digital technologies.

To conclude, the author wants to flag one challenge that can be adequately solved only through such broader structural mechanisms: the way divergent disciplinary, professional, and personal norms and expectations fundamentally shape how values are articulated and incorporated into design decisions, and how such disagreements can reflect a fundamental clash of values[42]. Although differences in social, professional, and epistemological norms comprised many of the roadblocks to collaboration during the course of the FIA project, at times project participants disagreed on an ontological level: about the fundamental values at issue in the development of network architecture and about which values are more or less important. Privacy is one clear example: some members of the computational teams were simply not convinced that human privacy as a value superseded others such as speed, user convenience, or network security.

These disagreements entail what Flanagan and Nissenbaum term "values trade-offs", or moments within the process of creation in which some values are prioritized over others[36]. While Apologos provides a mechanism for surfacing and highlighting such values trade-offs and their possible effects, it does not provide guidance *per se* on how to adjudicate between such tradeoffs. As Flanagan and Nissenbaum observe, "It is not surprising to find that design projects (particularly those with multiple requirements, goals, constituencies, and constraints) are rife with clashes and conflicts"[36]. And while Apologos does provide for identifying "negative" effects of potential design decisions as a means for some guidance, on what scale such negatives are judged remains at the discretion of the particular group of people doing the designing. Their clashes and conflicts, and the broader structural conflicts they represent, will be both unique to each situation and challenging in all cases[19, 42].

Nonetheless, Apologos has utility as one method among many in the broader conversation around how to account for human values and ethics in digital technologies. Lightweight exploratory methods like Apologos will ideally support space for broader conversations around novel engineering solutions[38], the necessity of participatory design across technical fields[25], and the necessity of refusal or non-deployment as an R&D option[47]. As sociotechnical analyses of pressing societal challenges become more urgent, and more frequent, we need ongoing focus on the work of interdisciplinary translation and design implementation as we navigate the ethics and values designed into and emerging from digital technologies.

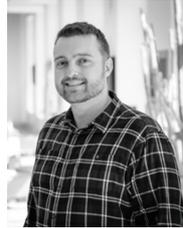

**Luke Stark** is an assistant professor in the Faculty of Information and Media Studies at Western University in London, ON, Canada. His work interrogating the historical, social, and ethical impacts of computing and AI technologies has appeared in journals including the *Information Society, Social Studies of Science, and New Media & Society*. He was previously a postdoctoral researcher in AI ethics at Microsoft Research, and a postdoctoral fellow in sociology at Dartmouth College; he received the PhD degree from Department of Media, Culture, and Communication at New York University in 2016, and the BA and MA degrees from University of Toronto in 2006 and 2008, respectively.

# Creating Technology Worthy of the Human Spirit

Aden Van Noppen*

**Abstract:** Spiritual caretakers have been present in every culture throughout human history. We know them as ministers, rabbis, lamas, shamans, imams, chaplains, gurus, and wise elders. In modern, secular times, they also include therapists, social workers, meditation teachers, and more. These caretakers support us through birth, death, and many of the most intimate and complex parts of the human experience. They use skills honed over many years that require paying radical attention to the humanity of others. Yet where is this expertise to be found in the creation of the digital technologies that have become portals through which we live, love, learn, grieve, and connect with our communities? Those who design and build digital technology must accept that we have become de-facto spiritual caretakers with the power to treat the well-being of humanity with care or with negligence. Unfortunately, caretaking is a role that computer science degrees do not prepare people for, few business models optimize for, and algorithms can not easily solve. This article outlines two concrete best practices that can help foster genuine responsibility and care on the part of technologists and technology companies. First, technologists must recognize that what we create is an expression of our own inner state. Our spiritual and emotional health is inextricably linked with our ability to build technology with responsibility and wisdom. Second, technologists must create an empowered seat at the table for those with the expertise and orientation needed to care for our souls, whether from a religious or secular lens.

**Key words:** spirituality; ethics; well-being; humane technology

## 1 Introduction

Spiritual caretakers have been present in every culture throughout human history. They support people through birth, death, and many of the most intimate and complex parts of the human experience that exist in between. Spiritual caretaking requires paying radical attention to the humanity of others. Yet its nature is shifting in dramatic ways in the Digital Age, when technology mediates many aspects of the human experience. When Siri and Alexa are on the receiving end of suicidal pleas[1] and vaccine misinformation spread on social media is killing tens of thousands of people, we live in a world in which spiritual care is frequently in the hands of algorithms. This means that the technologists who create them are de-facto spiritual caretakers of our world. Unfortunately, caretaking is a role that computer science degrees do not prepare people for, few business models optimize for, and algorithms can not easily solve.

Providing spiritual care has traditionally been among the most respected roles in a society. Spiritual caretakers include ministers, rabbis, lamas, shamans, imams, chaplains, gurus, wise elders, and more. In modern, secular times, they also include therapists, social workers, and meditation teachers. In most cases, they draw on tradition, training, and ritual that have been passed down for thousands of years. People in these roles deal with some of the most ineffable yet fundamental dimensions of the human experience, from our deepest grief to our greatest joy, and help us maintain a sense of connection to something larger than ourselves. This work, often referred to as "pastoral care" in Christian traditions, requires wise attention, compassion, and an understanding of the responsibility that comes with

● Aden Van Noppen is with Mobius, San Geronimo, CA 94963, USA. E-mail: Aden@mobi.us.org.
∗ To whom correspondence should be addressed.

accompanying people through existential questions of meaning, purpose, and our very existence. In 590 AD, Pope Gregory the Great wrote a manual of pastoral care that is still a foundational teaching text in seminaries and divinity schools around the world. In the manual, Gregory writes that "the care of souls is the art of arts"[2].

It is not enough for individual technologists to accept the spiritual implications of our work—the responsibility for spiritual care extends to the institutions within which we are housed. The interfaith Association of Professional Chaplains states, in reference to providing spiritual care in a hospital setting, "many persons both inside and outside traditional religious structures report profound experiences of transcendence, wonder, awe, joy, and connection to nature, self, and others as they strive to make their lives meaningful and to maintain hope when illness strikes... Institutions that ignore the spiritual dimension in their mission statement or daily provision of care increase their risk of becoming only 'biological garages where dysfunctional human parts are repaired or replaced' (Gibbons & Miller, 1989). Such 'prisons of technical mercy' (Berry, 1994) obscure the integrity and scope of persons".[3] Tech companies that ignore the spiritual dimensions of their work become like these hospitals: garages where superficial desires are met but the impacts of their products on our holistic well-being are overlooked.

Those individuals and institutions wishing to rise to the task of true spiritual care, which we must in order to thrive, will need to allow this commitment to lead us past our comfort zones. Rising to the task means seeing and accepting the suffering we cause ourselves and others by adhering to the status quo, and then taking brave action to change course at a crucial moment in the history of humanity and technology. It will challenge us to face our fears and the dark sides of human nature and capitalism. It may mean altering the underlying structures, belief systems, and assumptions that drive technologies, business models, cultures, and organizations as we know them.

Whether we realize it or not, technologists and technology companies are in a position to decide if we treat humanity with care or with negligence. Seriously accepting the responsibility of spiritual caretaking will require valuing care for human souls over care for profit. Choosing profit will have grave implications for the well-being of humanity and the planet.①

This article illustrates two concrete best practices that can help foster genuine responsibility and care on the part of technology companies. These suggestions are based on my personal experience working at the intersection of technology, ethics, and justice as a senior advisor to the US Chief Technology Officer in the Obama White House, as a resident fellow at Harvard Divinity School, and more recently, founding and leading Mobius. Mobius is a collective of technologists, entrepreneurs, scientists, spiritual teachers, artists, and organizers working together to create a more responsible, compassionate, and just tech ecosystem.

The two interventions I offer here are by no means a complete solution. Meaningfully addressing the harms of technology requires an ecosystem of interventions, including regulation, employee and consumer movements, values-oriented business models, empowered ethics teams inside companies, and addressing the toxicity of the underlying systems that gave rise to them in the first place. But all of these efforts will not create technology that is worthy of the human spirit—technology that shifts us from greed to generosity, from anxiety to ease, that heals us and brings us together—unless we broaden the frame. Curing what ails the tech sector also requires us to see the role of technologist through the lens of caretaking.

First, technologists must recognize that what we create is an expression of our own inner state. Our spiritual and emotional health is inextricably linked with our ability to build technology with responsibility and wisdom. Second, technologists must create an empowered seat at the table for those with the expertise and orientation needed to care for our spiritual and emotional well-being. Both practices have been key to spiritual caretaking for millennia. If adopted as part of a larger ecosystem of changes, they could help mitigate the harms of technology, and perhaps even lead to more technology that brings out the best in humanity.

## 1.1 A note on language

This article attempts to bridge between the spiritual and the technological. Despite the fact that these two domains are inextricably linked, they rarely speak to each other. This makes language inherently difficult.

① The planet is included here since digital technology so often disconnects humans from the natural world and makes it easy to "numb out" instead of seriously engaging with the realities of climate change, the implications of our treatment of the planet, and the action that is called for in response.

Words such as "soul" and "spirituality" can understandably be alienating in secular contexts, but one does not have to believe in God, associate with a religious tradition, or use this language to connect to the underlying concepts. When I say "spiritual well-being", I am referring to a healthy inner life, sense of wholeness, and connection to something larger than oneself. The nearly ubiquitous use of "well-being" in secular spaces refers to many of the same aspects of the human experience.

I also use "technologist" to refer to a wide range of roles and orientations. For the purposes of this piece, a technologist is anyone making decisions that influence technology products or services, regardless of their role. For this reason, I include myself in this category. Finally, I recognize that there are many kinds of technology. When I say "technology" in this article, I am referring primarily to consumer-facing digital technology.

## 1.2 How we got here

Accepting and meeting the responsibility of "care of souls" contain unprecedented challenges when mediated through technology built to succeed in the context of capitalism, an economic system that rewards greed, division, and competition. Barriers include incentives structures, societal norms and narratives, and the culture of the tech sector, to name a few. These interconnected systemic conditions give rise to an endlessly complex web of technologies that are integrated into the fabric of nearly every aspect of the human experience.

While there are many benefits to this integration—the democratization of information access, the spread of social movements, and the ability to connect with loved ones across continents—the dark side is also increasingly clear—hacking of elections, the spread of violent extremism via social media, fake news and the degradation of truth, and the mental health implications of tech addiction. When business models are built to maximize the time we spend engaging with technology, it is no wonder we become afraid, violent, polarized, and addicted. When selling our data is a primary revenue stream, it is no wonder we are exposed to highly targeted political ads and our democracy breaks down.

Yet we may be at a tipping point. There is a perfect storm that may create the conditions needed for greater alignment between technology and humanity. Journalists, academics, consumers, and tech employees are speaking out about the negative impacts of technology. Former and current tech executives are admitting to feelings of guilt over creating "tools that are ripping apart the fabric of how society works"[4]. All of these recognition and vocalization are leading to a reckoning in the tech sector with unprecedented levels of motivation and courage to address the negative impacts of tech on our well-being.

This makes Silicon Valley akin to a patient with a chronic illness in its first flare-up. Some are reacting by deflecting and denying, trying to prevent anyone from knowing we are sick[5]. Some are focused on treating the symptoms quickly and superficially to get through the crisis of the moment[6]. A third group wants to find cures. This group is growing and increasingly organized. We are made up of passionate consumers, academics, foundations, tech employees, and civil society organizations such as Data & Society, the UCLA Center for Critical Internet Inquiry, the Algorithmic Justice League, and Mobius, the organization that I lead. Together, we are addressing the challenge from a variety of angles and beginning to create change that seemed impossible until quite recently.

## 1.3 An alternative

As long as technologists build tools that touch nearly every aspect of our lives, rising to the task of spiritual care in the Digital Age will be an essential component not just of ethical and responsible design, but also of the larger systems change that is needed. I outline two powerful yet realistic strategies as places to start. They will not come close to shifting the direction of tech alone—they are intended to complement but not replace other regulatory, cultural, economic, and educational reforms to the tech industry.

First, technologists must recognize that our own spiritual and emotional states are inextricably linked with the ability to create responsible and humane technology. Systems theorist and senior lecturer in the MIT Sloan School of Management, Otto Scharmer, writes about a major blind spot in leadership theory, organizational development, and our everyday lives: we rarely recognize the importance of the inner state from which our actions, decisions, and creations originate. Scharmer writes that the "inner state of the intervener is perhaps the most important determinant of the intervention"[7]. Put another way by Wheatley, "without reflection, we go blindly on our way, creating more unintended consequences, and failing to achieve anything useful[8]." It is no wonder there are so many negative consequences of technology when we are surrounded by innovation created from states of anxiety, rushing, and greed.

I recognize that slowing down is exceedingly difficult in many tech companies, where company cultures, incentives, working conditions, and even job security rely on moving as quickly as possible. Even if it was easy, slowing down and bringing reflection, mindfulness, meditation, and other well-being practices into tech and entrepreneurial cultures are also not enough. While this can set important groundwork for shifting out of destructive inner states like anxiety and greed and into the thoughtful, clear, and compassionate states needed to responsibly design and build tech, it must be accompanied by an awareness that these very same tools can be dangerous when used primarily as coping mechanisms to feel less anxious and more productive at the individual or company level. In doing so, there is a risk that they become like numbing agents that actually keep the status quo in place. Their misuse can make it easier to ignore pain, including the pain caused by the products technologists build. True spiritual growth will actually lead a person to more uncomfortable places and support the clarity and strength needed to change course. Chögyam Trungpa, Tibetan Buddhist meditation master who played a major role in the dissemination of Buddhism in the West, wrote, "meditation is not a matter of trying to achieve ecstasy, spiritual bliss or tranquility, nor is it attempting to be a better person. It is simply the creation of a space in which we are able to expose and undo our neurotic games, our self-deceptions, our hidden fears and hopes."[9] This deeper work is required to create the spiritual and emotional states needed to build responsible and humane technology.

Second, technologists must create an empowered seat at the table for those with the expertise and orientation needed to care for our spiritual and emotional well-being. Dealing with the delicate territory of the soul requires knowledge, skills, and methods that are largely absent in tech companies. I am not saying that technologists need to be expert caretakers. In fact, it would be dangerous to assume we could be. We do not expect everyone to have the legal knowledge of a lawyer, but no major tech company would imagine creating a product without consulting one. Similarly, we need the humility to recognize the nuanced caretaking knowledge and wisdom that exists outside the walls of our companies and seek out that expertise. Their perspective should be embedded in product design and strategy at all levels.②

I offer these two strategies based on my experiences supporting tech leaders who are committed to taking the responsibility of spiritual caretaking seriously. I work with technology leaders who share the mission to put our individual and collective well-being at the center of what they are building. Some of these people are among the most influential in Silicon Valley: they control multibillion-dollar portfolios, oversee tens of thousands of employees, and influence the direction of technologies that affect billions of people globally. Yet, even with this mission and power, they are working within systems, incentive structures, and cultures that are designed to keep the status quo in place.③

Mobius supports these mission-aligned leaders in two overlapping ways. Each contributes to the shifts called for above. First, we bring these leaders together, across competitors, into a nurturing and supportive community that builds the trust needed to make their company-specific work bigger than their sum of its parts. Second, we curate groups of the world's leading experts on well-being and caretaking to advise on product and strategy. These experts have deep wisdom on how to care for our well-being. They span from senior spiritual teachers (such as Jack Kornfield and Roshi Joan Halifax) to prominent neuroscientists studying the development of compassion and empathy (such as Dr. Sará King and Dr. Emiliana Simon-Thomas), and scholars of racial justice and healing (such as Dr. Angel Acosta and john a powell). While some have previously been invited to

---

② This often requires bringing in people who are not currently on tech teams, but one must be careful of creating the false dichotomy that technologists cannot also be spiritual caretakers and spiritual caretakers cannot also be technologists. There are brilliant people who bridge that divide, but it is rare to find that combination in a single person or existing tech team.

③ Some may argue that senior leaders at the tech giants are inherently unethical and should not be supported. We choose to support these leaders because we believe that systemic change requires shifts from both inside and outside the major tech companies. We know firsthand that there are many people working at Google, Facebook, Twitter, and other big tech companies who are deeply concerned with the negative consequences of their technologies. Instead of being in denial, they are pushing for responsible strategies to change course. These employees are found at all levels of the companies, from the most junior employees to the C-suite. Mobius works with senior executives because of the scale of their influence, and we collaborate closely with other civil society organizations who are supporting mission-aligned tech employees throughout all levels of the companies.

visit a tech company to lead a meditation or give a talk, they are almost never in the rooms where products are designed. Mobius also weaves ancient practices such as meditation, reflection, and ritual into our facilitation in order to create inner states of compassion, clarity, and courage while decisions are being made about products. Each of these strategies is complex, and we recognize that there are potential unintended negative consequences of our work as well, including the possibility of "ethics washing" when tech companies are able to say they consulted with experts regardless of whether they integrate the recommendations.

While our work is far from a silver bullet, our hope is to help equip technologists to more responsibly take on the work of spiritual and emotional caretaking in tech's next chapter. This article is grounded in what I see as we work to support tech leaders and their teams to create technology that not only avoids harm, but also brings out the best in humanity.

From this perspective, this article takes a close look at technology's complex impact on our individual and collective well-being, followed by a deeper dive into each of the two interventions called for above and some of the promising interventions that are already or should be happening. Finally, I discuss what this all adds up to and how it fits into the growing ecosystem of changes that, even though we have a long way to go, are pushing the tech sector to value our shared well-being over the fastest route to profit.

## 2   The Status Quo

Some say we are in the midst of a "Fourth Industrial Revolution", driven by the rapidly growing and nearly ubiquitous integration of digital technology into all parts of society[10]. This is by no means humanity's first technological transformation, but never has a transformation been so intimately linked with nearly every aspect of our lives. Billions of people use technology as a primary portal through which to work, play, learn, and love. As a consequence, the direction of technology has profound and rapidly shifting effects on our individual and collective well-being.

### 2.1   The dark side of tech—implications of negligent "spiritual caretakers"

The dominant business models, cultures, and norms in the tech sector have led to technology that frequently and often consciously preys on the most vulnerable parts of human nature. We are surrounded by devices and platforms that hijack our attention and keep us from connecting deeply with ourselves, others, and the physical world around us. The negative implications of such technology are increasingly clear. Tech executives and their teams are facing one ethical quandary after the next, ranging from the spread of misinformation breaking apart our civic fabric, to the mental health implications of seventy two percent of teens in the United States feeling the need to immediately respond to notifications on their phones, to a steady stream of atrocities such as Facebook posts inciting genocide against the Rohingya Muslims[11].

Many problems stem from the mental and emotional effects of spending more time connected to our digital devices. Adults in the United States spend an average of eleven hours a day interacting with screens—nearly half our lives[12]. Netflix's CEO recently said that sleep is their biggest competitor[13]. I would argue that the health of our intimate relationships is a close second. As Turkle writes, "We have become accustomed to a new way of being 'alone together'. Technology-enabled, we are able to be with one another, and also elsewhere, connected to wherever we want to be."[14]

As just one example, millions of young people allow their friendships to hang in the balance of whether they maintain their Snapchat "streak", a feature that relies on friends sending direct snaps back and forth with each other every day. The longer one goes without breaking the chain of communication, the longer the streak and the "stronger" the friendship. Some Snap users manage hundreds of streaks simultaneously, and many go so far as to have their friends log into their accounts to maintain their streaks if their phone is taken away by parents[15]. This highly addictive feature preys on a wide swath of a psychologically vulnerable population—sixty nine percent of American teenagers use Snapchat[16].

#### 2.1.1   Designing for addiction

It makes sense that there are so many negative impacts when we look at the context within which these technologies are created. Engineers and designers are frequently driven to build highly addictive features because of the business models of the companies that employ them. "It is as if they are taking behavioral cocaine and just sprinkling it all over your interface and that is the thing that keeps you coming back and back and back", said Aza Raskin, former senior leader at Mozilla and Jawbone. Raskin invented the "infinite scroll" in 2006, an extremely common feature of apps that allows users to endlessly swipe down through content without extra click[17]. The infinite scroll was designed to be "maximally addictive … if you do not give your brain time to catch up with your impulses you just keep scrolling". This matters because "in order to get the next round of funding, in order to get your stock price up, the amount of time that people spend on your app has to go up", Raskin said. "So, when you put that much pressure on that one number, you are going to start trying to invent new ways of getting people to stay hooked." Raskin was ironically working at a tech company called "Humanized" when he invented the infinite scroll. In addition to the interventions and mind shifts discussed in this article, Raskin's point reinforces the importance of actions such as changes in policy, funding, and business models.

Raskin went on to cofound the Center for Humane Technology (CHT) in 2018 with former Googler Tristan Harris. CHT is part of a growing set of advocacy organizations that are building a movement to "realign technology with humanity". Raskin is among a relatively large community of technologists who admit feelings of guilt about the consequences of the tools they helped create and are working to shift the direction of the tech sector as former and current tech insiders.

Chamath Palihapitiya, Facebook's former vice president for User Growth, left the company in 2018 said he felt "tremendous guilt" over his role in creating "tools that are ripping apart the social fabric". He said, in reference not just to Facebook, but to the wider online ecosystem, that, "The short-term, dopamine-driven feedback loops that we have created are destroying how society works. No civil discourse, no cooperation, misinformation, mistruth. This is not about Russian ads," he added. "This is a global problem. It is eroding the core foundations of how people behave by and between each other"[4].

It is not just former big tech executives speaking out. There are countless scholars and activists who have been sounding the alarm bell for decades. As many point out, seemingly minor design choices, such as Snapchat's fire emoji that indicates whether a streak is still going, the buzz of the phone with each new email, or the infinite scroll that keeps us refreshing our feeds by swiping down on the screen, add up to a bigger picture with grave implications for our mental health and the health of our close relationships, civic fabric, and even our planet[18].

### 2.1.2 The fuel of toxic culture

In addition to the business models, the dominant culture of Silicon Valley drives people to create technology that treats the well-being of humanity with recklessness. This is true on the company and industry levels. "Move fast and break things" is not how pastoral care works. Even though Facebook founder Mark Zuckerberg has publicly changed this company motto, it is in their cultural DNA. Facebook structures their strategic planning and performance reviews in "halves", or six-month horizons[19]. The public pays the price of Facebook's short-term thinking. For example, algorithms designed to maximize our time on the site have numerous consequences, many of which can be avoided with scenario planning and foresight. One such consequence is that these algorithms separate us into "filter bubbles" within which we are primarily fed content that we already agree with, thus making our worlds smaller instead of bringing us together[20]. Moving fast and breaking things do not stay within Facebook's walls. It is indicative of a larger culture of "disruption" and the common belief that more and faster is always better. This orientation runs completely counter to acting with awareness, intention, and care.

This culture of speed and recklessness is not unique to the tech sector, or even to the private sector. It is pronounced across most industries and seeps into people's private lives by the nearly ubiquitous presence of our devices. People suffer information overload and the expectation that we are constantly plugged in and available. In his manual of pastoral care, Pope Gregory the Great warned about the impact of this fractured attention. When the minister distracts their heart "with a diversity of things, and as his mind is divided among many interests and becomes confused, he finds he is unfit for any of them and becomes so preoccupied during its journey as to forget what its destination was"[2]. Jack Kornfield, a well-known American Buddhist teacher and Mobius founding senior advisor, explains it another way. "We live in a society that almost demands life at double time, speed and addictions numb us to our own experience. In such a society, it is almost impossible to settle into our bodies or stay connected with our hearts, let alone connect with one another or the earth where we live"[21]. Even those who go into a tech company with a clear social mission are prone to forget the destination when they are swimming in rapid currents of short-term targets, emails and slack notifications, and rushing to release new features before the competition does.

The tech sector's bias toward speed and short-term thinking are also compounded by the nature of the ethical reckoning we are going through. Since it is like being diagnosed with an illness that has no simple cure and constantly evolving symptoms, there is understandably a new level of fear and overwhelm that puts many technologists into crisis mode, even as we try to work toward solutions. This means strategies to make things better are often created within the same short-term, quick-fix, and fearful approach that got us into this predicament in the first place.

In contrast, nearly every spiritual tradition teaches us that contemplative practices and slowing down to gather and focus attention are a necessary step towards responsible and wise action. Jews observe the Sabbath by taking a full day of rest, reflection, and prayer every week. Jews and non-Jews are putting a modern spin on Shabbat by observing "tech sabbath" as a sustained period of unplugging[22]. Observant Muslims perform ritualized prayer called "Salah" five times a day. This practice of stepping away at regular intervals is not only to connect with God, but also to "purify the heart", which in Islam is considered to be the center of all feelings, emotions, desires, remembrance, and attention. This practice of stopping, resting, reflecting, and reconnecting with the heart is a foil to the modus operandi of most tech companies.

## 2.2  The high side of tech—enabling the most positive human qualities

Digital technology can and often does enable us to live more connected lives of meaning. For example, there are transgender teens in rural America who develop their emerging queer identities online through social media affinity group[23]. Facebook introduced thoughtful memorialization features that recognize the complex emotions that are intertwined with the Facebook page of someone who passed away. Loved ones can activate a tribute page and new algorithms prevent memorialized profiles from showing up in "places that might cause distress", like event recommendations and birthday reminders[24]. Caring for someone's community when they die is a classic pastoral role. Not coincidentally the design of these features was led by a Buddhist chaplain who was trained in how to provide this care offline. All spiritual traditions have rituals and practices related to death, and Facebook's tribute page for the deceased is also reminiscent of the Jewish practice of sitting shiva. Family members observe seven days of mourning during which the community brings food and shares memories of the person who has died. Facebook's memorialization features are a concrete example of what it looks like to draw on offline ancient and sacred rituals to care for us online.

On the societal level, just as tech divides, polarizes, and dehumanizes, it also enables us to come together at unprecedented scales. Many of the most significant social movements of our time were fueled in part by hashtags. In July of 2020, shortly following the murder of George Floyd, the #BlackLivesMatter hashtag had been used 47.8 million times on Twitter from May 26th to June 7th, 2020. That is just under 3.7 million times per day[25]. Since its origin in a Facebook post after the 2012 shooting of 17-year-old Trayvon Martin, the hashtag has become a central unification and mobilization tool for the most widespread and visible racial justice movement since the 1960s[25]. From October 16th, 2017 until May 1st, 2018, #MeToo appeared an average of 61911 times per day on Twitter, dramatically shifting the conversation about sexual assault in the United States[26].

The above examples show that it is possible to use technology as a tool to bring out the best in humanity. What if technologists designed for that instead of designing to maximize the amount of time spent, and attention extracted? What if tech encouraged pausing, and approached every design decision with mindfulness and compassion? Above all, what if technologists valued deep expertise on how to care for our well-being as much as the expertise of great engineering and design? And what if we acted as if the care of our souls is more important than how easy it is to refresh our Twitter feeds?

## 3  Intervention

There is an increasing number of people looking for a cure to Silicon Valley's chronic illness. This includes policymakers, organizers and activists, tech employees, consumers, journalists, scholars, and former tech insiders speaking out about the implications of what they built. A true ethical transformation of the tech sector will require bold regulation, outside pressure, values-oriented business models, empowered ethics teams inside companies who are not reprimanded for speaking the truth, and humane company cultures. It necessitates lifting up leadership and perspectives that are often unrecognized by the mainstream technology sector and ensuring that a multitude of world views and skills are shaping its future. And long-term solutions rest on the herculean task of disentangling ourselves from the tentacles of an economic system fueled on greed.

That said, technology companies are not monoliths. They are made up of people with agency who are making decisions every day. Many of the individuals working inside the tech sector were drawn in part by the companies' stated values and missions, many of which we now know are dangerously idealistic and naive. Twitter's mission is to "give everyone the power to create and share ideas and information instantly without barriers". Facebook's mission is to "build community and bring the world closer together", and Steve Jobs' articulation of Apple's mission was to "make a contribution to the world by making tools for the mind that advance humankind". Capitalism, culture, and the complexity of the relationship between tech and humans have warped these missions at the expense of the long-term health of society. Yet, it is important to remember that the altruistic impulses of many of the people who make up the tech sector remain and can be the seeds of accepting the moral responsibility that comes with holding our spiritual well-being in their hands.

There are increasingly concrete examples of tech executives making unconventional choices that return to the original intentions behind their mission statements. For example, despite the fact that it clearly hurts short-term profits, Twitter's CEO Jack Dorsey, banned political ads in the leadup to the 2020 presidential election because "Internet political ads present entirely new challenges to civic discourse: machine learning-based optimization of messaging and micro-targeting, unchecked misleading information, and deep fakes. All at increasing velocity, sophistication, and overwhelming scale. These challenges will affect all internet communication, not just political ads. Best to focus our efforts on the root problems, without the additional burden and complexity taking money brings"[27]. If one reads between the lines, Dorsey is saying that Twitter's mission to "share information instantly without barriers" is not actually in the best interest of society. Twitter's vice president of Revenue and Content Partnerships, Matt Derella, also stated that "We want to make sure we don't create filter bubbles with this powerful ad system we have"[28]. There is a long way to go, but both Dorsey and Deralla are acknowledging the moral responsibility that comes with their power and they are taking action as a result.

Below I present two shifts that, while only part of the solution, are required for responsible spiritual care in the Digital Age, and they are often overlooked. First, technologists must pay closer attention to their own spiritual and emotional states, as that gives rise to the products we create. Second, we must make sure that those with the wisdom and expertise to care for our souls are helping to shape tech products and strategies.

### 3.1 Shifting the inner state of the intervener

Technologists must recognize that our own spiritual and emotional health is paramount, especially because of the ways that power and stress blind us. Gregory the Great warned ministers in 590 AD of the propensity for power to cloud the mind and the heart. "What else is power in a post of superiority but a tempest in the mind, wherein the ship of the heart is ever shaken by hurricanes of thought"[2]. Operating inside the clouds of power and privilege makes it even more important that technologists cultivate the awareness and spiritual fortitude to see clearly the implications of our decisions and to design from a place of wisdom and compassion. As the systems theorist and author Margaret Wheatley said, "without reflection, we go blindly on our way, creating more unintended consequences, and failing to achieve anything useful"[8].

#### 3.1.1 Spiritual bypassing

Ironically, much of the tech sector already embraces spiritual language and ancient practices, but often for self-serving ends that unwittingly disrespect the sanctity, depth, and intentions behind them. Entrepreneurs are using the South American ceremonial hallucinogen ayahuasca to come up with more creative business ideas[29], there are thousands of people on the waitlist for Google's two-day intensive mindfulness course[30], and whole startup teams are fasting for 36 hours to improve clarity[31].

In contrast, most spiritual and religious traditions include fasting as a sacred act of renunciation, atonement, or connection with God. Fasting during Ramadan is considered one of the five pillars of Islam. It is meant to reduce greed and increase empathy for those who are poor and hungry, thus encouraging acts of generosity and charity. Using fasting to increase profit is an offensive perversion of the altruistic intention behind the practice.

Applying ancient practices in modern, secular contexts is not negative in principle, but when these practices are primarily a way to feel less overwhelmed and more productive as individuals or companies, they risk becoming a numbing agent that makes it easier to ignore our own pain and the pain caused by our institutions. If spiritual work does not go beyond our own self-interest we risk engaging in a collective "spiritual bypass", the use of spiritual ideas and practices to avoid facing reality, especially if it involves feeling pain and discomfort[32].

There is a long history of spiritual bypassing and using spiritual practices to maintain destructive practices and institutions. The role of "chaplain" as we know it was established for the US Army in 1775, when Congress authorized one chaplain for each regiment of the Continental Army. Since then, the official mission of Army Chaplains has been to assess and boost the "spiritual fitness" of the soldiers. It is believed that spiritual fitness is a key component of "soldier readiness and force protection", and that it improves the soldier's ability to cope with the guilt of killing other people and the tragedy of losing their fellow soldiers[33]. It is undeniable that the mental health and spiritual well-being of soldiers is important—the traumas many soldiers experience are more extreme than most of us can imagine, and twenty veterans commit suicide every day[34]. But this focus on spiritual fitness puts band-aids on deep wounds long enough for soldiers to keep fighting, but without actually addressing their well-being in the long run. At the collective level, it helps keep a violent status quo in place even when there are countless moral and ethical reasons to question it.

Spiritual bypassing is built into the very fabric of our culture and economy. The Cherokee healer and psychologist Anne Wilson Schaef writes, "the best-adjusted person in our society is the person who is not dead and not alive, just numb, a zombie. When you are dead you are not able to do the work of society. When you are fully alive you are constantly saying 'No' to many of the processes of society, the racism, the polluted environment, the nuclear threat, the arms race ... Thus it is in the interest of our society to promote those things that take the edge off, keep us busy with our fixes, and keep us slightly numbed out and zombie-like. In this way our modern consumer society itself functions as an addict"[35].

The tech sector is no exception. Many tech employees are using meditation and mindfulness to increase productivity so they can build the tools that hijack our attention and make it harder for us to exist outside of the digital realm. There is deep hypocrisy in the fact that Mark Zuckerberg does not let his daughter use Facebook Messenger Kids, and Steve Jobs' children had strict limits on technology use at home[36]. The most sought-after private school in Silicon Valley, the Waldorf School of the Peninsula, bans technical devices for those under eleven and teaches the children of Google, Uber, Ebay, and Apple how to make go-karts, knit, and cook, saying that computers inhibit creative thinking, movement, human interaction, and attention spans. As Alice Thompson, an associate editor and weekly columnist for *The Times* in the UK said, "It is astonishing if you think about it: the more money you make out of the tech industry, the more you appear to shield your family of its effects".[37] This is akin to tobacco executives saying cigarettes have no harmful health effects while banning their own teenagers from smoking.

Yet it is easier to maintain cognitive dissonance than to reckon with the deep hypocrisy of choosing to build something that one knows is causing harm. As the Tibetan nun Pema Chödrön writes, "We can spend our whole lives escaping from the monsters of our minds", and the misuse of spiritual practices and rituals can be a powerful way to do this[38].

### 3.1.2   Moving from spiritual bypassing to wise action

Spiritual practices can also cultivate the courage and resilience to be with discomfort and look more honestly at the implications of one's actions. They can increase awareness of the rampant narratives and cultures that maintain the delusion of social benefit when the reality is far darker.

This can be difficult since humans are hardwired to run away from pain and seek pleasure. But moving beyond the use of spiritual practices purely for individual enhancement is a necessary step toward more ethical and compassionate technology. It means taking responsibility for the fact that our inner state shapes the decisions we make and what we create. Therefore, it is reckless not to cultivate awareness in service of a mission that is larger than oneself.

Few people articulate the relationship between one's inner state and what one creates better than the Quaker author and activist, Parker Palmer, in his explanation of the mobius strip, a surface with the mathematical property of being unorientable, causing it to appear double-sided even though it has only one side[39].

*"If you take your index finger and trace what seems to be the outside surface, you suddenly find yourself on what seems to be the inside surface. Continue along what seems to be the inside surface, and you suddenly find yourself on what seems to be the outside surface. What looks like its inner and outer surfaces flow into each other seamlessly, co-creating the whole. The first time I saw a Mobius strip, I thought, 'Amazing! That is exactly how life works!' Whatever is inside of us continually flows outward, helping to form or deform the world—depending on what we send out. Whatever is outside us continually flows inward, helping to form or deform us—depending on how we take it in. Bit by bit, we and our world are endlessly re-made in this eternal inner-outer exchange. Much depends on what we choose to put into the world from within ourselves—and much depends on how we handle what the world sends back to us…*

*Here's the question I've been asking myself ever since I understood that we live our lives on the Mobius strip: 'How can I make more life-giving choices about what to put into the world and how to deal with what the world sends back—choices that might bring new life to me, to others, and to the world we share?'"*

The connection between inner and outer states means that technologists have a moral responsibility to create company cultures that encourage reflection and compassion.

Palmer's discussion of the Mobius strip is the motivation behind my organization's name. Mobius' goal is to help tech leaders shape technology for the well-being of humanity, in part by helping them, as Palmer suggests, "make more life-giving choices about what to put into the world and how to deal with what the world sends back". In doing so, we aim to help technology leaders and their teams act ethically as they design the products that shape our experience of being human. We try to create the conditions for them to treat technology development as an act of pastoral care by "paying radical attention" to their humanity and the humanity of those who use their products.

Even moving beyond spiritual bypassing is not enough if the awareness that results does not influence product decisions. This requires responsibly integrating spiritual practices into the design process itself, moment to moment. This can feel uncomfortable in work settings, where culture often discourages merging the "spiritual" with the "professional". This is especially true in predominantly secular environments such as Silicon Valley. Seventy percent of adults in the San Francisco Bay Area, the heart of the tech industry in the US, are religiously unaffiliated, atheist, or agnostic. There are often appropriate reasons for separating religion and the workplace, especially with the risk of discomfort or discrimination based on religious beliefs. However, there are ways to sensitively bring the benefits of spiritual practices into the workplace without including the baggage that so often understandably accompanies it. It may sound insignificant in comparison to the scale of the challenge, and in many ways, it is, but inserting small moments of mindfulness that are explicitly connected to impact can shift the inner states of the people building technology, so we are more reflective and connected to our own intentions and the implications of our decisions. Given that tech companies are made up of individuals making decisions all day, this can have an outsized impact. And, even so, it is important to note the limitations. Simply being more reflective will not get us to where we need to go. That claim would ignore the realities of working within institutions that incentivize behavior that is often in direct contrast with ethical decisions.

However, a masterclass on the impact of contemplative practices supporting social change comes from the Leadership Conference of Women Religious, the leadership body of Catholic nuns in the US. In 2012, the nuns were being investigated by the Vatican for their feminist beliefs and political advocacy for LGBTQ and reproductive rights, which, they were told, ran counter to church doctrine. Their meetings began with thirty minutes of silent contemplation, a simple practice that bolstered their courage, resilience, and ability to act wisely while under fire[40]. Similarly, the Quaker practice of silent listening, followed by speaking when moved, arguably helped create the foundation of clarity and bravery that enabled Quakers to become some of the first White abolitionists. In the realm of physical design, traditional Chinese gardens build bridges according to Zen philosophy and teachings. The bridges proceed in right angles, not straight lines, such that the person walking needs to slow down and be mindful. Otherwise, they risk falling into the water. These are just three examples of what a culture of more deeply integrated mindful practice might look like.

Catholic nuns, Quakers, and Zen philosophers have understood for centuries how even small amounts of this kind of pause, especially amidst crisis and urgency, provide the clarity to take courageous and ethical action including in the fight for feminist rights and the abolition of slavery. This tipping point moment in the tech sector calls for similar levels of courage.

Mobius is witness to the power of small moments of mindful pause when we facilitate advising sessions inside tech companies. Thirty minutes of silence is ambitious in standard corporate settings, but even smaller moments of intentional pause and reflection can make a difference. Pauses, especially in the midst of overwhelming to-do lists and overflowing inboxes, increase the possibility of making more conscious choices. Especially if there is a deliberate effort to go beyond spiritual bypassing, these pauses can help set the foundation for transformation and changing course.

One example comes from Mobius' work with one of the largest tech companies to create more nuanced and responsible well-being metrics to understand how the platform affects peoples' mental and emotional health. What they find will inform product decisions across the company. Their definition of "well-being" will have a global impact. We facilitated a workshop that brought together outside experts, including spiritual teachers, with the company's well-being team. The meditation teacher, Jack Kornfield, began by leading a meditative reflection on the fact that, given their reach, influencing the company's definition of well-being directly impacts the well-being of humanity. He named that this is both a privilege and includes great responsibilities. We then led the team through a process of envisioning the impact they want to have on people and setting intentions. These efforts grounded the rest of the advising session in a sense of purpose that was much deeper than meeting their six-month targets. The moment of pause was simple, and yet we heard from the team that this was a radical act of slowing down in the context of a company culture that is dominated by rushing and anxiety about meeting performance metrics.

Integrating heartfelt reflection in that workshop did not change the course of the company. Advising tech companies on well-being has shown me over and over that, when the rubber hits the road, meaningful change requires making tradeoffs that value responsibility and care over core metrics of engagement, speed, and profit. Usually, these tradeoffs do not happen and the work becomes a band aid or is not sustained. However, if more pauses and guided reflection were built into the overall company culture and practice, people might be more likely to make those tradeoffs. These micro-interventions are a small piece of what is needed in the tech sector, but they help create conditions for more ethical and brave action in the moment and contribute to culture change over time.

Adopting practices like is difficult on one's own, regardless of the context. Community has always been key to the spiritual path. This is true of lay people who are part of religious congregations as well as of monks and nuns who support each other in lifelong commitments spiritual practice.

Mobius is also experimenting with how to meet this need in the tech sector by building an intimate community of mission-aligned tech leaders across companies. This is another method to shift the "interior condition of the intervener", counter the ways in which power and stress can blind well-intentioned people, and support people to move from good intentions to wise action. We host gatherings for senior leaders from across the major tech companies who share the mission to put our shared humanity at the center of their products and services. These gatherings are often hosted in someone's home and integrate spiritual practice in order to foster deeper connections to ourselves, each other, and a shared sense of purpose. The people who are part of the Mobius community work for competitors, so there are limits to what they can and will share with each other: they can rarely talk about specific product features. But there is an increasing desire to discuss common challenges, develop shared standards and principles, and envision new forms of industry-level responses.

We are certainly not the only community-builders in the ethical tech movement. The Trust and Safety Professional Association is a new entity to foster community and cross-company learning for those in Trust and Safety roles across the tech sector. New_Public is a community of people from a range of disciplines working to create healthier online spaces, and the list goes on.

The Mobius cross-company is particularly inspired by the Buddhist concept of the sangha, a community of Buddhists who gather consistently to practice together. Sanghas emphasize that members of the community are all walking a spiritual path together, even when not in the same physical space. This can create powerful levels of psychological safety to see the implications of one's actions and what it will take to change these actions.

As the Vietnamese Buddhist monk, activist and teacher Thich Nhat Hanh wrote:

*"The sangha is not a place to hide in order to avoid your responsibilities. The sangha is a place to practice for the transformation and the healing of self and society. When you are strong, you can be there in order to help society. If your society is in trouble, if your family is broken, if your church is no longer capable of providing you with spiritual life, then you work to take refuge in the sangha so that you can restore your strength, your understanding, your compassion, your confidence. And then in turn you can use that strength, understanding and compassion to rebuild your family and society, to renew your church, to restore communication and harmony. This can only be done as a community—not as an individual, but as a sangha."*[41]

Building a community among leaders is a radical act in the context of a sector that is usually allergic to collaboration. There are rare exceptions, such as the Global Network Initiative, a cross-tech industry coalition that was created to prevent human rights violations in response to the Chinese government finding and torturing political dissidents using data that it accessed from Yahoo. But as Thich Nhat Hahn explains, community has the power to bolster greater moral courage and provide the fortitude to do the difficult work of social transformation.

That fortitude is sorely needed in this case. While building community takes patience and requires trust, many of these leaders are lonely, overwhelmed, swimming upstream, and deeply hungry for like-minded individuals who share a commitment to responsibility and well-being. They are fighting against the strong forces of our economic system and how that translates into the incentives, structures, and cultures within which they are trying to create change. Locking arms in community can help provide the strength to see more clearly and act more radically in service of the larger whole.

### 3.2 Bringing the spiritual caretaker to the table

In his manual of Pastoral Care, Gregory the Great

implores those in power to maintain a "humility of office" that allows them to identify clouded perspectives, subconscious motivations, and blind spots.[2] In the tech sector, this humility needs to extend to a recognition that caring for the soul warrants expertise that rarely is present in tech companies. Whether in the form of a minister, Rabbi, Buddhist meditation teacher, or psychologist, these are experts on timeless questions about how to be healthy and whole human beings and communities.

Throughout much of human history, these roles have been accompanied by many different forms of preparation that include the cultivation of wisdom through deep spiritual practices that have been passed down for thousands of years. As such, it would be unrealistic and even dangerous to assume that everyone who touches product decisions could have the knowledge, skills, wisdom, and methods required to responsibly care for our souls—or that these people could acquire such expertise through a few meditation or spiritual retreats. We do not expect everyone to have the legal knowledge of a lawyer, but no major tech company would imagine shipping a product without consulting one. The same should apply to spiritual care when humans and technology are so intimately intertwined. It should not be acceptable to decide how Siri or Alexa talks a teenager out of a suicidal attempt without involving experts on nuanced and responsible spiritual care.

Tech companies are increasingly hiring the equivalent of chief ethics officers who, given the nature of the crises at hand, are scrambling to define their role, put out constant fires, and develop long-term ethical processes and principles[42]. Companies also bring in outside experts, mostly academics, to build their knowledge base about well-being. But these experts are often consulted in superficial and one-off ways rather than being deeply integrated into the design and strategy process. While these new ethics roles are important steps, they do not create the conditions for true pastoral care for the users of technology.

For example, Alexa is increasingly the only companion for many older people in a given day. Mobius convened a group of caretakers, meditation teachers, and neuroscientists to advise a team at Amazon that is exploring how Alexa might help alleviate loneliness and social isolation among the elderly. Alexa is suddenly "caring" for millions of older people around the world.

Alexa's engineers could either treat this as an interesting fact that is good for their business but does not influence how they define the success of their product, or they could accept the caretaker role with the responsibility it deserves. Thankfully, this particular Alexa team is taking their responsibility seriously. The experts we assembled worked with the Amazon technologists to imagine a world in which Alexa connects people via video to others who share their interests, collect stories and memories for their families by "interviewing" them over time (with consent), and helps people live in accordance with their values and goals for this stage of their lives. This workshop was early in the Alexa team's visioning process, so whether the ideas make their way into the product is yet to be seen. Regardless, this kind of intervention is unlikely to create sustainable change until expertise like this is present in the tech teams themselves or otherwise integrally woven into the decision-making process.

In the Alexa case, it is worth noting that being thoughtful about addressing loneliness most likely helps Amazon's bottom line. The real test is whether companies will make the necessary tradeoffs to value well-being over the fastest route to a profit. Meaningfully integrating caretaking expertise into product teams does not address that root cause, and it is important to be realistic about what that kind of intervention can and cannot accomplish without shifting what is incentivized and valued in the company.

The integration of such care could take a variety of forms, at the product and strategy levels. There could be resident chaplains who are part of product teams, cohorts of graduates from divinity schools and seminaries who are trained in tech and ethics and embedded in tech teams, engineers who attend tailored programs on spiritual care, or ethical councils that include faith leaders in addition to ethicists, lawyers, and tech policy experts. There are many strategies to explore, none of which should be one-off or treated as a silver bullet. They should be built into every part of the design, build, and launch process. It is only at the intersection of a wide range of wisdom, knowledge, skills, and life experiences that we can begin to create technology that is truly worthy of the human spirit.

## 4   Conclusion

The past few years were key to pointing out and naming the negative impacts of technology. We know there is an illness and the symptoms are undeniable. But now it is time to focus on a cure without succumbing to denial, band-aids, or purely putting out the latest fire. We need change at a greater depth and scale than any of the interventions discussed in this piece can create on their own. There is now a vibrant and growing ecosystem of individuals and organizations who are addressing this challenge from a myriad of angles. People are shifting business models, pushing for anti-trust regulation, increasing the diversity of the tech workforce, creating new ethical design principles and performance metrics, and organizing employee movements and walkouts. We need all of these efforts working in concert.

But if we fail to see solutions to tech's impact on humanity within the broader frame of care for souls, we will continue to create quick fixes and small interventions that are misaligned with the fact that technology is influencing nearly every aspect of the human experience. Thankfully, we are surrounded by wisdom that has a great deal to teach us about how to bring technology and humanity into alignment. We know what practices shift us from greed to compassion. We know how to create space for awareness and acceptance. We know how to provide pastoral care through the greatest joys and sorrows of life. Translating this into the digital world is not simple, but it is necessary.

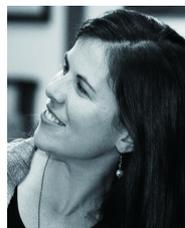

**Aden Van Noppen** received the BA degree from Brown University in 2009 and was a resident fellow at Harvard Divinity School in the 2017−2018 academic year. She was a senior advisor to the United States Chief Technology Officer during the Obama Administration and was a founding organizer of the Sanctuaries, the first interfaith arts organization in the United States. She is currently the founder and executive director of Mobius, a non-profit focused on creating a more responsible, compassionate, and just tech sector. Her works have been featured in the *New Yorker*, the *New York Times*, *Wired*, and elsewhere.

# Real Estate Politik: Democracy and the Financialization of Social Networks

Joanne Cheung*

**Abstract:** The power of social network platforms to amplify the scale, speed, and significance of everyday communication is increasingly weaponized against democracy. Analyses of social networks predominantly focus on design and its effects on politics. This article shifts the debate to their business model. Built as platform businesses, social networks are privately owned public spaces with structurally limited democratic affordances. Drawing from the history, theory, and practice of land use, I develop an analogy between the financialization of land by commercial real estate development and the financialization of attention by platform businesses. Historical policies, such as incentive zoning and exclusionary zoning, shed light on how platform businesses use systems of measurement and valuation to conflate users' roles, tokenize the incentives that drive behavior, and defer the ethical responsibilities businesses have to the public. While the real estate framing reveals social networks' structural flaws and colonial roots, lessons from urban planning, community land trusts, and Indigenous land stewardship can inform their regulation and reform. Building on the broader effort to embed ethics in the development of technology, I describe possibilities to steward social networks in the public interest.

**Key words:** social networks; social media; platform studies; financialization; urban planning; land use

## 1 Introduction

*While in the past there may have been difficulty in identifying the most important places (in a spatial sense) for the exchange of views, today the answer is clear. It is cyberspace—the "vast democratic forums of the Internet" in general, and social media in particular.*

– Anthony Kennedy, Former Associate Justice of the Supreme Court of the United States[1]

*The size of our user base and our users' level of engagement are critical to our success …We generate substantially all of our revenue from selling advertising placements to marketers.*

– Annual Report Pursuant to Section 13 or 15(d) of the Securities Exchange Act of 1934 for the fiscal year ended December 31, 2020 filed by Facebook, Inc.[2]

● Joanne Cheung is with the Haas School of Business, University of California, Berkeley, CA 94720, USA. E-mail: j@joannekcheung.com.
∗ To whom correspondence should be addressed.

*Cities have arisen through geographical and social concentration of a surplus product. Urbanization has always been a class phenomenon, since surpluses are extracted from somewhere and someone, while the control over their disbursement typically lies in a few hands.*

– David Harvey, "The Right to the City"[3]

In the Republic, Plato claimed the democratic affordance of the ideal city was measured by the distance of a herald's cry. In the "virtual city" of social network platforms, the speed and distance at which a user's voice travels are decoupled from physical constraints. Like providing a speaker with a hyper-visible soapbox and a hyper-amplified megaphone, social network platforms boost the political power of an individual's everyday conversations. While these platforms have the potential to expand democracy, their power instead has a growing dark side. From misinformation and polarization to hate speech and the incitation of violence, the power of social networks is increasingly weaponized against democracy.

The history of technology for communication is deeply intertwined with the political history of the empire[4–8]. With Meta Platforms (the parent organization of Facebook) passing $1 trillion market capitalization, social network platforms are effectively a contemporary form of empire. By linking individuals' communicative power with their spending power, the platform simultaneously extracts market information from individuals on the network and expands the financial market for the network itself. The historical origins of this model have been characterized by cultural theorist Nicholas Mirzoeff as the "colonial complex" (local surveillance of individuals) and the "imperial complex" (the control of "primitive" remote populations by a "cultured" centralized authority)[9]. As a platform designed, developed, and headquartered in Silicon Valley and deployed globally to its 2.91 billion monthly active users[10], Facebook resembles both complexes: the extent of its reach is planetary and the specificity of its surveillance is intimate.

If communication at a distance enabled the creation of empire, then the distance between the site of extraction (the colony) and the site of authority (the administrative center) is the basis of its power. The various nomenclatures used to address social network platforms conceal this distance. The term "social network platforms" itself collapses the linguistic distance between the site of extraction (the social network) and the site of authority (the platform), thereby subsuming the platform's business-facing dimension within discussions of public-facing social issues. The term "social media" similarly overly focuses on the consequences of user actions such as content moderation and information integrity[11–14] and behavioral implications of interface designs such as dark patterns, persuasive design, and technological seduction[15–17]. While these analyses are critical for understanding the symptoms and gravity of the problem, they are insufficient for exposing the mechanisms underlying the platform that are critical for their regulation and reform.

In order to expose the hidden mechanisms of platform power and their effects on democracy, I will first decouple "social network" from "platform" in analysis. "Platform" is a multi-sided marketplace business that develops and owns the technology infrastructure that creates social networks. Situating social networks in their current form within Lawrence Lessig's framework for regulation[18], I believe the market force bears direct responsibility for the systemic problems. Given the outsized influence of the financial dimension, I direct the critique from the platform's design and technical mechanisms to its financialization.

This article proceeds in three sections. The first section links democratic practice in public space to social networks, frames platform businesses as commercial real estate development, and explores their democratic affordances as "privately owned public space". The second section contextualizes the financialization of attention in terms of the financialization of land, historicizes how the platform business model encodes colonial assumptions into its management systems for measurement and valuation, and unpacks its mechanisms and systemic effects on democracy. Finally, the third section adapts lessons from urban planning and land justice practices to the context of social networks and proposes new possibilities for roles, incentives, and responsibilities to steward social networks in the public interest.

## 2 Democracy in Privately Owned Public Space

The health of democracy is sustained by communication in everyday life[19–21]. In *Democracy and Education,* philosopher John Dewey[22] described democracy as "more than a form of government; it is primarily a mode of associated living, of conjoint communicated experience". While American voters cast their ballot in the presidential elections every four years, the decision recorded in that instance forms over time. Through everyday interactions with their community, individuals deliberate political opinions that shape their democratic decision-making. Spaces beyond the voting booth—the community center, the neighborhood park, and the library—bear a democratic purpose: to enable a heterogeneous population to recognize and celebrate their differences[23–25]. They set the conditions for democratic life.

Social networks expand the space for the "conjoint communicated experience" defined by Dewey. Between 2010 and 2021, the percentage of Americans using platforms to regularly communicate increased from 47% to 72%[26]. Alongside this growth, a host of design, regulatory, and ethical challenges arise when the democratic affordance of physical space extends into the virtual realm. From Facebook groups acting as virtual assemblies to hashtag activism[27–29], social networks' unique communication features[30, 31] and digital architecture[32] introduce new dynamics and risks. While participation on social networks transformed the ways people engage politically, how they should be governed as political forums remains under debate. In 2017, the Supreme Court case *Packingham v. North Carolina* (*582 US __*) held that a North Carolina statute prohibiting sex offenders from accessing social network platforms violated the First Amendment to the United States Constitution. In the holdings of the case, Justice Kennedy described social networks as the "modern public square", drawing an analogy between access to online communication and access to public space. In 2021, *Knight Institute v. Trump* ruled that by blocking several users, President Trump had violated the First Amendment because comment threads on Twitter constituted a public forum; the case was later rendered moot in the ruling for *Biden v. Knight First Amendment Institute at Columbia University* (*593 U.S. __*) after Trump's presidency ended and his own Twitter access was terminated. In his opinion, Justice Thomas expressed the urgent need to regulate platforms that build social networks: "We will soon have no choice but to address how our legal doctrines apply to highly concentrated, privately owned information infrastructure such as digital platforms."

Platforms are hybrid entities: privately owned businesses that offer public service. Borrowing a term from urban planning[33], a platform is a privately owned public space. Although users may experience them as a "modern public square", their underlying economic incentive and legal constructs are much closer to that of commercial real estate development. The goals they serve are inherently dissonant: a public square exists to serve the public interest, while commercial real estate exists to generate return on capital. Platforms' private ownership structurally constrains their democratic affordance. When public and private interests clash, platforms' allegiances will favor the private over the public.

This dynamic played out during the Occupy Wall Street protests in Zuccotti Park, a privately owned public space in the Financial District of Manhattan, New York City. Zuccotti Park was constructed in 1972 alongside One Liberty Plaza, a 2.3-million-square-foot office tower currently valued at $1.55 billion. Both the tower and the park are owned by Brookfield Office Properties (with the park named after company chairman John Zuccotti). Brookfield Office Properties is a subsidiary of the commercial real estate company, Brookfield Property Partners, which itself is a subsidiary of one of the world's largest alternative asset management companies with over $625 billion of assets under management. Zuccotti Park is one of more than 500 privately owned public spaces in New York City created through a "Floor Area Bonus for a Plaza" regulation in the 1961 New York City Zoning Resolution. This regulation—commonly known as Incentive Zoning—incentivized the creation of open spaces in urban areas with high real estate value by permitting developers an additional ten square feet of built space in exchange for one square foot of "an open area accessible to the public at all times"[34]. "The equivalent of thirty average New York City blocks" was created as a result, "at no direct cost to the public"[35]. Privately owned public space was meant to be a win-win for both the public and the private sectors.

The Occupy Wall Street protest revealed the limits of the democratic affordances of the privately owned public space. Occupy Wall Street protesters turned Zuccotti Park into a makeshift village with tents and shared communal resources[36] and exercised consensus decision making. As the park's owners, Brookfield Office Properties maintained the power to amend the park's code of conduct—and they did. The new amendments banned "tents, sleeping bags, and lying down"[37], which were then used as grounds to evict Occupy protesters from the park. Zuccotti Park's democratic affordance was weakened by its private ownership, and a movement centered on financial inequality was ultimately evicted by the center of financial power.

Privately owned public spaces like Zuccotti Park exist not because of a direct investment in democratic public spaces. Rather, they are byproducts of high-profit skyscrapers developed during real estate market booms when the speculative value of the building soars above the material value of the land[38, 39]. The incentive behind the creation of the park therefore lies precisely in profit maximization. Born from the dissonance of extreme public and private interest, Zuccotti Park and One Liberty Plaza are inextricably linked; the park could not have existed without the tower. This dependency fundamentally weakens the park's democratic affordance. Public good is subsumed by the logic of financial capital and public interest is lodged within the most extreme expression of private interests.

Social networks and platform businesses have an interdependent relationship similar to Zuccotti Park and One Liberty Plaza. Social networks would not exist without platform businesses and their social benefits are intertwined with their profit-seeking purpose.

Businesses like Facebook operate under a platform business model that relies on the public to serve its private interests. They are multi-sided marketplaces with "many faces"[40]. Rather than creating value on a linear supply chain, a platform business generates revenue by connecting the multiple groups and brokering exchanges between them[41, 42]. Different from a traditional market, where the transaction occurs directly between the buyer and the supplier, exchanges between buyers of ads (advertisers) and suppliers (users) on a platform are indirect. Users not only supply attention for the ads, they also supply data about how they use their attention, which helps continuously improve the accuracy and value of Facebook ads. No real goods is exchanged on a multi-sided advertising market. Instead, advertisers who buy Facebook ads are buying the possibility to turn Facebook users into their future customers. Consumption begets more consumption.

Facebook principally mediates three layers of exchanges: (1) between users and their social groups, (2) between individuals and the platform, and (3) between the platform and its clients, third-party advertisers. The terms of exchange across these layers are not equal. In the user-facing layer, the exchange centers on everyday communication. On the client-facing side, the exchange centers on conversion: the moment the audience of the advertisement performs an action desired by the advertiser, such as discovering and purchasing a product. The platforms' objective is to reduce conversion time and increase the number of converted people. The second layer of exchange—between users and the platform—is the most opaque and hidden in black-box algorithms[43]. By positioning itself as a free service whose purpose is to enable users to connect and build community, Facebook turns non-financial exchanges (between users) into financial ones. In other words, the platform financializes everyday communication into sellable data, social relations into marketing channels, and users into consumers, while obscuring the terms of its financialization.

To disentangle the conflicting private and public interests in platforms, we must understand how financialization works. In the next section, I build on the analogy between platforms and real estate development and use the financialization of land to illuminate problematic mechanisms and their effects in the financialization of attention.

## 3 Financialization of Land and Attention

With its multifaceted dynamics, financialization has been characterized as cognitive capitalism[44], supply chain capitalism[45], racial capitalism[46], platform capitalism[47], surveillance capitalism[48], rentier capitalism[49, 50], technoscientific capitalism[51], and terror capitalism[52], as well as part of the growing field of "platform studies"[53, 54]. Due to its extractive effects, financialization of human cognitive capacities has been termed "immaterial labor"[55], "attention brokerage"[56], and the "subprime attention crisis"[57]. While financialization has extensively reconfigured the language, culture, and patterns of contemporary life[58−60], its mechanisms are not entirely new. Rather, they bear striking resemblance to colonial patterns of dispossession[61−67].

Both commercial real estate development and platform businesses financialize finite resources: *land* and *attention*. They do so through systems that measure and assign value, and in the process, reconstruct colonial myths in everyday life.

Measurement systems serve the purpose of the authority who institutes them. Scholars across philosophy, geography, anthropology, and more have termed this process "classification"[9], "the nomination of the visible"[68], "commensuration"[69], and "modular simplification"[70]. Measurement systems not only represent but also construct what they measure[71, 72]. As sociologist Donald MacKenzie writes, they are "not a camera, but an engine". The distinguishing feature of both land and attention measurement lies in how extensively the process subsumes otherwise non-financial entities[73−76] within its financial logic and as a consequence reduces the relevance of local contexts[77] in service of increasing the efficiency of the financial exchange.

Historical processes of land measurement encoded a colonial myth: before being made legible to the centralized authority, settlers of the American West declared that land existed in a "pristine" state, untouched by "humans". By discounting the humanity and stewardship of indigenous populations, the "pristine" myth helped justify the dispossession of Indigenous populations[78–83] and normalize the exploitation of "virgin land"[84] as monocultural fields optimized for commodity crops[85–88]. This myth sheds light on the extractive assumption that platforms make about participants on social networks: that people's attention, like land, is "unprocessed data"[89] ready to be converted into and exchanged as financial assets.

Once an entity has been converted into a financial asset, it is then assigned value by a centralized authority with the purpose of accruing it. The asset becomes sorted based on its perceived productivity—that is, its ability to generate profit. For example, the practice of scientific forestry optimized forests for lumber output—an asset that could generate profit—while excluding all other vegetation, which resulted in the systematic deterioration of soil health and ecosystem collapse[90]. Being embedded within the definition of productivity is a value judgement: the idea that certain entities are more valuable than others is an assumption and not a fact of nature, and the assumption is often made by a centralized authority about a site of extraction. The myth of productivity helps maintain the power of the centralized authority by intentionally obfuscating the subjectivity of its value system, rendering the systemic biases embedded within this judgment to appear normalized in practice. This myth, based on the colonial assumption that certain uses of land (agricultural cultivation) are more "productive" than others (Indigenous land use), was used to exclude Indigenous land and people[91].

The productivity myth is deeply ingrained into the financialization of attention today. "Highest and best use" is a framework commonly used in commercial real estate development to appraise the potential profit of a piece of land and decide on its use. Created by economist Irving Fisher, the framework assesses land use based on four criteria: the development must be (1) legally permissible, (2) physically possible, (3) financially feasible, and (4) maximally productive. The last criterion, "maximally productive", means that the chosen development should prioritize a type of use (for example, hotel over housing) that could generate maximum profit, disregarding the parcel's current purpose[92]. Uses that are not "maximally productive"—the balance of various types of commercial, civic, and residential programs, a diverse mix of residents, or the availability of transportation infrastructure that promotes active lifestyles—do not factor into this analysis because their benefits cannot be quantified as direct revenue.

Systems of measurement and valuation do not operate linearly, they reinforce one another iteratively. The history of land use demonstrates this self-reinforcing dynamic: measurement serves to progressively subdivide land while the value of the land progressively increases[93]. This dynamic also plays out in the financialization of attention by platform businesses. With the more granular subdivision of attention through user engagement, such as "like" and "share" and more accurate valuation of user behaviors, the value of attention increases in turn. When left unchecked, this dynamic complicates the roles, incentives, and responsibilities that are key to maintaining the health of democratic practice.

Using lessons from real estate and urban development, I expose three mechanisms platforms used to financialize attention: (1) conflate user roles in ways that undermine their agency, (2) tokenize the incentives behind everyday communication to drive up engagement, and (3) use proxy metrics to defer the social responsibility inherent in exclusionary practices.

## 3.1 Conflate roles

Platforms exploit the intersection of surveillance capitalism and identity politics. Individuals are valued for their authenticity while being asked to play multiple roles. Engagement metrics, such as "like" and "share", privilege the quantity and frequency of individuals' immediate reactions while reducing their agency in their actions. These mechanisms conflate roles in two ways. In an era of information overflow, authentic self-expressions—which is scarce by nature—has become a valuable commodity. How people express themselves reveals their preferences, interests, and connections and affirms their position as a member of their social network. However, these expressions are also the key input into platforms' mechanisms for increasing conversion—algorithmic ranking and personalized advertisements—that influence individuals' purchasing and political decisions. Herein lies platform extraction: using the authenticity of an individual's role *as a member of their social group* to categorize and predict their role *as a consumer (of commercial products and political advertising)*. The two roles that an individual is asked to play on a platform are not equally consensual. To an individual user, the platform markets itself as a provider of communication infrastructure and not as an advertising channel personalized based on their personal data. In blurring these two roles, the extent to which an individual's behavior in their first role (as a member of their social group) is influenced by their second role (as a consumer) becomes obscured as well.

Optimizing for engagement also conflates otherwise separate roles in the information supply chain: individual users are not only consumers—they are also producers and distributors. These separate roles typically enable the terms of transaction for each activity to be clearly delineated. However, interactions on the platform are designed to encourage all three types of activities at once. On Facebook, for example, all posts in the News Feed are followed with the "like" button, the "comment" button, and the "share" button. Furthermore, these engagement interactions are all reward-based. The quantity of "likes" given to a piece of content rewards producers with a sense of popularity. Badges such as "Top Fan"—awarded to the most active participants—turn communication into competition. By communicating through the platforms, individuals become unwitting contestants in the commodification of their authenticity.

### 3.2 Tokenize incentives

Similar to how casinos turn cash into token currencies in the form of gambling chips, platforms turn incentives driving social interactions into token currencies in the form of "likes" and "shares". Token currencies increase the psychological distance between the cost of consumption and the action of consumption, and as a consequence, they make it easier for users to consume more[94]. Management scientist and economist Drazen Prelec refers to token currencies as "hedonic buffers": "by buffering themselves between real money and consumption, they protect consumption from the moral tax"[95]. When purchasing a token currency, the consumer does not need to specify how the currency will be used. When the consumer spends the currency, they do not feel the need to evaluate the implications of the transaction as carefully as they would with a cash transaction.

Similarly, when a user posts on the platform, he/she does not need to specify his/her intended audience. Every interaction on Facebook—be it a post, a comment, or a "like"—is, by default, broadcast to the entirety of the user's social network. If a user is required to specify to whom they are speaking every time they write a post, they would be more likely to consider the immediate consequences of their action. By removing choice in one's audience, the psychological distance between the social cost of an interaction and the interaction itself widens, and user engagement increases as a result.

Considering current debates around the limits of social cognition online[96–98], tokenizing social incentives exploits and undermines the cognitive limits of individuals on social networks. Responses to this extractive pattern have themselves been subsumed by financialization. The rise of the social quantification sector[67] capitalizes on the extraction of attention as well as its conservation. In the last decade, "digital mindfulness"—from meditation apps to features like Screen Time—has become a billion-dollar business; in parallel, social network platforms feed emotions into the "outrage industry"[99]. Like the false dichotomy of land as either a pristine wilderness[78] or a site of extraction, seeing people's attention as either "protected" or "exploited" ultimately distracts from the extent of disempowerment caused by platforms and the fact that both result in the commodification of authenticity.

### 3.3 Defer responsibilities

Proxy metrics defer social responsibility to the technical implementations of the system. This form of obfuscation makes the values (such as racial discrimination or the relentless pursuit of profit) that fundamentally drive decisions shielded from direct critique.

In the context of cities, the ongoing struggle for segregation demonstrates the extent to which exclusionary practices have co-evolved with the history of urban development. Discrimination acts and persists, indirectly, through proxy metrics that encode bias. If measures to counteract discrimination are not proactively instituted, exclusionary practices will reinforce discrimination. In the early 1900s, White homeowners who perceived people of color as a threat to their property value began adopting racially restrictive covenants to bar people of color from home ownership in their neighborhoods. When the federal government created the Home Owners' Loan Corporation with the aim to expand home ownership opportunities as a part of the New Deal, rather than proactively mitigating the racial discrimination, government surveyors based their neighborhood ranking system on local officials' and bankers' racially charged risk assessments. In this way, they encoded racial discrimination into the value of land[100, 101], which resulted in racial segregation and concentrated poverty that still persist today[102–106]. Beyond the direct encoding of exclusion, single-family zoning ordinances conceal the discrimination behind proxy metrics like building density. Institutionalized in *Village of Arlington Heights v. Metropolitan Housing Development Corp.(1977)*, single-family zoning de facto separates lower-income populations—disproportionately racial minorities—from wealthier populations[107], perpetuating systemic disinvestments.

Proxy metrics for revenue used by platforms make the prioritization of private profit at the expense of other public good an unquestioned practice, and they underscore how extensively the entanglements between exclusionary practice and finance have been normalized. Instead of proactively integrating different perspectives, platforms by default algorithmically rank messages based on relevance, measured as "the number of comments, likes, and reactions a post receives"[108]. Algorithmic ranking prioritizes messages that support one's preexisting beliefs and exclude ones that may challenge those beliefs. Changes to the default sorting method, such as chronological sorting, must be manually selected by the user. Although Facebook's News Feed preferences proclaim to let individuals "take control and customize" the feed, the only way a user can make changes is to prune their News Feed: to add or remove up to 30 users to be prioritized to "see first". Individuals have no power to meaningfully change the exclusionary ranking mechanism that determines the value of what they see.

This systematic reinforcement of confirmation bias undermines a fundamental condition for a healthy democracy: a shared context that includes divergent beliefs, founded on a spirit of generosity rather than animosity. The efficacy of democratic practice lies in the collective ability to empathize, internalize, and reconcile differing opinions and beliefs. As anthropologist Elizabeth Povinelli writes, "The power of a particular form of communication to commensurate morally and epistemologically divergent social groups lies at the heart of liberal hopes for a nonviolent democratic form of governmentality".

Filtering one's interactions based on existing preferences and social connections narrows the context of one's preexisting beliefs. In *Liberalism and Social Action*, John Dewey[109] writes, "The method of democracy is to bring conflicts out into the open where their special claims can be seen and appraised, where they can be discussed and judged in light of more inclusive interests than are represented by either of them separately". The "meaningful inefficiencies"[110] inherent in the integration of diverse perspectives is foundation for democracy and yet is at odds with the platforms' exclusive focus on productivity. If social networks are to exist in service of democracy, then they need to proactively create the conditions for pluralism—to make it possible and desirable to reconcile differences rather than obscuring or exploiting them for profit.

In order to mitigate the systemic biases inherent in social networks and the detrimental effects of social exclusion, the first step is to question the status quo. Historically, land use that supports democracy has not been a given; it needed to be directly advocated for and formalized through law. The same expectations should be set for the democratic affordance of social networks. The Montgomery bus boycott, Freedom Rides, and many other protests of the civil rights movement were fights for African Americans to gain equal rights in public space. The Civil Rights Act of 1964 ended segregation in the public space, and the Fair Housing Act of 1978 made it illegal to write racially restrictive covenants into property deeds. A part of the work of changing the system is to expose its mechanisms. As Richard Rothstein advocated in *The Color of Law*, revealing how the mechanisms work creates opportunities for their reform. In the next section, I draw from the practice of urban planning and land justice movements to imagine new roles, incentives, and responsibilities for social networks.

## 4  Reclaim Social Networks from Financialization

Reclaiming social networks from financialization will require creating mechanisms that align the incentives of the platform with the public interest. This begins with recognizing the colonial underpinnings of American

 

democracy[111–114] and relinquishing the nostalgic vision of the colonial New England town halls. To create a collective space for an experimentalist democracy fit for our time, we need to embrace rather than obscure the "contingency of context"[115] of our globally connected society. Using lessons from urban planning, land justice, and Indigenous land stewardship, I propose three mechanisms to help reclaim social networks from financialization and reorient them to the public interest: (1) use urban planning to redefine roles that have been conflated by platforms, (2) use community land trusts to illustrate how public interest can be protected from market forces, and (3) use the practice of Indigenous land stewardship to inspire new thinking about the meaning of social responsibility.

### 4.1 Redefine roles: "urban planning"

Urban planning can serve as a model for a professional role that serves the public interest. As designers of public space, urban planners must wrestle with large private interests while they "continuously pursue and faithfully serve the public interest"[116]. In order to receive the licensure to practice—and to ensure that "the public interest" prevails in these negotiations—urban planners must follow Ethical Principles set by the American Planning Association's Institute of Certified Planners:

(1) Recognize the rights of citizens to participate in planning decisions;

(2) Strive to give citizens (including those who lack formal organization or influence) full, clear, and accurate information on planning issues and the opportunity to have a meaningful role in the development of plans and programs;

(3) Strive to expand choice and opportunity for all persons, recognizing a special responsibility to plan for the needs of disadvantaged groups and persons;

(4) Assist in the clarification of community goals, objectives, and policies in plan-making;

(5) Ensure that reports, records, and any other non-confidential information which is, or will be, available to decision makers is made available to the public in a convenient format and sufficiently in advance of any decision;

(6) Strive to protect the integrity of the natural environment and the heritage of the built environment;

(7) Pay special attention to the interrelatedness of decisions and the long-range consequences of present actions.

Given the similarity in challenges faced by urban planners and stewards of social networks, the American Planning Association's Ethical Principles seem eminently applicable to their roles. Values of inclusivity, fairness, and transparency are all values that should guide the design of a healthy digital democracy. As a thought experiment, what if platforms adopted the following ethical principles, based on the ones set forth by the American Planning Association?

(1) Recognize the rights of people to participate in platform design decisions;

(2) Strive to give people (including those who lack formal organization or influence) full, clear, and accurate information on product development issues and the opportunity to have a meaningful role in the design and development of the platform;

(3) Strive to expand choice and opportunity for all persons, recognizing a special responsibility to plan for the needs of disadvantaged groups and persons;

(4) Assist in the clarification of community goals, objectives, and policies in plan-making;

(5) Ensure that reports, records, and any other non-confidential information which is, or will be, available to decision makers is made available to the public in a convenient format and sufficiently in advance of any decision;

(6) Strive to protect the integrity of the digital public sphere;

(7) Pay special attention to the interrelatedness of decisions and the long-range consequences of present actions.

These ethical principles would encourage better decisions on the level of individual designers. However, though these principles reflect core democratic values, to substantively improve the business of the platforms, ethical principles are far from enough. Individuals' decisions and responsibilities correlate to their decision-making power and scope of accountability. A designer or engineer at the level of an "individual contributor" in a technology company may be responsible for their own output, such as a "share" button or refresh content controls. While their design decisions potentially shape the communication systems between billions of people, their social impact massively exceeds their power within the organization. Even if the designer or engineer adopted these ethical principles in the public interest, they would face enormous barriers in practice and would personally bear the risk of acting against the interest of their employers. Ethical principles must operate in a context that is greater than any individual designer or organization; they need to align with or shift the incentive structure of the business model.

In land use, urban planning and real estate development are different fields with distinct duties and ethics. Social network platforms, in their current formation, collapse incentives, roles, and responsibilities that help preserve meaningful checks and balances between the private and public interests. In the absence of an equivalent field of "urban planning" dedicated to the public interest for social networks, platforms are driven exclusively as commercial real estate development. Recognizing the different incentives behind these two professions is critical for discussions on technology ethics. Unlike urban planners, commercial real estate developers have no professional association nor explicit ethical principles. Codes of ethics are often found in fiduciary duty-defined relationships, which obligates a practitioner to act solely in the interests of their client—for example, doctor and patient. Real estate developers, on the other hand, do not have a fiduciary duty towards the users of buildings they develop. Instead, they act in accordance with the profit motives of their investors, whose interest in maximizing the bottom line is often at odds with the interest of users.

Similarly, social network platforms act in the best interest of their investors, shareholders, and clients (advertisers), which leads neither to the benefit nor protection of the participants in the network.

Because of this misalignment of individual and organizational values inherent in social network platforms, it will be critical to develop "public interest" roles for social networks, the equivalent of urban planners and land justice activists—professions with a fiduciary duty that aligns with their democratic responsibilities. In addition, beyond growing the field of public interest designers and technologists, institutions need to continue to create permanent positions for these roles to ensure their long-term viability.

### 4.2 Restore incentives: "community land trusts"

Current debates around individual data ownership apply property rights to address inequities in monetization, but this approach is limited. In *Colonial Lives of Property,* legal scholar Brenna Bhandar unpacks how colonial logics have shaped modern conceptions of property and created "the racial regimes of ownership"[117, 118]. Focusing on individual data ownership shifts the burden to individuals without addressing the commodification of their attention in the first place. Similarly, fixes that regulate individual user behavior—such as automatically limiting the time a user can spend on platforms (Social Media Addiction Reduction Technology Act 2019)—do not address the root of the problem. Real change requires creating alternatives to existing platforms that differentiate ownership from use and remove attention from the commodity market.

Community land trusts are nonprofit organizations that own and hold land in perpetuity in the permanent benefit of the communities they serve[119]. Robert Swann, who formalized the concept of the community land trust in *Community Land Trust: A Guide to a New Model of Land Tenure in America,* connects the concept to historical roots in Indigenous land stewardship: "American Indian tradition holds that the land belongs to God. Individual ownership and personal possession of land and resources were unknown"[120]. Community land trusts remove land from the commodity market, thereby buffering it from the booms and busts of the real estate market cycles. Crucially, a community land trust decouples the incentive of ownership from the incentive of use. Ownership is maintained in the public interest, while use allows for private interests through 99-year ground leases, the longest possible term of lease of real estate property. The community ownership of land aligns the incentives of the users and the owners; users have long-term access to affordable space, and the trust has a strong legal position to serve its mission and preserve affordability.

To develop an analogous mechanism for social networks that could incentivize the platform to serve its people in the long term, we must recognize how the community land trust is inextricably linked to place. While the legal arrangement can be replicated across geographies and adapted to suit local needs, a community land trust is anchored in its specific community. This usage of "community" is entirely different from the "community" used in Facebook's mission statement ("Facebook's mission is to give people the power to build community and bring the world closer together"). The community of a community land trust is defined by and bound by place, whereas the "community" of Facebook refers to its user base and is both decoupled from place and ever-expanding. Further, the residents in the community have voting power by holding board seats in the community land trust; Facebook users have no such power. Reclaiming social networks for the real benefit of communities means that a community, defined by place, should own the technology infrastructure and decision-making power in its use.

### 4.3 Reframe responsibilities: "Indigenous land stewardship"

Indigenous land stewardship is an example of collective stewardship that creates systems-level ecological benefits like biodiversity[121] and resilience[122−125]. The success of this practice depends on a mutually constitutive relationship between people and land. As Native American poet Paula Gunn Allen writes, "We are the land... The land is not really the place (separate from ourselves) where we act out the drama of our isolate destinies. It is not a means of survival, a setting for our affairs... It is rather a part of our being, dynamic, significant, real. It is ourself"[126]. The responsibility to care for the land and care for the self are one and the same. Learning from this practice, we can reorient social networks from financialization to care. This shift suggests a new approach to thinking about social responsibilities: from being external to being embodied.

Likewise, the technology that serves this community must not act from a distance; it must be co-designed. Laura Mannell, Frank Palermo, and Crispin Smith wrote in *Reclaiming Indigenous Urban Planning,* "A community plan cannot be developed from the outside looking in. It cannot be done for a community, it must be done with and by a community". [127] A community's social network, similarly, must be created with and by the community. Technology that supports social networks in the public interest begins with honoring the existing knowledge, capacities, and practices in a community as its starting point.

The fact that urban planning, community land trusts, and Indigenous land stewardship are all not-for-profit practices that exist to primarily serve social good brings up a natural question: can social network reforms based on lessons from these practices be achieved from within existing for-profit platform businesses? Fully addressing this question—which is fundamentally about transforming the political economy of data—is beyond the scope of this article. However, I do wish to highlight three entry points that are specifically relevant to businesses. First, businesses comprise groups of employees who hold a plurality of motivations and beliefs; these differences can and should be channeled towards social change. Second, profit and social good are not necessarily mutually exclusive; businesses with broad-based shared ownership and cooperative governance structures naturally align with democratic practice. Third, coloniality runs deep in the culture of technology; recognizing colonial inequities within organizational culture itself is a critical first step.

## 5   Conclusion

Social networks are now an undeniable public forum. However, their democratic potential has been undercut by the goals of platform businesses and their mechanisms of financialization. The incentives driving the platform set private and public interests in direct conflict. As publicly traded companies, platforms are ultimately accountable to their shareholders and must prioritize private interests—the health of the business, defined by its profitability and market share—over the public interest and the health of democracy. As privately owned public spaces in their current form, social networks' public-facing experiences, which purport to champion connection and community in practice obscure the extractive nature of their business model. Connection is exploited for its network power to expand the customer base; community is exploited as an input into a platform's advertising product. The language, interactions, and relationships of social networks have been coopted.

In order to reclaim social networks from financialization by platform businesses, we need to first expose the systems and mechanisms driving the process. As this article has shown, the financialization of land provides a critical lens for examining the systems that enable the financialization of attention and the constitutive role colonialism played in shaping them. Examples from land use also demonstrate possibilities for rethinking roles, incentives, and responsibilities, shifting social networks from extraction to mutualism, from expansion to place. Given this new understanding, I hope we can move from critique to creation and collaboratively build the theoretical frameworks, legal instruments, funding models, technical infrastructure, and social norms to steward social networks in the public interest.

## Acknowledgment

The author wishes to thank the Ethical Tech Working Group at the Harvard Berkman Klein Center for Internet & Society.

　　　　　　　　　　　　　　　　　　　　*Journal of Social Computing, December* 2021, 2(4): 323−336

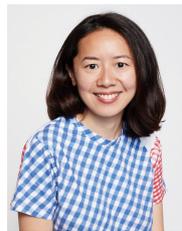**Joanne Cheung** is a lecturer at the Haas School of Business, University of California, Berkeley, USA. She has been a fellow at the Harvard Berkman Klein Center for Internet & Society and the American Association of University Women. She received her M.Arch degree from Harvard Graduate School of Design in 2018, her MFA degree from Bard College Milton Avery Graduate School of the Arts in 2013, her BA degree from Dartmouth College in 2009. Her current research focuses on climate justice and Indigenous data sovereignty.

# Algorithmic Silence: A Call to Decomputerize

Jonnie Penn*

**Abstract:** Tech critics become technocrats when they overlook the daunting administrative density of a digital-first society. The author implores critics to reject structural dependencies on digital tools rather than naturalize their integration through critique and reform. At stake is the degree to which citizens must defer to unelected experts to navigate such density. Democracy dies in the darkness of sysadmin. The argument and a candidate solution proceed as follows. Since entropy is intrinsic to all physical systems, including digital systems, perfect automation is a fiction. Concealing this fiction, however, are five historical forces usually treated in isolation: *ghost work*, *technical debt*, *intellectual debt*, the labor of algorithmic *critique*, and various types of *participatory labor*. The author connects these topics to emphasize the *systemic* impositions of digital decision tools, which compound entangled genealogies of oppression and temporal attrition. In search of a harmonious balance between the use of "AI" tools and the non-digital decision systems they are meant to supplant, the author draws inspiration from an unexpected source: musical notation. Just as musical notes require silence to be operative, the author positions algorithmic silence—the deliberate exclusion of highly abstract digital decision systems from human decision-making environments—as a strategic corrective to the fiction of total automation. Facial recognition bans and the Right to Disconnect are recent examples of algorithmic silence as an active trend.

**Key words:** technocracy; algorithmic silence; history; labor; artificial intelligence; AI ethics; automation; decomputerization

## 1 Introduction

In 1948, in an article in *Business Week*, a Vice President at the Ford Motor Company coined the term "automation" to promote the use of mechanized self-governance in manufacturing. Since entropy, error, and deterioration are intrinsic to all physical systems, including digital systems, perfect automation is a fiction. Even still, economists, industrialists, and technologists continue to invoke idealizations of "automation" in their influential visions of society. In this article, the author challenges the heightened rhetoric major technology companies and computer scientists have recently used to characterise the autonomous and predictive capabilities of advanced digital decision tools, the current vogue of the automated society. The author shows how reports of a looming "AI Revolution" misrepresent the complex ways in which such tools have been used, in practice, to preserve the political status quo in the United States and United Kingdom.[①] Yet this article is not just a critique. In pursuit of a harmonious balance between the use of such tools, the use of the non-digital decision systems they are meant to supplant, and the modes of administrative labor required for each, the author draws inspiration from an unexpected source: musical notation. Just as musical notes require silence in order to be operative, the author argues that societies must strategically emphasize—rather than simply seeking to displace—non-digital decision systems by limiting their use of digital alternatives. To crystallize this point, the author introduces the concept of algorithmic silence: the

• Jonnie Penn is with the Department of History and Philosophy of Science, University of Cambridge, Cambridge, CB2 3RH, UK, and also with the Berkman Klein Center, Harvard University, Cambridge, MA 02138, USA. E-mail: jnpenn@gmail.com.
* To whom correspondence should be addressed.

---

① The author reserves his comments to the two countries about which he has most expertise.

designation of a deliberate exclusion of highly abstract digital decision systems from human decision-making environments. Recent bans on facial recognition technologies are an example of algorithmic silence.

While the rise of digital automation has afforded tremendous opportunities for social transformation, it has also disguised growing administrative burdens. This underappreciated coupling is, by my account, a key reason to normalize algorithmic silence. As the cost of digital decision systems decreases globally and their use becomes more prolific, the accompanying need for diverse types of administrative labor will escalate, perhaps precipitously. To evidence this trend, the author connects five realms of scholarship usually treated in isolation: *ghost work, technical debt, intellectual debt,* the labor of algorithmic *critique,* and various types of *participatory labor.* The author emphasizes the systemic impositions that digital decision systems make on human beings not only as workers and members of different racial, class, or gender groups, as other scholars have shown, but also as consumers, citizens, parents, or any other number of identity frames. These obligations compound in idiosyncratic proportions depending on one's entangled identities, and their harms should be mitigated in respect to these differences. Yet, the author adds, the potential also exists to forge a cross-cutting form of solidarity that addresses broad exposures to the Kafkaesque cacophony of digital decision systems in oversupply. Modes of collective restraint, such acts of algorithmic silence, could help distance AI development from technocracy and align it with traditions of de-escalation, such as decomputerization and degrowth.

## 2   Disingenuous Rhetoric and "The AI Revolution"

In popular use today, the term "artificial intelligence" is a palimpsest: etched over the disciplines' mid-twentieth century origins, rife with theories of neural activity, is a radical ethos of imminent social transformation via automation.[②] AI is a catch-all not just for a branch of computer science and its subsets, but for myriad other digital automation techniques as well. Yarden Katz excavates this layering to reveal how, in the early 2010s, major American technology firms lent panache to sales of their data science and machine learning products and services by perpetuating the existence of "The AI Revolution"[1]. Their campaigns publicly consummated[2] the field's longstanding but underappreciated entanglements with institutional patrons intent on developing sophisticated tools for social analysis and control[3]. These interventions capitalized on tropes of imminent technological potential inherited through Western myth, science fiction, religion, economics, and popular culture[4–9]. *Blade Runner,* for example, which builds its narrative around the existence of synthetic human-like "replicants", is set on November 20, 2019, the rough date of this article's writing[10]. The future, it seems, is now.

The AI Revolution, like the computer revolution, is not a real revolution[11, 12].[③] Proponents do not seek to forcibly overthrow an existing social order. Far from it. As Katz shows, the AI Revolution is largely a conservative push to preserve and benefit from the political status quo, which, as this issue attests, is marked by historic levels of financial and informational inequality. A growing body of scholarship clarifies how such tool and services repackage and reinforce anti-black[13, 14], anti-poor[15], and chauvinist logics[16]—all under the pretense of progress and efficiency[11, 17–21]. The AI Revolution is thus genuinely political—just not in the ways it is made out to be[22].

Disingenuous rhetoric plays an important role in constructing civic imaginaries about the future. A critical audit of the evocative terminology used in and around AI research is long overdue[23–25]. A 1976 missive by an MIT AI engineer challenged the field's "contagious" use of wishful mnemonics: words that served as "incantations" for a desired result, rather than sober descriptions of a mechanism or function[23, 26–29].[④] A recent framing captures this trick in action. In 2018, a team at the Toronto Rotman School of Management cast AI as "a drop in the cost of prediction"[30]. As prediction became cheaper, the team reasoned, it would be used to solve problems that were not traditionally prediction problems, such as autonomous driving. This

---

② A palimpsest is a manuscript on which later writing has been superimposed on earlier writing. Thank you to Sarah Dillon for this metaphor.

③ See Hicks for a critical take on how the 1950−1970s computer "revolution" in the UK served to entrench existing gender inequalities. Summary in Ref. [12].

④ Naming conventions were judged to have warped researcher's relationship to the epistemic significance of their designs. Artificial intelligence is itself a wishful mnemonic, unique from chemistry and physics in that the name portrays an intention. See Garvey for a survey of AI critique over the second half of the twentieth century and Dreyfus for a glimpse into various eras of critique.

is an insightful observation, but not necessarily for the reasons its authors intended. The AI Revolution does not mark a genuine drop in the cost of prediction, but it may, instead, mark a meaningful drop in the cost to feign prediction. Stated differently, it is becoming trivially easy to manufacture the pretense of "predicting" an outcome in areas where prediction, in fact, defies natural law.

Critics clarify that, at a technical level, contemporary AI capabilities are closer in substance to Katz's account than to the account put forward by those at the Rotman School[24, 31]. Most so-called "predictive" analytics lack the necessary relation to causality to genuinely foretell an outcome in advance. "I have not found a single paper predicting a future result. All of them claim that a prediction could have been made; i.e., they are post-hoc analysis"[31]. The term is mistakenly used to describe novel statistical correlations after events have occurred, rather than identifying a determinate causal mechanism beforehand. One example is the recently debunked claim that AI can "predict" someone's sexual preference from their photograph[32]. Prediction implies prophecy, which is intimidating and inaccurate. At a technical level, argues Momin M. Malik, the term "detect" is more precise, if still not totally satisfying.[5]

The risks involved in indulging such prophetic rhetoric are compounded in cases in which a user's environment can be altered to make a product appear more "predictive" than it is[33–35]. For instance, it is far easier for a driverless vehicle to appear autonomous within the perpetually dry city grid of Phoenix, Arizona, than it would be for that same vehicle to navigate the wet, twisted lanes of Aberdeen, Scotland. Phoenix has fewer characteristic features, which makes changes easier for an "autonomous" vehicle to infer. Disingenuous rhetoric arises when results from a constrained environment (e.g., Phoenix) are treated as universally applicable (e.g., adequate to navigate all locales, including Aberdeen). These claims are covertly subjective not just because they overstate the competency of the algorithmic system in question under the guise of technological objectivity, but also because they treat the value of certain constraints (e.g., a city in a grid formation) as self-evident, as if worthy of mass reproduction along with the new autonomous technology. "Prediction" rhetoric fuses a model with the environment it is most successful in, incentivizing the recreation of those constrained environments to accompany propagation of those models[36]. This conservative push for the hegemonic standardization of human environments and behaviors is especially pernicious when deployed in value sensitive domains like healthcare.[6]

These dynamics are not new. The profundity of automatic manufacturing has long been a matter of training audiences' perspective to notice certain contributing features at the expense of others. In the nineteenth century London, recounts Stephanie Dick, Karl Marx criticized Charles Babbage for anthropomorphising cogs and gears while simultaneously failing to recognize the humanity of his own craftsmen[38, 39]. When the term "automation" was coined in 1948 by a Vice President at the Ford Motor Company, economists, industrialists, and unionists seized the term—under inconsistent definitions—to articulate their own competing visions of society[40]. In present day, Astra Taylor coins the term "fauxtomation" to provide a more accurate characterization of the concealed chains of labor that sustain contemporary modes of digital automation[41]. The notion of "autonomy" is a fiction concealed through the chronic underreporting and/or dehumanization of living contributors, argues Taylor. It is a horizon sought for but never reached, like an asymptote stretching hopelessly toward zero.

Having briefly considered how various rhetorical maneuvres distort civic imaginaries of automation both past and present, it is appropriate to ask what is, in fact, required to sustain pursuit of the endless horizon that is ubiquitous digital automation. In the section that follows, the author connects five labor trends usually treated in isolation: *ghost work*, *technical debt*, *intellectual debt*, the labor of algorithmic *critique*, and various types of *participatory labor*. The author's aim in connecting these threads is to emphasize the *systematic* nature in which different modes of digital automation extract and appropriate human labor simultaneously. The shadowy politics active in these systems are perhaps best recognized in cases of piecemeal low-pay tasks, as in the category of *ghost work*. Here, industrial actors dehumanize contingent workers to rationalize indecent conditions and maximize profits. Yet ghost work, on its own, is not fully illustrative of the broad spectrum of underappreciated impositions that digital automation makes upon human labor. The author explores four additional categories. As the author will show, *technical debt* and *intellectual debt* normalize poor craftsmanship and pseudoscience in the development of digital products and services, thereby offsetting an unspecified burden of maintenance and repair labor onto future generations. In a similar vein, the labor of *critique* and various modes of *participatory labor* help to sustain the acceptability and reliability of these products and services today. One wonders, in view of these labor trends: if software eats the world… who will digest it?

---

[5] Personal correspondence. Thank you to Momin for these critical readings.

[6] Rhetoric of this type has already been found to obscure the flawed scientific foundations of such tools[37] and to legitimize pseudoscience in areas like criminal justice, human resources, credit scoring and in medicine.

## 3 Performing "The AI Revolution" — A Taxonomy of Contingent Labor

### 3.1 Ghost work

The first category of labor to explore is *ghost work*, a phenomenon that reveals the banality of the AI Revolution in practice. Gray and Suri coined the term in 2019 to illuminate the opaque world of digital on-demand task fulfillment, in which online platforms aggregate piecemeal low-pay tasks and repackage them as the outputs of automation[42]. Examples of ghost work include rideshare driving and the search and categorization of micro tasks online. These platform systems emerged from decades of corporate led casualization and outsourcing, which normalized precarious modes of employment[43]. Their existence is critical to AI. For example, Fei-Fei Li's AI team at Stanford University estimated in 2007 that it would take nineteen years of undergraduate labor to create ImageNet, a large, gold-standard database of accurately labeled images. Using ghost work, the team accessed 49000 human contributors from 167 countries to produce the database in two and a half years[42]. ImageNet has been celebrated as a benchmark for computer vision algorithms; one that fueled a surge of media attention around AI techniques. Ghost workers, in contrast, remain "the AI revolution's unsung heroes"[42].

As the title suggests, ghost work is predicated on a status of tortured impermanence. Workers are hired as independent contractors rather than employees. This makes precise figures on the scale and nature of the phenomenon difficult to source. In 2017, the platform economy employed an estimated 70 million workers globally, with estimates for 2025 as high as 540 million (as cited in Ref. [44]). In the post-industrial economies of the US and UK, statistics indicate that ghost work is large and growing[42].⑦ Recent news around the poor performance of Facebook, Inc.'s platform content moderation algorithms provides a glimpse into how ghost work intersects with a well-funded and large-scale AI project. In this domain, content moderators are contracted to sort inappropriate content, often in conjunction with algorithmic systems. In 2009, Facebook was cited as paying twelve content moderators for its one hundred and twenty million users[45]. By 2017, this number allegedly grew to 4500 moderators. By 2019, it reached between 15000−20000 moderators for Facebook's two and a quarter billion users[46–48].⑧ Between 2009−2019 then, Facebook's content moderator-to-user ratio grew approximately sixty times.

Ghost work is core to the AI Revolution. Facebook is one of many corporations now intent on reconfiguring their business around AI and, consequently, precarious labor. In late 2017, YouTube LLC. declared it would hire 10000 content moderators for its 1.5−1.8 billion viewers, more than double the number of its current 5000-person employee base[49–51]. The most well-known ghost work platform is Amazon.com, Inc.'s Mechanical Turk (or MTurk) system, which provides businesses and consumers with structured access to a marketplace of low-cost and globally situated click workers. Between 2005−2016, MTurk grew five times, from approximately 100000 to 500000[42]. Amazon touts MTurk as "artificial artificial intelligence". In comparison, DefinedCrowd, one of many start-ups now competing with MTurk, claims eighty employees and 211468 click workers, more than the 163800 people working in oil and gas extraction across the United States[52–54].⑨ Sector analysts claim that the marketplace for third-party data labeling will grow six times by 2023

---

⑦ In 2016, twenty million workers were estimated to earn money via the completion of on-demand tasks in the United States. Estimates hold that analogous modes of semi "automation" could reconfigure 38 percent of US jobs by 2030. In developing countries, where much of ghost work is based, there are not even these figures.

⑧ In comparison, Facebook, Inc. reported 27705 employees in 2018.

⑨ At time of writing, competing outlets include: Alegion, Appen, Cape Start, Click Work, Cloud Factory, Cloud Sight, Data Pure, Defined Crowd, Figure8, Cloud AutoML Vision, hCaptcha, Gengo, Gems, Hive, iMerit, Labelbox, Lotus Quality Assurance, Micro Workers, MightyAI, OC Lavi, Playment, Reef, Scale, Superb, and TaskUs.

into a one billion dollar marketplace, with other estimates reaching as high as five billion dollars[55–57].

The federal government in the United States has yet to acknowledge or set labor protections for ghost workers, whose fight for recognition has only recently materialized into legislation in a handful of US states[58]. The job category "Content Moderator" remains unrecognized by the Bureau of Labor Statistics; it is also absent from the 21000 industry and 31000 occupational titles measured by the US Census[59, 60]. This uneasy status, along with the frequent lack of a shared worksite or uniform job title, deepens workers' precarity by adding friction to collective action and the protections it yields[42, 61].⑩

As in the era of Babbage, automation remains a matter of perspective. Regulators maintain a stubborn faith in narratives of imminent technological transformation. Despite the troubling size and character of the ghost work phenomenon, regulators fail to confront the possibility of its persistence, and thus fail to accept it as a site for reform. A 2015 World Bank report on online outsourcing claimed that forecasting beyond 2020 was "highly speculative" due to the sector's susceptibility to rapid technological change[62]. Gray and Suri challenge this idleness. They revisit how Microsoft leveraged Permatemp contracts as far back as the 1980s[42]. "We can not be sure if the 'last mile' of the journey toward full automation will ever be completed," they warn, adding that, "the great paradox of automation is that the desire to eliminate human labor always generates new tasks for humans"[42]. Even as technological boundaries change, workers' precarious status remains the same.

### 3.2   Technical debt

The second labor category to assess is *technical debt*. Technical debt is a form of delayed labor normalized through the acceptance of poor craftsmanship. In recent years, the programming community has used the term to characterize the compounding maintenance costs associated with poor design choices in program writing. Ward Cunningham coined the term in 1992, stating, "Shipping first time code is like going into debt. A little debt speeds development so long as it is paid back promptly with a rewrite... The danger occurs when the debt is not repaid. Every minute spent on not-quite-right code counts as interest on that debt."[63] Attempts at a framework for how to measure and monitor technical debt remain theoretical at best[64–69]. Estimates hold that in the development of machine learning systems, technical debt accrues at a rate comparable to that of a high-interest credit card[70, 71]. Researchers at Google, Inc. warn of compounding "correction cascades" in these fragile models, meaning hidden feedback loops, signal entanglements, and other technical challenges due to what they describe as the CACE principle, for "Changing Anything Changes Everything"[70].

Tomorrow's workers, both expert and not, will inherit the labor required to constantly repair and maintain this delicate infrastructure. That Facebook's moderator-to-user ratio increased sixty-fold between 2009−2019 speaks to the scope of the labor force required to algorithmically oblige evolving norms, customs, and laws in an ever-increasing number of overlapping domains. The European Commission, by analogy, employs a full-time "Protocol Service" to keep its human leadership tuned to ever-shifting cultural and political norms in national and regional contexts within that boundary[72].⑪ As the CACE principle distills, it is difficult to design AI systems that integrate a similarly fluid and complex set of concerns in real-time without human support. This difficulty rises further as developers attempt to model three dimensional environments. Sally Applin argues that software active in an "autonomous" vehicle must, in principle, seamlessly and unfailingly update across shifting municipal, city, regional, state/province, national, and international borders[73]. This software would also presumably register and integrate all relevant changes to the unfixed physical world (e.g., downed trees, new construction, etc.). These are Sisyphean undertakings. Narratives of an AI "revolution" belie the distribution of labor that make these performances of autonomy feasible at all.

### 3.3   Intellectual debt

As with technical debt, *intellectual debt* is a form of delayed labor. Zittrain uses the term to characterize the manner in which AI—and machine learning specifically—serve to "increase our collective intellectual credit line" by providing atomized solutions to problems without any clear explanation of the causal

---

⑩ Gray and Suri caution that no laws yet govern who counts as an "employer" or "employee" in this domain. Roberts explains that content moderators are also hired under the work titles "screener" or "community manager".

⑪ They are responsible to oversee appropriate gifts, actions, attire, and even choice in songs for events.

mechanisms involved[74]. In principle, access to this credit line could normalize widespread offsetting of theoretical explanation, where isolated decisions not to identify causal mechanisms accrue into a network of unchecked faith. Despite digital tools being the primary cause of this phenomenon, they are also held up as a primary solution, which fuels a feedback loop toward trained dependency and the centralization of power amidst cacophony. "A world of knowledge without understanding becomes a world without discernible cause and effect, in which we grow dependent on our digital concierges to tell us what to do and when"[74].

Influential figures in the American technology sector have extolled this horizon. In a 2008 article entitled "The End of Theory: The Data Deluge Makes the Scientific Method Obsolete", Chris Anderson, chief editor of *Wired* Magazine, called on his readers to reimagine science in the mold of Google's data-intensive advertising business. He celebrated an explanatory paradigm in which approximations to scientific truth follow from correlations found in massive stores of behavioral data, rather than from hypothesis and testing[75]. Also in 2008, Peter Norvig, Google's research director, advocated to update the statisticians' maxim "All models are wrong but some are useful", to "All models are wrong, and increasingly you can succeed without them"[75].[12] Weinberger, in a 2017 op-ed for *Wired*, reaffirmed Anderson's vision for a new decade, claiming, "Knowing the world may require giving up on understanding it."[76]

Intellectual debt is not unique to machine learning. As Zittrain notes, it is routinely accepted in areas of medicine. The drug Modafinil, for example, is sold with a disclaimer stating that its reasons for being effective are unknown. In the healthcare sector, however, such decisions face significant regulatory scrutiny and oversight. These burdens do not yet weigh as heavily on the tech establishment. Nor is mistrust of intellectual debt a guarantee that such heavy restrictions will naturally emerge over time. In the 1980s, automated and semi-automated document retrieval systems were met with a similar mistrust[77]. Indeed, the embrace of instrumentalist statistics in the United States can be traced back to the late nineteenth century[78]. Without regulatory oversights in place to ensure genuine social progress, the merits of which have already been overlooked by existing AI principles[79], this trend will likely burden tomorrow's workers with the mountain of tedious responsibilities that accompany navigating an experimental turn away from the reliability of causation.

### 3.4 Critique

A fourth category of labor is *critique*. This category is broad: it could feasibly encompass the labor required to investigate, identify, articulate, remedy, and/or reject the degenerative aspects of "autonomous" systems. This characterization provides a wide enough berth to encompass the work of theorists like, say, Langdon Winner, activists like those in the Carceral Tech Resistance Network, and those whose labor sustains movements of technological prohibition like Neo-Luddism. The ACM FAccT conference, which highlights engineering critiques of algorithmic systems, offers a window into the growth of at least one aspect of this broad domain: since the conference was formed in the late 2010s submissions have increased roughly two times annually, from 73 in 2018 to 290 in 2020.[13] While the growth of the AI industry is now regularly indexed by top universities and businesses[80], the growth of so-called AI Ethics, a contentious title for the body of criticism (as this issue conveys), is not as well understood.

Of note is that, at present, much of this labor is subsidized by the public. Of the seventy sets of recommendations on trustworthy AI produced between 2017−2019, industry produced roughly a fifth of submissions, and civil society and governments, together, roughly a half[81, 82]. Principled proposals for citizen juries and government-run data trusts extend, in their orientation, a similar expectation for the public to pay for the failures of automation. Zittrain, for instance, positions academia, along with public libraries, as the natural home for new modes of critique. He proposes that datasets and algorithms that meet a sufficiently broad level of public use could be tested by researchers to mitigate errors and vulnerabilities before they compound.

If adopted in tandem with structural reforms to labor standards, such proposals could bear fruit. Regrettably, most academic labor is now precarious and prone to exploitation. 73 percent of faculty in American higher education institutions work part-time or otherwise off the tenure track, which provides little job security[83]. 60 percent of higher education staff in UK universities struggle to make ends meet, with part-time and hourly paid teachers doing, on average, 45 percent of their work without compensation[84]. Meanwhile, in early 2020, Google, Inc.'s parent company Alphabet Inc. became the fourth US technology company to reach a market cap of over a trillion dollars, following Apple Inc., Amazon, and the Microsoft Corporation, with Facebook now close behind. The normalization of un- or low-paid critique thus threatens to normalize public responsibility for avoidable harms ill-managed by industry.

---

[12] The first maxim is commonly attributed to the statistician George Box.

[13] ACM FAccT (formerly FAT*) stands for Association for Computing Machinery's Conference on Fairness, Accountability, and Transparency in machine learning. Thank you to Christo Wilson for the figures.

### 3.5   Participatory labor

The final category of labor the author assesses defies reduction to a single classification. This cluster encompasses the surfeit of unpaid and often unrecognized tasks and offerings undertaken by consumers, users, and citizens when they engage, passively and actively, with digital modes of automation. This includes but is not limited to:

• Do-it-yourself economies (e.g., self-checkouts, self-check-ins, self-booking systems, solve-it-yourself customer service);

• Open-source software economies (e.g., pro-bono support of for-profit infrastructures);

• Inference economies (e.g., proprietary model training via auto-complete, CAPTCHA or service fulfillment, such as traffic patterns inferred from a driver's rideshare activity without fair compensation);

• *Digital labor* and *informational labor* economies (e.g., online community management, such as the labor volunteered by women of color in response to misogyny and racism on platform systems[85–87]);

• Covert agency economies (e.g., the unacknowledged workarounds users employ to modify or overcome limited affordances in an algorithmic system[88]);

• Dark pattern economies (e.g., design affordances that trick a user into signing up for something they do not want[89]);

• Reputation maintenance economies (e.g., labor undertaken to maintain one's standing when it is impacted by a system's shortcomings or outright failings[90]).

These diverse types of labor substantiate the "human infrastructure" required to integrate digital automation into daily life[91]. When deployed into structurally racist, sexist, and ableist societies, such structures tend to disproportionately penalize marginalized groups[92, 93].

These burdens are normalized through appeals to a neoliberal conception of consent, which assumes a base level "capacity for consent" that is unsubstantiated in reality[94]. When collective harms are framed as the responsibility of each individual to navigate, only those with power can afford to understand and overcome them. Others face exile or deprivation when they try to resist. Robust taxonomies and lines of solidarity are needed to map, connect, reform, or reject these entangled forms of labor, and to identify the toll of their collective impositions. These taxonomies might also be used to build toward renumeration and reparation structures that recognize and respond to each party's contingent inputs[95, 96].

This brief survey of *ghost work*, *technical debt*, *intellectual debt*, the labor of *critique* and *participatory labor* highlights the significant labor—both in the present and in the future—that organizations depend upon to further the sales friendly mythos of AI. "We are all system administrators now, whether we realize it or not," write Dick and Volmer, who assess user-supplied maintenance in relation to Microsoft's Windows platform[97]. Much of what the author has covered here reduces to the extended labor economies of error and anomaly management. Given this common source, it is worth noting that the earliest pioneers of computing had not anticipated that such labor would be necessary. They believed, wrongly, that computers would not have bugs. In his autobiography, Maurice Wilkes, who developed EDSAC, the first practical use stored-program digital computer, grappled with the realization that a good portion of the remainder of his life would be spent fixing errors in his own code[98]. "Debugging had to be discovered," he recalled[98].[⑭] In that era, and again with AI's maturation, the messy and irreducible complexities of material reality interrupt the principled but all too abstract aspirations of even the most accomplished computing engineers.

Since the development of EDSAC in the 1940s, the labor required to analyze, design, test, debug, and develop computer programs has become a recognized and deeply influential employment category known as, "Software Development and Programming". In the United States, it is one of the few employment categories to have emerged over the past century that employs a significant proportion of the population. As of 2010,

⑭ Emphasis mine.

there were thirty-five million computer experts employed around the globe, five orders of magnitude more than the initial group of scientists, engineers, and support staff working in the midcentury[99]. In 2016, 1.7 million were employed as software developers in the US alone, with an estimated 300000 expected to join in the decade to come[100]. Low-cost fauxtomation broadens this labor network even further, reaching into exploitational labor categories that remain to be taxonomized and acknowledged in the way that Software Development and Programming was during and after the 1960s.

Remaining to be seen, as responsibility for integrating these errors translates slowly into a tree of discernible job categories (e.g., content moderator, quality assurance officer for driverless vehicles), is the extent to which the accruing errors, harms, and sacrifices involved in adopting these systems should be absorbed by an already over-leveraged public. These impositions are particularly difficult to characterize, as is their chain of responsibility[101–103].⑤ By analogy, in 2016 analysts positioned medical errors as the third leading cause of death in the US[104, 105]. A 2018 report estimates that software bugs killed more than one thousand patients per year in the UK, with blame often passed on to doctors or nurses[90, 106, 107].⑥ A decade prior to the AI Revolution, the US Commerce Department estimated that computer users shared half the cost of the ＄22.2−59.5 billion lost annually as a result of inadequate software testing infrastructure[108]. These sacrifices—lost lives, lost wages, lost recognition, lost opportunity, lost insights, and lost time—are substantial, and they will grow larger still.

## 4  Automation's Impositions: A Structural View

The author's reason for connecting these threads is to draw attention to the outcomes of neglecting digital automation's systemic impositions, which entangle in ways that resist simple reduction. Notions of labor provide one lens into this change, as the prior sections demonstrates. Yet labor, alone, is not the only way to understand this change. As "predictive" technologies swell and rescript the logic of daily behaviors in healthcare, education, and beyond, competing automated systems will vie for citizens' finite time and encode their behavior with sophisticated interactivity[109]. Without adequate protections in place to monitor and/or meaningfully prohibit such impositions, low-cost decision systems will compound the public's digital obligations and slowly (or perhaps rapidly) sap their availability to non-digital systems. Existing terms of critique fail to capture the full character of this levy. Loss is treated in financial terms, as technical debt or intellectual debt, rather than a more profound loss of possibility. Ruha Benjamin subverts this trend when saying, in relation to technology's role in perpetuating anti-black logics, "Most people are forced to live inside someone else's imagination" (Ref. [110]; see also, in relation to critique of normative conceptions of time[111–113]).

An analogy is useful here as a means to characterize the scale of this type of systemic phenomenon and the related power that new vocabulary can have to communicate the complex reasons for an equally broad shift in course. The terms "global warming", "climate change", and "Anthropocene" introduced the public to the idea that local environmental harms, when taken in aggregate, amounted to a fatal error in cultural logic, one that now threatens the survival of our societies, with marginalized groups around the globe faced with the most dire risks[14]. These marquee terms speak to the sum-total harm caused by a complex web of operators whose default perspective was to treat carbon emissions as an acceptable negative externality. Emissions were considered someone else's problem—just as automation's impositions are now. "Global warming" and related terms interrupt that base assumption. They illuminate the inescapable hazards for everyone that accompany unrestrained material consumption.

That a climate crisis loomed in the late twentieth century was clear to many long before the invention of those aforementioned terms. In 1955, John Von Neumann, whose logical architecture laid the blueprint

---

⑤ Hobbyists, historians, and risk researchers maintain venues to catalogue and characterize the impact of poor error management in digital systems, but no sophisticated repository captures a broad picture of their aggregate toll, both economic and otherwise. For a moderated forum on the safety and security of computer and related systems see the Risk Digest. For a hobbyist's collection of serious or novel bugs see Huckle. For recent research on the role of error in the history of computing, see SIGCIS.

⑥ Elish calls this phenomenon of blame "the moral crumple zone" of automated systems. "Just as the crumple zone in a car is designed to absorb the force of impact in a crash, the human in a highly complex and automated system may become simply a component—accidentally or intentionally—that bears the brunt of the moral and legal responsibilities when the overall system malfunctions."

for the digital era, opined about this inflection point in an article entitled, "Can We Survive Technology?"[114]. During the first industrial revolution, he reasoned, "It was possible to accommodate the major tensions created by technological progress. Now this safety mechanism is being sharply inhibited; literally and figuratively, we are running out of room. At long last, we begin to feel the effects of the finite, actual size of the earth in a critical way."[114] John Von Neumann reckoned with technology's aggregate material implications. In this article, the author gestures to its aggregate temporal implications and administrative obligations.

As with climate change, the localized impositions of, in this case, low-cost decision systems, are dismissed by society at large as uncontentious in the short-term. Only once a ceiling asserts itself might this fleet of impositions be seen as degenerative and systemic. Regrettably, as with climate change, the existence of this ceiling is difficult to convey to the broader public—until it is not. Instead of fires, floods, and ecosystem collapse, temporal erosion may come to resemble, say, a latent denial-of-service (DoS) attack on a society's daily decision-making abilities. A DoS attack is a cyber-attack in which a communication pathway is flooded with enough superfluous requests to make it unavailable. By analogy, a poverty of time, caused by the proliferation of digital obligations and delights (deployed at low-cost), could hobble the public's collective capacity to consider or even imagine alternative modes of social organization, such as those that do not center on data, efficiency, or technological progress. Wood writes, from a related vantage, "Surely the most wretched unfreedom of all would be to lose the ability even to conceive of what it would be like to have the freedom we lack, and so dismiss even the aspiration to freedom, as something wicked and dangerous" (as cited in Ref. [92]).[⑰]

The difficultly of conveying this complex problem to the public is that time attrition is the product of a threatening system, not a threatening character or object. The harms of automation in oversupply are captured narratively in a folktale about The Sorcerer's Apprentice, in which an enchanted broom causes a flood by collecting and pouring out too much water for its new, inexperienced master. In the West, however, advanced automation is often personified, through characters like the Terminator, rather than being cast as infrastructural or distributed. These accounts of automation-as-individual, also captured in narratives about job losses to robots, distort the public's sensitivity to both the banality of the AI Revolution and its contingent harms. These stories convey a threat, but as with climate change, they may underemphasize the decentered nature of that threat.

Adding to the challenge of effective public communication of a world awash with low-cost decision systems is that skeuomorphs (i.e., features passed from one technology to another related technology, like the familiar "click" of a smartphone's shutter, which does not in fact exist or make a sound) have so far failed to preserve traditional prohibitory functions, such as those that ritualized natural limits and restraint. Digital automation techniques know no opening hours, holiday closures, snow days, sick days, periods of grievance, nor even strict regulatory limits on their collective impositions. These are the technological manifestations of the neoliberal attitudes that preceded them. Interventions in privacy law, labor law, consumer protections, and in the digital wellbeing movement add friction to select intrusions, as epitomized by worker's right to disconnect in France and Germany. Yet, as with climate change, reform is still often cast in relation to the individual, as if the potential to meter excess is somehow unavailable at the group level. This is a false restriction. Collective remedies, as always, remain viable.

The irony of this dilemma is that automation, at a certain level of proliferation, eventually fails to fulfill on its own celebrated purpose: to save time. The endless need to integrate different types of automation draws the ideal toward self-contradiction. Each new act of coordination creates a new labor requirement. This labor can be automated, but then that new automated system must be integrated, too. This feedback loop introduces new types of administrative obligations that, as the five labor trends outlined above adequately suggests, can be easily overlooked by those who benefit from their presence. As with climate change, marginalized peoples suffer these harms first. In the long run, however, as for the Sorcerer's Apprentice, a world awash with such obligations would presumably ensnare their elite creators as well by interweaving them in a society shaped by the same scripted logics they have used to control

---

⑰ Although this may sound alarmist, the emergence of light and sound pollution evidence how impacted parties can overlook what is lost amidst poor regulation. The author once met a child who had never seen the stars due to light pollution in his neighborhood.

others. The unrestrained use of low-cost decision systems would amount to death by a thousand paper cuts for a society callous to the compounding effects of such temporal pollutants.

By my account, the prolific use of digital decision systems, fueled by low marginal costs for proliferation and ascendant narratives of an imminent AI Revolution, marks a new stage in complex debates over the societal role(s) of automation. The characteristic the author seeks to denaturalize is the assumption that digital automation—by its own logic—merits recognition as a self-evident form of cultural progress. In the author's view, critics of automation who entertain this horizon (e.g., automation-as-progress) without also embracing acts of prohibition assume too readily that technical solutions can be found—eventually—and that, as a result, solutions should be labored toward. This endless-horizon narrative permits systemic harms to persist, with marginalized peoples bearing the brunt of tomorrow's maintenance. Acts of prohibition create decision making systems in which knowledge of such tools is not a prerequisite. With these spaces, critics endorse a growing distance between them and the non-expert communities they often aim to represent. Stated differently, advanced automation techniques may need to be resisted wholesale if tech ethics experts are to avoid becoming the technocrats they seek to displace.

## 5  On Formalization and Its Alternatives

One way to resist the encroachment of digital automation is to question the methodologies that clear a path for its use. One such methodology is the use of formalization to describe a system's presumed nature. In his introduction to Minsky's 1961 paper, "Steps Toward Artificial Intelligence", which laid out a research agenda for that discipline[115], guest-editor Harry T. Larson wrote, "When the practitioner has overcome his fear of the machine, and when the scientist and practitioner are communicating, the attack is relentless. The scientific mind has found an un-formalised field, and it cannot rest until it identifies, understands, and organizes basic elements of the field"[116]. Aspects of contemporary research on fairness, accountability and transparency in machine learning echo Larson's positivist dogma by implying that highly formalized engineering techniques will muster adequate solutions, rather than re-inscribing underlying harms or reifying ever more bureaucratization[78, 117].⑱ Intervening at the point at which attempts are made to formalize a social system helps to provide citizens the derivative economic or administrative relief needed to decide on a civic future for themselves. Operating this far upstream avoids their being automatically ensnared in debate over a decision tool or technique that continues ad nauseam.

To conclude, the author fosters a metaphor that he hopes will lend subtly to dialogue about how to reshape positivist inclinations in the automation space into something less brutal and domineering. In sheet music—indeed, in music composition generally—special notation is used to convey the role of a deliberative silence. These constructions build negative space purposefully, as a mode of art. Without rests, music would be cacophony. A recent wave of legal prohibitions on facial recognition technologies across American cities substantiate deliberative restraint in response to automation. US communities have opted to preserve what the author calls an algorithmic silence: the purposeful exclusion of highly abstract algorithmic methods from human decision-making environments. A silence of this type asserts that the value of such theory is worth more to the community when left unrealized. Such acts of prohibition leave room to incorporate holistic thinking about the myriad ways that advanced decision systems re-shape and bear upon human societies. Bans and moratoriums hold a space for reflection on the systemic burdens disguised by disingenuous rhetoric and incremental reformism. It provides the proverbial "frog" with the interruption necessary to recognize that it is in the proverbial "boiling pot".

Another benefit of this approach to resisting automation's impositions is that it reconfigures the distribution of labor involved in shaping the roles that digital decision systems ought to have in society. Algorithmic silence places the burden of proof on enthusiasts, rather than on critics, to prove why formal techniques and technological artifacts should be welcomed into a social system at all. Revoking entitlements to public goodwill reveals the actual toll of integrating such systems into daily life. Enthusiasts would need to prove ahead of time how their automated systems function without access to no-pay and low-pay surrogates to clean up the mess caused by piloting poor

---

⑱ Jones uses "data positivism" to describe this instrumentalist model of induction, which seeks functions that fit to the data, rather than functions that fit to a corresponding law of nature.

tech craftsmanship on the public. This tempts reflection on automation's full bill (and distribution) of costs, the nature of which transcend financial levees.

A third additional benefit to the normalization of prohibitions as a response to the excesses of an automated society is that this path would limit corporations' access to public coffers. By this route, universities and colleges would be spared reduction to the role of algorithmic custodians; history departments would need to be shuttered so that a new generation of scholars can find and resolve software errors on behalf of Facebook. Algorithmic silence asserts that the significant and underappreciated costs of experimenting with automation in the wild are paid for by the scientist and their patrons, rather than by the communities those groups treat as laboratories. Those who champion the horizon politics of automation, meaning the notion that decency will come "eventually" and that the status quo must remain until then, are handed responsibility for these "acceptable" burdens instead.

The motive power of a well-timed silence rings loudly. Rest, some forget, is its own vehicle. The ambience it creates is inhabitable and thus sacred. By this view, algorithmic silence is another safe road to progress. Sahlins—aware that declines in leisure time have been naturalized over centuries and can thus be denaturalized—famously memorializes hunter gathers as the original affluent society given that they toiled only three to five hours a day[118]. Via a far more theory-laden approach, Mejias introduces the term "paranode" to characterize the multitudes that lie beyond the network logics used in contemporary life to model and assimilate all that is social. A paranode is a place beyond the conceptual limits of networks[119]; a structural component that alters network outcomes but from outside the network's reach. An act of paranodality is one of disidentification with the logic of that network. Consider a broken URL, RFID (radio-frequency identification) blocker, or pirate radio. Each exists slightly beyond the validation of the networks designed to subsume it. By rejecting the hegemony of advanced decision systems, algorithmic silence fosters paranodality.

This account of paranodality from Mejias implies that those who resist disidentification from a network are more radical than those who cause it. By my account, those who reject algorithmic silence are tantamount to those who reject silence in music. This willingness to create cacophony is deeply political, since it is often not those enthusiasts who suffer its hazards. In response, these parties claim that acts of prohibition are antithetical to progress. This shaky platform would seek to undermine that silence is in fact co-constitutive of harmony; the two cannot exist apart. Writes musicologist Zofia Lissa, "In its symbiosis with sonority, silence is one of the structural elements of the sound fabric, though in itself silence is the very negation of a sound fabric."[120] Mejias, too, positions paranodality as intrinsic to a networks' structure. An attack on disidentification is thus an attack on the structure of the network.

At root, musical notation and network structures can be understood as metaphors for epistemic sovereignty in the face of technoscientific hegemony. Each makes a virtue of noncompliance. Algorithmic silence, likewise, provides an ambience that is, at first, epistemically nonhierarchical. What comes from this state, however, is unpromised. At best, respite from the perils of ubiquitous AI could provide a window into a way of knowing that colonialism has forcefully displaced; an occasion, per Nelson, to witness that "the human is not a problem to move beyond"[121]. Silence for the sake of silence constrains positivist technoscience by asserting arbitrary limits to its valorization of hyper rationalization and administration. It is an invitation to technocrats to stand outside of that rationalist bubble; to grieve, instead, the presumptuous fictions of progress and futurity. A chorus of algorithmic silences, the author wagers, could help to break the spell of AI by building harmony between its countless alternatives. Proponents of such techniques would arrive, instead, into the present, occupied as it is by the durability of imperialism[122] and the permanence of pollution[123]. Here, a different set of experts call the tune.

The growing ubiquity of advanced low-cost automation techniques has made strange bedfellows of those who seek the dangers of unrestrained automation. Military researchers, both in the US and India, have recently framed contemporary information flows as a growing *impediment* to their ideological aims rather than a cherished resource[124, 125]. "The desire to have maximum inputs for decision making is a tempting proposition but will have to be tempered with the necessity of giving a decision in time. As time pressures become more acute, we may well end up with 'information decoherence'."[126] This is a remarkable outcome given that the US military played a definitive role in pioneering modern information management techniques via the development of systems analysis, operations research, game theory, and digital computing and digital networking generally[127, 128]. For military researchers to insinuate the need to de-escalate information management is telling of the hazardous path dependencies of unrestrained automation. It speaks to a carrying capacity, or ceiling, after which even hardline proponents see diminishing returns from the logics behind mass automation. Cowan, similarly, debunks the popular myth that American domestic technologies saved domestic laborers time through automation. In fact, Cowan shows, such tools introduced more work for these laborers by upsetting the equitable models of labor distribution assumed in prior centuries[93].

In raising these critiques, and the unique possibilities afforded by the thoughtful use of prohibition amidst the rapid development of low-cost automated systems, the author seeks to emphasize the search for harmony in the development of digital automation regimes, particularly in the value-sensitive realm of democratic governance. It bears mention at this juncture that silence, on its own, is not harmonious, although the experience of it may be pleasing at times. Harmony, by definition, requires the thoughtful combination of positive expressions and their opposites, rather than simply the preservation of a dead signal or cacophony. The possibilities for proverbial harmony, in this regard, are vast[129]. In their 2020 book *Meaningful Inefficiencies*, for instance, Gordon and Mugar argue that public trust in civic organizations requires that such systems are designed *not* to be efficient[130].

In consideration of what precise balance to strike, it is worth considering that contemporary debates over acceptable levels of formalization and algorithmic management in a given context mirror a longstanding dilemma in American political theory about the appropriate balance between democratic representation and the agents who administer it. Herein lies a thorny trade-off: administrative decision makers in large-scale democracies, such as monetary experts, hold both the specialist knowledge to make an informed judgement and a capricious discretion over outcomes that no elected representative could ever hope to oversee. Sheer administrative complexity stifles democratic accountability by furnishing these experts with determinative rather than consultative capabilities[131]. Since there are too many experts for any elected representative to ever manage in these large systems, this group of specialists effectively skirt traditional modes of civic accountability.

The AI "revolution" teases this dilemma into new territory. As in industry, political administrators are easily tempted toward the presumed incentives of fauxtomation—efficiency, self-regulation, cost savings, etc.[79] This temptation leads them headlong toward a murky accounting of the contingent labor required to accomplish desired outcomes. The introduction of yet another layer of abstraction into state administration puts yet more distance between the public and their representatives[132, 133].⑨ Worse, Kafkaesque modes of administrative accountability fatigue the public's sensitivity to their civic entitlements. "Decision-making structures become systems of domination", warn Downey and Simons about the failings of contemporary pre-automated democratic procedures, "Nobody appears to have responsibility for the reproduction of injustice over time: not elected representatives, delegated agencies or private corporations"[131]. As in the American and Indian military contexts referenced above, complexity has exhausted the system's potential for capacity.

The promise (or specter) of automation is that it can resolve complex administrative tradeoffs in a seemingly rational fashion. Regrettably, as demonstrated in the opening to this article, disingenuous rhetoric around the true capabilities of such techniques distorts a clear appraisal of their worth. Confusion over this accounting becomes, in the process, its own powerful form of deflection. When questioned by the US Congress and Senate about Facebook's content moderation architecture in 2018, for instance, Mark Zuckerberg made frequent appeals to the efficacy of "artificial intelligence" to solve known problems[134], despite the

---

⑨ Lanius introduces how statistical technologies distort expectations about evidence amongst black and white communities. Hill shows how access to evidence from sophisticated analytical tools privileges those in the criminal justice system but penalizes marginalized individuals. Literature on the digital divide substantiates other disparities caused by the politics of digitization, such as the fact that the majority of content on the internet is in English, which alienates people who speak other languages, and that this content is most often developed for haptic interfaces on computers and smartphones, which alienates people with disabilities.

efficacy of such methods remaining untested. From this perspective, Zuckerberg's call for patience is in fact a call for the public to subsidize the status quo; to absorb the costs of his failure indefinitely in the hopes of an imminent technological solution—a simple expression of horizon politics in action. In the process, technical and intellectual debts continue to accrue, along with the social costs of abuse, harassment, and misinformation that traffic on his channels.

While Zuckerberg and Facebook can, for the moment, sustain this violent charade, it is less clear that a genuine large-scale democracy can do so as well. Consider the right to a public defender. This right is made trivial if that defender is too overburdened to adequately fulfil the duty, as is now the case in areas in the United States[135]. In this instance, a failure in due process negates the possibility to assert hard-won democratic principles; justice delayed is justice denied. While new technologies are held up as solutions as such problems, their total compounded administrative costs remain unclear at best, as the author has argued. At worst, sophisticated digital architecture is a known hazard to accountability. In an indicative case-study, Dick and Volmar capture what is called "dependency hell" in the use of Microsoft's infrastructure[97]. In this hell, individual components function precisely as intended but systemic failure results, nonetheless. "Who ultimately 'owns' a failure in a system like this?" they ask, "More importantly, who fixes it?"[97]

Algorithmic silence tempts these obscure politics into the light. The term connects acts of restraint that might otherwise be read as dissimilar. If ubiquitous automation is liable for its burdens and not just it promises, then bans on facial recognition technologies can be understood as of a kind with, say, the EU's Working Time Directive (2003/88/EC) and Right to Disconnect, which set out minimum requirements for rest in relation to telework. Each intervention imposes regulatory limits on the prospect of algorithmic optimization. Whether or not the human workplace or the human face is pliable to such techniques is made moot. Regulators, following public pressure, preserve the relatively intimate (if imperfect) modes of accountability permitted by human-to-human scale interaction.

The need to protect time and space from the AI Revolution echoes in literature on AI and medicine. Topol speculates that the core benefit of advanced decision systems will be time savings gained by experts moving away from automation[136]. US doctors currently face a degenerative cycle; more than 50 percent suffer burnout and 25 percent suffer depression—pressures that beget additional medical errors and strain, which exacerbate suffering and can lead to suicide[136]. Topol positions protections on time as a promising line of resolution to this feedback loop, not just for clinician's work/life balance, but also for patient outcomes. A study of 60000 caregiver visits identified the provision of additional patient-to-expert time as the most reliable path to decreasing hospital readmissions, as other studies support[136].[20]

In medicine, human-to-human accountability regimes led to improved outcomes. Summarizing one of several such studies, Topol writes, "Taking the computer out of the exam room and supporting doctors with human medical assistants led to a striking reduction in physician burnout, from 53 percent to 13 percent."[136] This solution is not new. On the contrary, Topol's thesis echoes the sentiment of William Osler, co-founder of John Hopkins Hospital, who wrote in 1895, "A sick man cannot be satisfactorily examined in less than half an hour."[136] Indra Joshi, Digital Health and AI Clinical Lead for NHS England, agrees. Joshi describes the experience of waiting in the journey for treatment—for results, a specialist, or a bed—not as a process, but as a state of being, "A feeling of being neither here nor there"[137]. This is the same torturous state of being that Zuckerberg, Facebook, and other influential proponents of ubiquitous digital automation advocate for and enforce through the tact they take to technological development[138]. Just hold on, the story goes, we are almost there.

To interrupt this rhetoric, critics must adequately diagnose its charm. Crucially, Zuckerberg and peers assume no finite constraints on time. This is their faux reality. Such appeals benefit from at least three levels of illocution[139] (Garvey characterizes the history of AI as a string of illocutionary acts or promises). Within AI, as the author has introduced, technical terms like "predict" describe a desired end state, not a procedure in time. The term "artificial intelligence" is an exemplar of this trend; a vague yet seemingly prophetic sign of a movement yet to come. Reckless critique overlooks this folly. It accepts

---

[20] Giving a patient an additional minute with an expert reduced their probability to being readmitted by 18%, or 13% in the case of nurses. A separate study found that additional time with experts reduced hospitalizations by twenty percent.

AI rhetoric without scrutiny and diverts attention from wishful mnemonics to "wishful worries", Brock's term for "problems that it would be nice to have, in contrast to the actual agonies of the present"[140]. Meanwhile, a fleet of human contributions, both paid and unpaid, perform, unknowingly and knowingly, a broad array of discreet tasks that, if overlooked as systemic and connected, might lend AI an air of legitimacy and imminence. Like the Church-Turing Thesis, AI provides a tantalizing and multifaceted escape from the existence of time and space, but only for a privileged few.

To interrupt this wishful cycle, critics must situate AI within the post digital era, meaning the period in which, "The revolutionary phase of the information age has surely passed"[141]. Cut off from the ability to escape time or make vague appeals to imminent transformation, AI advocates would be pressed to justify their interventions on alternative grounds. One option the author has championed here is to audit the labor required to develop, deploy, maintain, critique, and use such tools. If this was a norm, a clearer picture of AI's proffered impact on labor could begin to emerge. More likely, expert-led calls for algorithmic accountability would be met with a charge akin to "Luddite!". The author, for one, fears that the history of Luddism is too disanalogous to today to accommodate the paradoxes of contemporary automation, replete as it is with the compounding intersectional realities of gender, race, class, coloniality, and globalization[21, 142]. Digital tools embody opportunities and risk across many layers simultaneously; their treatment deserves more nuance.

Enter algorithmic silence. If unburdened by the accumulated labor required to perform the AI Revolution ad infinitum, citizens would gain the incremental derivative economic or administrative relief needed to decide on a civic future for themselves. Their reliance on technocrats posturing as AI ethicists would be diminished in proportion to the nonproliferation of faux automation systems, since—in principle—the civic space in which they operate would be relatively less influenced by unrestricted impositions on their finite time. Algorithmic silence provides a content agnostic framework for solidarity across settings, be it restraint for workers, consumers, parents, prisoners, women, youth, etc. The prospect of solidarity across these contexts is, in principle, broad enough to answer orthogonal pressures from data science. Ribes, for example, shows how the term "domain" presupposes a role for computing in areas of life not yet conscripted into such methods[143–145]. For solidarity to emerge across countercultures, interventions must evidence a larger movement, whatever it may be called. Algorithmic silence is a step toward that end.

As critics mobilize against automation's harms, they must confront the possibility of achieving a Pyrrhic victory. Clearly articulated ethical principles would indeed be a positive result, but their enshrinement into law remains only half the battle (see also Ref. [146], this issue). Commitments to due process must also be considered, articulated, enacted, and enforced, or hard-won principles will be a farce, as is witnessed with overworked public defenders and caregivers. The politics of procedure and promise of automation merit deep contemplation in a moment when indigenous leaders and scholars in particular reaffirm ancient notions of accountability to place, planet, and people that stand to exceed the shortcomings of liberal democratic imaginaries[147–149]. Transformation is possible, but likely not via appeasement. By continuing to normalize the presumption that automation can be refined and improved—that satisfactory tech ethics can be articulated—those in the realm of automation development and critique point to a loadstar that either misguides them, or makes real a system of politics that, in fact, they endorse but have not yet been held accountable for.

# 6 Conclusion

Arthur C. Clarke's popular Third Law About the Future boasts, "Any sufficiently advanced technology is indistinguishable from magic."[150] This literary "law" is often cited in salesmanship that surrounds the AI Revolution. It is used to paint a boundary between those who create technology and those who merely witness it. In this article, the author has questioned that boundary by exploring the ways in which groups who experience the "magic" of digital automation is often made into co-managers of that performance via ghost work, technical debt, intellectual debt, the labor of critique, participatory labor, or some combination therein. The author questions how the experience of advanced technologies changes as onlookers participate in an increasing number of performances simultaneously, day after day, week after week, without structured relief to their expected vigilance. Clarke's "law" claims to speak to the performative aspects of a new technology. Yet, tellingly, it speaks not at all to experience of those performers whose labor substantiates the act.

Given the need for public awareness around the structural impositions caused by an automated society, as well as the risk of paternalism that accompanies unchecked faith in a technocratic expert-led resistance, it is worthwhile to question which vocabularies adequately capture the character of the phenomenon the author has engaged herein. Algorithmic silence resists the tradition of highly formalized and positivist articulations of social dynamics that prefigure and inform contemporary forms of digital automation. The concept, instead, reifies the virtues of deliberate relief from these types of knowing. At best, it affords collective freedoms from the onslaught of formalisms and encoded behaviors that are sure to accompany the prolific use of low-cost automation. Algorithmic silence treats rest as its own dignified vehicle to progress—one that could surface lines of solidarity across otherwise divisive relationships changed by a rising torrent of discrete obligations. With each passing day, the global community awakens to the reality that, as Dick and Volmar suggest, we are all system administers now (or will be, eventually). Servicing the need for spaces untouched by algorithmic enclosure would allow civic communities the distance to reflect on and shape this unfolding phenomenon for themselves—or at least see that it is occurring.

Acts of wholesale prohibition such as that which the author distills as algorithmic silence tempt reflection on the ethos of entitlement that sustains contemporary myths about digital automation and a looming AI Revolution. If judged in relation to time and space, as opposed to the timelessness of an endless horizons, AI fits more neatly into the post-digital era in which no significant change to the existing social order is to be expected. At a superficial level, this reappraisal of rhetoric could help to steer AI development in line with existing traditions of de-escalation, such as decomputerization and degrowth, although the nuances of this proposal merit closer consideration (since algorithmic silence could also be abused). Those who address the environmental toll of machine learning systems, however, have made similar calls for decomputerization[151, 152]. Such acts of relief color the edges of what could become a powerful deindustrial revolution: a transformation equal in magnitude to the fabled AI Revolution but led, instead, by communities rather than corporate needs.

## Acknowledgment

Special thanks to Sarah T. Hamid, Sarah Dillon, Stephanie Dick, Richard Staley, Helen Anne Curry, Momin M. Malik, Mustafa Ali, Mary Gray, Sean McDonald, William Lazonick, Ernesto Oyarbide-Magaña, Ben Green, and attendees of the 2020 Istanbul Privacy Symposium.

**Jonnie Penn** received the doctorate and MPhil (First Class Honors) degrees in philosophy from the Department of History and Philosophy of Science at the University of Cambridge in 2020 and 2015, respectively. He also received the BA degree from McGill University in 2013. He serves as an affiliate at the Berkman Klein Center at Harvard University, a research fellow at St. Edmund's College, and as an associate fellow at the Leverhulme Centre for the Future of Intelligence. He currently teaches masters on AI ethics and society at the University of Cambridge (2021−2024), where is also a postdoctoral research fellow in the Department of History and Philosophy of Science and co-organizer of the Mellon Foundation Sawyer Seminar on "Histories of Artificial Intelligence: A Genealogy of Power".

# Connecting Race to Ethics Related to Technology:
# A Call for Critical Tech Ethics

Jenny Ungbha Korn*

**Abstract:** Critical tech ethics is my call for action to influencers, leaders, policymakers, and educators to help move our society towards centering race, deliberately and intentionally, to tech ethics. For too long, when "ethics" is applied broadly across different kinds of technology, ethics does not address race explicitly, including how diverse forms of technologies have contributed to violence against and the marginalization of communities of color. Across several years of research, I have studied online behavior to evaluate gender and racial biases. I have concluded that a way to improve technologies, including the Internet, is to create a specific type of ethics termed "critical tech ethics" that connects race to ethics related to technology. This article covers guiding theories for discovering critical tech ethical challenges, contemporary examples for illustrating critical tech ethical challenges, and institutional changes across business, education, and civil society actors for teaching critical tech ethics and encouraging the integration of critical tech ethics with undergraduate computer science. Critical tech ethics has been developed with the imperative to help improve society through connecting race to ethics related to technology, so that we may reduce the propagation of racial injustices currently occurring by educational institutions, technology corporations, and civil actors. My aim is to improve racial equity through the development of critical tech ethics as research, teaching, and practice in social norms, higher education, policy making, and civil society.

**Key words:** race; gender; ethics; tech; bias; equity; society; policy

## 1 Sociocultural Introduction

It is July of 2020, and I am writing this article during a time of racial unrest and personal loss. A few months ago, George Perry Floyd Jr, a Black man, was killed by a White police officer, Derek Chauvin, who knelt on Floyd's neck for nearly eight minutes. Like many activists, I participated in protests to draw attention to continued racism, police brutality, and racial injustice. Located in Alabama, I marched in my hometown's Black Lives Matter protest, where police snipers were stationed at the tops of buildings, poised to shoot ordinary citizens of different races engaged in peaceful activism. George Floyd's murder occurred during the COVID pandemic, which attacked both of my parents, causing them both to be hospitalized and intubated. My daddy died within a week of George Floyd's death.

I provide details about this particular sociocultural moment to make the point that the time for a closer inspection of how race relates to ethics and technology has arrived. Over the past few months, I have received dozens of emails from companies and organizations stating their condemnation of racism, promotion of equality, and support of inclusion. Entire associations are now stating that Black Lives Matter. They are stating that anti-Asian racism and medical racism related to COVID, which my family experienced[1], are wrong. This country is examining racial injustice across a variety of contexts, including sports, crime, politics, medicine, and technology. If there is a time to call for critical tech ethics, it is most assuredly right now.

● Jenny Ungbha Korn is with the Berkman Klein Center for Internet and Society, Harvard University, Cambridge, MA 02138, USA. E-mail: jkorn@cyber.harvard.edu.
∗ To whom correspondence should be addressed.
    Manuscript received: 2021-06-22; revised: 2021-11-22; accepted: 2021-11-25

## 2 Article Outline

Critical tech ethics is my call for action to influencers, leaders, policymakers, and educators to help move our society towards centering race, deliberately and intentionally, to tech ethics. For too long, when "ethics" is applied broadly across different kinds of technology, ethics does not address race explicitly, including how diverse forms of technologies have contributed to violence against and the marginalization of communities of color. Across several years of research[2−11], I have studied online behavior to evaluate gender and racial biases. I have concluded that a way to improve technologies, including the Internet, is to create a specific type of ethics termed critical tech ethics that connects race to ethics related to technology.

This article covers:

• Guiding theories for discovering critical tech ethical challenges;

• Contemporary examples for illustrating critical tech ethical challenges;

• Institutional changes across business, educational, and civil society actors for teaching critical tech ethics and encouraging the integration of critical tech ethics with undergraduate computer science.

## 3 Guiding Theories

The theories that inform this article are all drawn from critical theory, including critical race theory, intersectional feminist theory, and critical race feminist theory. In this section of the article, I briefly highlight significant components of all three related theories to demonstrate how their contributions on race, gender, and diverse axes of identity intersect with digital media ethics to create critical tech ethics.

Both critical race theory and intersectional feminist theory emerged in the 1980s[12, 13]. Immediately, critical race theory became popular within academia, especially in the fields of law and education. In contrast, though intersections of race with gender had been highlighted by prominent feminists of color in the 1980s[12], intersectional feminist theory was slower in its adoption, not gaining widespread recognition until the 1990s[14].

Critical race theory has at least three tenets that are relevant directly to critical tech ethics[13]:

• Concepts that are held as "race-neutral" are tied to White supremacy and racism.

• Racism is acknowledged as ordinary, fundamental, and embedded within American society.

• Awareness of examples of hegemonic Whiteness should lead practitioners of critical race theory to create and support interventions to transform social structures and advance social justice.

Intersectional feminist theory informs this article by stressing the concurrent ways that axes of identity are activated in their oppressions[12]. Specifically, the applications of intersectional feminist theory used in analyses for this article are:

• Race alone and gender alone are not adequate ways to analyze the results of the inputs and outputs related to online behavior.

• The intersection of race with gender lends important insights into understanding the inputs and outputs related to online behavior.

Finally, critical race feminist theory, as the name implies, combines components of critical race theory with intersectional feminist theory[15]. In fact, key advocates of both critical race theory and intersectional feminist theory have helped to form critical race feminist theory. Since the mid-1990s, critical race feminist theory has been forming adherents, but it lags in popularity behind critical race theory and intersectional feminist theory.

The key reminders from critical race feminist theory most applicable to this work are[15]:

• The socially-constructed categories of race with gender should not be reduced to essentialism. In other words, women of color, men of color, and people of color who do not identify with binary gender experience the world differently from one another across genders, which is a presumption that is different from earlier forms of critical race theory that lumped men and women of color together under the umbrella term of race.

• Women and people of color who do not identify with binary gender are not monolithic. Perceived differences across racial and ethnic divides influence concepts of what it is to be Indigenous, African American/Black, Asian/Asian American, Latina/Latine/Latinx, Caucasian/White, and so on. The presence of one woman of color online is not representative of all women of color, particularly across ethnicities.

Critical race theory[16], intersectional feminist theory[17], and critical race feminist theory[15] encourage special attention be paid to race and gender. Following such traditions set by scholars of color, I use this article to illustrate why and how critical tech ethics should be developed as an area that connects criticality around race and gender with technology ethics, including digital media. Indigenous scholar Ess[18] has defined digital media ethics as addressing the moral principles related to activities conducted via computing technologies and online systems. A data practice that I challenge is how acritical and supportive of the status quo "ethics" in artificial intelligence, computing technologies, and online systems has been. Digital media ethics has been heralded as a way to consider the social good, producing tech conferences devoted to the combination of ethics with artificial intelligences. Ethics is the current buzzword for the funding of grants for civil and academic artificial intelligence projects[19]. But how often does tech ethics explicitly engage with racial equity? I explore answers to this question here.

After presenting real-life examples illustrating ethical challenges that are not race-neutral, I advocate institutional changes for teaching critical tech ethics and marketplace changes for encouraging the integration of critical tech ethics into undergraduate computer science education. Though intersectional feminist theory does not include an action component in its application, critical race theory does emphasize interventions as part of analyses. I use the latter sections of this article to critique the algorithms that control so much of our online behavior and highlight interventions that could empower future technology builders to create a healthier Internet for all.

## 4　Online Images

Images influence our conceptions of the world[20, 21], and yet, they are often overlooked in examinations of computer-mediated communication[22, 23]. I use online image searches to highlight the reflexivity between society and technology in (re)producing contemporary American socioeconomic politics, while concurrently shaping attitudes, decisions, and actions about race and gender.

I am a woman of color in the academy. When I entered the keyword of "professor" in an online image search, the algorithm produced a screen full of thirty images[4]. Using critical visual discourse analyses, I examined the presence and absence of diverse embodiments for professors in images from online searches. Of those results from the screen, 87% of the images were highly biased in terms of age, race, gender, and appearance. Twenty-six images were variations of elderly, White men that wore glasses or laboratory coats or appeared in front of chalkboards in a conflation of "professor" with laboratory scientist[24]. The background of a chalkboard matches the emoji suggestion for professor made by iPhones running Apple's iOS 10 for an emoji of a White man standing in front of a chalkboard[25]. Men were shown as bedecked with grey or white hair that stuck out from the head in a hairstyle that has become associated with Dr. Albert Einstein[26, 27], who was a well-known and highly-regarded professor of physics. The visual images of "professor" tended to showcase individuals as rational and scientific, which has been an enduring perspective on the appearance of a professor since the late 1960s[24, 28, 29]. The embodiment of a professor that is normalized through these online image results is intertextual and upholds that a professor is expected to be White, male, and weight-proportionate[30]. Representations of professors within images influence students and their preconceived notions of whom to expect as authoritative and expert in the classroom, which lead to significant implications on student evaluations of teaching.

Another example of how image searches are biased and have real-life consequences is an online query for medical conditions related to skin. Those of us that have experienced bumps or dry patches on our skin might turn to the Internet for images to figure out what might be ailing us. Unfortunately, like other mass media representations[31], the images that result in online searches nearly always reinforce a dominance of White and male. In 2021, image examples of "bumps" or "hives" yield 100% pale skin as examples. When race and gender are rendered invisible in images online, the outcome may be classified as color-blind and gender-blind. Color- and gender-blindness, often under the guise of neutrality, maintain White racial and male gender domination by normalizing White men as the standard[13, 32, 33]. Pictures of diseases related to skin tend to be on white skin in medical textbooks, physical and online, which leads to the perpetuation of biases in health care, limitations on health diagnoses, and inequities in medical training related to allergies and diseases of the skin, by professionals and amateurs alike[34, 35]. Omissions from these online images results become othered[36]: White men are legitimated as professors, and white skin is validated as the foundation for visible detection of skin conditions.

Algorithmic othering is happening. As algorithmic systems become commonplace, we should be represented in algorithmic results. Examples of biases along race and gender extend beyond search results. For example, facial recognition and covert surveillance technologies have been used by those in power to oppress communities of color to unjust outcomes affecting employment, prosecution, and more. I choose examples of representation online because representation continues to matter and because their results often go unquestioned by acritical search users that believe online searches yield neutral findings. An education in critical tech ethics would behoove the individuals that program and impact the creation of the algorithms that increasingly construct our online world and the individuals that casually and critically use and benefit from such algorithms.

## 5 Institutional Changes

A reason to promote critical tech ethics is to ensure that race and gender are prioritized within digital media ethics. Earlier in this article, I questioned how often tech ethics explicitly engages with racial equity. One domain that provides empirical data on how ethics might connect with race lies within the university system of the United States.

As part of my keynote for Mozilla in 2019[37], to gain better understanding about the primacy of race within undergraduate computer science education, I analyzed a public, online listing from 2018 of the names of crowdsourced courses identified as ethics related to technology[38]. As part of the listing, instructors could opt to share their syllabus. With syllabi as my units of analyses, I used curricula by faculty to analyze how ethics was defined by the individuals that were teaching self-identified courses in ethics. What does "ethics" mean in praxis, not in theory, when ethics is taught to undergraduate students? And how often does such education in ethics intersect with issues related to race? To focus on how race is construed in the context of an American computer science department, I audited when and how the topic of race was explicitly referenced by faculty that used English as the primary language in their education of ethics to undergraduate students in computer science in the United States.

Using thematic and critical discourse analyses on the results of the ten syllabi whose entire contents were available publicly online, from undergraduate ethics courses taught in computer science in the United States, eight syllabi did not list race explicitly as a topic of focus for any class of the entire school term, leaving only two syllabi that featured race specifically for a class session. While stating race at all makes the faculty that created those two syllabi exemplary, it was unfortunate that the topic of race was constructed to fill only a single class session, as opposed to having race in tech ethics as a regular part of an ongoing discussion across all class sessions. Each of the two syllabi construed "race" in two different contexts: one syllabus defined race in terms of improving the racial diversity of employees in the field of computer science[39]. Another syllabus identified race as a factor for influencing, and being influenced by, algorithmic data[40]. Outside of those two syllabi that included a class session on race, four syllabi included links to supplemental readings that were aligned with the latter definition of race, namely, algorithmic bias in terms of racist outcomes against the Black community[41−44]. One syllabus mentioned "algorithmic fairness" as a topic, but race was never introduced; instead, ethical considerations about algorithmic fairness were defined in terms of the extinction of humanity by robots and the attachment of emotions related to robots. In other words, the ethics of robotics was considered a priority by faculty, but the ethics of race was rendered irrelevant for this undergraduate course: robots appeared as an ethical issue in artificial intelligence on this syllabus, but race as an ethical issue related to artificial intelligence did not materialize in the syllabus for this course. I provide these results as a snapshot in time of how tech ethics is such a broad area that the topic of race may be rendered invisible.

As a topic for teaching, research, and discussion, race may be more uncomfortable, and therefore more challenging, for those that are untrained in critical race theory or for those whose lived experiences represent the institutionally-dominant White community in the United States. For the vast majority of undergraduate computer science classes taught about ethics, ethics is acritical and supportive of the status quo. While tech ethics might be heralded as a way to consider the social good[45], tech ethics training tends not to engage explicitly with racial equity. In practice, across the training and education of civil society organizations, ethics tends to rely upon the work of heterosexual White cis male philosophers and does not address intersectional justice across races, genders, and sexualities.

Computer science ethics classes are often taught by computer science faculty that have minimal training in ethics, let alone any training in critical race theory. This lack of training is a systemic issue that reflects biases in what expertise is seen as valuable: computer science ethics is taught by technical scholars who have self-studied some ethics, rather than people with deep expertise in ethics and race. As I have advocated in my public scholarship[7, 37], universities supportive of critical tech ethics should seek to hire faculty with training in and whose scholarship promotes critical race theory, intersectional feminist theory, or critical race feminist theory connected to ethics related to technologies.

# 6   Critical Tech Ethics

Ethics without intentional criticality results in a panacea for people with the power to influence computer science, digital systems, and artificial intelligence. Ethics devoid of critical race training is incomplete and deleterious. I am concerned about a responsibility gap between decisions made by people designing algorithms and people experiencing algorithmic biases. I position accountability for racial fairness upon existing business, educational, and civil society institutions that train and hire individuals and upon established organizations that design and manage algorithms. A way to guide better interactions between artificial intelligence and diverse humans is to provide improved academic and social instruction related to racial equity to creators and users of technologies for academic communities, technology organizations, and civil society actors. Rather than present ethics as race-neutral, reflecting a philosophy of color-blindness[10], I seek to institutionalize considerations of racial equity through the establishment of critical tech ethics.

Technology is not neutral. Algorithms have embedded values. The question then is whose truth is reflected and whose truth is omitted in the design and use of algorithms. Algorithmic bias happens because values are implicit within the programming and design of the algorithms for online behavior[46−48]. Algorithms fit with and help advance a single race as the dominant culture in the United States[48]. Critical tech ethics makes explicit the implicit ways that Whiteness is hegemonic to the detriment of other races. Critical tech ethics is based on critical race training that offers both intellectual and political responses to challenge racial power and change American society. I encourage readers to engage in digital acts of racial realism, as described by Bell Jr[49] to "challenge principles of racial equality" and to use "social mechanisms" to "have voice and outrage heard".

Critical tech ethics is an area of study and application that includes:

(1) **Institutionalizing critical tech ethics through mandating racially-aware standards for reviewing research, awards, grants, and funding**: Specifically, I seek to construct "racial implications of this proposal" into a critical tech ethics standard for civil society organizations because downstream uses of artificial intelligence should be part of the intellectual rigor that is valued for judging work in reviews[8]. In doing so, organizations and companies that mandate considerations of racial implications in their applications signal that racial awareness is a significant factor in awarding funding and awards, which, in turn, encourages participants to reflect upon how their work is impacted by and imbricated with race, racism, and racial equity.

(2) **Setting expectations for teaching critical tech ethics centered on racial equity**: Required training in critical race theory would help those creating our technological worlds to understand better about ethical considerations related to race. Specifically, such education should be informed by critical race theory to change norms and demonstrate how computer science, digital systems, and artificial intelligence have played a role in the episteme and techne of racism[8]. In doing so, critical tech ethics actively builds in discussions of race, racism, and racial justice to minimize the reproduction and hegemony of Whiteness by those in programming, coding, computer science, engineering, and open source communities.

(3) **Establishing critical tech ethics practices for improving industry norms aligned with the goal of improving racial justice**: Due to the influence of capitalism upon choices made by students for profitable careers and choices made by universities to supply employees for in-demand occupational niches, industry must also be part of the equation to establish critical race thinking as part of everyday computer science education in the United States. To encourage institutions to mandate the addition of critical tech ethics, employers will need to update the requirement section of their job ads to state the desirability of hiring individuals with training in considering the racial implications of artificial intelligence[7, 8, 10]. In doing so, technology corporations may take a step towards contributing towards racial justice, which involves tactics, actions, and attitudes that challenge racial power, resulting in more equitable opportunities and outcomes[50].

## 7 Conclusion

Included within this article is a call for action to influencers, leaders, and policymakers to take note to help move our society towards greater justice for everyone, particularly communities of color. To combat racism and sexism[51, 52], changes to existing curricula must occur. Students themselves acknowledge that critical race thinking should be taught more frequently than they are available currently[53]. Across leading institutions globally, a lack of inclusion of race, gender, intersectionality, and power leads to an enactment of ethics education lacking in justice. For too long, the rhetoric of diversity has been unaccompanied by institutional change. We must recognize and address that computer science departments in the United States have overlooked how the technologies on which they are training future programmers are impacted by and imbricated with race, gender, sexuality, religion, and other axes of identity. Presumptions about the neutrality of algorithms have resulted in the biases we see today in the inputs and outputs of various technologies[46−48]. Countering those biases through critical tech ethics will be helpful in reducing unfair and unjust outcomes based on algorithms. Rendering diversity in race and gender as visible is a process that will take greater commitment by those producing the algorithms and those using the algorithms because online data are a social enterprise[23]. Critical tech ethics has been developed with the imperative to help improve society through connecting race to ethics related to technology, so that we may reduce the propagation of racial injustices currently occurring by educational institutions, technology corporations, and civil actors. We should live in a world in which the responsibilities for racial equity do not fall on people of color only, but are borne by everyone that influences and is influenced by algorithms. My project is to improve racial equity through the development of critical tech ethics as research, teaching, and practice in social norms, higher education, policy making, and civil society.

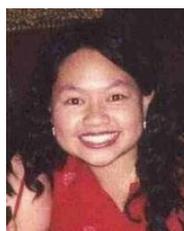 **Jenny Ungbha Korn** is a feminist activist of color for social justice, a ciswoman scholar of race and gender in mass media, digital life, and artificial intelligence, and a member of Mensa, the high intelligence quotient (IQ) society. She is an affiliate and the founding coordinator of the Race+Tech+Media Working Group at the Berkman Klein Center for Internet and Society at Harvard University. She is the founding director of Princeton Diversity Discussions, a free, public, ongoing series since 2014 of in-person and digital group gatherings to share personal opinions and lived experiences focused on race, racism, and racial justice that meets weekly across the sponsorship of 30+ associations in the United States. She is the author of numerous publications, she has won awards from the Carl Couch Center for Social and Internet Research; the Association for Information Science and Technology; the Philosophy of Communication Division and the African American Communication and Culture Division of the National Communication Association; the Minorities and Communication Division and the Communication Theory and Methodology Division of the Association for Education in Journalism and Mass Communication; and the Organization for the Study of Communication, Language, and Gender. She has given over one hundred talks as invited keynote presentations, university guest lectures, interactive community education, and refereed conference presentations. As a public scholar, she has been quoted in interviews with *NPR*, *CNN*, *SXSW*, *Bustle*, *Colorlines*, *Fox News*, *Forbes*, *Mashable*, *Reader's Digest*, *U.S. News & World Report*, *Washington Post*, and more. Drawing on critical race and intersectional feminist theories, she explores how Internet spaces and artificial intelligences influence, and are influenced, by assemblages of race and gender and how online producers-consumers have constructed inventive digital representations and computer-mediated communications of identity.

# Critical Technical Awakenings

Maya Malik and Momin M. Malik*

**Abstract:** Starting with Philip E. Agre's 1997 essay on "critical technical practice", we consider examples of writings from computer science where authors describe "waking up" from a previously narrow technical approach to the world, enabling them to recognize how their previous efforts towards social change had been ineffective. We use these examples first to talk about the underlying assumptions of a technology-centric approach to social problems, and second to theorize these awakenings in terms of Paulo Freire's idea of *critical consciousness*. Specifically, understanding these awakenings among technical practitioners as examples of this more general phenomenon gives guidance for how we might encourage and guide critical awakenings in order to get more technologists working effectively towards positive social change.

**Key words:** critical technical practice; critical consciousness; perspective transformation; education; machine learning

## 1 Introduction

In 1997, then-UCLA professor Philip E. Agre published a remarkable essay, entitled "Towards a Critical Technical Practice: Lessons Learned in Trying to Reform AI"[1]. In it, Agre describes his experience as a doctoral student in AI at MIT in the 1980s undergoing a crisis of faith in his discipline and looking to other disciplines for answers. Agre writes (bold emphasis added):

"As an AI practitioner already well immersed in the AI literature, I had incorporated the field's taste for technical formalization so thoroughly into my own cognitive style that **I literally could not read the literatures of nontechnical fields at anything beyond a popular level**. The problem was not exactly that I could not understand the vocabulary, but that **I insisted**

**on trying to read everything as a narration of the workings of a mechanism**.

"My first intellectual breakthrough came when, for reasons I do not recall, it finally occurred to me to stop translating these strange disciplinary languages into technical schemata, and instead simply to learn **them on their own terms**.

"**I still remember the vertigo I felt during this period**; I was speaking these strange disciplinary languages, in a wobbly fashion at first, without knowing what they meant—without knowing what *sort* of meaning they had... **in retrospect this was the period during which I began to 'wake up', breaking out of a technical cognitive style that I now regard as extremely constricting**."

In this paper, we use Agre's essay as a foil to discuss what we call *critical technical awakenings*: when people from technical disciplines, previously committed to a narrow technical view of the world, "wake up" from that perspective to what we identify as seeing the world through a critical, constructivist lens.

Other articles in this special issue do a fantastic job of analyzing the political economy of tech ethics[2, 3]. While recognizing that structural change at this level is our ultimate goal, our focus here is in taking up a specific slice of how to achieve this: what makes certain technical

• Maya Malik is with the School of Social Work, McGill University, Montréal, H3A 0G4, Canada. E-mail: maya.malik@mail.mcgill.ca.
• Momin M. Malik is with the Institute in Critical Quantitative, Computational, & Mixed Methodologies, Johns Hopkins University, Baltimore, MD 21218, USA. E-mail: momin.malik@gmail.com.
∗ To whom correspondence should be addressed.
  Manuscript received: 2021-05-20; revised: 2021-12-07; accepted: 2021-12-08

practitioners come to care about understanding this larger context, and how do some individuals become committed to working towards structural change? We do not mean to imply that "ethics" are a problem at the level of individuals; but, as we will argue, individual-level awakenings play a central role in building communities that effectively work towards positive structural change, and so are crucial to consider.

Our goal is not necessarily to convince people purely within a "technical perspective" that they should change (indeed, we argue that rational argumentation alone is insufficient to cause change), but rather to speak to people who are in the process of undergoing, or who have recently undergone, the type of awakening we identify. Awakenings can be a lonely and confusing process, but need not be. By pointing to existing examples and theorizing this process, and by providing guidance about how to productively channel and shape awakenings, we hope to make it less difficult to go through an awakening, and thereby encourage and contribute to growing a community of critical technical practitioners within modern data practice and technology design.

Specifically, we aim to:

• Review the existence of different ways of approaching the world and their different underlying assumptions (in Section 2);

• Identify what is initially compelling about a "technical perspective", but how and why some of its adherents rightly come to see this perspective as insufficient (in Section 3);

• Draw on Paulo Freire's idea of *critical consciousness* and subsequent theory from adult education[4], in order to theorize critical awakenings more broadly (in Section 4);

• Present a specific view of ethics and argue that this should be the goal of critical technical awakenings (in Section 5);

• Examine potential shortcomings of existing examples of critical technical awakenings in light of adult education's prescriptive positions on what makes a "complete" awakening, and by advocating for a care-based ethical code which the examples do not seem to have arrived at (in Section 6).

As a note, the awakenings we discuss are not technical in nature. Perhaps "critical-technical awakenings", "critical awakenings in tech", or "critical sociotechnical awakenings" would be more appropriate; we use the phrase "critical technical awakenings" to emphasize a connection to Agre's critical technical practice and, in contrast to other examples of people writing about "critical awakenings"[5, 6] to emphasize the awakenings in question being experienced by people in *technical* fields.

## 2 Paradigms of Social Research

Training in social research includes, as a basic part of any research methods course, an introduction to different research paradigms. For people who carry out social research from a technical background, this may not be something they have been exposed to; but even if it is, the abstract layout of different paradigms may not be meaningful. To set up the remainder of the discussion, we first present our take on the standard view of the contours of social research in Table 1, with further descriptions in a glossary Appendix, and try to point out how it relates to a technical perspective versus what people might awake to.

The rows correspond to subfields of philosophy, but here more specifically and narrowly represent types of assumptions within that philosophical domain, respectively about the nature of things (ontology), how we *can* know things (epistemology), and how we actually *go about* knowing things (methodology). While not always present in charts like this one, *axiology* is an additional branch of philosophy that contains ethics (what is good) and aesthetics (what is beautiful). Within this, we specifically care about *normative ethics*, which are choices of codes of conduct to which we should adhere (which are how we go about being ethical), as opposed to, say, descriptive ethics (descriptions of what certain people believe to be ethical).

The columns represent different paradigms of social research, and the cells are the assumptions that each paradigm makes. These assumptions are fundamental and foundational, and cannot be debated, justified, or refuted through empirical means (since, among other things, these assumptions are about the very possibility, reliability, and even definition of empirical evidence). In the Appendix, we provide a glossary with extensive descriptions of these columns and some specific terms that appear in the cells.

Neither the rows nor the columns are cleanly separated or singular; positions can bleed into one another, and a

**Table 1   Assumptions of social research paradigms. Based on Guba and Lincoln's "Basic beliefs (metaphysics) of alternative inquiry paradigms"[7]. See Appendix for details.**

| Issue | Positivism | Postpositivism | Critical theory | Constructivism | Participatory |
|---|---|---|---|---|---|
| Ontology (assumptions about the nature of things) | Naïve realism. Reality is independent of and prior to human conception of it, and apprehensible. | Critical realism: Reality is independent of and prior to human conception of it, but only imperfectly and approximately apprehensible. | Disenchantment theory: there is a reality, shaped by social, political, cultural, economic, ethnic, and gender values and solidified over time, but it is secret/hidden. | Relativism: multiple realities and experiences of truth, constructed in history through social processes. | Participative: multiple realities, each co-constructed through interactions between specific people and environments. |
| Epistemology (assumptions about how can know things) | Reality is knowable through reason and observation. It is possible to have findings that are singular, perspective-independent and neutral, atemporal, and therefore universally true. | Findings are provisionally true; multiple descriptions can be valid but are probably equivalent; findings can be affected/distorted by social and cultural factors. | The truth of findings is mediated by their value; how we come to know something, or who comes to know something, matters for how meaningful it is. | Relativistic: there is no neutral or objective perspective from which to adjudicate competing perspective or truth claims; truth is relative to a given perspective. | We come to know things, and create new understandings that can transform the world, by involving other people in the process of inquiry. |
| Methodology (how we go about trying to know things) | Experimental/ manipulative (hypothetico-deductive); hypotheses can be verified as true. Chiefly quantitative methods, and mathematical representation. | Modified experimental/ manipulative; *falsification* of hypotheses; primacy of quantitative methods, but may include qualitative and mixed methods. | Dialogic (through conversation and debate) or dialectical (through a process of thesis, antithesis, and a synthesis which becomes a new thesis) | Dialetical, or hermeneutical (a process of reading sources "against themselves" to identify inconsistencies, underlying assumptions, or implicit messages, and thereby interpret meaning). | Collaborative, action-focused; flattening researcher/ participant hierarchies; engaging in self- and collective reflection; jointly deciding to engage in individual or collective action. |
| Axiology (ethics; values; who matters, who is important, who has standing) | Knowledge achieved through hypothetico-deductive means is more valuable than other knowledge. The people who can carry out such investigation have privileged access to the truth, and thus have a special role and importance (and potentially a special responsibility). | Knowledge achieved through hypothetico-deductive is more valuable, but can be distorted by social/cultural factors, and this can sometimes only be uncovered by qualitative means and insight. Qualitative methods can provide checks and context, or raw material for quantification. | Marginalization is what is most important; experience of marginalization provides unique insights, and the knowledge of the marginalized is more valuable than the knowledge of dominant/legitimate paradigms. | Understanding the process of construction is what is valuable; value (including valuing understanding the process of construction) is relative to a given perspective. | Everyone is valuable. Reflexivity, co-created knowledge, and non-western ways of knowing are valuable and combat erasure and dehumanization. |

single column can cover a variety of irreconcilable different perspectives (for example, *logical* positivism tries to remove the ontological assumptions of realism from positivism's quantitative empirical commitments, and conversely, mathematical realism often disdains empiricism). We identify the purest form of a "technical perspective" as falling squarely within the "positivism" column, but the perspective we discuss is more

specifically about the power of technology to effect social change.①

These columns are not exhaustive or mutually exclusive, but represent useful clusters. But, even

---

① This includes the perspective of *technological determinism*, a position largely rejected in social science that holds that given technology inherently effects certain causal changes, independent of context. See Green's article in this special issue[8] for details. A softer version allows for context as a moderator, but still sees technology as having inherent causal power.

beyond this, as individuals we human beings can be inconsistent or even contradictory in the sets of assumptions we make (crossing multiple columns at different times or even at once), and we may not even be self-aware of the underlying assumptions we are making. Technical disciplines in particular are frequently positivist without realizing that it is a specific position, or that it is not the only way to see the world. Part of undergoing a critical awakening is coming to be aware that a technical perspective is only one way of looking at the world, and starting to recognize its core underlying assumptions—and reject them.

## 3    The Technical Perspective

One piece of Agre's argument is about the importance of taking AI seriously:

"The central practice of the field of AI, and its central value, was technical formalization. Inasmuch as they regarded technical formalization as the most scientific and the most productive of all known intellectual methods, the field's most prominent members tended to treat their research as the heir of virtually the whole of intellectual history. I have often heard AI people portray philosophy, for example, as a failed project, and describe the social sciences as intellectually sterile. In each case their diagnosis is the same: lacking the precise and expressive methods of AI, these fields are inherently imprecise, woolly, and vague. **Any attempt at a critical engagement with AI should begin with an appreciation of the experiences that have made these extreme views seem so compelling.**"

The target of Agre's critique (and the focus of the first half of his essay) is the AI that existed in the 1980s and 1990s, a very specific and peculiar field (seeing itself as seeking to understanding mechanisms of cognition, in contrast to the machine learning of today which is instrumentally focused on achieving specific tasks and effectively unconcerned with cognition; see Ref. [9]). But the same logic remains: we begin a critical engagement with an appreciation of the experiences that make extreme technical views seem so compelling.

The specific "technical perspective" we refer to here is a position around computation and digital technology and has been identified and critiqued under a series of related terms: Morozov's "tech solutionism"[10]; Toyama's "tech commandments"[11]; Broussard's "tech chauvinism"[12]; and Green's "tech goggles"[13]. These labels emphasize something about the arrogance and absolutism of the technical perspective, and all authors emphasize how adherents are dazzled by the apparent ability of technology (or, if engaging more with the intellectual content than the material artifacts, being dazzled by the apparent power of formalizing goals, operations, and human concepts into mathematical and/or software abstractions) to control and change the world.

As we noted above, at their purest, technical perspectives fall purely within the "positivism" column. We first review the overall appeal of positivism, before focusing specifically on its tech solutionist variety.

A statement by physicists Jean Bricmont and Alan Sokal[14] provides a pure expression:

"In the same way that nearly everyone in his or her everyday life disregards solipsism and radical skepticism and spontaneously adopts a 'realist' or 'objectivist' attitude toward the external world, scientists spontaneously do likewise in their professional work. Indeed, scientists rarely use the word 'realist,' because it is taken for granted: *of course* they want to discover (some aspects of) how the world really is! And *of course* they adhere to the so-called correspondence theory of truth (again, a word that is barely used): if someone says that it is true that a given disease is caused by a given virus, she means that, in actual fact, the disease is caused by the virus.

"We would not even call it a 'theory'; rather, we consider it a *precondition for the intelligibility* of assertions about the world."

This captures something about the aesthetic appeal of positivism and specifically its realist ontology: the world is fundamentally *knowable*. Furthermore, the technical person experiences the satisfaction of having command of the sole means by which to achieve that knowledge.

While, as suggested in this quote, this perspective is widespread in the natural, mathematical, or "hard" sciences, "positivism" was actually coined as an aspiration for social science in the 19th century (see Appendix). Past that period, Porter[15] describes post-WWII behavioralists adopting quantitative methodologies in social science in pursuit of "*liberating essence* of a proper objective methodology" that could "*rise above stubborn tradition and invisible culture*" (emphasis added). That is, they pursued a vision where it is possible to know how the social world "really is", such that it is possible to have intelligible assertions about it (rather than "stubborn tradition and invisible culture" getting in the way of intelligibility).

This idea of liberation through science leads to a view where quantification and formalization are not only *practically* superior, but *morally* superior as well. Everything else in the world is anecdotal evidence, naïve heuristics, and armchair philosophy—shackles of ignorance either useless for accomplishing concrete goals and characterized only by failure, or achieving success only through sheer luck or cheap trickery. That is, even if there is a case where technical approaches are not practically superior (like, for example, convincing climate change deniers), there is a view that they are *morally* superior: even if attempting to understand or intervene in the world through means other than abstraction (i.e., through means like through rhetoric, or narrative) may succeed, those alternatives are dishonest, unprincipled, or otherwise somehow ignoble and compromise our moral integrity.

In addition to this intrinsic moral superiority, positivism seems to comport well with a basis for morality. An observer-independent external world also justifies universal morality—a standard which we can hope to define, and then appeal to for solving moral questions. Indeed, in the so-called "science wars" of the 1990s, when some scientists (initially led by Alan Sokal) took up arms against what they saw as the "fashionable nonsense" of science and technology studies (and related areas), those scientists also bemoaned that while they and the "postmodernists" seemed to share progressive political goals of greater justice and equity, the postmodern perspective was undermining the basis for pursuing that goal and the basis of forming coalitions.

Even worse than getting in the way, "postmodern" arguments are in fact deployed in support of[16, 17] and by climate change deniers, creationists, and all sorts of religious nationalists and right-wing movements across the world. These reactionary elements of society seek to undermine the legitimacy of science in pursuit of a regressive political agenda, and while they clearly believe in a single reality (corresponding to their own beliefs), they co-opt language around plurality and relativism to prevent critique. One of the more forceful arguments around this is by Nanda[18], who argues how Enlightenment beliefs in universality are what we need to defend against perspectives like those of Hindu nationalists, whose weaponization of science studies she documents.

The computation- and technology-focused variety of positivism discussed by Morozov[10], Toyama[11], Broussard[12], and Green[13] is not necessarily about understanding the world, but about acting within it. Toyama discusses (before undergoing what seems like an awakening) thinking technology addresses "real problems"; that both means that the problems are prior to and independent of the perspective of the technologists, and that technology in itself can actually address and solve those problems. Morozov lists examples of Silicon Valley rhetoric about technology changing the world and solving global problems. He summarized the implicit technologist vision of the future in a satirical prediction:

"If Silicon Valley had a designated futurist, her bright vision of the near future... would go something like this: Humanity, equipped with powerful self-tracking devices, finally conquers obesity, insomnia, and global warming as everyone eats less, sleeps better, and emits more appropriately. The fallibility of human memory is conquered too, as the very same tracking devices record and store everything we do. Car keys, faces, factoids: We will never forget them again...

"Politics, finally under the constant and far-reaching gaze of the electorate, is freed from all the sleazy corruption, backroom deals, and inefficient horse trading. Parties are disaggregated and replaced by Groupon-like political campaigns, where users come together—once—to weigh in on issues of direct and immediate relevance to their lives, only to disband shortly afterward. Now that every word—nay, sound— ever uttered by politicians is recorded and stored for posterity, hypocrisy has become obsolete as well. Lobbyists of all stripes have gone extinct as the wealth of data about politicians—their schedules, lunch menus, travel expenses—are posted online for everyone to review...

"Crime is a distant memory, while courts are overstaffed and underworked. Both physical and virtual environments—walls, pavements, doors, and log-in screens—have become 'smart.' That is, they have integrated the plethora of data generated by the self-tracking devices and social-networking services so that now they can predict and prevent criminal behavior simply by analyzing their users. And as users don't even have the chance to commit crimes, prisons are no longer needed either. A triumph of humanism, courtesy of

Silicon Valley."

This is a synthetic caricature, but we can use it to discuss what might be compelling in the perspective that Morozov identifies and critiques. There is a view that technology is practically superior, in that it will succeed where stubborn tradition and invisible culture have failed. But also, tradition and culture are the cause of social problems in the first place; technology is not compromised by their failings, and thus to approach social problems with technology rather than society is a morally superior and more responsible move.

There is an ignominious aspect of the appeal of this technical perspective as well, which Broussard shows. She argues that technologists, who are frequently white, male, and upper-class, fixate on technology as a way to try and solve social problems traditionally managed by people who are Black, women, and/or poor. These men seek to use technology to avoid engaging with the complex and messy labor and understandings these groups have mobilized to manage and address social problems. That is, part of the appeal to the technical perspective is a chauvinistic one: of providing a means to distance oneself from the knowledge, labor, and even existence of devalued people who are women and/or non-white. If we just invent the right device, formalization, or processes, the thinking goes, we can avoid needing to deal with all the ambiguities, nuances, and emotional labor with which, say, Black women social workers engage.

These are the appeals of a technical perspective. What, then, leads people away from it? In awakenings, a common theme seems to be a precipitating event or moment that put the sleeper into a moment of crisis. For Agre, what he described is fairly abstract and intellectual: when trying to decide on a dissertation topic, he found that "Every topic I investigated seemed driven by its own powerful internal logic into a small number of technical solutions, each of which had already been investigated in the literature". In his description, it was his search for a novel topic led him to read the literatures of other disciplines.

Agre does allude to a "large and diverse set of historical conditions" beyond what he presents in the essay. But as he does not elaborate on this, we turn to two other examples of described awakenings, respectively from Kentaro Toyama and Phil Rogaway.

First, we consider Kentaro Toyama, who rejected a technical perspective in a rather "scientific" way. In his book *Geek Heresy: Rescuing Social Change from the Cult of Technology*[11], he describes working after his PhD on "ICT4D"-type projects (Information and Communication Technologies for Development) for Microsoft in India. His position involved expanding technology products' audiences beyond the educationally advantaged Indian middle class to try and help those in poverty. But he repeatedly found his attempted interventions failing.

"In the course of five years, I oversaw at least ten different technology-for-education projects. We explored video-recorded lessons by master teachers; presentation tools that minimized prep time; learning games customizable through simple text editing; inexpensive clickers to poll and track student understanding; software to convert PowerPoint slides into discs for commonly available DVD players; split screens to allow students to work side by side; and on and on. Each time, we thought we were addressing a real problem. But while the designs varied, in the end it didn't matter—technology never made up for a lack of good teachers or good principals. Indifferent administrators didn't suddenly care more because their schools gained clever gadgets; undertrained teachers didn't improve just because they could use digital content; and school budgets didn't expand no matter how many 'cost-saving' machines the schools purchased. If anything, these problems were exacerbated by the technology, which brought its own burdens.

"These revelations were hard to take. I was a computer scientist, a Microsoft employee, and the head of a group that aimed to find digital solutions for the developing world. I wanted nothing more than to see innovation triumph, just as it always did in the engineering papers I was immersed in. But exactly where the need was greatest, technology seemed unable to make a difference."

This was "scientific" in the sense that Toyama was open to evidence by which he tested his assumption that technical tools can circumvent the messiness of society. But the fact that he was even able to recognize that he had such foundational assumptions is not a given; Toyama contrasts his insights to the perspective of a prominent technologist, One Laptop Per Child founder Nicholas Negroponte:

"I was once on a panel at MIT with Negroponte where I outlined my hard-won lessons about technology for education. He didn't like what I said, and he went on the offensive. But he did it with such confidence and self-assurance that, as I listened, I felt myself wanting to be persuaded: Children *are* naturally curious, aren't they? Why *wouldn't* they teach themselves on a nice, friendly laptop?

"As I heard more of the technology hype, however, I realized that it didn't engage with rigorous evidence. It was empty sloganeering that collapsed under critical thinking."

That is, many scientists and technologists are not, in this sense, open to a particular type of empirical evidence. This is not inherently bad or even "unscientific"—work in the history, sociology, and philosophy of science points out that interpretations of empirical evidence require layers of theories and assumptions[19], including the idea that evidence can be erroneous due to human error, issues with instrumentation, or natural variability. Indeed, skepticism of evidence that challenges established theory is an important part of science: but this is all to say, evidence alone is not enough to change minds, such as in an awakening. Kuhn[20] famously theorized that one-off failures in experimental science seldom affect theory, but strings of failures can precipitate a *crisis*, potentially leading to a *paradigm shift* in understanding and defining basic scientific concepts differently (and, conversely, it takes a crisis and not simply routine failures to produce a paradigm shift).

Second, we look at the account of cryptographer Phil Rogaway in his essay, "The Moral Character of Cryptographic Work"[21]. For Rogaway as well, there was a discrete empirical event that led to his identifying and rethinking some fundamental assumptions, but here the challenge posed was a moral one rather than one of assumptions about how the world works not fitting evidence.

"Most academic cryptographers seem to think that our field is a fun, deep, and politically neutral game—a set of puzzles involving communicating parties and notional adversaries. This vision of who we are animates a field whose work is intellectually impressive and rapidly produced, but also quite inbred and divorced from real-world concerns. Is this what cryptography *should* be like? Is it how we *should* expend the bulk of our intellectual capital?

"For me, these questions came to a head with the Snowden disclosures of 2013. If cryptography's most basic aim is to enable secure communications, how could it not be a colossal failure of our field when ordinary people lack even a modicum of communication privacy when interacting electronically? Yet I soon realized that most cryptographers didn't see it this way. Most seemed to feel that the disclosures didn't even implicate us cryptographers."

Also noteworthy is how both Rogaway and Toyama (and Agre as well) describe resistance from their peers to their crisis of faith, and how the experience that led to their transformation did not succeed in triggering others. This contrast again emphasizes that evidence, or external triggers, are not sufficient to cause an awakening; they are only catalysts for already-existing potential.

These accounts do not reflect on what made their authors different from their peers. But understanding these accounts through the lens of adult education and specifically work on critical consciousness (see Appendix), below, will help fill in key answers.

We can also contrast these descriptions to others who, while recognizing the limitations of purely technical approaches, remain within a positivist paradigm (or, at most, soften to a post-positivist one).

Physicist and applied mathematician turned sociologist Duncan Watts[22] wrote that "many of the ideas and metrics of the 'new' science of networks have either been borrowed from, or else rediscovered independently of, a distinguished lineage of work in mathematics, economics, and sociology", acknowledging sociological contributions but reading them in an essentially positivist light. Another person trained in physics and working in network science, César Hidalgo[23], wrote about realizing why "social and natural scientists fail to see eye to eye": "Social scientists focus on explaining how context specific social and economic mechanisms drive the structure of networks and on how networks shape social and economic outcomes. By contrast, natural scientists focus primarily on modeling network characteristics that are independent of context, since their focus is to identify universal characteristics of systems instead of context specific mechanisms". This again positions social science's role by reference to the task of finding universal and objective truths, rather than understanding that (at least some) social science rejects the idea that there could be universal characteristics.

A more personal potential example is Hannah

Wallach's viewpoint, "Computational Social Science ≠ Computer Science + Social Data"[24]. In this she writes, "Despite all the hype, machine learning is not a be-all and end-all solution. We still need social scientists if we are going to use machine learning to study social phenomena in a responsible and ethical manner." A dilemma was only hinted at:

"When I first started working in computational social science, I kept overhearing conversations between computer scientists and social scientists that involved sentences like, 'I don't get it—how is that even research?' And I could not understand why. But then I found this quote by Gary King and Dan Hopkins—two political scientists—that, I think, really captures the heart of this disconnect: 'computer scientists may be interested in finding the needle in the haystack—such as... the right Web page to display from a search—but social scientists are more commonly interested in characterizing the haystack.'

"In other words, the conversations I kept overhearing were occurring because the goals typically pursued by computer scientists and social scientists fall into two very different categories... models for prediction are often intended to *replace* human interpretation or reasoning, whereas models for explanation are intended to *inform* or *guide* human reasoning."

But what she describes overall only goes so far as to recognize the importance of *quantitative* social science—areas of economics like econometrics and game theory, and political science, all of which build formal models for the task of causal understanding. There is no mention of "thick" disciplines that do not use quantitative modeling, such as cultural anthropology, critical sociology, critical race studies, human geography, critical gender studies, media studies, or cultural studies, let alone any mention of experiential ways of knowing outside of academic disciplines.

Like with Agre, from this piece alone it is impossible to know if this encapsulates Wallach's understandings, or if it is rhetorical strategy (indeed, in a later piece, Wallach[25] seems to go beyond post-positivism in recognizing that the notions of "objectivity" are both ill-defined and not desirable, as well as acknowledging positionality[②] [see Appendix]). After all, it is much

② "Will these changes of always having a sociotechnical lens make machine learning less fun? Maybe, for some people. But that is their privilege talking about their ethical debt. Machine learning has never been all that fun for people who are involuntarily represented in datasets or subject to uncontestable life-altering decisions made by machine learning systems."

easier to convince computer scientists of the value of the formalism- and data-heavy discipline of economics than of interpretive disciplines like cultural studies, or of knowledge that comes from lived experience.

## 4 Critical Awakenings

Earlier, we mentioned Kuhn's idea of paradigm shifts. Recognizing that this may be too simple a model for scientific development[26], Mezirow[27, 28] offers a similar model but instead describing individual psychosocial development, which he called *perspective transformation*. More immediately, Mezirow's idea comes from the work of Paulo Freire and his idea of *critical consciousness* (see Appendix), and has a robust body of follow-up work investigating the idea empirically[29] and developing it theoretically[30−32]. We will also draw on subsequent work that has noted shortcomings in Mezirow's theory not going far enough in considering context, other cultural settings, and the significance of interpersonal relationships[32].

Perspective transformation came from Mezirow's study with women who re-entered college programs mid-life. He identified the ultimate value of such programs as being in the personal transformation that took place among the women, rather than any material outcomes. He theorized 10 stages of this process:

"(1) A disorienting dilemma;

"(2) Self-examination with feelings of fear, anger, guilt, or shame;

"(3) A critical assessment of assumptions and a sense of alienation from taken-for-granted social roles and expectations;

"(4) Recognition that one's discontent and the process of transformation are shared and that others have negotiated a similar change;

"(5) Exploration of options for new roles, relationships, and actions;

"(6) Planning a course of action;

"(7) Acquiring knowledge and skills for implementing one's plans;

"(8) Provisional trying of new roles;

"(9) Building competence and self-confidence in new roles and relationships;

"(10) A reintegration into one's life on the basis of conditions dictated by one's new perspective."

These ten stages are somewhere between descriptive and normative. They are descriptive, insofar as they describe a process undergone by the subjects of Mezirow's study, but normative, insofar as Mezirow identified perspective transformation as something valuable and possibly aided by knowing about this sequence in advance and following it (following Freire, and the idea of critical consciousness as a normative goal). While this alone does not necessarily shed light on *who* would experience a dilemma as disorienting and change in response (since Mezirow encountered women already pursuing a change), it does point to how this change does not happen in isolation, and indeed how connecting with others who have negotiated a similar change is key for shaping awakenings towards productive ends. But Mezirow[33] does provide an answer for the question of what is needed beyond evidence, observing that an additional condition is that a person *reflect* about assumptions and beliefs that structured how they understood an experience (or evidence).

Also noteworthy are the examples of disorienting dilemmas: they included "the death of a husband, a divorce, the loss of a job, a change of city of residence, retirement, an empty nest, a remarriage, the near fatal accident of an only child, or jealousy of a friend who had launched a new career successfully". In comparison, the *dilemmas* of Agre, Toyama, and Rogaway are decidedly elite and privileged experiences. Still, we can identify critical technical awakenings as a specific form of a much more general phenomenon of critical consciousness, thus making it appropriate to theorize with perspective transformation.

There are several lessons to draw from this connection. The first is how critical technical awakenings may relate to critical consciousness (CC) overall. Jemal[34] notes that much work on critical consciousness has deliberately excluded privileged populations, but argues this exclusion "...may inadvertently support the proposition that oppression is a problem for the oppressed to solve. When, in essence, CC is important for members of privileged groups who have greater access to resources and power and can operate as allies privileged by the system of social injustice, unfair distribution of resources and opportunities, and inequity, be able to recognize unjust social processes and acquire the knowledge and skills needed for social change."

Drawing from Freire, she continues：

"It is imperative that those who may be privileged by the system of social injustice, unfair distribution of resources and opportunities, and inequity, be able to recognize unjust social processes and acquire the knowledge and skills needed for social change... CC would help individuals understand their role in a system of oppression, as members of either the privileged or stigmatized groups. Liberation requires true solidarity in which the oppressor not only fights at the side of the oppressed, but also takes a radical posture of empathy by 'entering into the situation of those with whom one is solidary'.[35] Thus, CC, with the goal of liberation, has the radical requirement that the oppressor, those who deny others the right to speak their word, and the oppressed, those whose right to speak has been denied, must collaborate to transform the structures that beget oppression.[35]"

The second is that all of the descriptions of possible critical technical awakenings do not recognize "that one's discontent and the process of transformation are shared and that others have negotiated a similar change". From the perspective of Mezirow's theory, this means they fall short. Indeed, our article here is an attempt to directly address the fragmentary nature of narratives of critical technical awakenings, and to draw connections between people's experiences. We can also continue the normative route, and note that in order to fully achieve the potential for social change from critical technical awakenings, we should try to see how to continue past stage (5) and on to stages (6)−(10).

What might new roles (stages (5)−(9)) be, in which technical practitioners should build competence and self-confidence, and make provisional efforts? We suggest that one role might be in opposing gatekeeping. It is rare even for qualitative researchers to have a seat at the table of technological adoption, let alone communities affected by it. But by leveraging the social standing that comes with quantitative legitimacy, and translating concerns into terms that are (more) acceptable for technical audiences as a first step, technical practitioners can help bring others into the processes of technology development—whether to participate, or to oppose development and deployment that does not empower those communities.

The relationships that come with those roles would be with allies outside of technical disciplines and sectors, and particularly through learning from and working with communities affected by technology (whether directly, by a technology itself, or indirectly, such as in gentrification resulting from real estate expansions by the tech industry or of universities who receive influxes of tech money). These would be new roles not only for the technical practitioners, but indeed new social roles, and would require weathering all the difficulties of negotiating roles outside of recognized categories.

Drawing on the follow-up work to Mezirow, we also draw attention to the importance of looking at perspective transformations outside of frames of self-realization[31], and indeed outside of depicting the process as a deeply *rational* one in molds of western rationality. One example is a study that identifies disorienting dilemmas among women in Botswana that led to questioning assumptions, but with the value of the outcome being oriented towards the spiritual, community responsibility and relationships, and gender roles[36]. Indeed, acknowledging other ways of knowing that are not expressed in the language of rationality makes perspective transformation far less novel. Johnson-Bailey[37], coming from the perspective of a Black woman, writes about "transformational learning as the only medium in which we exist, learn, and teach. Since it is the air we breathe, maybe we just take it for granted and didn't attend to or claim it sufficiently." This is also an example of a more general issue; in "The Race for Theory", Barbara Christian[38] wrote, "people of color have always theorized—but in forms quite different from the Western form of abstract logic... our theorizing is often in narrative forms, in the stories we create, in riddles and proverb, in the play with language, since dynamic rather than fixed ideas seem more to our liking. How else have we managed to survive with such spiritedness the assault on our bodies, social institutions, countries our very humanity? ...My folk, in other words, have always been a race for theory".

The third is in looking at recommendations from adult education about how we might encourage perspective transformations. Unfortunately, as Taylor and Snyder[32] note, work has focused on support based around assumptions from Mezirow, "such as creating a safe and inclusive learning environment, focusing on the individual learner's needs, and building on life experiences". One strand of work that does go beyond Mezirow's assumptions looks at how the significance of spontaneous action depends on social recognition. That work finds that what would otherwise be a spontaneous action becomes personally meaningful when others point it out and provide positive feedback about it.

Combining these strands together, we can say: those who have undergone a critical technical awakening should think about relationships with others in which we create safe and inclusive learning environments, facilitate opportunities for experience, serve as guides who can give focus to specific learning needs, and give positive feedback around disorienting dilemmas and other opportunities for reflecting and questioning assumptions.

While these principles were developed in opposition to existing formalized education, there may be opportunities to incorporate them into formal education as well. Trbušić[39] argues for integrating critical methods into engineering education as a way of making ethics more than a superficial part of training. She specifically suggests using Augusto Boal's technique of *Theatre of the Oppressed*[40] (itself based on the work of Freire, with whom Boal was friends), using improvisation and role-playing to encourage critical consciousness. Incorporating role-playing with scenarios where engineering students are put into ethical dilemmas could encourage taking an active stance, trying different roles, and stimulating reflection in a way that presenting formal models of ethics would not.

Especially insofar as critical technical awakenings may fall short more than other types of critical consciousness, there is also a task for how to deepen our own awareness and practice. Taylor and Snyder[32] identify work about "social accountability", where a moral underpinning is an outcome of transformative learning. More specifically, "the outcome of transformative learning involves recognizing the reasons why, for what purpose, and for whom a new identity was constructed", especially as an essential component of trusting relationships[41]. Having transformations be ethically grounded for what kind of world we want to see and work towards, and making this a focus of interpersonal relationships and community-building, can also help achieve more complete and powerful transformations.

## 5 Ethics

Earlier, we raised reasons why it seems like positivism is compelling as a basis for ethics. But Rogaway and Toyama's accounts, in particular, get at how positivism and technical disciplines are harmful in the consequences of their epistemological assumptions: if quantitative forms of knowledge are superior, then other forms of knowledge are inferior. Consequently, those who do not hold quantitative knowledge do not have anything to offer.

De Sousa Santos[42, 43] discusses the interconnection of ecologies of knowledge and how people are valued. When knowledge is put in hierarchies, it also places people into hierarchies. Sylvia Wynter, in her landmark work on "No Humans Involved"[44], has a stark presentation of this idea. Her title refers to a term used by the Los Angeles Police Department to classify police encounters where they enacted violence on young Black men who were jobless in the inner city: by saying that these encounters did not involved "humans", the department excused themselves from documenting their use of force and gave them a license to continue. The literal, administrative category reflected metaphorical dehumanization: there is no brutality or injustice if the targets are not human.

Critical, constructivist, and participatory paradigms link epistemology and axiology, saying: how do we value people, if we do not value their knowledge? Even post-positivism is insufficient; we can see calls for "Human-Centered AI", or "Human-Centered Machine Learning", or "Human-Centered Data Science" as fitting into a post-positivist frame, where we pursue objective knowledge and "real" technology that is focused around the figure of the human and its subjectivity. But human-centeredness does not address dehumanization, who gets recognition as being in the category of "human", and how exclusion happens (e.g., being "human" is reserved for people who look, talk, think, act, and exist in certain ways). Any form of human-centered computing that takes the category of "human" for granted will not undo the status quo of what Wynter calls "narrative condemnation". Participatory approaches, in particular, start with the proposition that everyone is valuable, and then derive knowledge from there.

As in the premise of critical theory, the Enlightenment led to or at least did not prevent the atrocities of the Holocaust, to which we can also add the atrocities of indigenous genocides in the Americas and Australia, the brutality of colonialism like in the anthropogenic Bengal famine or the atrocities in Congo Free State, and especially the trans-Atlantic slave trade. Science was a weapon to dehumanize and make exclusionary standards for moral standing throughout history[45]. It was utilized as a tool to control otherized populations, alienate them from the public sphere, and remove them from societal participation. Pretending these things did not happen, or pretending as though they were aberrations from the natural course of science, does nothing to prevent them from happening in the future. Atrocity and oppression cannot happen without devaluing entire groups of people, and excluding them from belonging to the same sort of category of being; this is the only way we can apply different standards, for example, of surveillance or accountability or resource distribution or violence to people based on different labels (e.g., criminal, immigrant, welfare beneficiary, and foreign citizen). Then, instead of making universal morality the basis of our ethics, we should seek to dismantle knowledge hierarchies. We should valorize knowledge creation that resisted and persisted through dehumanization[46] through empirical but also artistic, narrative, and cultural means, and see these as no lesser than quantitative forms of knowledge.

We advocate specifically for the *ethics of care* from Black feminist frameworks[47−50]. Traditionally, descriptive ethics have linked recognition, belonging, and moral standing: normatively, the way to be ethical, and achieve justice, is to extend recognition, equal standing, and the protection of rights to people who have been marginalized and excluded (such as by bringing marginalized people into full participation in the public sphere, or by policies framed around safeguarding human or civil rights). In contrast, the ethics of care is a normative ethical position that reacts to the ethics of recognition and how it descriptively concedes to "recognition" as being an acceptable basis for treatment. This ethical position is found in a long history of the labor of Black women (including potentially not under the explicit label of "ethics of care"[51]), specifically in Black feminist circles and in value-based social services disciplines[52] like social work, thinking about how to have ethical and holistic interpersonal relationships, and focusing on care for marginalized people[53−55]. Instead of recognition, the basis of these ethics is empathy, love, and connection, coming from non-Eurocentric world-views, and advocating treating every living being with care. Scaling up interpersonal care to systems creates a principle that systems must serve the most marginalized and disadvantaged, rather than those people needing to fit into systems or gain social capital before they are respected or considered important.

## 6 Traps

A critical technical awakening destabilizes a positivist worldview, opening up the possibility of a perspective transformation that leads to people working with deliberation and awareness towards a better world. But it is not sufficient. In a reflection of the language of Selbst et al.[56] who talk about five "traps" of the (positivist) formalisms of computer science, we discuss two traps in critical technical awakenings that reject positivism but may fail to achieve genuine transformation. There are other traps as well, for example co-option, as discussed in other articles in this special issue[2, 3], but here we discuss *incomplete awakenings*, and *technical abandonment*.

The first and most important trap is of *incomplete awakenings*, where one's perspective only widens somewhat, and specifically does not get past knowledge hierarchies. We have sketched out a particular normative path for an awakening, with this dismantlement as the goal. But none of the critical technical awakenings we identify necessarily get this far. Agre's characterization of his awakening, for example, seemed more like it was about intellectual fulfillment, and (at least from the description) did not engage with positionality. What he describes is coming to see some other forms of elite knowledge, namely those from the humanities and social sciences, as superior to his former narrow technical worldview.

The blindness Broussard[12] identifies of technologists to other forms of knowledge from experience is not ever recognized or addressed in Agre's work. Again, the work may not reflect the full extent of Agre's experience, and it may do so in a particular rhetorical strategy of not trying to overturn positivism *and* knowledge hierarchies all at once; but, this is a theme across the other descriptions of awakenings as well. In none of them is there a recognition of the existence and value of other very different forms of knowledge, or the value of the people who hold those other forms of knowledge.

The second trap is a more subjective one: that of *abandonment*. There is a temptation, upon having an awakening and becoming disillusioned, to abandon technical work entirely. We argue this is bad for two reasons. The first is a strategic one: at the risk of reifying quantification and technology, we believe that there is a role for those trained in these methods to push back and develop critiques in "internal" terms that can be intelligible to those still in a technical mindset (and perhaps even leading others to having their own critical technical awakening). These are some of the potential "new roles", as in Mezirow's ten steps, we explore above. This temptation is parallel to how, upon recognizing and becoming disillusioned with privilege, one temptation is to attempt to reject that privilege; but, such attempts do not actually erase the privilege one has benefitted from in the past. Finding ways to engage with and leverage this privilege is the more responsible course.

The second reason we argue against abandonment is more abstract and speculative. Just as modern qualitative research originated in the oppressive project of colonial anthropology but has since worked to reform on grounds of being reflexive and pursuing justice, so too might quantitative research move away from positivism[57].

Given that quantification is about abstraction[56], and abstraction flattens meanings[15], it is difficult to imagine quantitative knowledge that can be reflexive and acknowledge other forms of knowledge, but is worth exploring. Agre's own suggestion of a "critical technical practice" is itself a call to continue creating technical knowledge, but through a critical lens. What that might mean or how it might look is unclear from Agre's work or the handful of subsequent works that have taken on that label, but the development of technical knowledge on something other than a realist ontology and a hierarchical axiology can be seen as a worthwhile challenge.

## 7 The Path Forward

Despite being a powerful expression of a profound shared experience, Agre's call for "critical technical practice" has largely languished for the past two decades. For personal reasons, Agre himself has not been active in academia[58] to continue exploring and developing this idea himself. Critical technical practice has been continued by a few people, like Phoebe Sengers[59], but even that has been mostly within design and Human-Computer Interaction[60, 61], rather than in more formal mathematic and technical areas where critical and constructivist approaches are most alien.

As discussed before, one key missing element from Agre's narrative and those of others is Mezirow's stage (4), "Recognition that one's discontent and process of transformation are shared and that others have negotiated a similar chang." While it is hard to say why critical technical practice failed to take hold—Agre no longer being active in academia? The original essay not having any clear statement of what, exactly, critical technical practice is or looks like? Critical technical practice not being a good way to productively channel awakenings? There not being enough awakenings to form a critical mass? Agre simply being ahead of his time[62]?—building community and coalitions seems to be a critical missing step.

Some of what we detail in sources of awakening suggest ways that we can try to encourage more people with a technical perspective to undergo critical awakenings: exposure to anti-positivist and anti-realist ideas, putting them in contact with non-technical individuals, and finding ways to attack compartmentalization (as is done in other articles in this collection like those of Green[8], and in the design method that Stark[63] offers). Or, if these were integrated in technical education sufficiently early on[25, 39], perhaps people would never develop a distinctly technical perspective and would not need (as abrupt of) an awakening, in a topic that also relates to the article in this special issue by Korn[64]. This article (as well as that of Hu[3]) also partially take the form of personal reflections, which are central in critical awakenings; while we have chosen, primarily for reasons of length and coherence, to make this essay a primarily informational and analytic one rather than discuss our experiences, we cite these articles as examples of how we should seek to create more opportunities for technical practitioners to, respectively, engage in their own personal reflection as a technical practitioner[3] and with the experiences of others[64].

Seeking out perspectives from others, both contemporary and historical, is one way to break through ossified visions. In "Informatics of the Oppressed", Ochigame describes in English for the first time two Latin American informatics projects[65]. First, Cuban librarians and computer scientists in the 1980s, facing US embargoes, set up an alternative information indexing and retrieval system whose mathematical model, among other features, adjusted readership-based indexes by the number of librarians in recognition of the "author-reader social communication that happens in libraries". Second, liberation theologists in Brazil resisting the post-1964 military dictatorship set up a print and mail-based "intercommunication network" to solicit and internationally distribute writings by those most subjected to domination, in a vision of advancing Freire's project past a need for intermediaries and towards "'inter-conscientization' between the oppressed". Ochigame notes that these projects were, like libertarian fantasies coming out of California, overly optimistic in what technology (alone) would achieve; but these visions were still valuable in the alternative they offered to ranking based only on productivity or popularity (in Cuba), and in justifying and structuring dissemination not just in terms of free speech or in the politics of "whether one is free to speak, but whose voices one can hear and which listeners one's voice can reach" (in Brazil). We can take inspiration from these alternative visions, and seek out others that have similarly been silenced and pushed aside (indeed, Ochigame's discovery of these projects came through personal meetings, and not online searches). Those of us trained in technology development and quantitative forms of knowing should try to build on these, and explore alternative visions. We hold that the potential value of quantitative knowledge *outside* of its connection to and role in upholding power, hierarchy, and privileged access to truth have yet to be fully explored.

Another key part of any path forward is to *build community* to encourage, support, and guide critical technical awakenings, and channel those who undergo such awakenings towards developing a critical technical practice. Here, we can point to conference workshops[60, 61, 63, 66], networks like the one formed from the Ethical Tech Working Group that generated this special issue, fellowship cohorts, and mentorship as paths forward. But as a caveat, while community-building aimed at reaching technical practitioners will most likely need to operate within institutional elitism (indeed, like the Ethical Tech Working Group being at Harvard), this should only be one part of larger community-building. After all, during his exile under Brazil's 21-year military dictatorship, Freire also spent a year as a visiting professor at Harvard; but he eventually returned to Brazil and continued to develop both theory and practice, including serving as a municipal Secretary of Education.

But questions remain. What is the value of quantitative approaches outside of knowledge hierarchies? As Bricmont and Sokal suggest[14], are quantitative and technical approaches to the world only valuable if they are getting at a single universal truth? If we reject positivism, and choose participatory paradigms and the ethics of care, must we reject technical approaches? Or even if not, how can we integrate the ethics of care into technology to achieve "doing no unintended harm", and not further marginalizing resource-deprived communities? What sorts of technical practices might emerge not from an *elite* critical stance, but from a critical *pedagogical* stance?

It seems daunting, but qualitative research also was once positivist and hierarchical, for example, in seeing the role of a colonial anthropologist as providing neutral description about colonized or imperialized peoples to better facilitate control.

Lastly, we hope this article has served as an orientation, encouragement, and guidance for those who are undergoing the kind of vertigo that Agre described. The technical variant of critical consciousness is a profound and important experience, just like critical consciousness in general. But if it happens in isolation, it may be unnecessarily painful, and more importantly may not overcome the most pernicious part of positivism: creating and defending hierarchies of knowledge that structure the ways we approach the world, value ideas, and treat other beings. We hope that this article points to how this experience is not isolated, and gives support towards building community, overcoming knowledge hierarchies, adopting an ethics of care, and taking action towards more liberated ways of being.

## Appendix

### Glossary of Key Terms

**Realism** is the belief in a single underlying reality that exists independent of and prior to human conception of it. A specific form of this was articulated by Plato, where mathematical forms are immutable and that invariance what determines what is "real". Confusingly but perhaps more appropriately, this is sometimes also called "idealism", since reality is associated with *ideas* rather than perception.

**Positivism** was coined by Auguste Comte in philosophical writings around 1830−1842. It was an application of methodology from natural sciences to study human behavior and social phenomena. Comte articulated positivism[57] in terms of a premise that universal truths exist for human behavior and social phenomenon (i.e., a realist ontology), and that empirical observations through scientific measurement can discover these universal truths (i.e., an empiricist epistemology and methodology).

Positivism now describes any research paradigm that holds that a singular truth exists and can be uncovered by empirical observation, and covers natural sciences as well as social and behavioral sciences. There are versions of positivism that try to avoid the realist commitment, and there can also be realism without empiricism (such as in pure mathematics) but the key point of either realism or positivism as compared to other sets of assumptions is *belief in an external world that takes primacy over actors' interpretations and renegotiations of it*[67].

**Post-positivism** is a softening of positivism, and held by people who still find positivism aesthetically compelling, but acknowledge that contingent and malleable (and non-scientifically measurable) history, society, and culture can come in the way of our ability to discover universal truths through observation, and so must be accounted for (potentially through qualitative means).

**Critical theory** is a type of philosophy often viewed as originating from a specific group of European intellectuals based in Frankfurt in the period between the World Wars. Against the prevailing view that the Enlightenment had led to constant social improvement, this Frankfurt School and their successors sought to theorize how the Enlightenment led to, or at least failed to prevent, World War I, the rise of anti-Semitism, and other forms of oppression (eventually leading to the Holocaust) in liberal capitalist societies. Of course, earlier major atrocities—such as the trans-Atlantic slave trade, or colonial genocide of indigenous populations—tellingly did not lead to similar soul-searching among European intellectuals about the consequences of the Enlightenment. Still, the Frankfurt School represented when a major European philosophical school caught up to people in the colonized world in acknowledging marginalization as a central philosophical question. For example, Rabaka[68] argues that Martinique-born psychiatrist and philosopher Frantz Fanon (discussed more below), built on prior work from the colonized world and went far beyond the Frankfurt school in analyzing the nature of the racism and exploitation of settler colonialism.

It is from the Frankfurt School's use of "critical" that the term is applied to theories that dispute prevailing assumptions about social development needing only continue along its current course to eventually result in the end of forms of oppression, e.g., around gender, race, sexuality, disability, etc.

A good definition of what makes a "critical social science" is in Fay's *Critical Social Science: Liberation and its Limits*[69]. Fay conceives of critical social science as a type of "estrangement theory". This is a view of the world that holds that there is a manifest/ordinary sphere in which most people live, but this keeps them trapped from what is best in life, which exists in a hidden/extraordinary sphere. Specifically, critical social science is a *humanist* variant of estrangement theory, that locates the hidden/extraordinary sphere not in a religious or spiritual plane (like religious and mystical traditions do), but in the social plane. He additionally theorizes that a complete critical theory includes a *theory of false consciousness* (identifying certain understandings and explaining how they are false and/or incoherent, and how they come to be and are maintained), a *theory of crisis* (how a society is in a crisis from felt dissatisfactions that threaten social cohesion and cannot be resolved within existing social organization and self-understandings), a *theory of education* (the necessary and sufficient conditions for overcoming the false consciousness), and a *theory of transformative action* (identifying what needs to change, and a plan of action for who are "carriers" of anticipated social change and how they will go about achieving it).

Note that positivism (or realism) can have an estrangement aspect as well, where there is a hidden truth that reality is apprehensible through the language of mathematics and/or experimental methods, leading to liberation. Indeed, Plato's parable of the cave, and Platonism (as well as the neo-Platonism of mystic cults throughout the Mediterranean and West Asia centuries after Plato) sees universal abstract mathematical forms as the truth from which the masses are estranged. But the estrangement aspect of positivism need not be present, whereas it is an essential part of any critical theory.

**Relativism** is a stance that potentially spans ontology, epistemology, and axiology. Ontological (or conceptual) relativism holds that there is no observer-independent reality, and that an observer creates their own reality. Epistemically, relativism holds that there is no neutral frame in which we can arbitrate whether claims are "true" or "false". This can be understood empirically (rather than normatively)[③]: for example, speaking purely empirically, there is no frame of reference to which a Biblical creationist and an evolutionary biologist would agree for arbitrating their competing claims about the origin of biological diversity. Each would insist on their own frame being the "neutral" or superior one, and any logical or empirical basis for deciding between the frames would itself rely on agreement over what counts as logical or empirical. Moral relativism holds that there is no neutral frame in which we can decide what is good or bad. Similar to epistemic relativism, moral relativism may be a descriptive rather than a normative position, built on the observation that people have genuine disagreements about morality that cannot be logically resolved by an appeal to universal underlying principles. That is, a relativist can have their own (non-relativist) normative morality that they believe is correct, alongside a relativist ontology and/or relativist epistemology that they also believe is correct, but they recognize that there is not necessarily any deeper universal principle to which to appeal and *logically* convince others. As a corollary, we can account for people with perspectives we find bizarre or moral codes that we find abhorrent who cannot be convinced through logical means, rather than needing to dismiss them as insane.

Relativism represents a break from a singular truth, and can be deeply uncomfortable and threatening for those accustomed to the pursuit of certainty and finality. Worse, when every possible position and action *can* be critiqued, relativized, destabilized, and once we know how to do this, it can be debilitating. See below for how participatory paradigms provide a way out of this.

**Constructivism** is built on relativism, and describes

---

③ Barnes and Bloor[70] have a relatively simple response to the frequent initial objection that relativism is paradoxical or self-refuting (i.e., if all perspectives are equally valid, then by its own admission relativism concedes to non-relativism): relativism is not saying we cannot hold our own perspective, or we cannot condemn those of others or say they are wrong (whether morally, or in terms of knowledge); relativism can be just the recognition that others can and will reject our views or condemnations, and that our condemnations alone will not convince them otherwise. Of course, it is possible to interpret relativism in such a way as to defend the right of regressive perspectives to exist, but that treats relativism as a standard to which to aspire, rather than a description of how things are. And, relativism is self-referential and can create paradoxes, but we believe that accepting these paradoxes as axiomatic is enormously insightful.

the process by which multiple "truths" come to exist. It is an idea coming out of the sociology of knowledge that holds that our experiences of the world, and knowledge, are not references to or reflections of an underlying external reality, but are the product of historical, cultural, and material forces that, had they been different, would have built something different. Note that saying something (like scientific knowledge) is "constructed" does not mean that it is not real, or not solid, or not robust; a metaphor used to illustrate this perspective[71] is that of a house, which is perfectly "real" but it came to exist at a certain point in time, and was built in one specific way out of specific materials out of many alternatives. We can come to understand this building process without claiming the building is anything other than solid and durable. However, other versions of constructivism stress the fluidity of things like scientific knowledge, rejecting the idea of knowledge as hierarchical structures anchored to, if not a solid underlying reality, then to society and history; these versions of constructivism instead see knowledge as ungrounded webs of mutual reference. Then, the task of inquiry is to understand the construction and maintenance of these webs of mutual reference (with the inquiry being itself a part of the webs it considers).

There is a tension between critical theory and constructivism[72] in how *critical* perspectives can end up holding that there is an external world, just that it is something different than what most people think it is. So, for example, Fay offers the Marxist-humanist model of political revolution as an example of a critical theory, where there is a "true nature" that bourgeoisie oppressors derive power from the self-understandings of the oppressed working classes.

However, they frequently appear together. Hacking[73] points out how looking at how things are put together also gives people grounds to see how they come apart, and *deconstruct* them. A crucial part of a critical toolbox is in showing the historical construction of ideas, forms of knowledge, institutions, and cultural forms, thereby demonstrating that they are not inevitable, and letting us imagine and advocate for alternatives.

For example, in critical race studies and critical gender studies, there is a "false consciousness" of thinking that the categories by which people are marginalized are based on biological traits or even cultural ones. But there is no such thing as biological race or gender, let alone inferiority by them (and the "value" of cultures, like European culture versus indigenous cultures, come from how they are valued, and not something intrinsic). Instead, such categories and their value are socially constructed by and maintained through power relationships. Going further, there is a second layer to the false consciousness, of holding marginalized people individually responsible for their suffering and deprivation. Once categories are so constructed, those that fall within the marginalized categories like women of color and others with individual or intersecting marginalized identities are treated as inferior, in ways often enacted on an interpersonal level but structurally and culturally encouraged and permitted. The result is marginalized people face greater mental and physical suffering, and material deprivation, entirely apart from their individual "effort", yet over which they are held responsible. Even holding those who *enact* the double standards individually responsible (i.e., seeing racism or sexism as an interpersonal problem), rather than seeing the larger structure, is a false consciousness. Only by recognizing the true nature of modern civilization as fundamentally structured on white supremacy, patriarchy, colonialism, and other forms of domination can we effect change and improve human life.

Indeed, Agre's[74] idea of "critical" is actually more about constructivism (and unfortunately he sets it up using ableist language). In one entry from his Red Rock Eater Newsletter (a listserv over which Agre sent out writings that has been cited as a precedent for blogs), he wrote:

"I finally comprehended the difference between critical thinking and its opposite. Technical people are not dumb [sic], quite the contrary, but technical curricula rarely include critical thinking in the sense I have in mind. **Critical thinking means that you can, so to speak, see your glasses. You can look at the world, or you can back up and look at the framework of concepts and assumptions and practices through which you look at the world.**"

Agre continues: "Not that critical thinking makes you omniscient: you're still wearing glasses even when you're looking at your glasses." That is, there is no perspective without any glasses, no "view from nowhere".④ The experience of "seeing one's glasses" is different than just replacing one's glasses; it opens the

---

④ Ludwig Wittgenstein, another figure who underwent a transformation in his basic beliefs and how he saw the world, also used this metaphor much earlier[75]: "The ideal, as we think of it, is unshakable. You can never get outside it; you must always turn back. There is no outside; outside you cannot breathe.—Where does this idea come from? It is like a pair of glasses on our nose through which we see whatever we look at. It never occurs to us to take them off."

path to understanding endless contingency in ideas, structures, institutions, and frameworks.

**Critical consciousness** is a theory that came out of political mobilization and community development, also known as popular education, in the Global South[76–78], and specifically from the work of Brazilian educator, philosopher, and politician Paulo Freire (1921–1997).

Freire worked in the 1960s with populations like marginalized sugarcane harvesters with no access to formal education. He started education programs for political mobilization in conjunction with them, and used that mobilization to get the Brazilian government to financially support the programs they had created. He challenged a "banking" conception of education that assumed he was more of a knowledge holder and knowledge creator than the farmers he worked with, and that placed more value on him as a teacher, because he had access to formal education. He inverted the hierarchy to say that the marginalized are valuable because of their response to marginalization, their resilience, and how the experience of marginalization showed larger societal structures in a way that Freire, with his privilege, had not seen. He theorized how to unseat the teacher or researcher as the expert, and sought to develop a model where we all bring something to the table and learn from each other, and understanding emerges from our interactions.

Another key input, on whose work Freire drew, was Frantz Fanon (1925–1961). Fanon was hired as a psychiatrist by the French colonial government in Algeria to treat mental illness in colonial subjects. There, Fanon realized that his patients were not having mental health crises, but reacting to oppression, and the French government did not understand that their reaction was the most logical response to being otherized, dehumanized, and oppressed. Building on his previous work theorizing his own experience being treated as a French colonial subject[79], in interacting with his patients in Algeria he learned about his own position in a larger oppressive system and how it was causing harm to others[80, 81]. From this, he wrote about working with marginalized populations, unlearning harmful frames, and mobilizing for revolution and equity, himself joining the Algerian National Liberation Front to support Algeria's War of independence from France.

Freire gave the name *conscientizaçao* to the transformative process of interacting with other individuals and other communities[82, 83], translated as critical consciousness, or more literally as "conscientization", and sometimes as consciousness-raising[84].⑤ From there, others have continued to systematically develop tools, strategies, and methods for critical consciousness, including dialogue and critical reflection, reflective questioning, psychosocial support, co-learning, group processes, civic engagement and sociopolitical action, and identity development[34]. Critical consciousness has inspired a field within education known as critical pedagogy[86] which has been carried forward particularly in adult education[4] and has had a large impact on the development of Participatory Action Research[87] and Community Based Participatory Research[88].

**Positionality** is awareness and discussion of ones' social and institutional position with regards to research, particularly of power imbalances, and limitations the researcher may have because of differences in lived experience.

**Reflexivity** is the process of "turning back on" and reflecting on experience and our positionality. For example, in anthropology, this is researchers being explicit about their emotions and how they related to research subjects[89]. Positivism, in particular, does not and cannot engage in reflexivity[90], since it holds that knowledge is independent of the knowledge-holder.

**Participatory** paradigms address an important moral aspect lacking in both critical theory and constructivism. Certain streams of critical theory frequently have a condescending aspect to them: that people are unaware of their own oppression, and it is the role of the critical theorist to educate them. On the other hand, constructivism does not account for experiential knowing[90]. Building explicitly from the ideas of Freire, participatory paradigms value and highlight experience, following a methodology that challenges hierarchies between teacher and student, or researcher and subject, and seeks to construct knowledge collectively. Its methodology and axiology prioritize understanding and improving the world by changing it through collective, reflexive inquiry[91].

This paradigm has a relativistic component in seeing knowledge as malleable and multiple rather than absolute and singular; by locating value in others and their experiences, rather than seeing the status of

---

⑤ Consciousness-raising also appears, without reference to Freire, in US feminist movements in the 1960s[85].

knowledge as the most important thing in life, the instability of knowledge does not become a reason to be nihilistic.

## Acknowledgment

Thanks to two anonymous reviewers for great feedback and pointing us to some relevant literature, and Ben Green both for fantastic continuous comments and for corralling and managing this special issue.

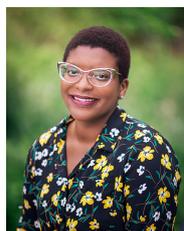

**Maya Malik** received the BA degree in psychology from Warren Wilson College in 2014 and the MS degree in Social Work from Columbia University's School of Social Work in 2017, focusing on international social welfare and rights for immigrants and refugees through program design, research, and evaluation. They are currently a doctoral student at the School of Social Work, McGill University, where they are researching how to utilize arts-based Youth-Led Participatory Action Research (YPAR) methods to work with Queer Black American youth who have been justice-involved to improve educational intervention programs. They are also a researcher in the Participatory Axis of the McGill Global Child Research Group, and a research contributor at the Youth and Media project at the Berkman Klein Center for Internet & Society at Harvard University. They are also a member of an interdisciplinary public health team to address the possible impact of reparations on quality of life and mortality for Black Americans.

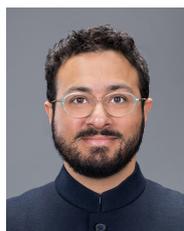

**Momin M. Malik** received the AB degree in History and Science from Harvard University in 2009, the MSc degree in Social Science of the Internet from the University of Oxford in 2012, and the MS degree in machine learning and the PhD degree in societal computing from Carnegie Mellon University both in 2018. Previously, he was the Data Science Postdoctoral Fellow at the Berkman Klein Center for Internet & Society at Harvard University, and the Director of Data Science at Avant-garde Health. He is currently a Senior Data Scientist for AI Ethics at the Center for Digital Health, Mayo Clinic, a fellow at the Institute in Critical Quantitative, Computational, & Mixed Methodologies, and a lecturer for the School of Social Policy & Practice at the University of Pennsylvania. He is a Senior Program Committee Member for the International Conference for Web and Social Media.